\documentclass[seceq]{ptptex}

\usepackage{graphicx}
\usepackage{wrapft}

\def\mbf#1{\hbox{\boldmath $#1$}}
\def\eq#1{Eq.\ (\ref{#1})}

\def\bS{{\mbf S}}

\def\bS{{\mbf S}}

\def\ba{{\mbf a}}
\def\bb{{\mbf b}}
\def\be{{\mbf e}}

\def\bn{{\mbf n}}
\def\bp{{\mbf p}}
\def\bq{{\mbf q}}

\def\bfsigma{{\mbf \sigma}}

\def\CP{{\cal P}}

\def\CS{{\cal S}}

\def\CY{{\cal Y}}

\def\H{{\scriptstyle \frac{1}{2}}}
\def\3H{{\scriptstyle \frac{3}{2}}}
\def\5H{{\scriptstyle \frac{5}{2}}}
\def\7H{{\scriptstyle \frac{7}{2}}}

\title{
Quark-Model Baryon-Baryon Interaction Applied to the Neutron-Deuteron
Scattering (II)
}

\subtitle{Polalization Observables of the Elastic Scattering
}

\author{Kenji \textsc{Fukukawa} and Yoshikazu \textsc{Fujiwara}
}

\inst{
Department of Physics, Kyoto University, Kyoto 606-8502, Japan
}

\abst{
The neutron-deuteron ($nd$) scattering is solved in the Faddeev formalism, 
employing the energy-independent version of the quark-model baryon-baryon
interaction fss2. The differential cross sections and the spin polarization
of the elastic scattering up to the neutron incident energy $E_n=65$ MeV
are well reproduced without reinforcing fss2 with the three-body force.
The vector analyzing-power of the neutron, $A_y(\theta)$,
in the energy region $E_n \leq 25$ MeV is largely improved in comparison
with the predictions by the meson-exchange potentials, thus yielding
a partial solution of the long-standing $A_y$-puzzle owing to the
nonlocality of the short-range repulsion produced by the quark-model
baryon-baryon interaction. The large Coulomb effect in the vector and
tensor analyzing-powers in $E_n \leq 10$ MeV is also analyzed based on
the Vincent and Phatak method and recent detailed studies by other authors. 
}

\PTPindex{205}  

\begin{document}

\maketitle

\section{Introduction}

The QCD-inspired spin-flavor $SU_6$ quark model (QM) for the
baryon-baryon interaction, developed by the Kyoto-Niigata group,
has achieved accurate description of available nucleon-nucleon ($NN$) data,
comparable with the modern meson-exchange potentials.\cite{PPNP}
The naive three-quark structure of the nucleon is incorporated 
in the microscopic framework of the resonating-group method (RGM) for two
three-quark clusters, leading to the well-defined nonlocality 
and the energy dependence of the $NN$ interaction, 
inherent to the RGM framework. In particular, the short-range repulsion
of the $NN$ interaction is mainly described by the quark-exchange
kernel originating from the color-magnetic term 
of the quark-quark interaction.
The energy dependence of the interaction is eliminated
by the standard off-shell transformation,
utilizing a square root of the normalization kernel.\cite{renRGM}
This procedure yields an extra source of nonlocality, 
whose effect was examined in detail for the
three-nucleon ($3N$) bound state and for the hypertriton.\cite{ren}
The QM baryon-baryon interaction thus constructed is expected
to give quite different off-shell properties from the
standard meson-exchange potentials.
It is therefore interesting to examine predictions by the QM $NN$ interaction
to the $3N$ scattering, especially in this renormalized framework 
with no explicit energy dependence.

In a previous paper,\cite{ndsc1} referred to as I hereafter, we have
developed a new algorithm to solve the Alt-Grassberger-Sandhas (AGS) 
equations \cite{AGS} with the deuteron singularity,
employing the Noyes-Kowalski method.\cite{NO65,KO65}
Another notorious moving singularity of the free three-body Green function 
is treated by the standard spline interpolation technique developed 
by the Bochum-Krakow group.\cite{spline82,Wi03,PREP,Liu05}
The AGS equation is solved in the momentum representation, 
using the off-shell RGM $t$-matrix obtained from
the energy-independent renormalized RGM kernel.
The total cross sections derived from the optical theorem
and $nd$ elastic differential cross sections are well reproduced
up to the neutron incident energy $E_n=65$ MeV in the laboratory
system. It was found that the predicted elastic differential cross sections
have larger diffraction minima than the experiment in the energy
region $E_n=35\,\hbox{-}\,65$ MeV, in contrast with the predictions
by the meson-exchange potentials. This is in consistent with the fact
that fss2 reproduces nearly correct triton binding energy
without the three-body force.
In these calculations, we have used the Gaussian 
nonlocal potential constructed from the original fss2 
interaction.\cite{apfb08} This potential preserves not only 
the on-shell properties of the $t$-matrix with the accuracy 
of less than 0.1 degree for the phase shift parameters, 
but also the nonlocality of the interaction completely,
which was confirmed by recalculating the triton binding energy 
with this potential.\cite{scl10}  
The calculation of the $S$-wave $nd$ scattering length by this
potential yields almost correct values $\hbox{}^{2}a_{nd}=0.66~\hbox{fm}$
for the spin doublet scattering length and
$\hbox{}^{4}a_{nd}=6.30~\hbox{fm}$ for the quartet scattering length,
without the three-body force and the charge dependence
of the $NN$ force.\cite{scl10} 

In this paper, we extend the study of the $nd$ elastic scattering 
in I and Ref.\,\citen{scl10} to various types
of spin polarization observables,
and compare them with the experimental data and predictions by 
other theoretical calculations. 
In the low-energy $nd$ and $pd$ scattering, there is a long-standing
nucleon analyzing-power puzzle, which implies the failure
of rigorous $3N$ calculations to account for the magnitude
of the measured analyzing-power $A_y(\theta)$.\cite{To98a,To98b,To02,To08}
We find that there is no serious discrepancy in $A_y(\theta)$ at
the low energies $E_n \leq 25$ MeV, although the difference
of about $15\%$ still remains for the magnitude of maximum peaks.
We have analyzed the low-energy $nd$ and $pd$ eigenphase shifts below
the deuteron breakup threshold
and found that the main origin of this improvement is the more 
attractive feature of the $\hbox{}^2S_{1/2}$ phase shift 
in our model, originating from the nonlocal description
of the short-range repulsion.\cite{scl10}
The comparison of tensor-type deuteron analyzing-powers
with experiment is blurred with the Coulomb effect, since all
the experimental data are for the $pd$ or $dp$ scattering.
We here apply the Vincent and Phatak method\cite{Vi74,apfb05}
using the cut-off Coulomb potential to the $pd$ scattering,
and show some preliminary results in comparison
with recent detailed studies by other authors 
in the $E_p \leq 10$ MeV region.\cite{Be90,Al02,Ki96,Ki01,De05,De05c,Is09}
We find that this method works well to reproduce
the characteristic behavior of the forward angular distributions
for the $dp$ tensor analyzing-powers.
  
The organization of this paper is as follows.
In $\S\,2.1$, we first develop a general framework to calculate
the spin polarization observables for the $nd$ elastic scattering,
using the Blatt and Biedenharn technique.\cite{Bl52}
The applications to the analyzing-power and polarization
of the nucleon and deuteron, the polarization transfer for the nucleon,
the nucleon to deuteron polarization transfer, 
and the spin correlation coefficients are made in
subsections 2.2, 2.3, 2.4, and 2.5, respectively.
The definition of the deuteron spin operators, used in this paper,
are summarized in Appendix A, together with the relationship
between various different notations.
The results of polarization observables are shown in $\S\,3$.
In $\S\,3.1$, we discuss the nucleon analyzing-power, focusing
on the analyzing-power puzzle.
The vector and tensor analyzing-powers 
of the deuteron are discussed in $\S3.2$, where the Coulomb effect
is important at $E_p \le 10$ MeV.
In $\S\,3.3$, we extend the analysis of the low-energy observables
below the deuteron breakup threshold, developed in Ref.\,\citen{scl10},
by employing the present approach to the Coulomb force.
Various types of polarization transfer and spin correlation 
coefficients are discussed in $\S\,3.4$ and $\S\,3.5$, respectively.
The last section is devoted to a summary.

\section{Formulation}

\subsection{General framework}

In this subsection, we derive general formulas for the spin polarization
observables for the $nd$ elastic scattering 
by using the Blatt and Biedenharn technique.\cite{Bl52}
This method is used in the previous paper I to calculate
the differential cross sections and the extension to the various 
spin observables is rather straightforward.
We calculate $I=Tr\,\{f \CS_i f^\dagger \CS_f\}$, where $f$ is the $nd$ 
scattering amplitude defined by the partial-wave 
amplitudes $f^J_{(\ell^\prime S^\prime_c), (\ell S_c)}$ in
Eq.\,I(2.92)\footnote{In the following, we cite equations
of the previous paper I with adding I in front of the equation number.}
as
\begin{eqnarray}
f(\widehat{\bq}_f, \widehat{\bq}_i)
= 4\pi \sum_{(\ell^\prime S^\prime_c)(\ell S_c)J}
f^J_{(\ell^\prime S^\prime_c), (\ell S_c)}
\sum_{J_z} \CY_{(\ell^\prime S^\prime_c)JJ_z}(\widehat{\bq}_f; {\rm spin})
~\CY^{\ *}_{(\ell S_c)JJ_z}(\widehat{\bq}_i; {\rm spin})\ ,\nonumber \\
\label{sec4-5-1}
\end{eqnarray}
where $\CY_{(\ell S_c)JJ_z}(\widehat{\bq}; {\rm spin})
=\left[Y_{\ell}(\widehat{\bq}) \chi_{S_c}(12;3)\right]_{JJ_z}$ is
the angular-spin wave function.
The spin wave function is defined 
by $\chi_{S_c S_{cz}}(12;3)=[\chi_1(12) \chi_{\H}(3)]_{S_c S_{cz}}$.
The spin operators $\CS_i$ and $\CS_f$ are in general expressed as
a tensor operator defined by
\begin{eqnarray}
\CS^{(f)}_{f_z}=\left[\sigma^{(s)} S^{(\lambda)}\right]^{(f)}_{f_z}\ ,
\label{sec4-5-2}
\end{eqnarray}
where $s=0$ or 1 and $\lambda=0$, 1, or 2. The neutron spin operators
$\{1, \bfsigma\}$ and the deuteron spin operators $\{1, \bS, S^{(2)}\}$
form $4 \times 9=36$ independent matrix components in the spin space
spanned by $\chi_{S_c S_{cz}}(12;3)$ with $S_c=1/2$ and 3/2.
The deuteron spin operators $\bS$ and $S^{(2)}$ are summarized
in Appendix A, together with the relationship between various notations.
The standard expression for the $nd$ scattering is in the Cartesian
representation such as $\sigma_\alpha S_{\beta \gamma}$, which are 
expressed by the linear combination of the tensor coupled form
in \eq{sec4-5-2}. 
To calculate $I=Tr\,\{f \CS_i f^\dagger \CS_f\}$,
we use \eq{sec4-5-1} and take the trace for the spin variables. Then, we obtain
\begin{eqnarray}
I & =  & Tr\,\{f \CS_i f^\dagger \CS_f\}
\nonumber \\
 & = & (4\pi)^2 \sum_{(\ell^\prime S^\prime_c)(\ell S_c)J}
\sum_{(\widetilde{\ell}^\prime \widetilde{S}^\prime_c) 
(\widetilde{\ell} \widetilde{S}_c)\widetilde{J}}
f^J_{(\ell^\prime S^\prime_c) (\ell S_c)}
\left\{f^{\widetilde{J}}_{(\widetilde{\ell}^\prime \widetilde{S}^\prime_c)
(\widetilde{\ell} \widetilde{S}_c)}\right\}^* \nonumber \\
& & \times \sum_{J_z \widetilde{J}_z}
\langle \CY_{(\ell S_c)JJ_z}(\widehat{\bq}_i; {\rm spin}) |\CS_i|
\CY_{(\widetilde{\ell} \widetilde{S}_c)\widetilde{J} \widetilde{J}_z}
(\widehat{\bq}_i; {\rm spin}) \rangle_{\rm spin} \nonumber \\
& & \times
\langle \CY_{(\widetilde{\ell}^\prime \widetilde{S}^\prime_c)
\widetilde{J} \widetilde{J}_z}(\widehat{\bq}_f; {\rm spin}) |\CS_f|
\CY_{(\ell^\prime S^\prime_c)J J_z}
(\widehat{\bq}_f; {\rm spin}) \rangle_{\rm spin}\ ,
\label{sec4-5-25}
\end{eqnarray}
where the last two matrix elements are taken only for the spin 
variables. We assume that the spin operators $\CS_i$ and $\CS_f$ are
tensor operators in \eq{sec4-5-2} and calculate 
\begin{eqnarray}
I(\lambda_i \mu_i; \lambda_f \mu_f)
=Tr\,\{f \CS^{(\lambda_i)}_{\mu_i} f^\dagger \CS^{(\lambda_f)}_{\mu_f}\}\ .
\label{sec4-5-26}
\end{eqnarray}
We first calculate
\begin{eqnarray}
J_{\lambda \mu}(\widehat{\bq}_i)
=\langle \CY_{(\ell S_c)J J_z}
(\widehat{\bq}_i; {\rm spin}) | \CS^{(\lambda)}_{\mu} |
\CY_{(\widetilde{\ell} \widetilde{S}_c)\widetilde{J} \widetilde{J}_z}
(\widehat{\bq}_i; {\rm spin}) \rangle_{\rm spin}\ ,
\label{sec4-5-27}
\end{eqnarray}
in the standard recoupling technique.
It is rather easy to derive
\begin{eqnarray}
J_{\lambda \mu}(\widehat{\bq}_i)
& = & \frac{1}{\sqrt{4\pi}}\sum_{LM} (-)^{J-S_c-L}
\frac{1}{\widehat{J}}~Y^*_{LM}(\widehat{\bq}_i)
\sum_{\kappa \nu} \langle LM \lambda \mu|\kappa \nu \rangle
~\langle \widetilde{J} \widetilde{J}_z \kappa \nu |J J_z \rangle
\nonumber \\
& & \times Z^{(\lambda)}_\kappa(\ell J \widetilde{\ell} \widetilde{J};
S_c \widetilde{S}_c L) \ ,
\label{sec4-5-39}
\end{eqnarray}
where the recoupling
coefficient $Z^{(\lambda)}_\kappa(\ell J \widetilde{\ell} \widetilde{J};
S_c \widetilde{S}_c L)$ is defined by
\begin{eqnarray}
Z^{(\lambda)}_\kappa(\ell J \widetilde{\ell} \widetilde{J};
S_c \widetilde{S}_c L)=(-)^{\ell+J-S_c}
\,\widehat{\widetilde{\ell}}\,\widehat{J}
~\langle \ell 0 \widetilde{\ell} 0 | L 0 \rangle
\left[ \begin{array}{ccc}
\widetilde{\ell} & \widetilde{S}_c & \widetilde{J} \\ [2mm]
L & \lambda & \kappa \\ [2mm]
\ell & S_c & J \\
\end{array} \right]
\langle \chi_{S_c}|| \CS^{(\lambda)}||
\chi_{\widetilde{S}_c} \rangle_{\rm unc}\ ,
\nonumber \\
\label{sec4-5-38}
\end{eqnarray}
using the unitary form of the 9-$j$ coefficient
and the unconventional reduced matrix element
$\langle \chi_{S_c}|| \CS^{(\lambda)}||
\chi_{\widetilde{S}_c} \rangle_{\rm unc}$.
We can use this formula both for $\CS^{(\lambda_i)}_{\mu_i}$ and
$\CS^{(\lambda_f)}_{\mu_f}$ in \eq{sec4-5-25}, 
changing $\widehat{\bq}_i$ to $\widehat{\bq}_f$ etc.
Then the sum over $J_z$ and
$\widetilde{J}_z$ can be taken. We further use the symmetry property
of $Z^{(\lambda)}_\kappa(\ell J \widetilde{\ell} \widetilde{J};
S_c \widetilde{S}_c L)$ in \eq{sec4-5-38}:
\begin{eqnarray}
Z^{(\lambda)}_\kappa(\widetilde{\ell} \widetilde{J} \ell J;
\widetilde{S}_c S_c L)=(-)^{\lambda+\kappa+L+\widetilde{S}_c
-S_c}\,Z^{(\lambda)}_\kappa(\ell J \widetilde{\ell} \widetilde{J};
S_c \widetilde{S}_c L)\ ,
\label{sec4-5-40}
\end{eqnarray}
by assuming ${\CS^{(\lambda)}_\mu}^\dagger=(-)^\mu \CS^{(\lambda)}_{-\mu}$.
After all, we obtain for \eq{sec4-5-26}
\begin{eqnarray}
& & I(\lambda_i \mu_i; \lambda_f \mu_f)
=\sum_{(\ell^\prime S^\prime_c)(\ell S_c)J}
\sum_{(\widetilde{\ell}^\prime \widetilde{S}^\prime_c) 
(\widetilde{\ell} \widetilde{S}_c)\widetilde{J}}
(-)^{S^\prime_c-S_c}~f^J_{(\ell^\prime S^\prime_c)(\ell S_c)}
\left\{f^{\widetilde{J}}_{(\widetilde{\ell}^\prime \widetilde{S}^\prime_c)
(\widetilde{\ell} \widetilde{S}_c)}\right\}^* \nonumber \\
& & \times \sum_{LM, L^\prime M^\prime}
(-)^{\mu_f+L+L^\prime}~Y^*_{LM}(\widehat{\bq}_i)
~Y_{L^\prime M^\prime}(\widehat{\bq}_f)
\sum_{\kappa \nu}\frac{4\pi}{(\widehat{\kappa})^2}
~\langle LM \lambda_i \mu_i|\kappa \nu \rangle
~\langle L^\prime M^\prime \lambda_f -\mu_f|\kappa \nu \rangle
\nonumber \\
& & \times 
Z^{(\lambda_i)}_\kappa(\ell J \widetilde{\ell} \widetilde{J};
S_c \widetilde{S}_c L)
~Z^{(\lambda_f)}_\kappa(\ell^\prime J \widetilde{\ell}^\prime
\widetilde{J}; S^\prime_c \widetilde{S}^\prime_c L^\prime)\ .
\label{sec4-5-41}
\end{eqnarray}
We write this expression as
\begin{eqnarray}
\ \hspace{-10mm} & & I(\lambda_i \mu_i; \lambda_f \mu_f)
=\sum_{(\ell^\prime S^\prime_c)(\ell S_c)J}
\sum_{(\widetilde{\ell}^\prime \widetilde{S}^\prime_c) 
(\widetilde{\ell} \widetilde{S}_c)\widetilde{J}}
(-)^{S^\prime_c-S_c}~f^J_{(\ell^\prime S^\prime_c)(\ell S_c)}
\left\{f^{\widetilde{J}}_{(\widetilde{\ell}^\prime \widetilde{S}^\prime_c)
(\widetilde{\ell} \widetilde{S}_c)}\right\}^* \nonumber \\
\ \hspace{-10mm} & & \times \sum_{L, L^\prime, \kappa}
C^{L, L^\prime}_\kappa(\lambda_i \mu_i,\lambda_f \mu_f; \widehat{\bq}_f,
\widehat{\bq}_i)
~Z^{(\lambda_i)}_\kappa(\ell J \widetilde{\ell} \widetilde{J};
S_c \widetilde{S}_c L)
~Z^{(\lambda_f)}_\kappa(\ell^\prime J \widetilde{\ell}^\prime
\widetilde{J}; S^\prime_c \widetilde{S}^\prime_c L^\prime)\ .
\label{sec4-5-42}
\end{eqnarray}
The spatial part $C^{L, L^\prime}_\kappa(\lambda_i \mu_i,\lambda_f \mu_f; 
\widehat{\bq}_f,\widehat{\bq}_i)$ can be written in various forms:
\begin{eqnarray}
\ \hspace{-10mm} & & C^{L, L^\prime}_\kappa(\lambda_i \mu_i,\lambda_f \mu_f; 
\widehat{\bq}_f,\widehat{\bq}_i)
\nonumber \\
\ \hspace{-10mm} & & = \sum_{M M^\prime \nu} (-)^{\mu_i}
\,\frac{4\pi}{\widehat{L}\,\widehat{L}^\prime}
\langle \kappa \nu \lambda_i -\mu_i|LM \rangle
\langle \kappa \nu \lambda_f \mu_f|L^\prime M^\prime \rangle
~Y_{L^\prime M^\prime}(\widehat{\bq}_f)
Y^*_{LM}(\widehat{\bq}_i)
\nonumber \\
\ \hspace{-10mm} & & = (-)^{\lambda_i+\lambda_f+L+L^\prime-\kappa} 
\,4\pi \sum_{f f_z} \langle \lambda_i \mu_i \lambda_f \mu_f|f f_z \rangle
\left\{ \begin{array}{ccc}
L & \lambda_i & \kappa \\
\lambda_f & L^\prime & f \\
\end{array} \right\}
~Y_{(L^\prime L)f f_z}(\widehat{\bq}_f, \widehat{\bq}_i),
\label{sec4-5-43}
\end{eqnarray}
but the separation into the invariant part and the non-invariant part
is not easy except for the $f=0$ and 1 cases.
We should note that the $L$ and $L^\prime$ sum in \eq{sec4-5-42} is
only for $|L-L^\prime|=$even because of the parity conservation.
We use a simplified notation 
$C^{L, L^\prime}_\kappa(\lambda_i \mu_i,\lambda_f \mu_f) 
=C^{L, L^\prime}_\kappa(\lambda_i \mu_i,\lambda_f \mu_f; 
\widehat{\bq}_f,\widehat{\bq}_i)$ in the following and
choose a special coordinate system with $\widehat{\bq}_f=(\theta, 0)$
and $\widehat{\bq}_i=\be_z=(0, 0)$, where the $z$-axis
is the beam direction and $\theta$ is the scattering angle in the
center-of-mass (cm) system. We use
\begin{eqnarray}
& & Y_{\ell m}(\be_z)=Y_{\ell m}(0, 0)=\delta_{m,0}
\frac{\widehat{\ell}}{\sqrt{4\pi}}\ ,\nonumber \\
& & Y_{\ell^\prime m^\prime}(\theta, 0)
=(-)^{\frac{m^\prime+|m^\prime|}{2}}\frac{\widehat{\ell}^\prime}{\sqrt{4\pi}}
\sqrt{\frac{(\ell^\prime-|m^\prime|)!}{(\ell^\prime+|m^\prime|)!}}
~P^{|m^\prime|}_{\ell^\prime}(\cos\,\theta)\ ,
\label{sec4-5-44}
\end{eqnarray}
which yield a basic symmetry property
\begin{eqnarray}
& & Y_{(\ell^\prime \ell)\kappa \nu}(\widehat{\bq}_f, \be_z)
=(-)^{\kappa+\nu}
~Y_{(\ell^\prime \ell)\kappa -\nu}(\widehat{\bq}_f, \be_z)
\nonumber \\
& & =(-)^{\ell^\prime}~\frac{\widehat{\kappa} \widehat{\ell}^\prime}
{4 \pi} \langle \ell^\prime \nu \kappa \, -\nu|\ell 0 \rangle
\sqrt{\frac{(\ell^\prime-\nu)!}{(\ell^\prime+\nu)!}}
~P^\nu_{\ell^\prime}(\cos\,\theta) \quad \hbox{for} \quad \nu \geq 0\ .
\label{sec4-5-45}
\end{eqnarray}
If we use \eq{sec4-5-44} in the first expression of \eq{sec4-5-43},
we find
\begin{eqnarray}
& & C^{L, L^\prime}_\kappa(\lambda_i \mu_i,\lambda_f \mu_f)
=(-)^{\lambda_i+\mu_i+\lambda_f+\mu_f}
C^{L, L^\prime}_\kappa(\lambda_i, -\mu_i,\lambda_f, -\mu_f)
\nonumber \\
& & =(-)^{\mu_f}
\langle \kappa \mu_i \lambda_i -\mu_i|L0 \rangle
\langle \kappa \mu_i \lambda_f \mu_f|L^\prime \mu_i+\mu_f \rangle
\sqrt{\frac{(L^\prime-\mu_i-\mu_f)!}{(L^\prime+\mu_i+\mu_f)!}}
~P^{\mu_i+\mu_f}_{L^\prime}(\cos\,\theta)
\nonumber \\
& & \hspace{10mm} \hbox{for} \quad \mu_i+\mu_f \geq 0\ .
\label{sec4-5-46}
\end{eqnarray}
If we set here $\lambda_f \mu_f=0 0$, then we obtain
\begin{eqnarray}
& & C^{L, L^\prime}_\kappa(\lambda_i \mu_i,0 0)
=(-)^{\lambda_i+\mu_i}
C^{L, L^\prime}_\kappa(\lambda_i, -\mu_i, 0 0)
\nonumber \\
& & =\delta_{\kappa, L^\prime}
\langle L^\prime \mu_i \lambda_i -\mu_i|L 0 \rangle
\sqrt{\frac{(L^\prime-\mu_i)!}{(L^\prime+\mu_i)!}}
~P^{\mu_i}_{L^\prime}(\cos\,\theta) \quad
\hbox{for} \quad \mu_i \geq 0\ .
\label{sec4-5-47}
\end{eqnarray}
For $\lambda_i \mu_i=0 0$, we find
\begin{eqnarray}
& & C^{L, L^\prime}_\kappa(0 0 ,\lambda_f \mu_f)
=(-)^{\lambda_f+\mu_f}
C^{L, L^\prime}_\kappa(0 0, \lambda_f, -\mu_f)
\nonumber \\
& & =\delta_{\kappa, L} \frac{\widehat{L}^\prime}{\widehat{L}}
\langle L^\prime \mu_f \lambda_f -\mu_f|L 0 \rangle
\sqrt{\frac{(L^\prime-\mu_f)!}{(L^\prime+\mu_f)!}}
~P^{\mu_f}_{L^\prime}(\cos\,\theta) \quad
\hbox{for} \quad \mu_f \geq 0\ .
\label{sec4-5-48}
\end{eqnarray}
If we further set $\lambda_i \mu_i=00$, we gain
\begin{eqnarray}
C^{L, L^\prime}_\kappa(0 0 , 0 0)
=\delta_{\kappa, L} \delta_{L, L^\prime}~P_L(\cos\,\theta)\ ,
\label{sec4-5-49}
\end{eqnarray}
which yields the differential cross sections in Eq.\,I(3.5)
by using the reduction
\begin{eqnarray}
Z^{(0)}_\kappa(\ell J \widetilde{\ell} \widetilde{J}; S_c
\widetilde{S}_c L)
=\delta_{\kappa, L} \delta_{S_c, \widetilde{S}_c}
~Z(\ell J \widetilde{\ell} \widetilde{J}; S_c L)\ .
\label{sec4-5-50}
\end{eqnarray}
The reduced matrix element of $\CS^{(\lambda)}$
in \eq{sec4-5-38} is given by
\begin{eqnarray}
& & \langle \chi_{S_c}||\left[\sigma^{(s)} S^{(\lambda)}\right]^{(f)}||
\chi_{\widetilde{S}_c} \rangle_{\rm unc}
\nonumber \\
& & =(-1)^{s+\lambda-f}
\left[ \begin{array}{ccc}
1 & \H & \widetilde{S}_c \\ [2mm]
\lambda & s & f \\ [2mm]
1 & \H & S_c \\
\end{array} \right]
\langle 1 ||S^{(\lambda)}||1 \rangle_{\rm unc}
\langle \H ||\sigma^{(s)}||\H \rangle_{\rm unc} \ ,
\label{sec4-5-51}
\end{eqnarray}
with $\langle \H||\bfsigma||\H \rangle_{\rm unc}=\sqrt{3}$,
$\langle 1||\bS||1 \rangle_{\rm unc}=\sqrt{2}$, and
$\langle 1||S^{(2)}||1 \rangle_{\rm unc}=\sqrt{5/3}$ (see \eq{sec4-5-10}).

\bigskip

\subsection{Analyzing-power and polarization}

The analyzing-power is characterized by $\lambda_f \mu_f=00$ and
the spatial function in \eq{sec4-5-47} is used.
The nucleon vector analyzing-power is defined with $\CS_i=\bfsigma$.
The expression in \eq{sec4-5-47} with $\lambda_i=1$
\begin{eqnarray}
& & C^{L, L^\prime}_\kappa(1 \mu,0 0)
=(-)^{1+\mu}
C^{L, L^\prime}_\kappa(1, -\mu, 0 0)
\nonumber \\
& & =\delta_{\kappa, L^\prime}
\langle L^\prime \mu \, 1 -\mu|L 0 \rangle
\sqrt{\frac{(L^\prime-\mu)!}{(L^\prime+\mu)!}}
~P^{\mu}_{L^\prime}(\cos\,\theta) \quad
\hbox{for} \quad \mu \geq 0\ ,
\label{sec4-5-52}
\end{eqnarray}
is transformed to the Cartesian representation by the spherical
vector rule in \eq{sec4-5-5}.
If we write these as $C^{L, L^\prime}_\kappa(\alpha)$, we find
that $x$ and $z$ components are zero because of the parity
symmetry in \eq{sec4-5-52} and $\langle L0 10|L0 \rangle=0$,
respectively.
(Note that we only have $L^\prime=L$ for $\lambda_i=1$.)
The $y$-component is from $\langle L1 1 -1|L0 \rangle=1/\sqrt{2}$
with $L \geq 1$
\begin{eqnarray}
C^{L, L^\prime}_\kappa(y)
=\delta_{\kappa, L} \delta_{L, L^\prime}
~i~\frac{1}{\sqrt{L^\prime (L^\prime+1)}}
~P^1_{L^\prime}(\cos\,\theta) \ .
\label{sec4-5-53}
\end{eqnarray}
The nucleon analyzing-power $A_y(\theta)$ is usually defined
through
\begin{eqnarray}
\frac{Tr \left\{ f \bfsigma f^\dagger\right\}}{Tr \left\{f f\dagger\right\}}
=A_y(\theta)~\widehat{\bn}\ ,
\label{sec4-5-54}
\end{eqnarray}
where $\bn=[\bq_i \times \bq_f]$. Since $\widehat{\bn}=\be_y$,
we find that $A_y(\theta)$ is calculated from $I_y/I_0$ with \eq{sec4-5-53},
where $I_0=Tr \left\{f f^\dagger\right\}$. The reduced matrix element 
for $Z^{(1)}_L(\ell J \widetilde{\ell} \widetilde{J}; S_c
\widetilde{S}_c L)$ is 
\begin{eqnarray}
\langle \chi_{S_c}||\bfsigma||\chi_{\widetilde{S}_c} \rangle_{\rm unc}
=(-1)^{S_c+\H} \sqrt{6}~\widehat{\widetilde{S}_c}
\left\{ \begin{array}{ccc}
\widetilde{S}_c & \H & 1 \\
\H & S_c & 1 \\
\end{array} \right\}\ .
\label{sec4-5-55}
\end{eqnarray}

The vector analyzing-power of the deuteron is similarly calculated,
by setting $\CS_i=\bS$. Only the reduced matrix element is different,
which is explicitly given by
\begin{eqnarray}
\langle \chi_{S_c}||\bS||\chi_{\widetilde{S}_c} \rangle_{\rm unc}
=(-1)^{\widetilde{S}_c+\H} \sqrt{6}~\widehat{\widetilde{S}_c}
\left\{ \begin{array}{ccc}
1 & 1 & 1 \\
\H & S_c & \widetilde{S}_c \\
\end{array} \right\}\ .
\label{sec4-5-56}
\end{eqnarray}
From the definition of $iT_{11}$ in \eq{sec4-5-24}, we should multiply
the factor $\sqrt{3}/2$, namely, $iT_{11}=(\sqrt{3}/2)(I(y)/I_0)$.

The tensor analyzing-power of the deuteron is obtained from the $\lambda_i
\mu_i=2 \mu$ case in \eq{sec4-5-47}. From the parity conservation
embodied in \eq{sec4-5-47}, the two components with $2 \mu$ and
$2, -\mu$ give the same contributions. 
Thus the factor $\sqrt{3}$ in \eq{sec4-5-18}
yields $T_{2\mu}=\sqrt{3}(I(2\mu)/I_0)$, where the reduced matrix
element for $Z^{(2)}_{L^\prime}(\ell J \widetilde{\ell} \widetilde{J}; S_c
\widetilde{S}_c L)$ (in this case, $L^\prime=L,~L\pm 2$) is given by
\begin{eqnarray}
\langle \chi_{S_c}||S^{(2)}||\chi_{\widetilde{S}_c} \rangle_{\rm unc}
=(-1)^{\widetilde{S}_c+\3H} \sqrt{5}~\widehat{\widetilde{S}_c}
\left\{ \begin{array}{ccc}
2 & 1 & 1 \\
\H & S_c & \widetilde{S}_c \\
\end{array} \right\}\ .
\label{sec4-5-57}
\end{eqnarray}
The spatial functions $I(2\mu)=I(2\mu,00)$ are calculated from
\begin{eqnarray}
& & C^{L, L^\prime}_\kappa(2 \mu,0 0)
=(-)^\mu~C^{L, L^\prime}_\kappa(2, -\mu, 0 0)
\nonumber \\
& & =\delta_{\kappa, L^\prime}
~\langle L^\prime \mu~2, -\mu|L 0 \rangle
\sqrt{\frac{(L^\prime-\mu)!}{(L^\prime+\mu)!}}
~P^{\mu}_{L^\prime}(\cos\,\theta) \quad
\hbox{for} \quad \mu=0,~1,~2\ .
\label{sec4-5-59}
\end{eqnarray}

It is convenient to write the final results for the analyzing-power
in the form similar to the differential cross sections in Eq.\,I(3.5)
(We take the sum over $S^\prime_c$ and $S_c$ inside.)
\begin{eqnarray}
I_0 & = & Tr\,\{f f^\dagger\}=6\,\sigma_0=\sum_L B_L~P_L(\cos\,\theta)\ \ ,
\nonumber \\
I_y & = & Tr\,\left\{f \left(\begin{array}{c}
\sigma_y \\
S_y \\
\end{array}\right)f^\dagger \right\}
=\sum_L \frac{1}{\sqrt{L(L+1)}} B^{(1)}_L~P^1_L(\cos\,\theta)
\ \ ,\nonumber \\
I(2m) & = & Tr\,\left\{f S^{(2)}_m f^\dagger \right\}
=\sum_{L^\prime,L}~B^{(2)}_{L, L^\prime}
~\sqrt{\frac{(L^\prime-m)!}{(L^\prime+m)!}}
\nonumber \\
& & \times \langle L^\prime m~2, -m|L 0 \rangle~P^m_{L^\prime}(\cos\,\theta)
\qquad \hbox{for} \qquad m=0,~1,~2\ \ ,
\label{sec4-5-60}
\end{eqnarray}
where $P^m_L(x)$ ($m=0,~1,~2,\cdots$) are the Legendre by-polynomials
and the $L^\prime$ sum in $I(2m)$ is only for $L^\prime=L,~L\pm 2$
from the parity conservation.
The coefficients, $B_L$, $B^{(1)}_L$ and $B^{(2)}_{L, L^\prime}$, are
given by 
\begin{eqnarray}
\ \hspace{-10mm} \left. \begin{array}{c}
B_{L^\prime} \\ [2mm]
B^{(1)}_{L^\prime} \\ [2mm]
B^{(2)}_{L, L^\prime} \\
\end{array} \right\}
& = & \sum_{(\ell^\prime S^\prime_c)(\ell S_c)J}
\sum_{(\widetilde{\ell}^\prime \widetilde{S}^\prime_c) 
(\widetilde{\ell} \widetilde{S}_c)\widetilde{J}}
(-)^{S^\prime_c-S_c}~f^J_{(\ell^\prime S^\prime_c)(\ell S_c)}
\left\{f^{\widetilde{J}}_{(\widetilde{\ell}^\prime \widetilde{S}^\prime_c)
(\widetilde{\ell} \widetilde{S}_c)}\right\}^* 
\nonumber \\
\ \hspace{-10mm} & & \times \left\{ \begin{array}{c}
\delta_{S_c, \widetilde{S}_c}
\,Z(\ell J \widetilde{\ell} \widetilde{J}; S_c L^\prime) \\ [2mm]
i~Z^{(1)}_L (\ell J \widetilde{\ell} \widetilde{J}; 
S_c \widetilde{S}_c L^\prime) \\ [2mm]
Z^{(2)}_{L^\prime}(\ell J \widetilde{\ell}
\widetilde{J}; S_c \widetilde{S}_c L)  \\
\end{array} \right\}
~\delta_{S^\prime_c, \widetilde{S}^\prime_c}
\,Z(\ell^\prime J \widetilde{\ell}^\prime \widetilde{J}; 
S^\prime_c L^\prime)\ .
\label{sec4-5-61}
\end{eqnarray}
The explicit expressions of the CG coefficients allow us to
express $I(2m)$ ($m=0,~1,~2$) in \eq{sec4-5-60} as
\begin{eqnarray}
\ \hspace{-10mm}
I(2m)=Tr\,\left\{f S^{(2)}_m f^\dagger \right\}
=\sum_{L \geq m} B^m_{L}
\left\{ \begin{array}{c}
1 \\ [2mm]
\frac{1}{\sqrt{L(L+1)}} \\ [2mm]
\frac{1}{\sqrt{(L-1)L(L+1)(L+2)}} \\
\end{array}\right\}
~P^m_{L}(\cos\,\theta)\ ,
\label{sec4-5-62}
\end{eqnarray}
for $m=0,~1,~2$ with
\begin{eqnarray}
B^0_{L}
& = & -\sqrt{\frac{L(L+1)}{(2L-1)(2L+3)}}~B^{(2)}_{LL}
+\sqrt{\frac{3(L+1)(L+2)}{2(2L+1)(2L+3)}}~B^{(2)}_{L+2,L}
\nonumber \\
& & +\sqrt{\frac{3(L-1)L}{2(2L-1)(2L+1)}}~B^{(2)}_{L-2,L}\ ,\nonumber \\
B^1_{L}
& = & \sqrt{\frac{3}{2(2L-1)(2L+3)}}~B^{(2)}_{LL}
+\sqrt{\frac{L(L+2)}{(2L+1)(2L+3)}}~B^{(2)}_{L+2,L}
\nonumber \\
& & -\sqrt{\frac{(L-1)(L+1)}{(2L-1)(2L+1)}}~B^{(2)}_{L-2,L}\ ,\nonumber \\
B^2_{L}
& = & \sqrt{\frac{3(L-1)(L+2)}{2(2L-1)(2L+3)}}~B^{(2)}_{LL}
+\frac{1}{2}\sqrt{\frac{(L-1)L}{(2L+1)(2L+3)}}~B^{(2)}_{L+2,L}
\nonumber \\
& & +\frac{1}{2}\sqrt{\frac{(L+1)(L+2)}{(2L-1)(2L+1)}}~B^{(2)}_{L-2,L}\ .
\label{sec4-5-63}
\end{eqnarray}

The polarization of the outgoing particles is calculated from
the $\lambda_i \mu_i=00$ case in \eq{sec4-5-48}.
In the cm system, the vector polarization
of the nucleon or the deuteron is the same as
the analyzing-power. Namely, we find
$C^{L, L^\prime}_\kappa(00,y)=C^{L, L^\prime}_\kappa(y,00)$.
For the tensor polarization of the deuteron, we find
\begin{eqnarray}
C^{L, L^\prime}_\kappa(00,2 \mu)
=\frac{\widehat{L}^\prime}{\widehat{L}}~C^{L, L^\prime}_\kappa(2 \mu,00)\ .
\label{sec4-5-64}
\end{eqnarray}

\bigskip

\subsection{Polarization transfer for the nucleon}

This is the case with $\CS_i=\bfsigma$ and $\CS_f=\bfsigma$.
Setting $\lambda_i=1$ and $\lambda_f=1$ in \eq{sec4-5-46},
we find
\begin{eqnarray}
& & C^{L, L^\prime}_\kappa(1 \mu_i,~1 \mu_f)
= (-)^{\mu_i+\mu_f}\,C^{L, L^\prime}_\kappa(1, -\mu_i;~1, -\mu_f)
\nonumber \\
& & =(-)^{\mu_f}
\langle \kappa \mu_i 1 -\mu_i|L0 \rangle
\langle \kappa \mu_i 1 \mu_f|L^\prime \mu_i+\mu_f \rangle
\sqrt{\frac{(L^\prime-\mu_i-\mu_f)!}{(L^\prime+\mu_i+\mu_f)!}}
~P^{\mu_i+\mu_f}_{L^\prime}(\cos\,\theta)
\nonumber \\
& & \hspace{10mm} \hbox{for} \quad \mu_i+\mu_f \geq 0\ .
\label{sec4-5-65}
\end{eqnarray}
From the symmetry of the parity conservation, the independent 
types of the polarization transfer are very restricted.
We use a simplified notation $C_{\mu, \mu^\prime}
= C^{L, L^\prime}_\kappa(1 \mu,~1 \mu^\prime)$ and express this
symmetry as $C_{\mu, \mu^\prime}=(-1)^{\mu+\mu^\prime}
C_{-\mu, -\mu^\prime}$ from \eq{sec4-5-65}.
For the fixed values of $\mu_i+\mu_f=f_z=0,~1,~2$,
we have the following five combinations of $\mu_i$ and $\mu_f$:
$(\mu_i, \mu_f)=(0,~0),~(1, -1)$ for $f_z=0$, (10), (01) for $f_z=1$,
and $(1, 1)$ for $f_z=2$. The correspondence to the Cartesian
representation is found, for example, as
\begin{eqnarray}
K_{x,x} \sim \sigma_x \sigma_x & = & \frac{1}{2}
(-\sigma_1+\sigma_{-1})(-\sigma_1+\sigma_{-1})
\sim \frac{1}{2}(C_{11}-C_{1,-1}-C_{-1,1}+C_{-1,-1})
\nonumber \\
& = & \frac{1}{2}(C_{11}-C_{1,-1}-C_{1,-1}+C_{1,1})
=C_{11}-C_{1,-1}\ ,\nonumber \\
K_{x,y} \sim \sigma_x \sigma_y & = & i\,\frac{1}{2}
(-\sigma_1+\sigma_{-1})(\sigma_1+\sigma_{-1})
\sim i\,\frac{1}{2}(-C_{11}-C_{1,-1}+C_{-1,1}+C_{-1,-1})
\nonumber \\
& = & 0\ ,\nonumber \\
K_{x,z} \sim \sigma_x \sigma_z & = & \frac{1}{\sqrt{2}}
(-\sigma_1+\sigma_{-1}) \sigma_0
\sim \frac{1}{\sqrt{2}}(-C_{10}+C_{-1,0})
=-\sqrt{2}~C_{10}\ ,\nonumber \\
\cdots & & \ .
\label{sec4-5-66}
\end{eqnarray}
We obtain
\begin{eqnarray}
& & K_{z,z} \sim C_{00}\ \ ,\qquad K_{x,x} \sim C_{11}-C_{1,-1}\ \ ,\qquad
K_{y,y} \sim -(C_{11}+C_{1,-1})\ ,\nonumber \\
& & K_{x,z} \sim (-\sqrt{2})~C_{10}\ \ ,\qquad
K_{z,x} \sim (-\sqrt{2})~C_{01}\ ,\nonumber \\
& & K_{x,y}=K_{y,x}=K_{y,z}=K_{z,y}=0\ .
\label{sec4-5-67}
\end{eqnarray}
After all, five independent polarization transfers of the nucleon
are calculated from the following spatial integrals:
\begin{eqnarray}
& & K_{z,z} \sim C^{L, L^\prime}_\kappa(10, 10)
=\langle \kappa 0 1 0|L0 \rangle
\langle \kappa 0 1 0|L^\prime 0 \rangle ~P_{L^\prime}(\cos\,\theta)
\ ,\nonumber \\
& & K_{x,z} \sim (-\sqrt{2}) C^{L, L^\prime}_\kappa(11, 10)
=-\sqrt{2}\langle \kappa\,1\,1,-1|L0 \rangle
\langle \kappa 1 1 0|L^\prime 1 \rangle
\frac{1}{\sqrt{L^\prime(L^\prime+1)}} P^1_{L^\prime}(\cos\,\theta)
\ ,\nonumber \\
& & K_{z,x} \sim (-\sqrt{2}) C^{L, L^\prime}_\kappa(10, 11)
=\sqrt{2} \langle \kappa 0 1 0|L0 \rangle
\langle \kappa 0 1 1|L^\prime 1 \rangle
\frac{1}{\sqrt{L^\prime(L^\prime+1)}}~P^1_{L^\prime}(\cos\,\theta)
\ ,\nonumber \\
& & \left. \begin{array}{c}
K_{x,x} \\ [2mm]
K_{y,y} \\
\end{array} \right\}
\sim \left\{ \begin{array}{c}
C^{L, L^\prime}_\kappa(11, 11)-C^{L, L^\prime}_\kappa(11; 1,-1) \\ [2mm]
-C^{L, L^\prime}_\kappa(11, 11)-C^{L, L^\prime}_\kappa(11; 1,-1) \\
\end{array}\right\} \nonumber \\
& & =\langle \kappa\,1\,1,-1|L 0\rangle
\left[ \mp \langle \kappa 1 1 1 |L^\prime 2 \rangle
\sqrt{\frac{(L^\prime-2)!}{(L^\prime+2)!}} P^2_{L^\prime}(\cos\,\theta)
+ \langle \kappa\,1\,1,-1|L^\prime 0 \rangle 
P_{L^\prime}(\cos\,\theta) \right].\nonumber\\
\label{sec4-5-68}
\end{eqnarray}
The reduced matrix elements of the two $Z$-factors are for $\bfsigma$,
given in \eq{sec4-5-55}.

We should note that the polarization transfers in \eq{sec4-5-68}
are for the cm system. These are however usually defined 
in the lab system and the notation ${K_\alpha}^{\beta^\prime}$ with
the upper $\beta^\prime$ is used to specify the $\beta$-axis 
in the laboratory system. 
We have to rotate $K_{\alpha, \beta}$ with respect to the
coordinate $\beta$ for the outgoing particle.
In the laboratory system, the scattering angle of the nucleon,
$\theta$ in \eq{sec4-5-44} is transformed into $\theta_1$, 
which can be calculated from
\begin{eqnarray}
{\rm tan}\,\theta_1=\frac{\sin\,\theta}{\gamma \left(\cos\, \theta
+\frac{\beta}{\beta^*}\right)}\ .
\label{sec4-5-68-1}
\end{eqnarray}
Here, $\gamma$ and $\beta$ are the usual relativistic factors
for the cm to lab transformation,
and $\beta^*$ is the $\beta$ factor of the outgoing nucleon
in the cm system. For the neutron-incident $nd$ elastic scattering
these are calculated from the ``kinetic'' energy of the incident
nucleon $T_{\rm lab}=E_{\rm lab}-M_N c^2$. ($M_N$ is the nucleon
mass.) If we assume the masses of the nucleon and
deuteron as $m_1=M_N$ and $m_2=2M_N$, the necessary factors
in \eq{sec4-5-68-1} are calculated to be
\begin{eqnarray}
\gamma=\frac{1+\frac{x}{3}}{\sqrt{1+\frac{4}{9}x}}
\ \ ,\qquad \frac{\beta}{\beta^*}
=\frac{1+\frac{2}{3}x}{2+\frac{2}{3}x}\ ,
\label{sec4-5-68-2}
\end{eqnarray}
with $x=T_{\rm lab}/(M_N c^2)$.
The rotation of the Cartesian frame $(x, y, z)$ in the Madison
convention around the $y$-axis by $\theta_1$ yields the transformation
of the unit vectors
\begin{eqnarray}
& & \be^\prime_z= \be_z \cos\,\theta_1+\be_x \sin\,\theta_1\ ,\nonumber \\
& & \be^\prime_x= -\be_z \sin\,\theta_1+\be_x \cos\,\theta_1\ ,\nonumber \\
& & \be^\prime_y=\be_y\ .
\label{sec4-5-68-3}
\end{eqnarray}
The components of the spin vector $\bfsigma$ of the outgoing nucleon
are therefore transformed to
\begin{eqnarray}
& & \sigma_{z^\prime}= \sigma_z \cos\,\theta_1
+\sigma_x \sin\,\theta_1\ ,\nonumber \\
& & \sigma_{x^\prime}= -\sigma_z \sin\,\theta_1
+\sigma_x \cos\,\theta_1\ ,\nonumber \\
& & \sigma_{y^\prime}=\sigma_y\ ,
\label{sec4-5-68-4}
\end{eqnarray}
since $\sigma_{\alpha^\prime}=(\bfsigma \cdot \be^\prime_\alpha)$
and $\sigma_\alpha=(\bfsigma \cdot \be_\alpha)$. 
In order to write down the explicit expression for $K_\alpha^{\beta^\prime}$,
we note that $K_{\alpha, \beta}$ in the cm system is actually
functions of $\widehat{\bp}_f=(\theta, \varphi)$ under the assumption
of $\widehat{\bp}_i=\be_z=(0, 0)$. We therefore understand
$K_{\alpha, \beta}$ in \eq{sec4-5-68} are actually
$K_{\alpha, \beta}=K_{\alpha, \beta}(\theta, 0)$, using
the notation $K_{\alpha, \beta}(\theta, \varphi)$.
Taking these into consideration, we obtain the following results.
\begin{eqnarray}
& & {K_z}^{z^\prime}=K_{z,z}\,\cos\,\theta_1
+K_{z,x}\,\sin\,\theta_1\ ,\nonumber \\
& & {K_z}^{x^\prime}=-K_{z,z}\,\sin\,\theta_1
+K_{z,x}\,\cos\,\theta_1\ ,\nonumber \\
& & {K_x}^{z^\prime}=K_{x,z}\,\cos\,\theta_1
+K_{x,x}\,\sin\,\theta_1\ ,\nonumber \\
& & {K_x}^{x^\prime}=-K_{x,z}\,\sin\,\theta_1
+K_{x,x}\,\cos\,\theta_1\ ,\nonumber \\
& & {K_y}^{y^\prime}=K_{y,y}\ ,
\label{sec4-5-68-5}
\end{eqnarray}
where $K_{\alpha, \beta}=K_{\alpha, \beta}(\theta, 0)$ and $\theta_1$
is calculated from Eqs.\,(\ref{sec4-5-68-1}) and (\ref{sec4-5-68-2}).

In fact, we further need small relativistic corrections
for the spin directions, as discussed in the $NN$ spin transfer
coefficients. (See for example Ref.\,\citen{Fu98} and references therein.)
We however neglect these, since we deal with the low-energy scattering
for the time being.

\bigskip

\subsection{Nucleon to deuteron polarization transfer}

The vector-type nucleon to deuteron polarization transfers 
in the cm system are calculated from
the same spatial functions as in \eq{sec4-5-68}, but with a different 
reduced matrix element of $\bS$ given in \eq{sec4-5-56} for the second
rank-one $Z$ factor. The same notation $K_{\alpha, \beta}$ is used
for these polarization transfer coefficients with the vector-type deuteron
polarization. However, we should be careful with some
complications originating from the fact that we are now detecting
the recoil particle. The scattering angle $\theta_2$ of the recoil
deuteron is given by
\begin{eqnarray}
{\rm tan}\,\theta_2=\frac{1}{\gamma} {\rm tan}\,\left(
\frac{\pi}{2}-\frac{\theta}{2}\right)\ ,
\label{sec4-5-68-6}
\end{eqnarray}
namely, $\theta_2=\pi/2$ - 0 with $\pi/2$ corresponding to $\theta=0$
and 0 to $\theta=\pi$. The direction of $\theta_2$ is opposite
to $\theta_1$, so that the $y^\prime$ axis is not equal to 
the original $y$ axis but to the opposite to it.
($\be^\prime_y=-\be_y$). If the nucleon comes out on the left-hand side
of the beam direction, the deuteron turns to the right-hand side.
This changes the sign of $\be^\prime_y$. This definition seems
to be inconvenient to the experimentalists, since the outgoing deuteron 
is further turned around by the spectrometer magnet
to measure the polarization. To circumvent this difficulty, we rotate
the whole system around the original $z$-axis by $\pi$. Then
the $y$-direction becomes the original $\be_y$ and
the deuteron comes out to the left-hand side (the same direction 
as the nucleon before). We can use
\begin{eqnarray}
& & S_{z^\prime}= S_z \cos\,\theta_2
+S_x \sin\,\theta_2\ ,\nonumber \\
& & S_{x^\prime}= -S_z \sin\,\theta_2
+S_x \cos\,\theta_2\ ,\nonumber \\
& & S_{y^\prime}=S_y\ ,
\label{sec4-5-68-7}
\end{eqnarray}
for the deuteron spin operator $\bS$, similarly to \eq{sec4-5-68-4},
but we should be careful to use $K_{\alpha, \beta}(\theta, \pi)$
and not $K_{\alpha, \beta}(\theta, 0)$. We therefore find
\begin{eqnarray}
& & {K_z}^{z^\prime}=K_{z,z}(\theta, \pi)\,\cos\,\theta_2
+K_{z,x}(\theta, \pi)\,\sin\,\theta_2\ ,\nonumber \\
& & {K_z}^{x^\prime}=-K_{z,z}(\theta, \pi)\,\sin\,\theta_2
+K_{z,x}(\theta, \pi)\,\cos\,\theta_2\ ,\nonumber \\
& & {K_x}^{z^\prime}=K_{x,z}(\theta, \pi)\,\cos\,\theta_2
+K_{x,x}(\theta, \pi)\,\sin\,\theta_2\ ,\nonumber \\
& & {K_x}^{x^\prime}=-K_{x,z}(\theta, \pi)\,\sin\,\theta_2
+K_{x,x}(\theta, \pi)\,\cos\,\theta_2\ ,\nonumber \\
& & {K_y}^{y^\prime}=K_{y,y}(\theta, \pi)\ .
\label{sec4-5-68-8}
\end{eqnarray}
The $\pi$-rotation around $z$-axis in the polarization transfer coefficients
$K_{\alpha, \beta}$ (and also ${K_\alpha}^{\beta^\prime}$) are
related to the even-odd symmetry for $\theta$ as
shown in Ohlsen's paper. \cite{Oh72} See Sec. 5-3-2 in page 751.
In our expression, this symmetry appears in \eq{sec4-5-67} 
as the fact that the coefficient $C_{\mu_i, \mu_f}$ changes 
the sign when $\mu_i+\mu_f=\hbox{odd}$. This is because
the $Y_{\ell^\prime m^\prime}(\theta, \pi)$ in \eq{sec4-5-44}
gives an extra factor $(-1)^{\mu_i+\mu_f}$ from the
wave function $\Phi_{m^\prime}(\varphi)=e^{i m^\prime \varphi}
/\sqrt{2\pi}$. For the polarization transfer coefficients this phase
factor is the same as the even-odd character of $N_x+N_y$,
where $N_x$ and $N_y$ are the number of times $x$ and $y$ appear.
Namely, from \eq{sec4-5-67}, we find that only $K_{x,z}$ and
$K_{z,x}$ change the sign, since $\mu_i+\mu_f=\hbox{odd}$.
In other words, $K_{x,z}(\theta,\pi)=-K_{x,z}(\theta,0)$
and $K_{z,x}(\theta,\pi)=-K_{z,x}(\theta,0)$, the others no phase change.
If we use the simplified 
notation $K_{\alpha, \beta}=K_{\alpha, \beta}(\theta, 0)$ as 
in \eq{sec4-5-68-5}, we finally obtain
\begin{eqnarray}
& & {K_z}^{z^\prime}=K_{z,z}\,\cos\,\theta_2
-K_{z,x}\,\sin\,\theta_2\ ,\nonumber \\
& & {K_z}^{x^\prime}=-K_{z,z}\,\sin\,\theta_2
-K_{z,x}\,\cos\,\theta_2\ ,\nonumber \\
& & {K_x}^{z^\prime}=-K_{x,z}\,\cos\,\theta_2
+K_{x,x}\,\sin\,\theta_2\ ,\nonumber \\
& & {K_x}^{x^\prime}=K_{x,z}\,\sin\,\theta_2
+K_{x,x}\,\cos\,\theta_2\ ,\nonumber \\
& & {K_y}^{y^\prime}=K_{y,y}\ .
\label{sec4-5-68-9}
\end{eqnarray}
Note the phase change in addition to $\theta_1 \rightarrow \theta_2$.

For the tensor-type deuteron polarization, we consider
the matrix elements for $\CS_i=\sigma_{\mu_i}$ and $\CS_f=S^{(2)}_{\mu_f}$.
Since $\lambda_i=1$ and $\lambda_f=2$, we deal with the spatial
integrals of the type
\begin{eqnarray}
& & C^{L, L^\prime}_\kappa(1 \mu_i,~2 \mu_f)
= (-)^{1+\mu_i+\mu_f}\,C^{L, L^\prime}_\kappa(1, -\mu_i;~2, -\mu_f)
\nonumber \\
& & =(-)^{\mu_f}
\langle \kappa \mu_i 1 -\mu_i|L0 \rangle
\langle \kappa \mu_i 2 \mu_f|L^\prime \mu_i+\mu_f \rangle
\sqrt{\frac{(L^\prime-\mu_i-\mu_f)!}{(L^\prime+\mu_i+\mu_f)!}}
~P^{\mu_i+\mu_f}_{L^\prime}(\cos\,\theta)
\nonumber \\
& & \hspace{10mm} \hbox{for} \quad \mu_i+\mu_f= 0,~1,~2,~3\ .
\label{sec4-5-69}
\end{eqnarray}
We again use the notation $C_{\mu, \mu^\prime}
= C^{L, L^\prime}_\kappa(1 \mu,~2 \mu^\prime)$ and express the parity
symmetry as $C_{\mu, \mu^\prime}=(-1)^{1+\mu+\mu^\prime}
C_{-\mu, -\mu^\prime}$ from \eq{sec4-5-69}.
From the parity conservation $(-)^{L+L^\prime}=1$, we only have 
$L^\prime=L,~L\pm2$.
We can calculate independent non-zero coefficients in a way similar 
to the nucleon polarization transfer.
Namely, we express the tensor-type nucleon to deuteron polarization transfer 
coefficients $K^{\beta, \gamma}_\alpha$ from the spatial matrix elements
for the corresponding spin operator $\sigma_\alpha \CP_{\beta, \gamma}$.
The transformation from $\CP_{\beta, \gamma}$ to $S^{(2)}_\mu$ is
given by Eqs.\,(\ref{sec4-5-21}) and (\ref{sec4-5-15}).
Again, from parity conservation, some of the coefficients are zero.
In particular, we have a relationship
$K^{x,x}_\alpha+K^{y,y}_\alpha+K^{z,z}_\alpha=0$ for 
all $\alpha=x,~y,~z$, since $\CP_{x,x}+\CP_{y,y}+\CP_{z,z}=0$.
Because of this symmetry, only seven coefficients are independent.
These are 
\begin{eqnarray}
& & K^{x,y}_x \sim i\,\frac{3}{\sqrt{2}}\,(C_{12}-C_{-1,2})\ ,\nonumber \\
& & K^{y,z}_x \sim -i \frac{3}{\sqrt{2}}\,(C_{11}+C_{1,-1})\ ,\nonumber \\
& & K^{x,x}_y \sim i\,\left(-\sqrt{3} C_{10}
+\frac{3}{\sqrt{2}}C_{12}+\frac{3}{\sqrt{2}}C_{-1,2}\right)\ ,\nonumber\\
& & K^{y,y}_y \sim i\,\left(-\sqrt{3} C_{10}
-\frac{3}{\sqrt{2}}C_{12}-\frac{3}{\sqrt{2}}C_{-1,2}\right)\ ,\nonumber\\
& & K^{z,z}_y = -\left(K^{x,x}_y+K^{y,y}_y \right)
\sim i\,2\sqrt{3}~C_{10}\ ,\nonumber\\
& & K^{x,z}_y \sim -i\,\frac{3}{\sqrt{2}}\,(C_{11}-C_{1,-1})\ ,\nonumber \\
& & K^{x,y}_z \sim i\,(-3)\,C_{02}\ ,\nonumber \\
& & K^{y,z}_z \sim i\,3\,C_{01}\ .
\label{sec4-5-70}
\end{eqnarray}
The correspondence also yields
\begin{eqnarray}
K^{x,x}_z=K^{y,y}_z=-\sqrt{\frac{3}{2}}\,C_{00}\ \ ,\qquad
K^{z,z}_z=(-2)\,K^{x,x}_z=\sqrt{6}\,C_{00}\ .
\label{sec4-5-71}
\end{eqnarray}
However, these are all zero since $C_{00}=(-)^{1+0+0}C_{00}=0$
from the parity conservation.
The independent seven spatial functions are the linear combinations of
$C_{1,-1}$ for $f_z=0$, $C_{10},~C_{01},~C_{-1,2}$ for $f_z=1$,
$C_{02},~C_{11}$ for $f_z=2$, and $C_{12}$ for $f_z=3$.
After all, we calculate seven independent tensor-type nucleon to deuteron
polarization transfer coefficients by the spatial integrals
\begin{eqnarray}
& & K^{x,y}_x \sim i\,\frac{3}{\sqrt{2}} \left[
C^{L,L^\prime}_\kappa(11,22)-C^{L,L^\prime}_\kappa(1,-1, 22)\right]
\ ,\nonumber \\
& & K^{y,z}_x \sim -i\,\frac{3}{\sqrt{2}}
\left[ C^{L,L^\prime}_\kappa(11,21)+C^{L,L^\prime}_\kappa(1,1,\,2,-1)\right]
\ ,\nonumber \\ [2mm]
& & K^{x,x}_y \sim i\,\left[ -\sqrt{3} C^{L,L^\prime}_\kappa(11,20)
+\frac{3}{\sqrt{2}} C^{L,L^\prime}_\kappa(11,22)
+\frac{3}{\sqrt{2}} C^{L,L^\prime}_\kappa(1,-1,22) \right]
\ ,\nonumber\\
& & K^{y,y}_y \sim i\,\left[ -\sqrt{3} C^{L,L^\prime}_\kappa(11,20)
-\frac{3}{\sqrt{2}} C^{L,L^\prime}_\kappa(11,22)
-\frac{3}{\sqrt{2}} C^{L,L^\prime}_\kappa(1,-1,22) \right]
\ ,\nonumber\\
& & K^{z,z}_y = -\left(K^{x,x}_y+K^{y,y}_y\right)
\sim i\,2\sqrt{3} C^{L,L^\prime}_\kappa(11,20)\ ,\nonumber\\ [2mm]
& & K^{x,z}_y \sim -i\,\frac{3}{\sqrt{2}} 
\left[ C^{L,L^\prime}_\kappa(11,21)-C^{L,L^\prime}_\kappa(1,1,\,2,-1)\right]
\ ,\nonumber \\ [2mm]
& & K^{x,y}_z \sim i\,(-3)\,C^{L,L^\prime}_\kappa(10,22)\ ,\nonumber \\ [2mm]
& & K^{y,z}_z \sim i\,3\,C^{L,L^\prime}_\kappa(10,21)\ .
\label{sec4-5-72}
\end{eqnarray}
The reduced matrix element for the second rank-two $Z$-factor is
given by \eq{sec4-5-57}.

The transformation to the laboratory system can be carried out 
in a way similar to ${K_\alpha}^{\beta^\prime}$.
We first use $\CP_{\alpha \alpha}=3 {S_\alpha}^2-2$ and
\eq{sec4-5-17}, and derive spin rotation rule according to 
\eq{sec4-5-68-7} (we here use the simplified notation $\theta$
for $\theta_2$ in \eq{sec4-5-68-7}.):
\begin{eqnarray}
& & \CP^\prime_{xx} = \CP_{xx} (\cos\,\theta)^2
+\CP_{zz} (\sin\,\theta)^2-\CP_{xz} (\sin\,2\theta)\ ,\nonumber \\
& & \CP^\prime_{yy} = \CP_{yy} ,\nonumber \\
& & \CP^\prime_{zz} = \CP_{xx} (\sin\,\theta)^2
+\CP_{zz} (\cos\,\theta)^2+\CP_{xz} (\sin\,2\theta)\ ,\nonumber \\
& & \CP^\prime_{xy} = \CP_{xy} \cos\,\theta
-\CP_{yz} \sin\,\theta\ ,\nonumber \\
& & \CP^\prime_{yz} = \CP_{xy} \sin\,\theta
+\CP_{yz} \cos\,\theta\ ,\nonumber \\
& & \CP^\prime_{xz} = \left(\CP_{xx}-\CP_{zz}\right) 
\frac{1}{2} (\sin\,2\theta)
+\CP_{xz} (\cos\,2\theta)\ .
\label{sec4-5-72-1}
\end{eqnarray}
The $\pi$-rotation around the $z$-axis yields
\begin{eqnarray}
& & {K_x}^{x^\prime, y^\prime} = K^{x,y}_x(\theta,\pi) \cos\,\theta_2
-K^{y,z}_x(\theta,\pi) \sin\,\theta_2\ ,\nonumber \\
& & {K_x}^{y^\prime, z^\prime} = K^{x,y}_x(\theta,\pi) \sin\,\theta_2
+K^{y,z}_x(\theta,\pi) \cos\,\theta_2\ ,\nonumber \\
& & {K_z}^{x^\prime, y^\prime} = K^{x,y}_z(\theta,\pi) \cos\,\theta_2
-K^{y,z}_z(\theta,\pi) \sin\,\theta_2\ ,\nonumber \\
& & {K_z}^{y^\prime, z^\prime} = K^{x,y}_z(\theta,\pi) \sin\,\theta_2
+K^{y,z}_z(\theta,\pi) \cos\,\theta_2\ ,\nonumber \\
& & {K_y}^{x^\prime, z^\prime} = \left[ K^{x,x}_y(\theta,\pi)
-K^{z,z}_y(\theta,\pi)\right]\frac{1}{2}(\sin\,2\theta_2)
+K^{x,z}_y(\theta,\pi) (\cos\,2\theta_2)\ , \nonumber \\
& & {K_y}^{x^\prime, x^\prime} = K^{x,x}_y(\theta,\pi) (\cos\,\theta_2)^2
+K^{z,z}_y(\theta,\pi) (\sin\,\theta_2)^2
-K^{x,z}_y(\theta,\pi) (\sin\,2\theta_2)\ ,\nonumber \\
& & {K_y}^{y^\prime, y^\prime} = K^{y,y}_y(\theta,\pi)\ ,\nonumber \\
& & {K_y}^{z^\prime, z^\prime} = K^{x,x}_y(\theta,\pi) (\sin\,\theta_2)^2
+K^{z,z}_y(\theta,\pi) (\cos\,\theta_2)^2
+K^{x,z}_y(\theta,\pi) (\sin\,2\theta_2)\ .\nonumber \\
\label{sec4-5-72-2}
\end{eqnarray}
If we further use the odd-even character 
of $K^{\beta, \gamma}_\alpha(\theta,0)$, we find that $K^{x,y}_x$,
$K^{x,x}_y$, $K^{y,y}_y$, $K^{z,z}_y$, and $K^{y,z}_z$
(namely, $N_x+N_y=\hbox{odd}$) change the sign from \eq{sec4-5-70}.
We therefore obtain the following final result.
\begin{eqnarray}
& & {K_x}^{x^\prime, y^\prime} = -K^{x,y}_x \cos\,\theta_2
-K^{y,z}_x \sin\,\theta_2\ ,\nonumber \\
& & {K_x}^{y^\prime, z^\prime} = -K^{x,y}_x \sin\,\theta_2
+K^{y,z}_x \cos\,\theta_2\ ,\nonumber \\
& & {K_z}^{x^\prime, y^\prime} = K^{x,y}_z \cos\,\theta_2
+K^{y,z}_z \sin\,\theta_2\ ,\nonumber \\
& & {K_z}^{y^\prime, z^\prime} = K^{x,y}_z \sin\,\theta_2
-K^{y,z}_z \cos\,\theta_2\ ,\nonumber \\
& & {K_y}^{x^\prime, z^\prime} = \left[-K^{x,x}_y+K^{z,z}_y\right]
\frac{1}{2}(\sin\,2\theta_2)
+K^{x,z}_y (\cos\,2\theta_2)\ , \nonumber \\
& & {K_y}^{x^\prime, x^\prime} = -K^{x,x}_y (\cos\,\theta_2)^2
-K^{z,z}_y (\sin\,\theta_2)^2
-K^{x,z}_y (\sin\,2\theta_2)\ ,\nonumber \\
& & {K_y}^{y^\prime, y^\prime} = -K^{y,y}_y\ ,\nonumber \\
& & {K_y}^{z^\prime, z^\prime} = -K^{x,x}_y (\sin\,\theta_2)^2
-K^{z,z}_y (\cos\,\theta_2)^2
+K^{x,z}_y (\sin\,2\theta_2)\ ,
\label{sec4-5-72-3}
\end{eqnarray}
where $K^{\beta, \gamma}_\alpha=K^{\beta, \gamma}_\alpha(\theta, 0)$ 
and $\theta_2$ is calculated from Eqs.\,(\ref{sec4-5-68-6}) 
and (\ref{sec4-5-68-2}). Note that $K^{y,y}_y$ changes the sign.

\bigskip

\subsection{Spin correlation coefficients}

In this case, $\CS_i=[\sigma S]^{(\lambda)}_\mu$ or
$\CS_i=[\sigma S^{(2)}]^{(\lambda)}_\mu$, together with
$\CS_f=1$. The exchanged case, $\CS_i \leftrightarrow \CS_f$ is
considered in a similar way (except for the phase correction
of the spin reduced matrix elements related to the symmetry
in \eq{sec4-5-40}).
We define the matrix element $I(\lambda \mu)$, which is a special case
of \eq{sec4-5-42} with $\lambda_i \mu_i=\lambda \mu$ and
$\lambda_f \mu_f=0 0$:
\begin{eqnarray}
& & I(\lambda \mu)=I(\lambda \mu;00)
=Tr\,\{f \CS^{(\lambda)}_{\mu} f^\dagger\}=(-)^{\lambda +\mu}
I(\lambda, -\mu) \nonumber \\
& & =\sum_{(\ell^\prime S^\prime_c)(\ell S_c)J}
\sum_{(\widetilde{\ell}^\prime \widetilde{S}^\prime_c) 
(\widetilde{\ell} \widetilde{S}_c)\widetilde{J}}
(-)^{S^\prime_c-S_c}~f^J_{(\ell^\prime S^\prime_c)(\ell S_c)}
\left\{f^{\widetilde{J}}_{(\widetilde{\ell}^\prime \widetilde{S}^\prime_c)
(\widetilde{\ell} \widetilde{S}_c)}\right\}^* \nonumber \\
& & \times \sum_{L, L^\prime}
C^{L, L^\prime}_{L^\prime}(\lambda \mu, 0 0)
~Z^{(\lambda)}_{L^\prime}(\ell J \widetilde{\ell} \widetilde{J};
S_c \widetilde{S}_c L)
~\delta_{S^\prime_c, \widetilde{S}^\prime_c}
Z(\ell^\prime J \widetilde{\ell}^\prime
\widetilde{J}; S^\prime_c L^\prime)\ .
\label{sec4-5-73}
\end{eqnarray}
Here, the spatial functions $C^{L, L^\prime}_{L^\prime}(\lambda \mu, 0 0)$
are given in \eq{sec4-5-49} for $\lambda \mu=00$, in \eq{sec4-5-52}
for $\lambda=1$, and in \eq{sec4-5-59} for $\lambda=2$. The reduced matrix
element for the first $Z^{(\lambda)}_{L^\prime}$ function is
given by \eq{sec4-5-51}.

Let us first consider $\CS^{(\lambda)}_\mu=[\sigma S]^{(\lambda)}_\mu$ case.
The reduced matrix element from \eq{sec4-5-51} is
\begin{eqnarray}
\langle \chi_{S_c}||\left[\sigma S\right]^{(\lambda)}||
\chi_{\widetilde{S}_c} \rangle_{\rm unc}
=(-1)^{\lambda} \sqrt{6}
\left[ \begin{array}{ccc}
1 & \H & \widetilde{S}_c \\
1 & 1 & \lambda \\
1 & \H & S_c \\
\end{array} \right]\ .
\label{sec4-5-74}
\end{eqnarray}
From the parity conservation, we find only 5 coefficients are 
independent; namely, we find the correspondence
\begin{eqnarray}
& & [\sigma S]^{(0)}_0 \sim I(00)\ ,\nonumber \\
& & [\sigma S]^{(1)}_1 + [\sigma S]^{(1)}_{-1}
\sim 2 I(11)\ \ ,\qquad [\sigma S]^{(1)}_1 - [\sigma S]^{(1)}_{-1} 
\sim 0\ \ ,\qquad [\sigma S]^{(1)}_0 \sim 0\ ,\nonumber \\
& & [\sigma S]^{(2)}_2 + [\sigma S]^{(2)}_{-2}
\sim 2 I(22)\ \ ,\qquad [\sigma S]^{(2)}_2 - [\sigma S]^{(2)}_{-2}
\sim 0\ ,\nonumber \\
& & [\sigma S]^{(2)}_1 + [\sigma S]^{(2)}_{-1} \sim 0\ \ ,\qquad
[\sigma S]^{(2)}_1 - [\sigma S]^{(2)}_{-1} \sim 2 I(21)\ ,\nonumber \\
& & [\sigma S]^{(2)}_0 \sim I(20)\ .
\label{sec4-5-75}
\end{eqnarray}
The transformation to the Cartesian representation is carried out 
by using the formula
\begin{eqnarray}
[\sigma S]^{(0)}_0=-\frac{1}{\sqrt{3}}(\bfsigma \cdot \bS)\ \ ,\qquad
[\sigma S]^{(1)}_\mu=i\,\frac{1}{\sqrt{2}} [\bfsigma \times \bS]_\mu\ ,
\label{sec4-5-76}
\end{eqnarray}
and the formula in \eq{sec4-5-14}. We find the correspondence
\begin{eqnarray}
& & [\sigma S]^{(0)}_0 = -\frac{1}{\sqrt{3}}(\sigma_x S_x
+\sigma_y S_y +\sigma_z S_z)\ ,\nonumber \\
& & [\sigma S]^{(1)}_1 + [\sigma S]^{(1)}_{-1}
=[\bfsigma \times \bS]_y=\sigma_z S_x - \sigma_x S_z\ ,\nonumber \\
& & [\sigma S]^{(1)}_1 - [\sigma S]^{(1)}_{-1}
=(-i) [\bfsigma \times \bS]_x
=(-i)(\sigma_y S_z - \sigma_z S_y)\ ,\nonumber \\
& & [\sigma S]^{(1)}_0
=i\,\frac{1}{\sqrt{2}} [\bfsigma \times \bS]_z
=i\,\frac{1}{\sqrt{2}}(\sigma_x S_y - \sigma_y S_x)\ ,\nonumber \\
& & [\sigma S]^{(2)}_2 + [\sigma S]^{(2)}_{-2}
=\sigma_x S_x - \sigma_y S_y\ \ ,\qquad
[\sigma S]^{(2)}_2 - [\sigma S]^{(2)}_{-2}
=i\,(\sigma_x S_y + \sigma_y S_x)\ , \nonumber \\
& & [\sigma S]^{(2)}_1 + [\sigma S]^{(2)}_{-1}
=-i\,(\sigma_y S_z + \sigma_z S_y)\ \ ,\qquad
[\sigma S]^{(2)}_1 - [\sigma S]^{(2)}_{-1}
=- (\sigma_x S_z + \sigma_z S_x)\ , \nonumber \\
& & [\sigma S]^{(2)}_0
=\frac{1}{\sqrt{6}}(2\sigma_z S_z - \sigma_x S_x - \sigma_y S_y)\ .
\label{sec4-5-77}
\end{eqnarray}
If we combine \eq{sec4-5-75} and \eq{sec4-5-77} and express $C_{\alpha,\beta}$
by $I(\lambda \mu)$, we obtain the following results.
\begin{eqnarray}
& & C_{x,x} \sim -\frac{1}{\sqrt{3}} I(00)-\frac{1}{\sqrt{6}} I(20)
+I(22)\ ,\nonumber \\
& & C_{y,y} \sim -\frac{1}{\sqrt{3}} I(00)-\frac{1}{\sqrt{6}} I(20)
-I(22)\ ,\nonumber \\
& & C_{z,z} \sim -\frac{1}{\sqrt{3}} I(00)+\sqrt{\frac{2}{3}} I(20)
\ ,\nonumber \\ [2mm]
& & C_{z,x} \sim I(11)-I(21)\ \ ,\qquad
C_{x,z} \sim -I(11)-I(21)\ ,\nonumber \\ [2mm]
& & C_{y,z}=C_{z,y}=C_{x,y}=C_{y,x}=0\ .
\label{sec4-5-78}
\end{eqnarray}

Let us move to the $\CS^{(\lambda)}_\mu=[\sigma S^{(2)}]^{(\lambda)}_\mu$ type
spin correlation coefficients. The reduced matrix element in this case
is
\begin{eqnarray}
\langle \chi_{S_c}||\left[\sigma S^{(2)}\right]^{(\lambda)}||
\chi_{\widetilde{S}_c} \rangle_{\rm unc}
=(-1)^{1+\lambda} \sqrt{5}
\left[ \begin{array}{ccc}
1 & \H & \widetilde{S}_c \\
2 & 1 & \lambda \\
1 & \H & S_c \\
\end{array} \right]\ .
\label{sec4-5-79}
\end{eqnarray}
We again use the same notation $I(\lambda \mu)$ of \eq{sec4-5-73}
with this reduced matrix element. Another convenient representation is
decoupled representation
\begin{eqnarray}
& & I_{\sigma \mu}=Tr\,\left\{f \sigma_\sigma S^{(2)}_\mu f^\dagger \right\}
=(-)^{1+\sigma +\mu} I_{-\sigma, -\mu} \nonumber \\
& & =\sum_{\lambda m} \langle 1 \sigma 2 \mu|\lambda m \rangle
~Tr\,\left\{f [\sigma S^{(2)}]^{(\lambda)}_m f^\dagger \right\}
=\sum_{\lambda m} \langle 1 \sigma 2 \mu|\lambda m \rangle 
~I(\lambda m)\ .
\label{sec4-5-80}
\end{eqnarray}
The symmetry, $I_{-\sigma, -\mu}=(-)^{1+\sigma+\mu}
I_{\sigma \mu}$, is the same as that of \eq{sec4-5-69}
for the nucleon to deuteron polarization transfer. This allows us
to use the relationship in \eq{sec4-5-70} by just replacing
$K^{\beta \gamma}_\alpha$ and $C_{\sigma \mu}$ by 
$C^{\beta \gamma}_\alpha$ and $I_{\sigma \mu}$, respectively:
\begin{eqnarray}
& & C^{x,y}_x \sim i\,\frac{3}{\sqrt{2}}\,(I_{12}-I_{-1,2})\ ,\nonumber \\
& & C^{y,z}_x \sim - i\,\frac{3}{\sqrt{2}}\,(I_{11}+I_{1,-1})\ ,\nonumber \\
& & C^{x,x}_y \sim i\,\left(-\sqrt{3}\,I_{10}
+\frac{3}{\sqrt{2}}I_{12}+\frac{3}{\sqrt{2}}I_{-1,2}\right)\ ,\nonumber\\
& & C^{y,y}_y \sim i\,\left(-\sqrt{3}\,I_{10}
-\frac{3}{\sqrt{2}}I_{12}-\frac{3}{\sqrt{2}}I_{-1,2}\right)\ ,\nonumber\\
& & C^{z,z}_y = -\left(C^{x,x}_y+C^{y,y}_y \right)
\sim i\,2\sqrt{3}~I_{10}\ ,\nonumber\\
& & C^{x,z}_y \sim - i\,\frac{3}{\sqrt{2}}\,(I_{11}-I_{1,-1})\ ,\nonumber \\
& & C^{x,y}_z \sim i\,(-3)\,I_{02}\ ,\nonumber \\
& & C^{y,z}_z \sim i\,3\,I_{01}\ .
\label{sec4-5-81}
\end{eqnarray}
We need seven independent components, 
$I_{01},~I_{02},~I_{10},~I_{11},~I_{1, -1},~I_{12},~I_{-1,2}$, 
which are related to $I(20),~I(11),~I(21),~I(31),
~I(22),~I(32),~I(33)$ through the CG coefficients.
Note that $I(10)=I(30)=0$, since $I(\lambda 0)=(-)^{\lambda+0} I(\lambda 0)$.
Equation (\ref{sec4-5-80}) is explicitly written as
\begin{eqnarray}
I_{12} & = & I(33)\ ,\nonumber \\
I_{-1,2} & = & \sqrt{\frac{3}{5}}I(11)-\frac{1}{\sqrt{3}}I(21)
+\frac{1}{\sqrt{15}}I(31)\ ,\nonumber \\
I_{11} & = & \frac{1}{\sqrt{3}} I(22)+ \sqrt{\frac{2}{3}}I(32)\ ,\nonumber \\
I_{1,-1} & = & \frac{1}{\sqrt{2}}I(20)\ ,\nonumber \\
I_{1 0} & = & \frac{1}{\sqrt{10}}I(11)+\frac{1}{\sqrt{2}}I(21)
+\sqrt{\frac{2}{5}} I(31)\ ,\nonumber \\
I_{02} & = & -\sqrt{\frac{2}{3}} I(22) + \frac{1}{\sqrt{3}} I(32)
\ ,\nonumber \\
I_{0 1} & = & -\sqrt{\frac{3}{10}} I(11)-\frac{1}{\sqrt{6}}I(21)
+2\sqrt{\frac{2}{15}} I(31)\ .
\label{sec4-5-83}
\end{eqnarray}
If we use these in \eq{sec4-5-81}, we eventually obtain
\begin{eqnarray}
& & C^{x,y}_x \sim i\,\left[ \frac{3}{\sqrt{2}} I(33)
-3\sqrt{\frac{3}{10}}I(11)+\sqrt{\frac{3}{2}}I(21)-\sqrt{\frac{3}{10}}
I(31) \right]\ ,\nonumber \\
& & C^{y,z}_x \sim i\,\left[ -\sqrt{\frac{3}{2}}I(22)-\sqrt{3} I(32)
-\frac{3}{2} I(20) \right]\, \nonumber \\
& & C^{x,x}_y \sim i\,\left[\frac{3}{\sqrt{2}} I(33)
+\sqrt{\frac{6}{5}}I(11)-\sqrt{6}I(21)-\sqrt{\frac{3}{10}}I(31) \right]
\ ,\nonumber \\
& & C^{y,y}_y \sim i\,\left[-\frac{3}{\sqrt{2}} I(33)
-2\sqrt{\frac{6}{5}}I(11)-3\sqrt{\frac{3}{10}}I(31) \right]
\ ,\nonumber \\
& & C^{z,z}_y = -\left(C^{x,x}_y+C^{y,y}_y \right)
\sim i\,\left[ \sqrt{\frac{6}{5}}I(11)+\sqrt{6} I(21)
+2\sqrt{\frac{6}{5}}I(31)\right]\ ,\nonumber\\
& & C^{x,z}_y \sim i\, \left[ -\sqrt{\frac{3}{2}} I(22)-\sqrt{3} I(32)
+\frac{3}{2} I(20) \right] \ ,\nonumber \\
& & C^{x,y}_z \sim i\,\left[\sqrt{6} I(22) - \sqrt{3} I(32) \right]
\ ,\nonumber \\
& & C^{y,z}_z \sim i\,\left[-3\sqrt{\frac{3}{10}} I(11)
-\sqrt{\frac{3}{2}} I(21)+2\sqrt{\frac{6}{5}} I(31) \right]\ .
\label{sec4-5-84}
\end{eqnarray}

\bigskip

\section{Results and discussion}


\subsection{Vector analyzing-power of the nucleon}

\begin{figure}[htb]
\begin{center}
\begin{minipage}{0.48\textwidth}
\includegraphics[angle=0,width=55mm]
{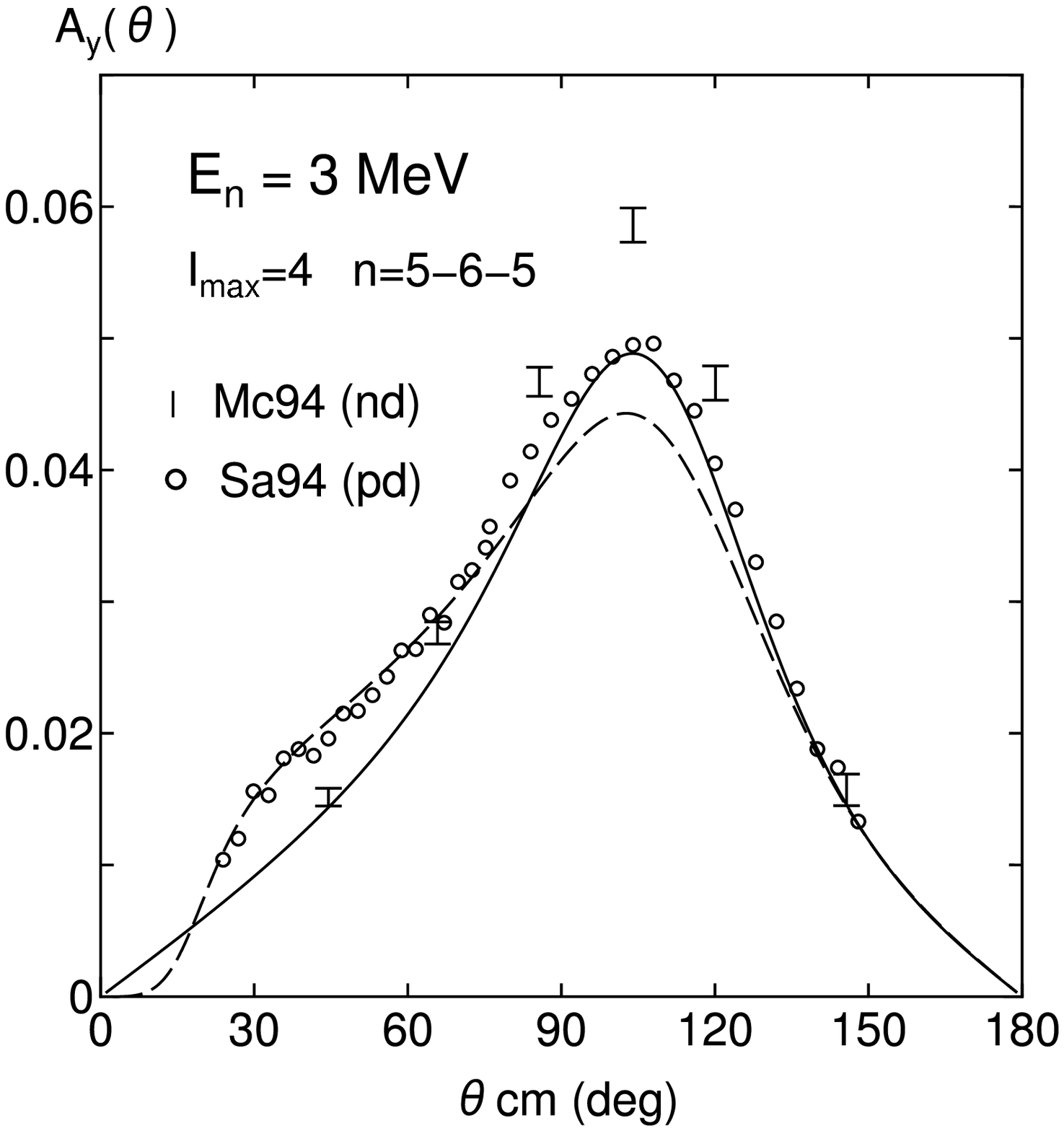}

\includegraphics[angle=0,width=55mm]
{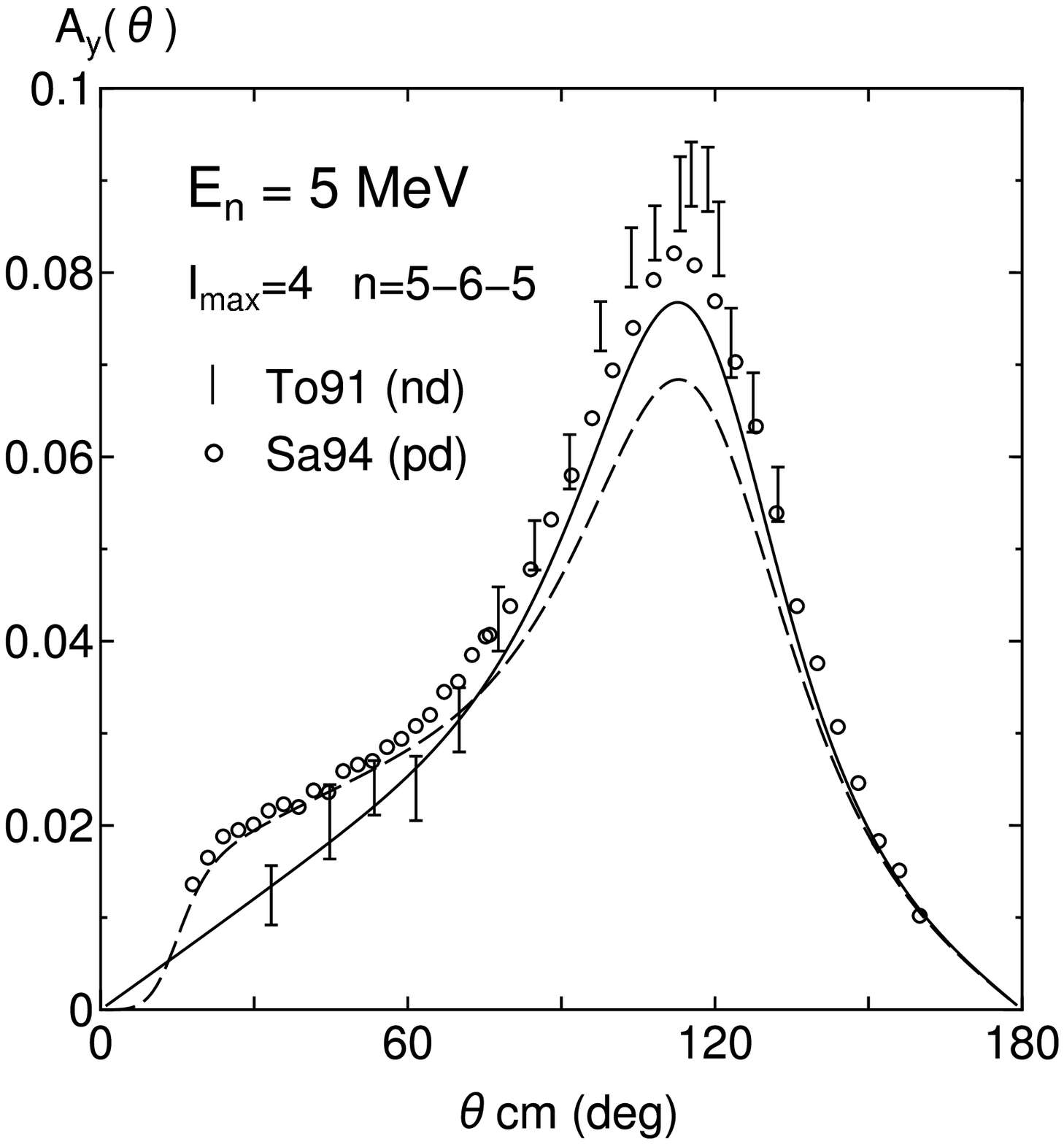}

\includegraphics[angle=0,width=55mm]
{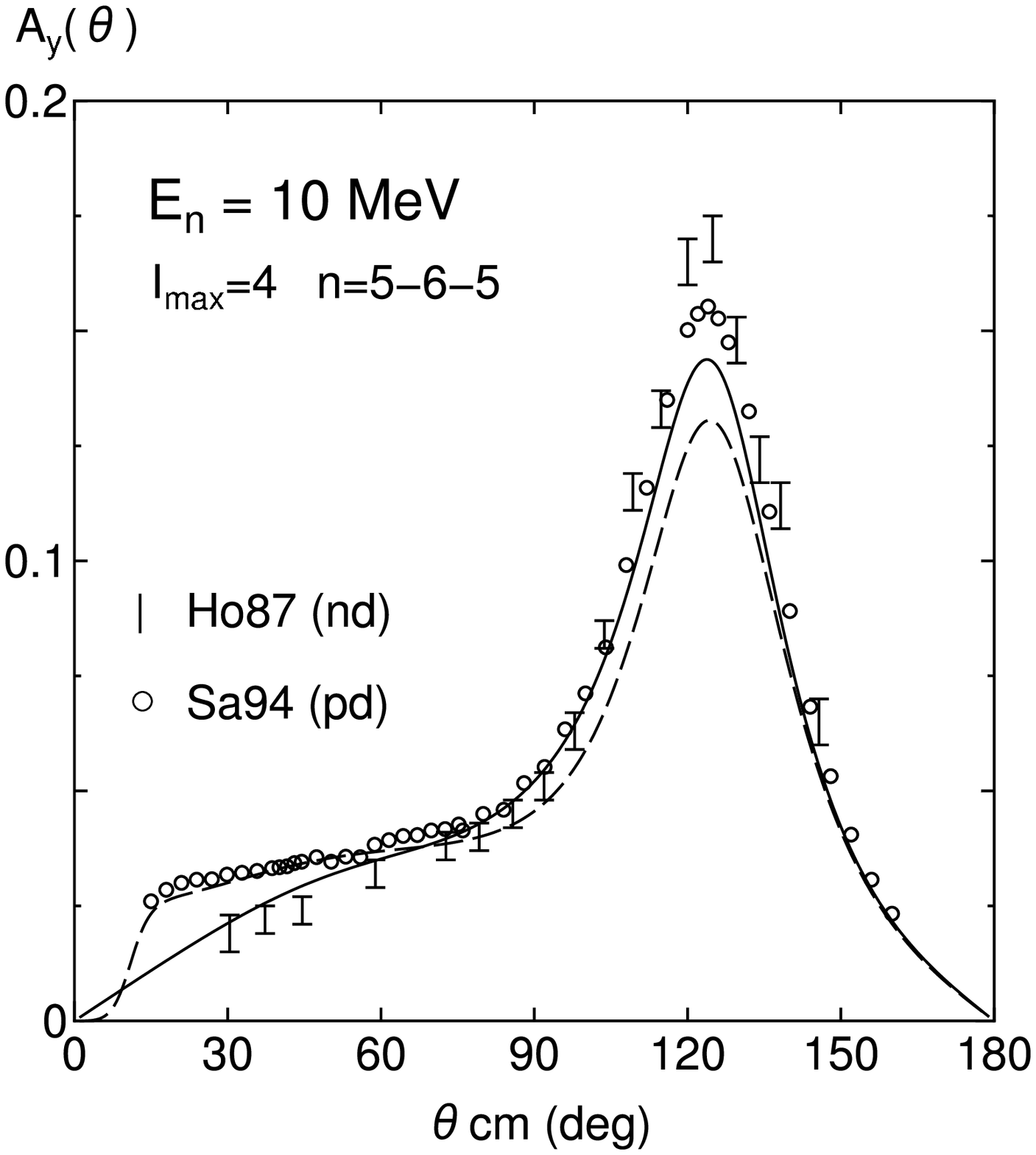}
\end{minipage}~%
\hfill~%
\begin{minipage}{0.48\textwidth}
\includegraphics[angle=0,width=55mm]
{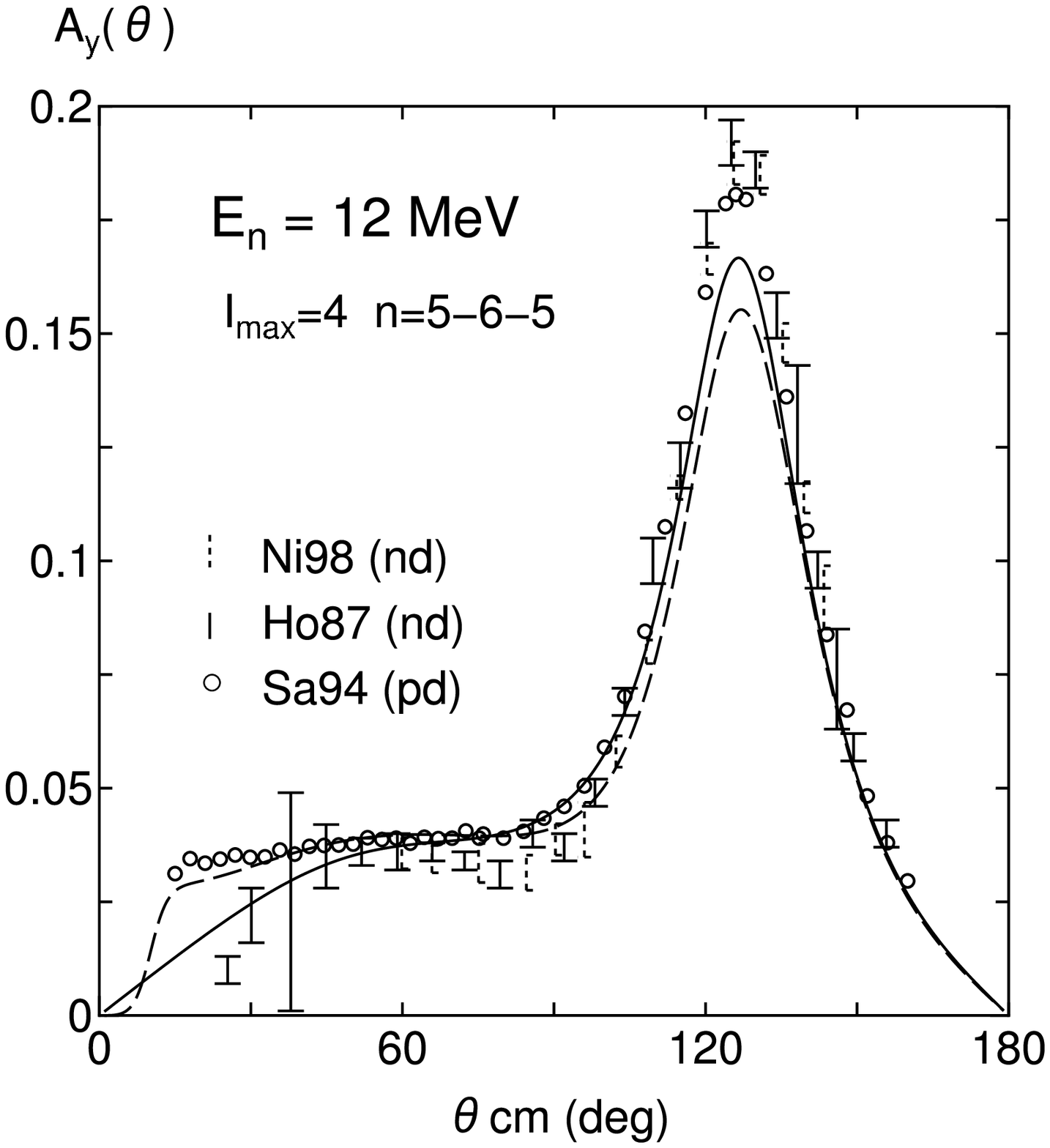}

\includegraphics[angle=0,width=55mm]
{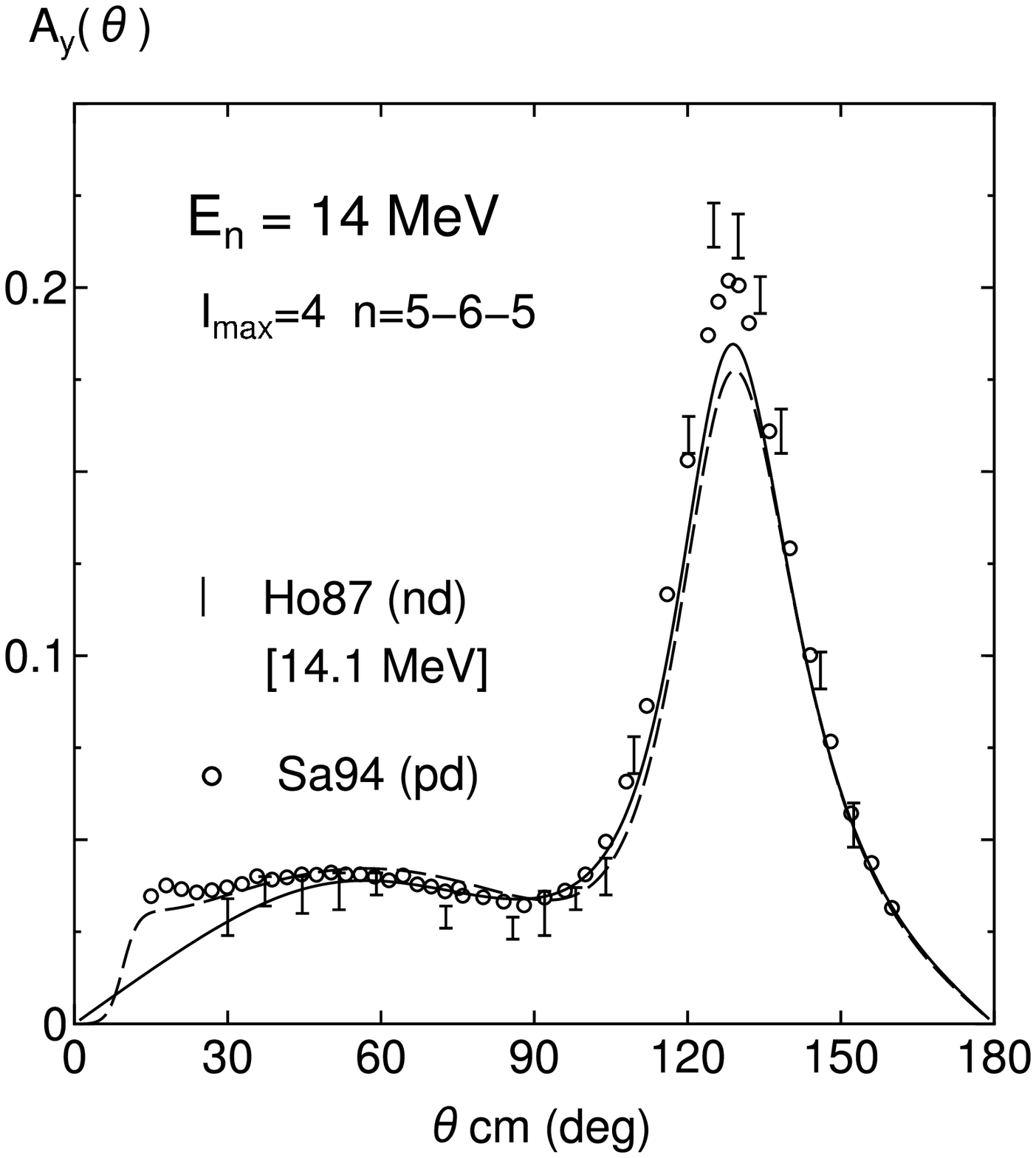}

\includegraphics[angle=0,width=54mm]
{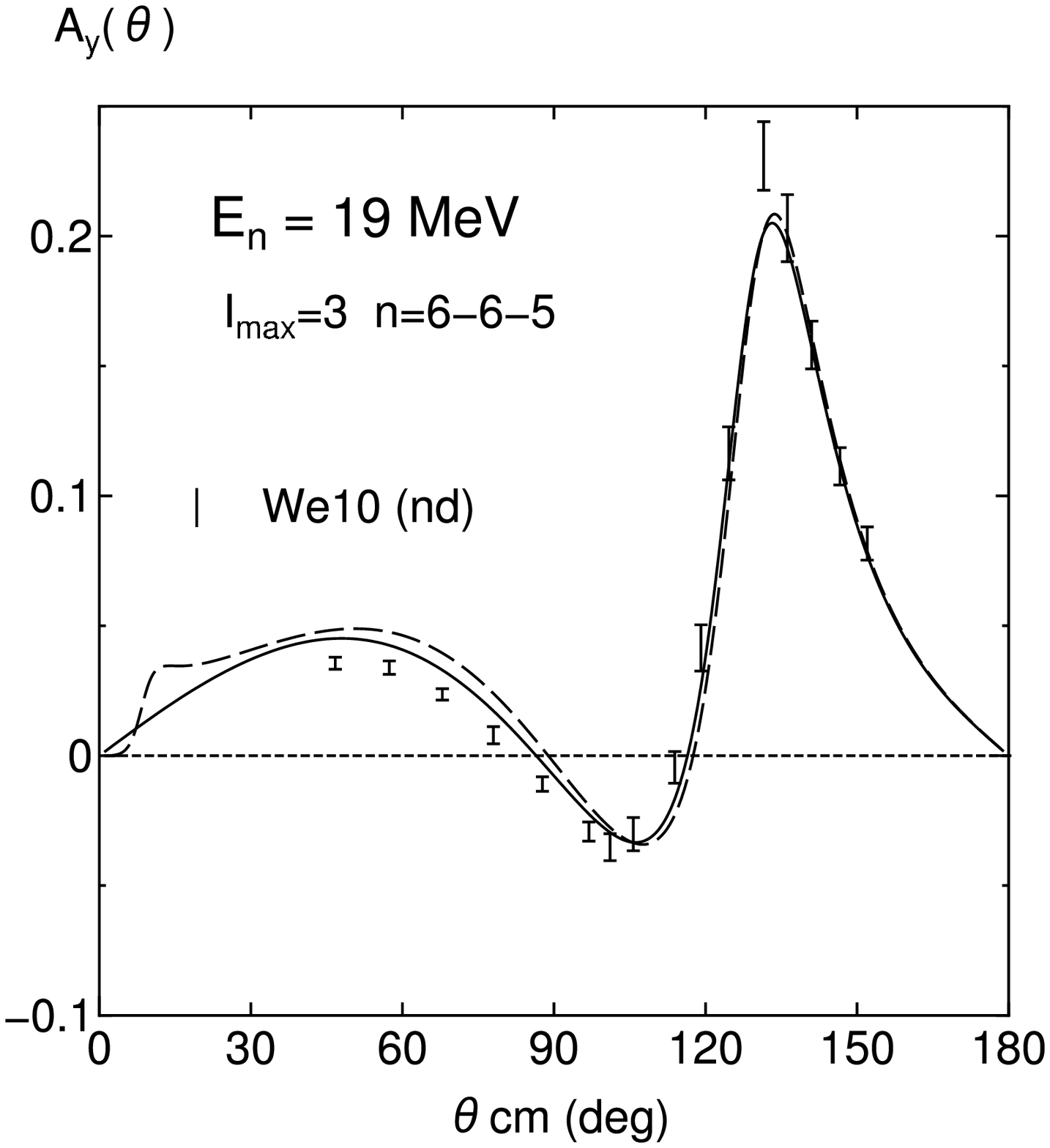}
\end{minipage}
\end{center}
\caption{
The nucleon analyzing-power $A_y(\theta)$ of the $nd$ elastic
scattering (solid curve) from $E_n=3$ to 19 MeV,
compared with the experiment (bars).
The $pd$ results with the cut-off Coulomb force are also shown
by dashed curves, which should be compared with empty circles.
The experimental data are taken from Refs.\,\citen{Mc94} for Mc94 ($nd$),
\citen{Sa94} for Sa94 ($pd$), \citen{To91} for To91 ($nd$),
\citen{Ho87} for Ho87 ($nd$), \citen{Ni98} for Ni98 ($nd$), 
and \citen{We10} for We10 ($nd$).
}
\label{Ay1}       
\end{figure}

\begin{figure}[htb]
\begin{center}
\begin{minipage}{0.48\textwidth}
\includegraphics[angle=0,width=56mm]
{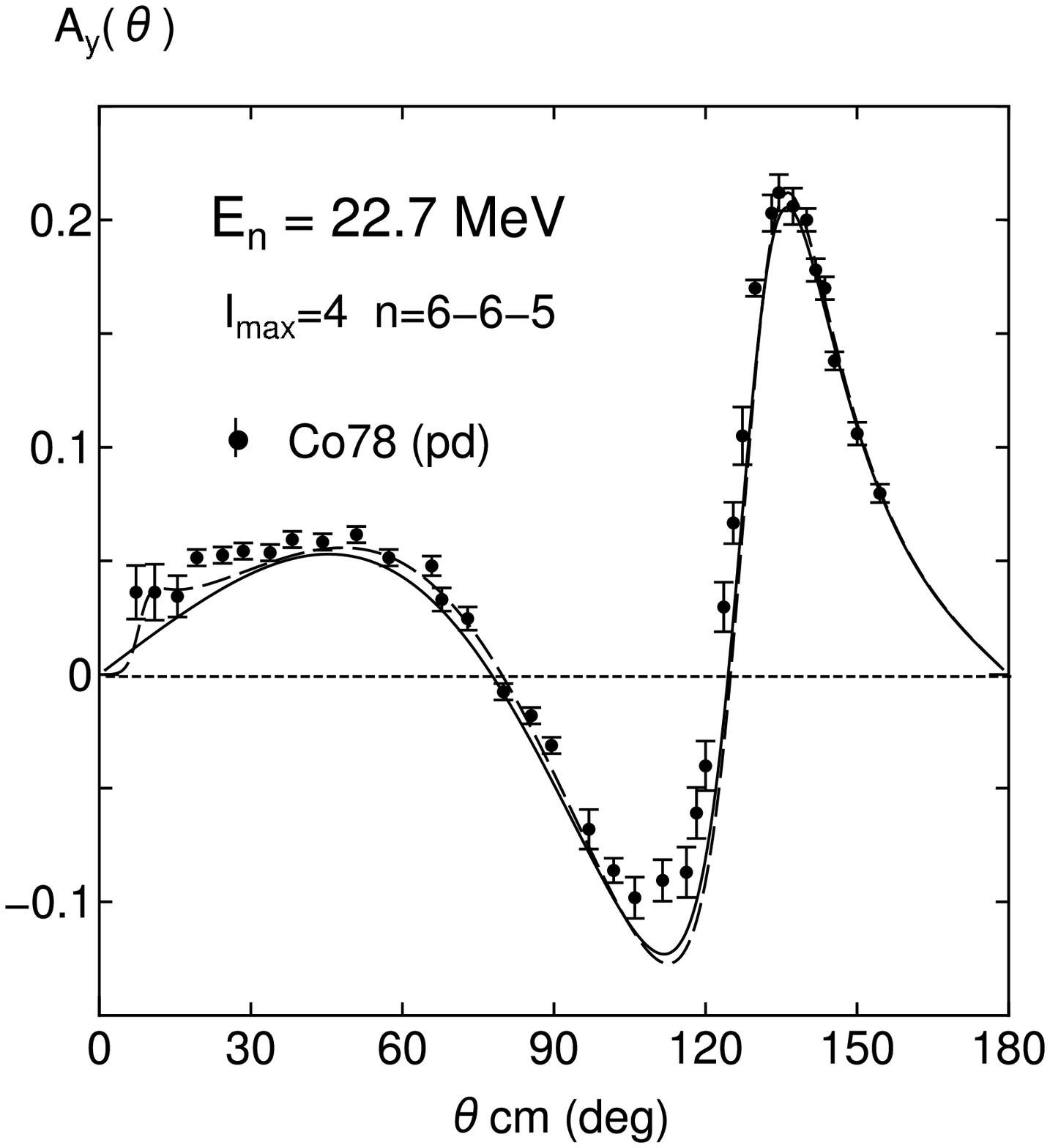}

\includegraphics[angle=0,width=56mm]
{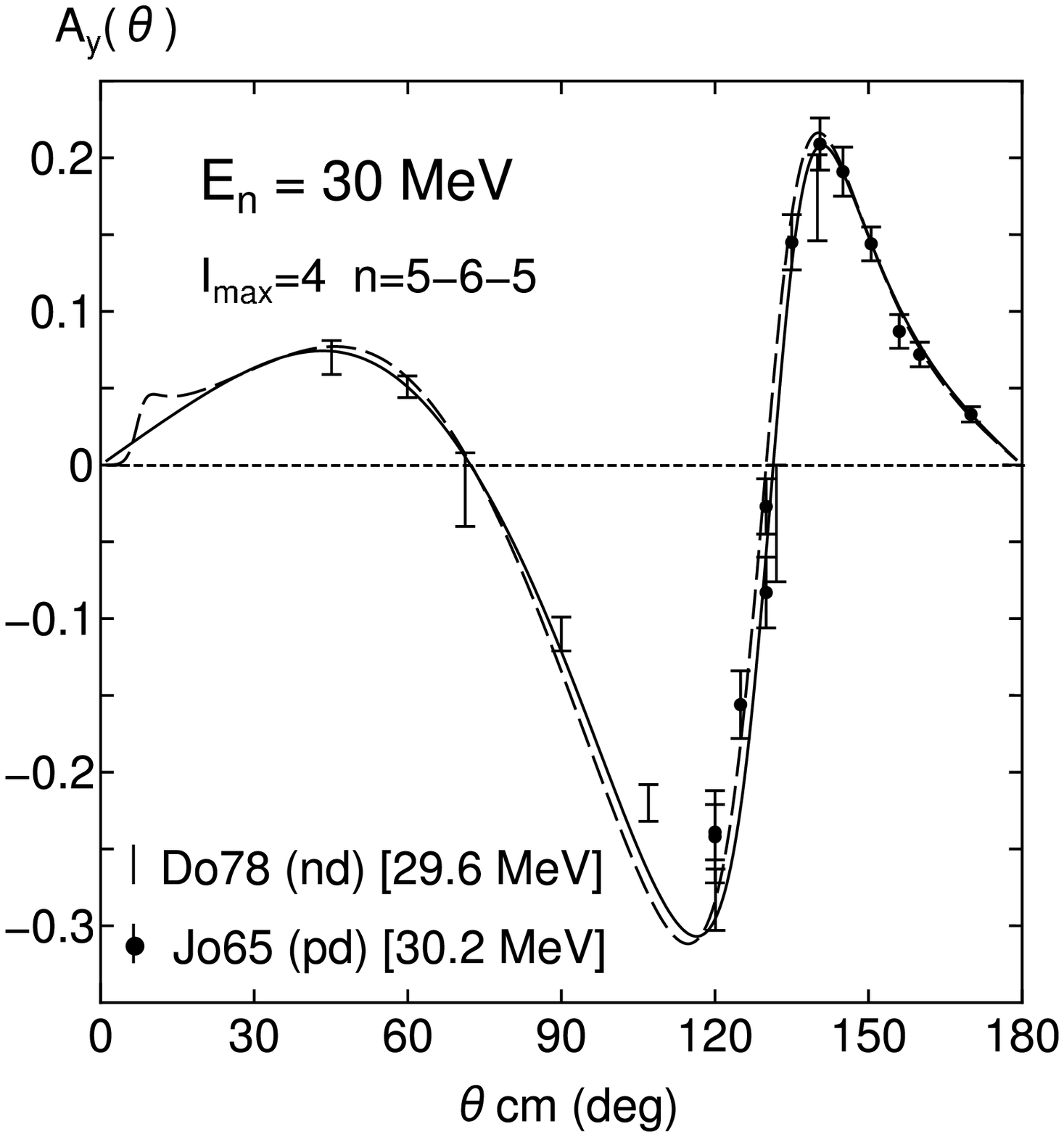}

\includegraphics[angle=0,width=56mm]
{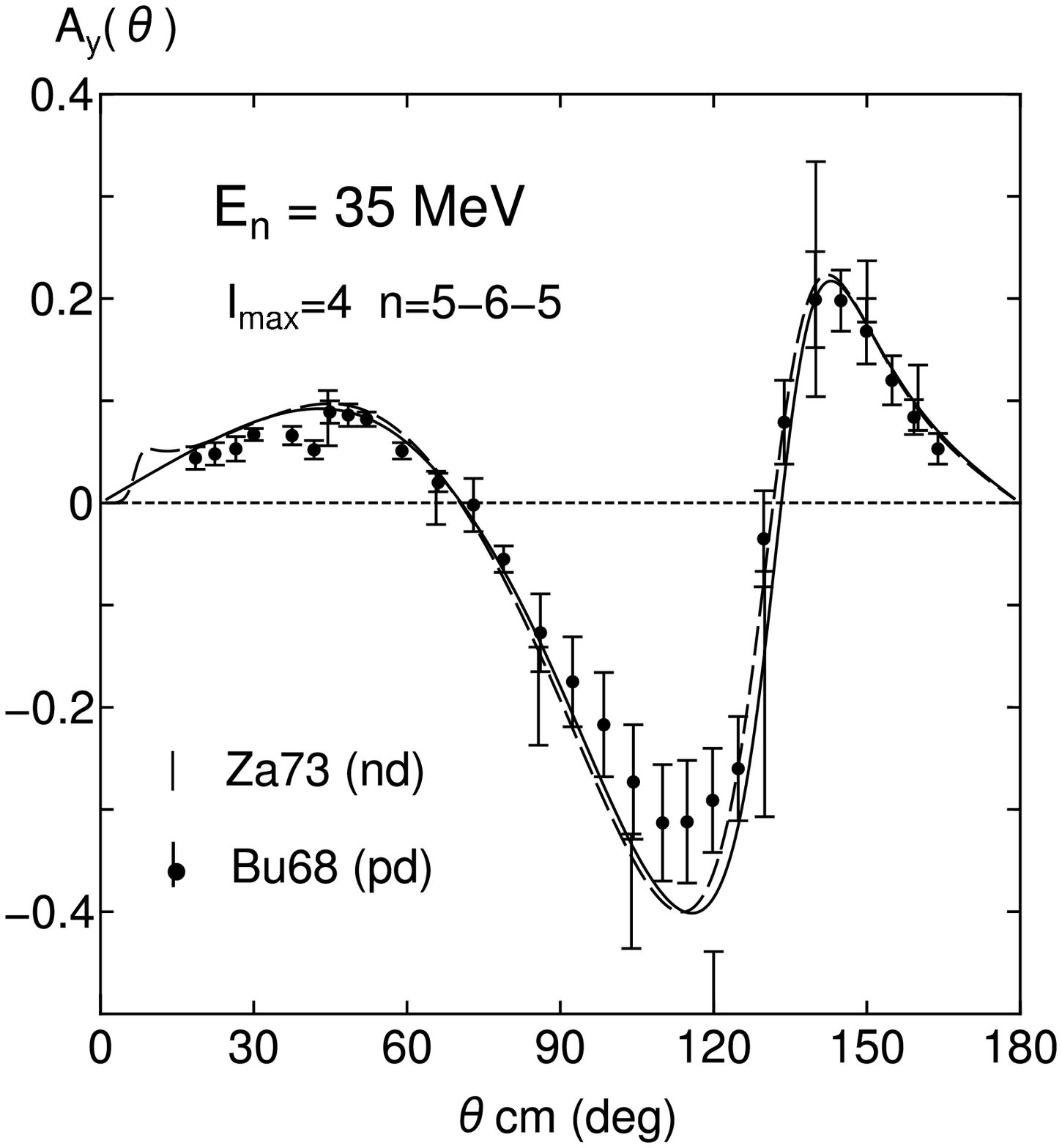}
\end{minipage}~%
\hfill~%
\begin{minipage}{0.48\textwidth}
\includegraphics[angle=0,width=56mm]
{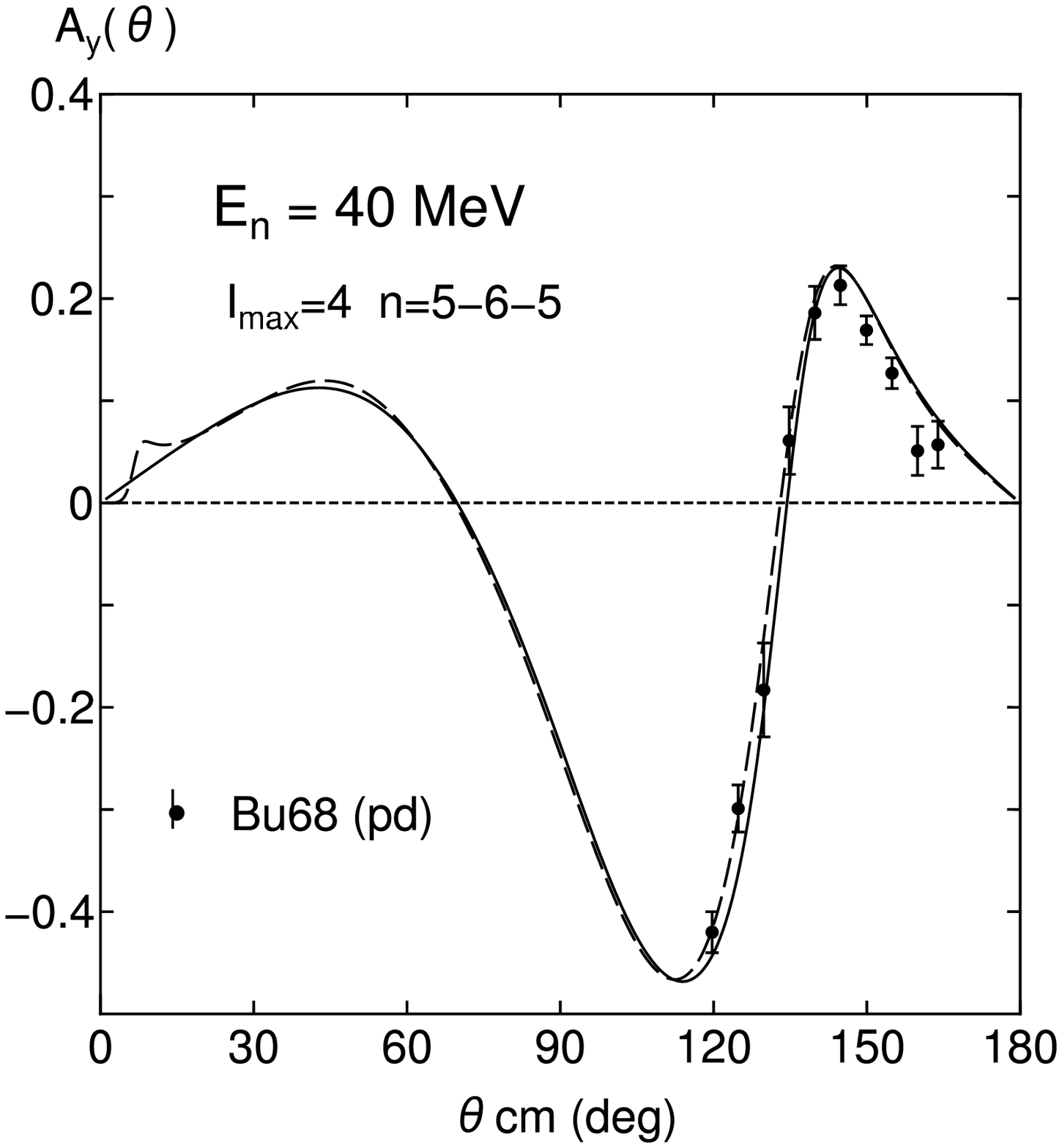}

\includegraphics[angle=0,width=56mm]
{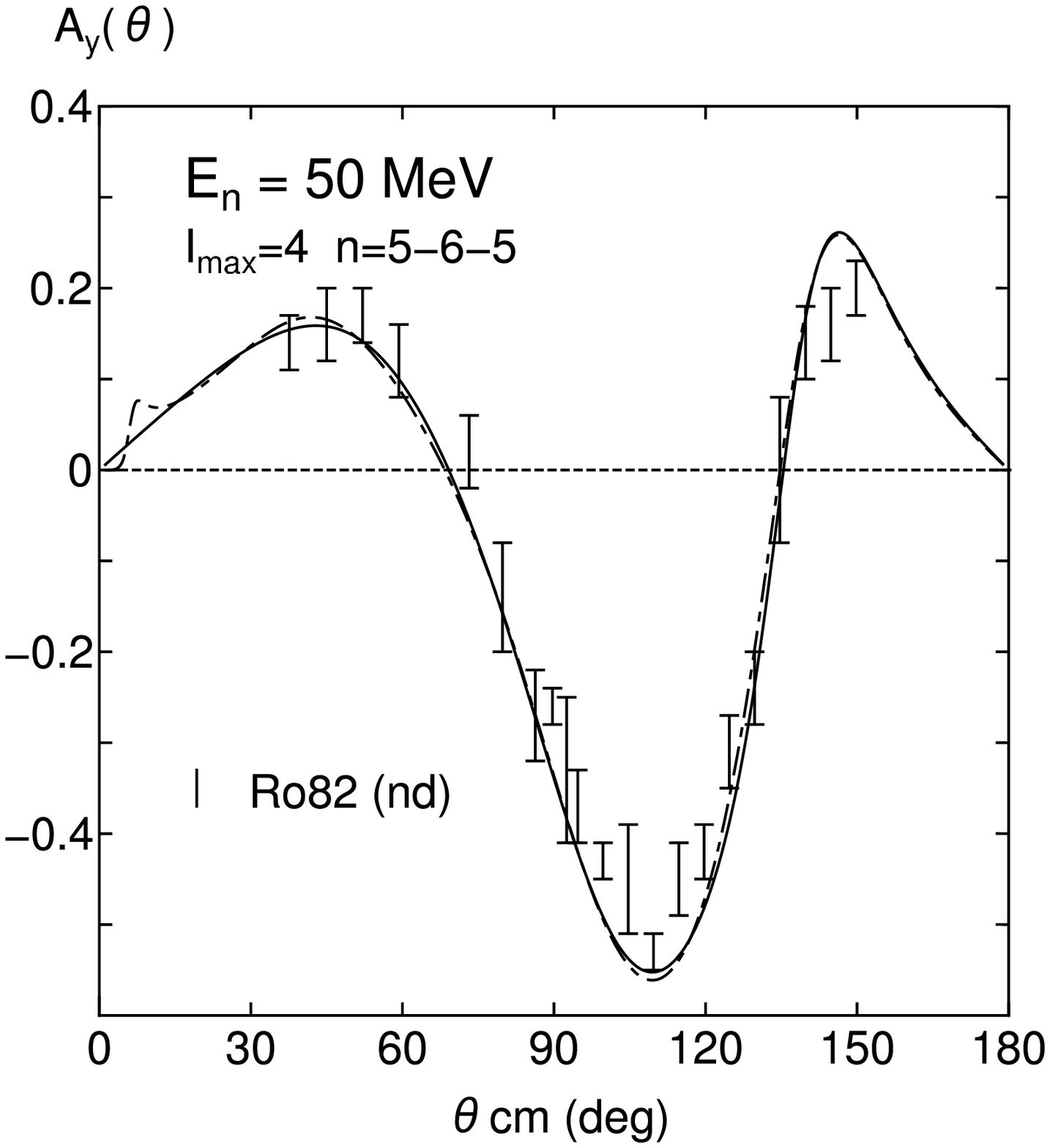}

\includegraphics[angle=0,width=56mm]
{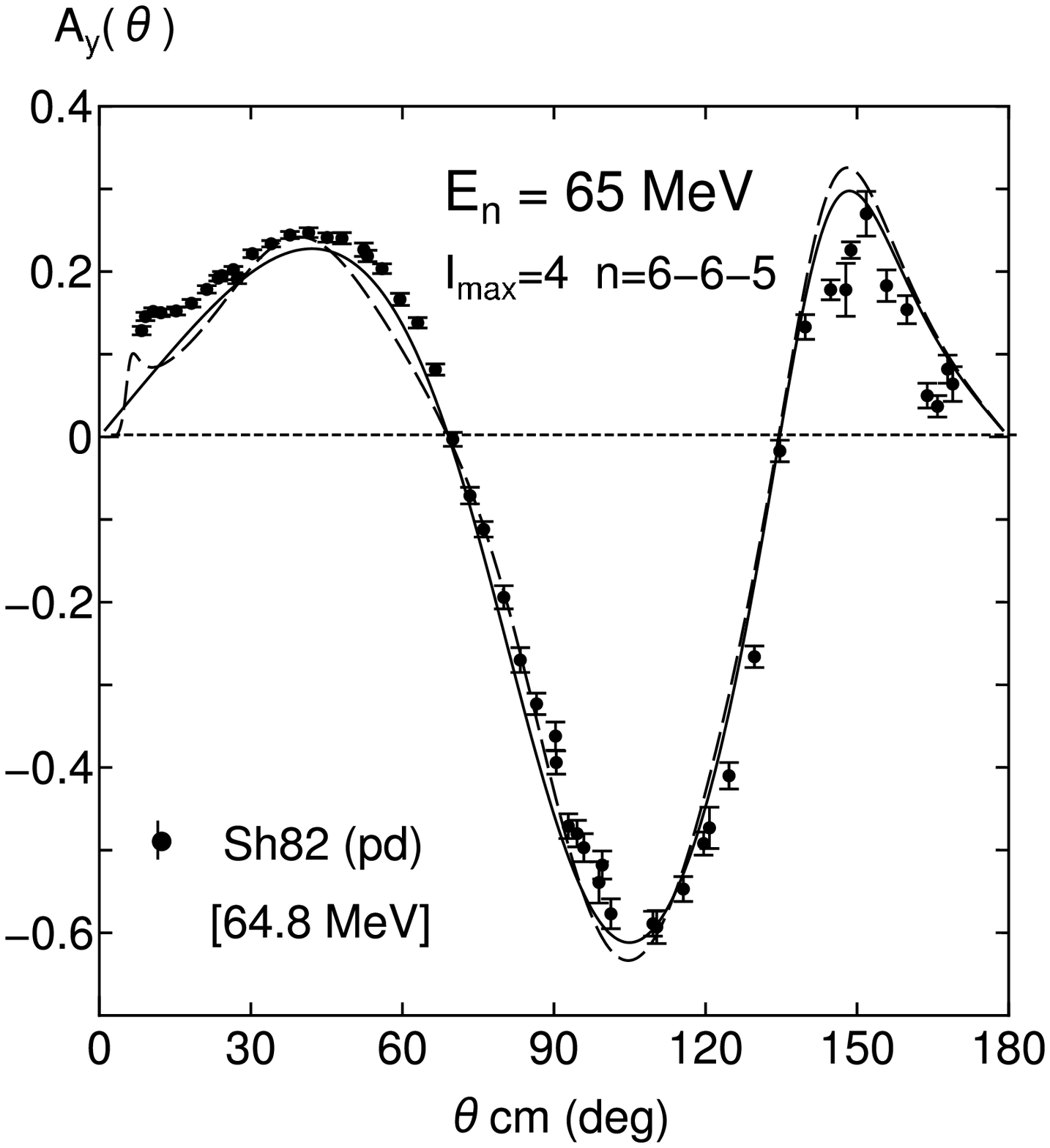}
\end{minipage}
\end{center}
\caption{
The same as Fig.\,\ref{Ay1}, but for the energies $E_n=22.7$ to 65 MeV.
The $nd$ data are shown by bars and the $pd$ data by filled circles
with bars.
The experimental data are taken from Refs.\,\citen{Co78} for Co78 ($pd$),
\citen{Do78} for Do78 ($nd$), \citen{Jo65} for Jo65 ($pd$),
\citen{Za73} for Za73 ($nd$), \citen{Bu68} for Bu68 ($pd$),
\citen{Ro82} for Ro82 ($nd$), and \citen{Sh82} for Sh82 ($pd$).
}
\label{Ay2}       
\end{figure}

The vector analyzing-powers of the nucleon, predicted by model fss2,
are shown in Figs.\,\ref{Ay1} and \ref{Ay2} for the neutron incident
energies $E_n=3$ to 65 MeV.
The calculations in this paper were carried out
using the maximum angular momentum for the $NN$ system, 
$I_{\rm max}=3$ or 4, and the momentum mesh-points $n=5$-6-5 or 6-6-5, 
in the definition defined in I.
For the energies $E_n \leq 3$ MeV, the partial waves up to $I_{\rm max}=3$
are good enough. We should note that the polarization observables are
more sensitive to the truncation of the model space than the differential
cross sections. They are also very sensitive to the Coulomb effect 
in the present low-energy region. Almost all polarization
data are for the $pd$ or $dp$ scattering and the detailed comparison with
the experiment requires the introduction of the Coulomb force.
Here we show with dashed curves preliminary results obtained
by applying the Vincent and Phatak method\cite{Vi74} to the
$pd$ scattering.\footnote{The details of this approach
to the $pd$ scattering will be reported in a separate paper. 
See also Ref.\,\protect\citen{apfb05}.}
The $pd$ calculations were made using the cut-off Coulomb potential with
the the cut-off radius $R_c=9$ fm (for $E_p \leq 3$ MeV)
or $R_c=8$ fm (for $E_p \geq 5$ MeV),
together with $I_{\rm max}=3$ and $n=6$-6-5.
The $nd$ data shown with bars should therefore be compared with the 
solid curves, while the $pd$ data with circles and others
correspond to the dashed curves.
The cited paper for these experimental data is specified by
two letters and double figures, which are short for the first author's
name and the publication year, respectively.
In the forward angular region with $\theta_{\rm cm} \leq 30^\circ$, 
the enhancement of $A_y$ for the $pd$ data is almost correctly reproduced 
by the Coulomb effect, although the reduction in $\theta_{\rm cm}
=60^\circ$ - $120^\circ$ makes the agreement with the experiment worse
at the energy region $E_p=3$ - 14 MeV.
We find that the difference between the solid curves and the dashed curves
gradually diminishes for higher energies except for the forward angles.

\begin{table}[htb]
\caption{
Comparison of the maximum values of $A_y$ and their positions
$\theta_{\rm cm}$ with the $nd$ and $pd$ experimental data. 
The ratio between the theory and experiment is also shown.
}
\label{table1}
\renewcommand{\arraystretch}{1.12}
\setlength{\tabcolsep}{3mm}
\begin{center}
\begin{tabular}{cccccccc}
\hline
$E_N$ & system & \multicolumn{2}{c}{fss2}& ratio 
    & \multicolumn{2}{c}{experiment} & Ref. \\
(MeV)    &       & $\theta_{\rm cm}$ (deg) & ${A_y}_{\rm max}$ 
&   & $\theta_{\rm cm}$ (deg) & ${A_y}_{\rm max}$ & \\
\hline
2   & $nd$ &  95 & 0.0341 & 0.81  &  95 & 0.042  & \citen{Ne03} [1.9 MeV] \\
    & $pd$ &  92 & 0.0320 & 0.973 &  95 & 0.0329 & \citen{Sa94} \\ [2mm]
2.5 & $nd$ & 100 & 0.0412 & $-$   &     &        &      \\
    & $pd$ &  98 & 0.0387 & 0.921 &  99 & 0.0420 & \citen{Sa94} \\ [2mm]
3   & $nd$ & 104 & 0.0487 & 0.81  & 104 & 0.060  & \citen{Mc94} \\
    & $pd$ & 103 & 0.0443 & 0.899 & 104 & 0.0493 & \citen{Sa94} \\ [2mm]
5   & $nd$ & 116 & 0.0726 & 0.80  & 115 & 0.091  & \citen{To91} \\
    & $pd$ & 113 & 0.0684 & 0.840 & 113 & 0.0814 & \citen{Sa94} \\ [2mm]
7   & $nd$ & 118 & 0.104  & $-$   &     &        &       \\
    & $pd$ & 119 & 0.0930 & 0.830 & 119 & 0.112  & \citen{Sa94} \\ [2mm]
9   & $nd$ & 122 & 0.131  & 0.87  & 119 & 0.15   & \citen{To91} [8.5 MeV] \\
    & $pd$ & 123 & 0.118  & 0.831 & 122 & 0.142  & \citen{Sa94} \\ [2mm]
10  & $nd$ & 124 & 0.144  & 0.85  & 126 & 0.17   & \citen{To82} \\
    & $pd$ & 124 & 0.131  & 0.845 & 124 & 0.155  & \citen{Sa94} \\ [2mm]
12  & $nd$ & 127 & 0.167  & 0.88  & 128 & 0.19   & \citen{Ni98} \\
    & $pd$ & 127 & 0.155  & 0.856 & 126 & 0.181  & \citen{Sa94} \\ [2mm]
14  & $nd$ & 129 & 0.185  & 0.84  & 130 & 0.22   & \citen{To83} [14.1 MeV] \\
    & $pd$ & 129 & 0.177  & 0.876 & 128 & 0.202  & \citen{Sa94} \\ [2mm]
18  & $nd$ & 133 & 0.203  & $-$   &     &        &       \\
    & $pd$ & 133 & 0.205  & 0.928 & 132 & 0.221  & \citen{Sa94}  \\ [2mm]
19  & $nd$ & 133 & 0.205  & 0.89  & 132 & 0.23   & \citen{We10} \\
    & $pd$ & 134 & 0.208  & $-$   &     &        &              \\
\hline
\end{tabular}
\end{center}
\end{table}

Our results in Fig.\,\ref{Ay1} imply that the long-standing $A_y$-puzzle
for the large discrepancies between the theory and experiment in the
energy region $E_n \le 25$ MeV is not so serious 
as the AV18 potential,\cite{To98a,To98b,To02,To08}
although the maximum peak height of $A_y$ is still too low.
We compare in Table \ref{table1} the theoretical peak heights
with the $nd$ experimental data and with the $pd$ data 
given in Ref.\,\citen{Sa94}.
The comparison of the no-Coulomb calculations implies
that our results are about 80 - 90$\%$ of the observed $nd$ data, 
although exact evaluation of the ratio is not easy
due to the experimental errorbars.
A similar amount of discrepancy is also seen in the comparison of
the $pd$ data, but it has an apparent minimum around $E_p=7$ MeV
of about 83$\%$.
As an average in this energy region,
the shortage of about 15$\%$ is a fair estimate both for the $nd$ and
$pd$ data, which is better than the rather constant
discrepancy by AV18 potential at the $25\%$ level
up to about 25 MeV.\cite{To08} (See Fig.\,2 of Ref.\,\citen{To08}.)
Looking back to the old predictions by realistic
separable potentials, $A_y(\theta)$ at some energies are very well
reproduced.\cite{Do82,Sa83,Ko87a,Ko87b,Ra88}
Our QM $NN$ interaction fss2 and the separable potentials are both
considered to have quite different off-shell properties
from the meson-exchange potentials, that are characterized
by the strong nonlocality of the interaction in the configuration space. 
For the energies $E_p=30$ - 65 MeV, there is no clear discrepancy
of $A_y$ between the theory and experiment.

We should mention that the improvement of $A_y(\theta)$ is not achieved
by the modification of the $\hbox{}^3P_J$ phase shifts
of the $NN$ interaction, as claimed in early studies of this problem.
Although $A_y(\theta)$ is very sensitive to these phase shifts,
an artificial modification (of almost $10\%$) is far beyond acceptable
from the modern phase shift analysis of the $NN$ interaction.\cite{To98a,To98b}
Our $NN$ model fss2 reproduces the empirical $\hbox{}^3P_J$ phase shifts
within the accuracy of one degree at the energies
less than 300 MeV. (See Fig.\,1 of Ref.\,\citen{PPNP}.)
From the detailed phase shift analysis of the $pd$ scattering
up to $E_p=10$ MeV, the authors of Ref.\,\citen{To02} claim that
the $\hbox{}^4P_{1/2}$ and $\varepsilon_{{3/2}^-}$ eigenphase shifts
should be affected by the $3N$ force, in order to improve $A_y(\theta)$.
In fact, the phase shift analysis of the $pd$ scattering is very
ambiguous in this energy region, since so many complex parameters are
involved. In Ref.\,\citen{scl10}, we have compared the $nd$ phase shifts
with the phase shift analysis\cite{Ki96} at much lower energies
less than 3 MeV and found that the desirable feature
to reproduce the sufficient
binding energy of the triton and the spin-doublet $nd$ scattering
length $\hbox{}^2a_{nd}$ is achieved by the more 
attractive feature of the $\hbox{}^2S_{1/2}$ phase shift of fss2
than the AV18 potential, originating from the nonlocal description 
of the short-range repulsion.
We can conjecture that a similar situation is taking place even
for the higher energies, although it is difficult to pinpoint
some particular eigenphase shifts.
 
\bigskip

\subsection{Vector and tensor analyzing-powers of the deuteron}

\begin{figure}[htb]
\begin{center}
\begin{minipage}{0.48\textwidth}
\includegraphics[angle=0,width=58mm]
{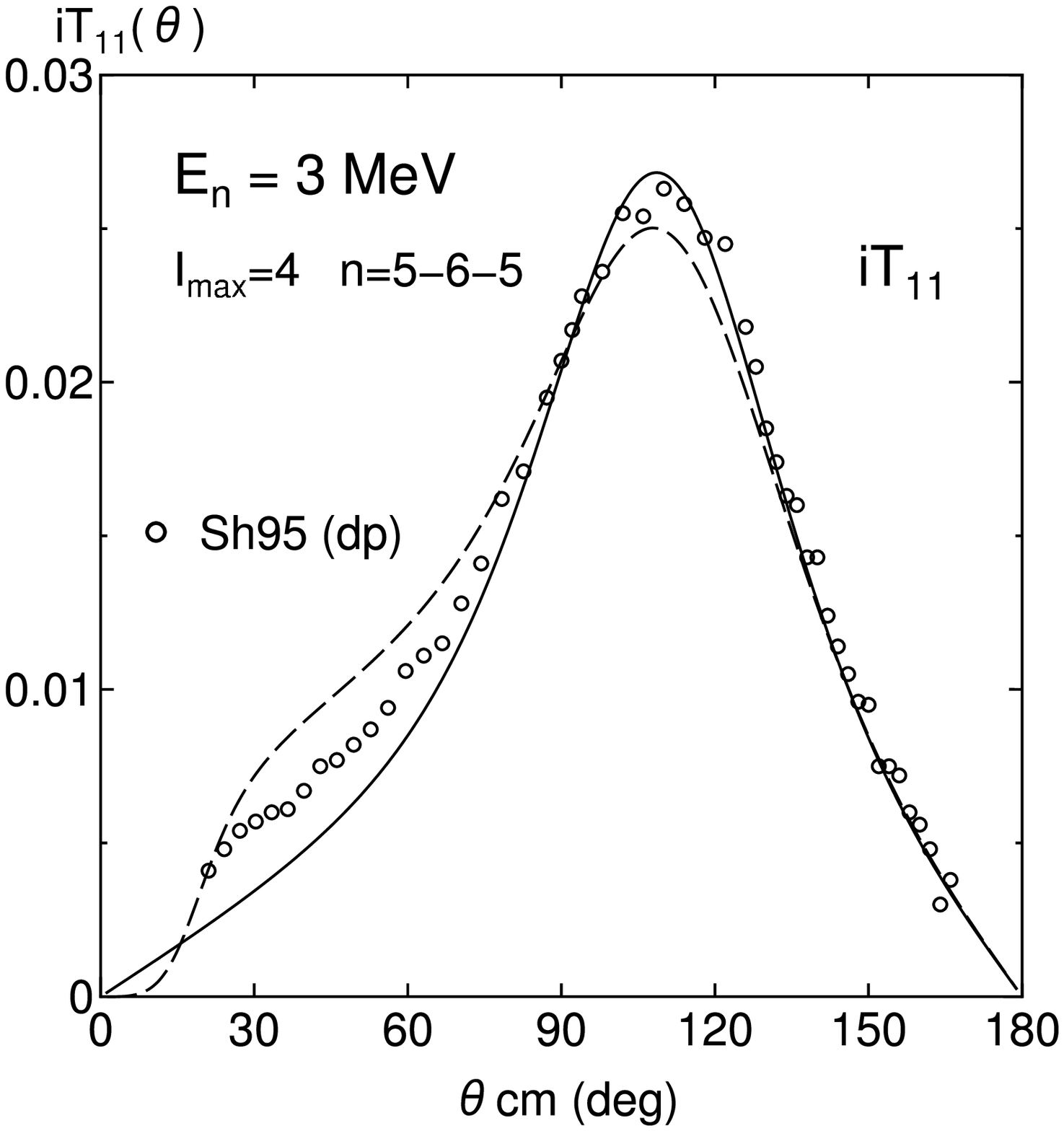}

\includegraphics[angle=0,width=58mm]
{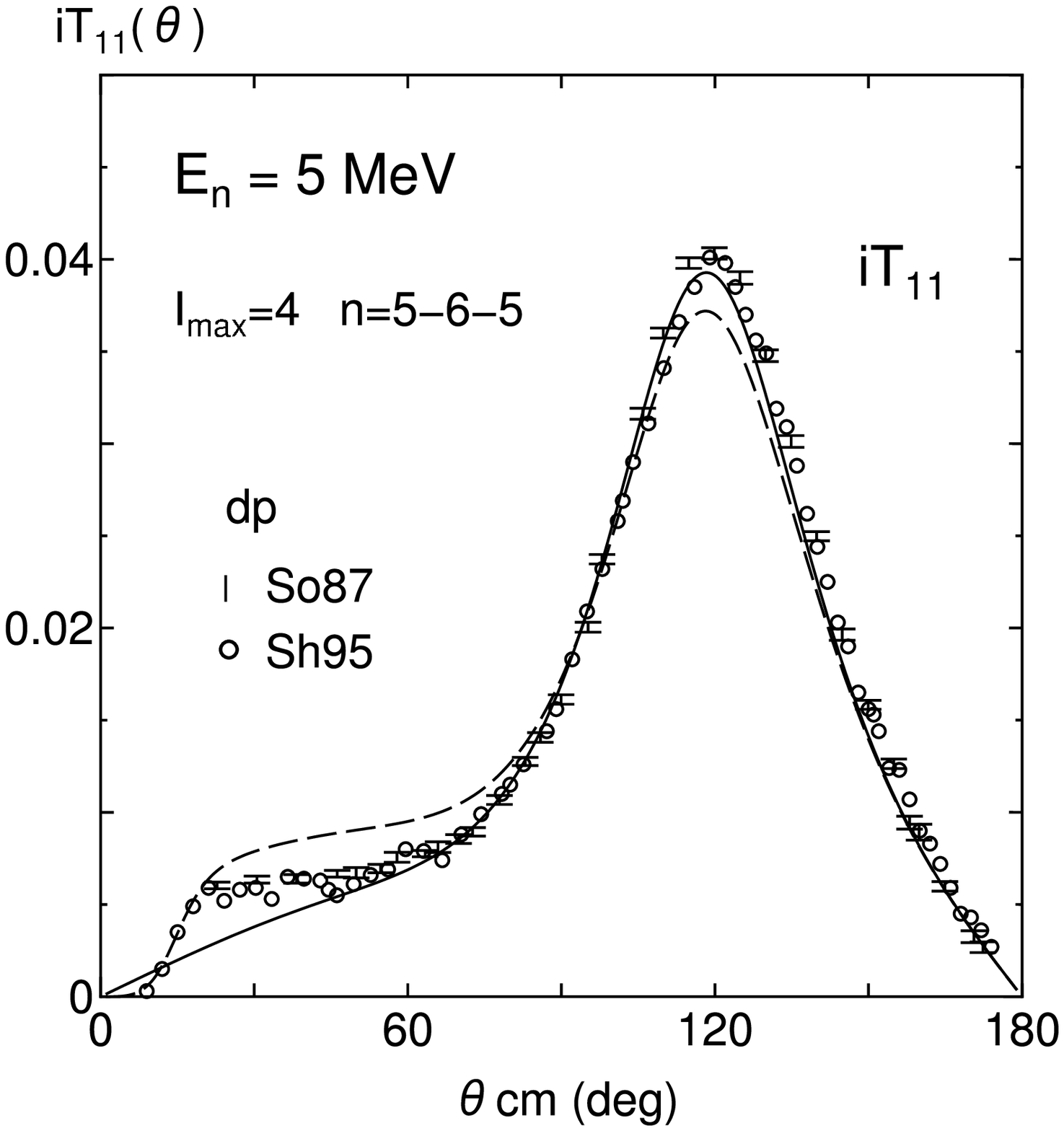}

\includegraphics[angle=0,width=58mm]
{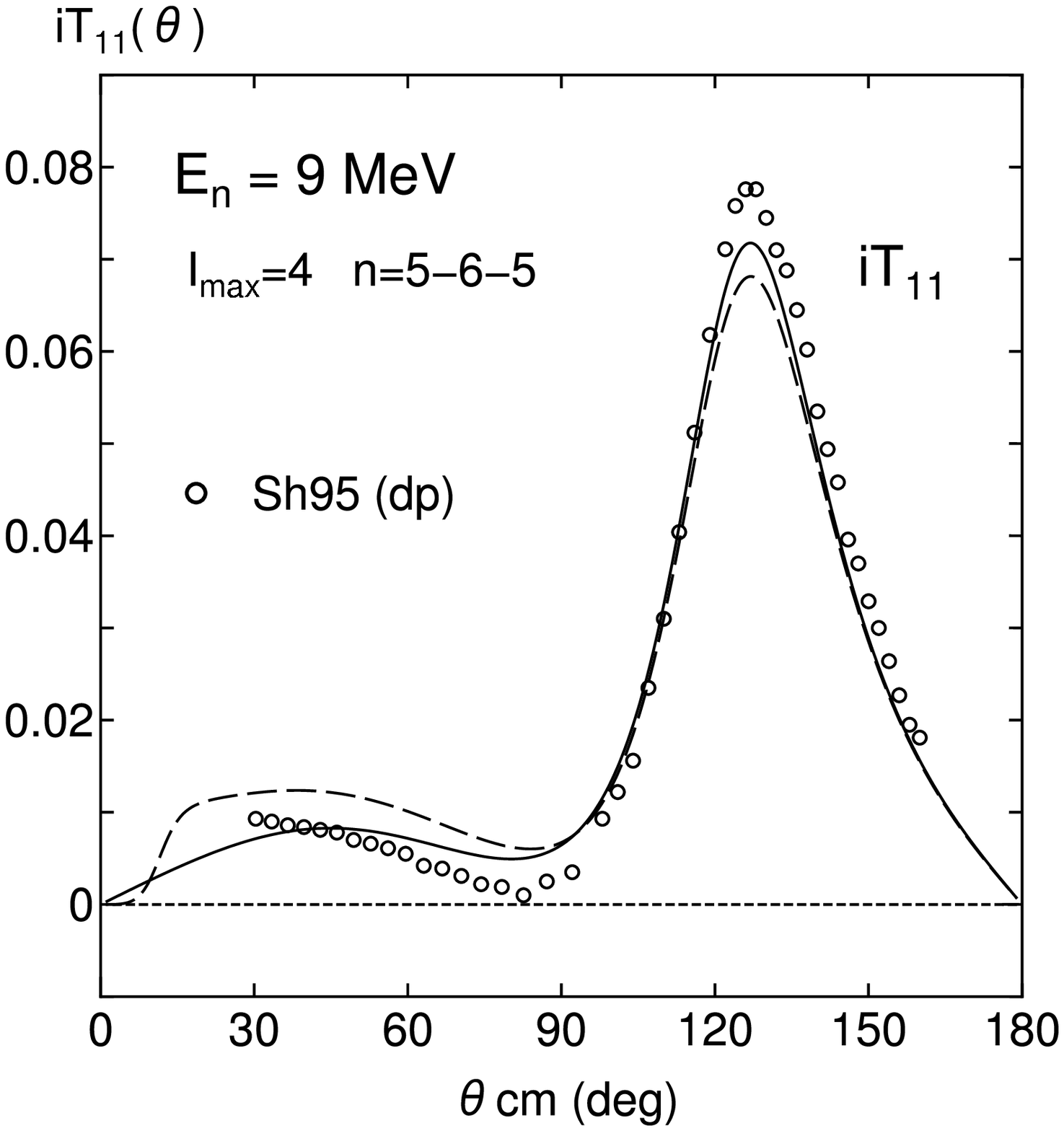}
\end{minipage}~%
\hfill~%
\begin{minipage}{0.48\textwidth}
\includegraphics[angle=0,width=58mm]
{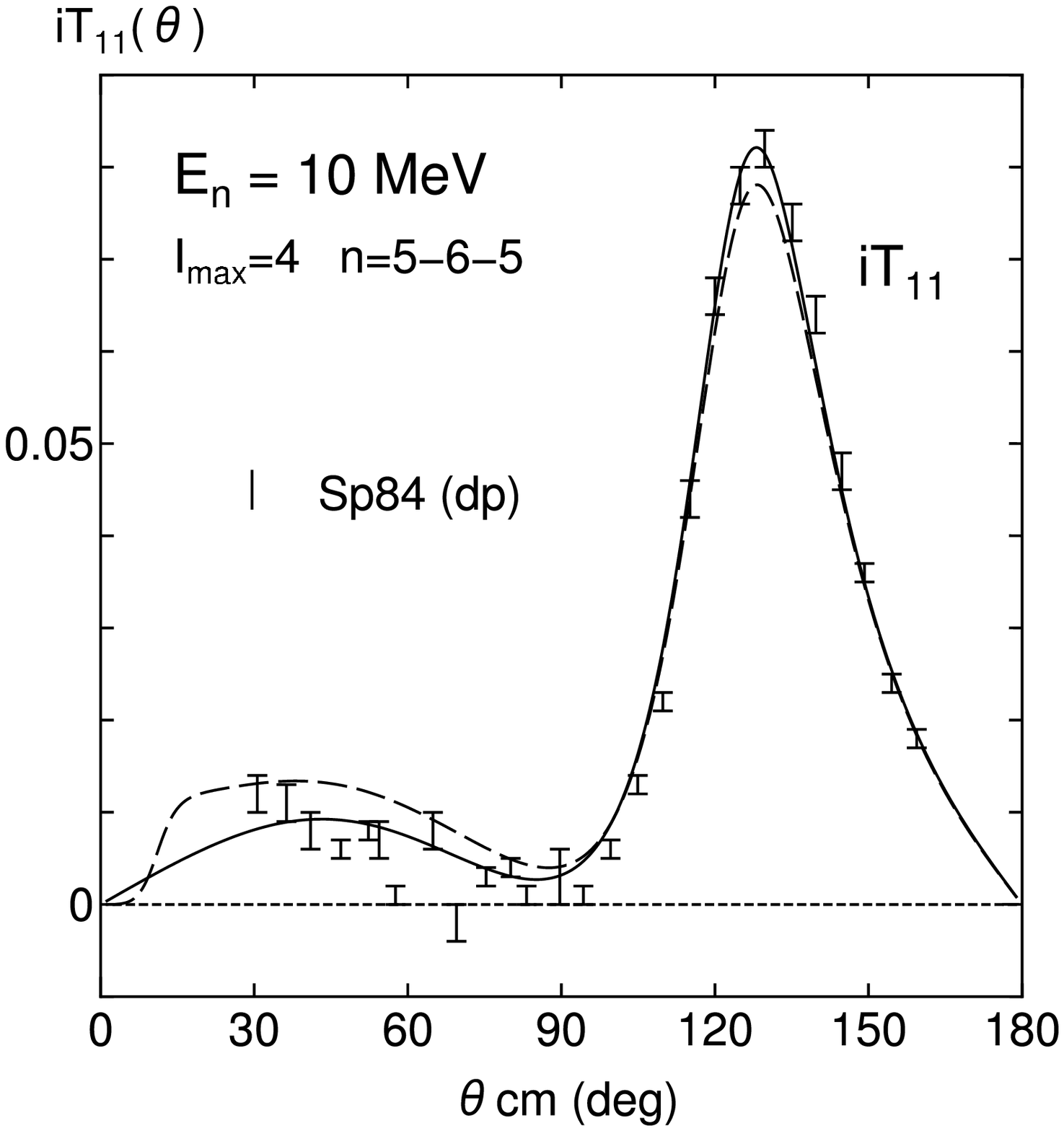}

\includegraphics[angle=0,width=58mm]
{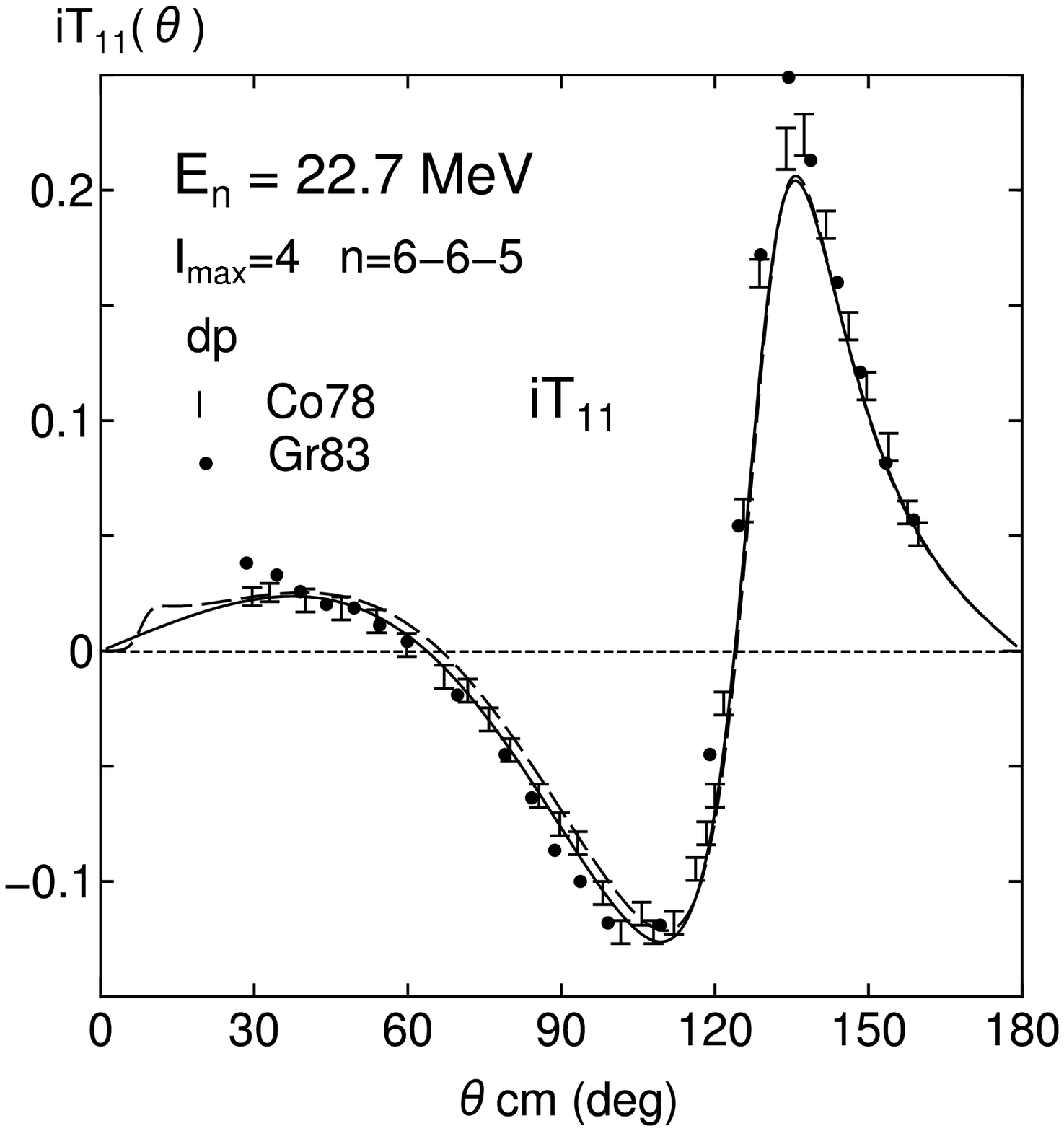}

\includegraphics[angle=0,width=58mm]
{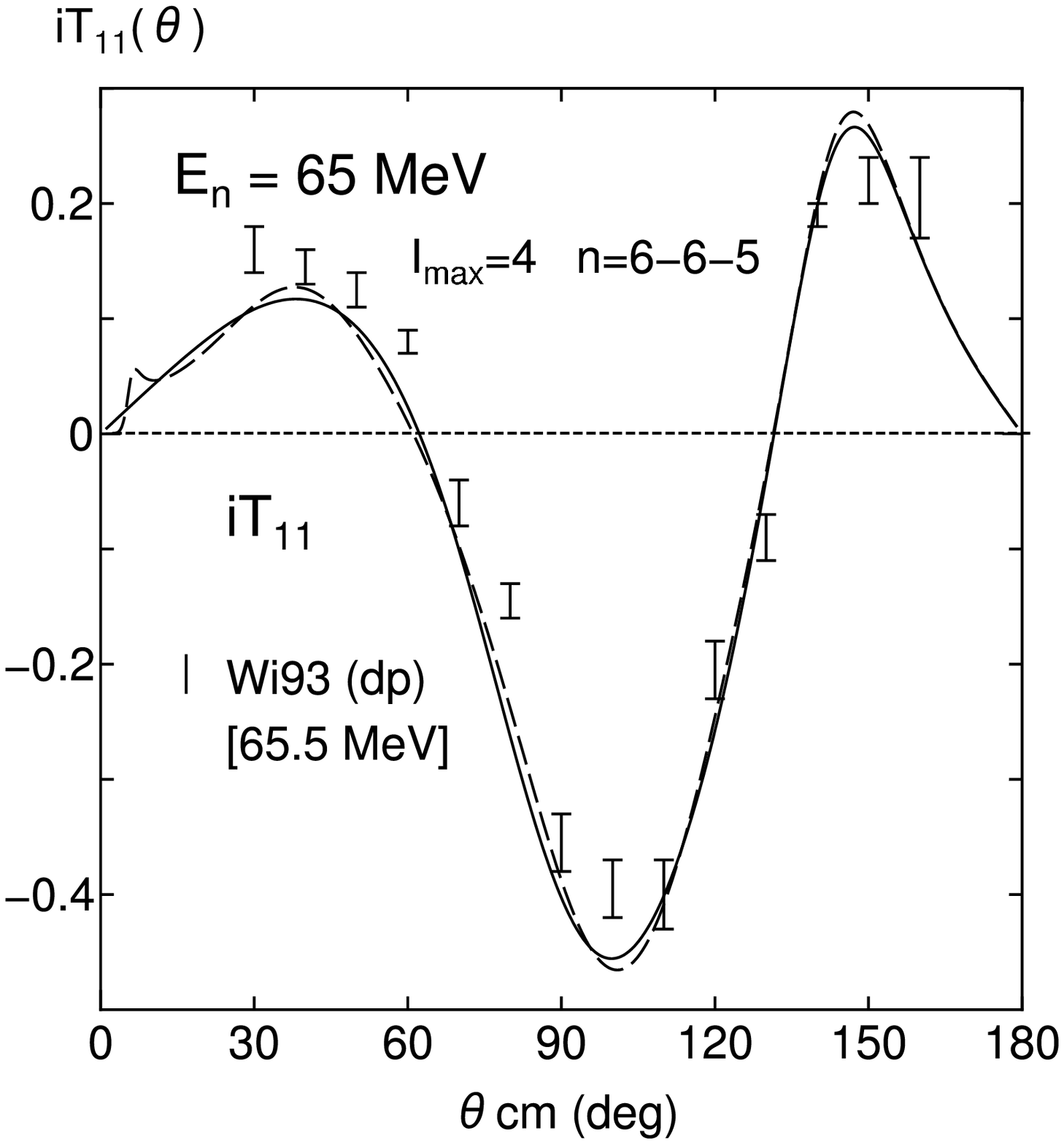}
\end{minipage}
\end{center}
\caption{
The vector-type deuteron analyzing-power $i\,T_{11}(\theta)$ 
for the $nd$ elastic scattering (solid curves) from $E_n=3$ to 65 MeV.
All the experimental data are for the $dp$ scattering, for which
the $pd$ results with the cut-off Coulomb force are also shown
by dashed curves.
The experimental data are taken from Refs.\,\citen{Sh95} for
Sh95, \citen{So87} for So87, \citen{Sp84} for Sp84,
\citen{Co78} for Co78, \citen{Gr83} for Gr83, and \citen{Wi93} for Wi93.
}
\label{T11}       
\end{figure}

\begin{figure}[htb]
\begin{center}
\begin{minipage}{0.48\textwidth}
\includegraphics[angle=0,width=60mm]
{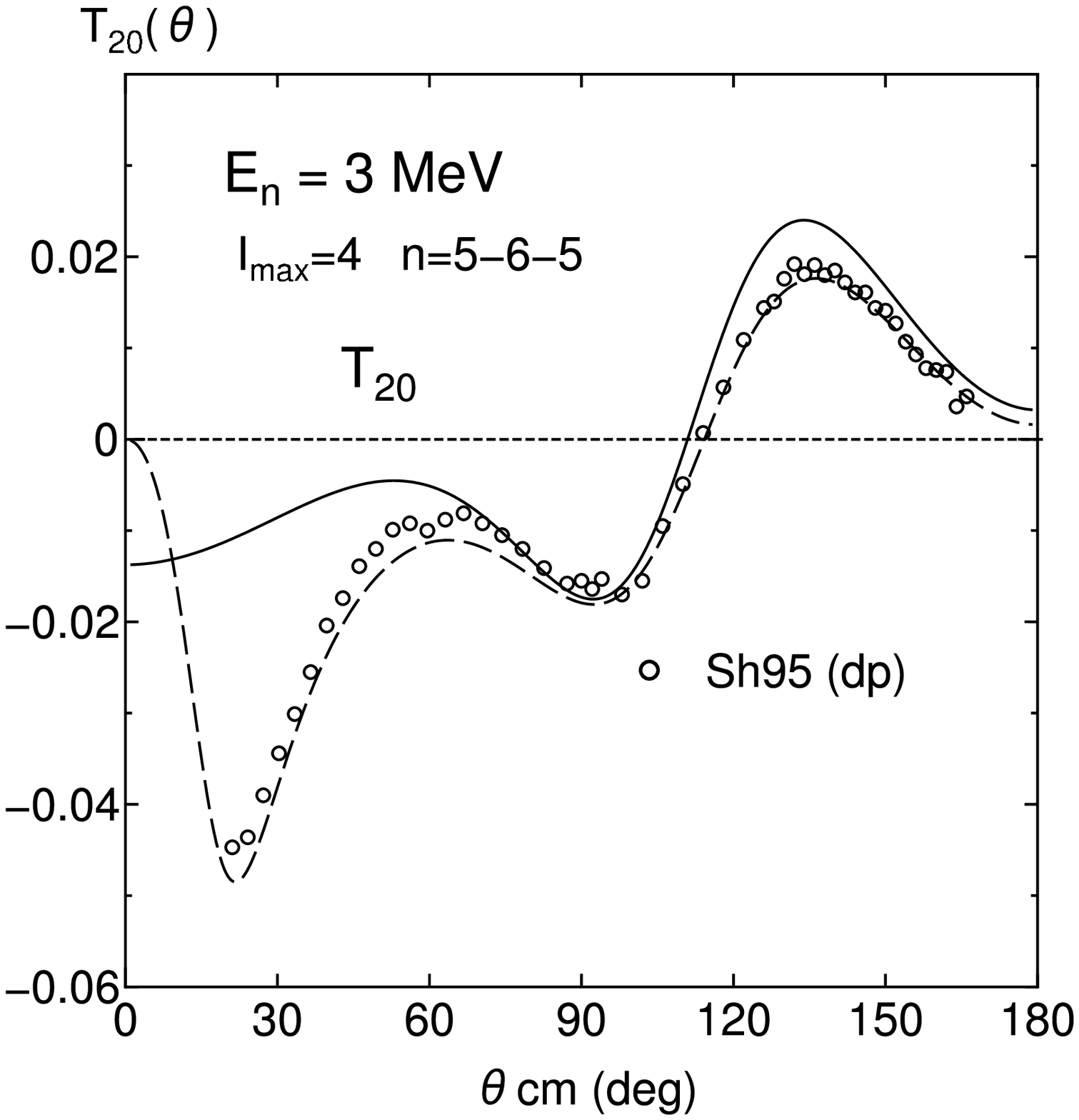}

\includegraphics[angle=0,width=60mm]
{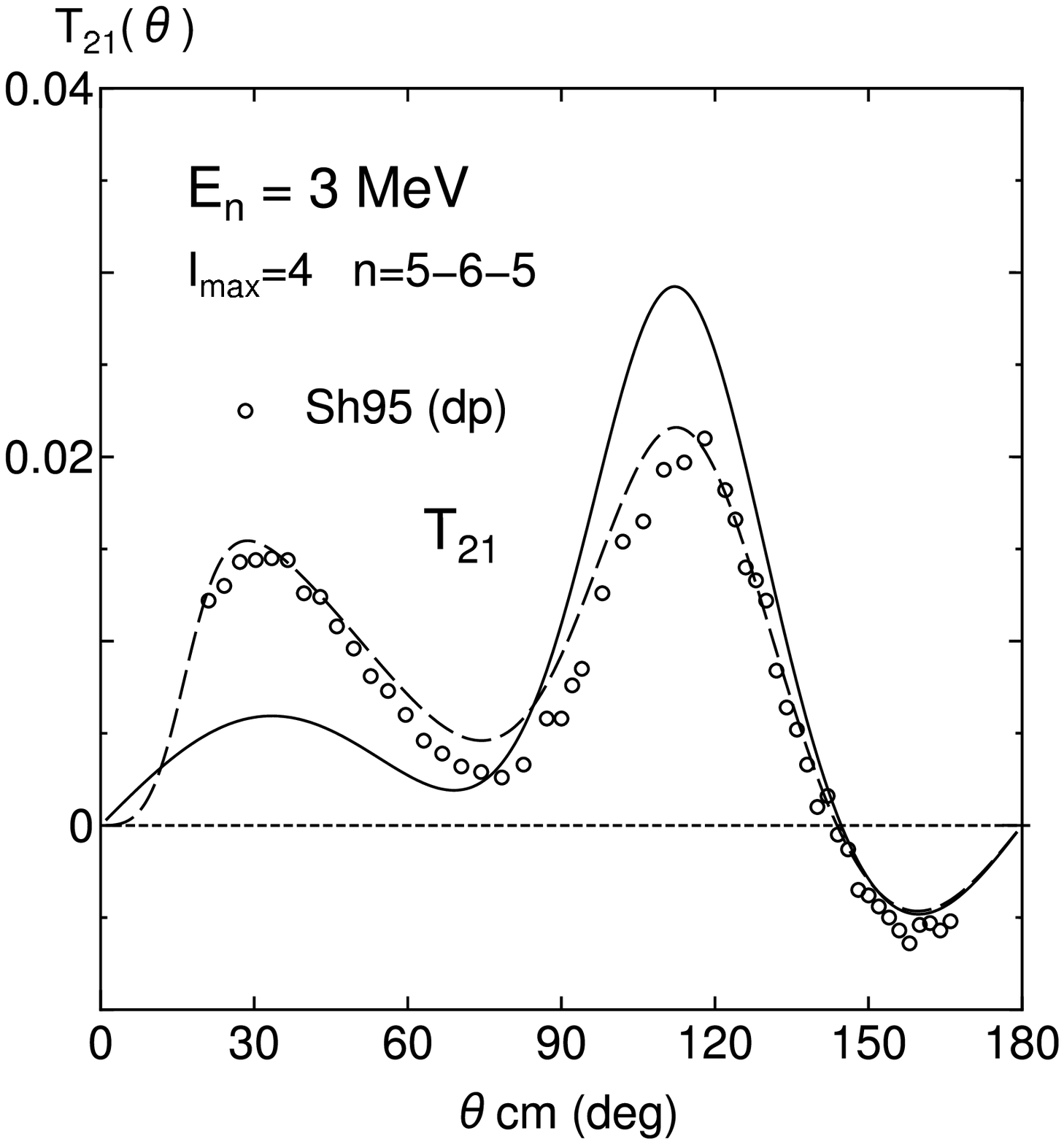}

\includegraphics[angle=0,width=60mm]
{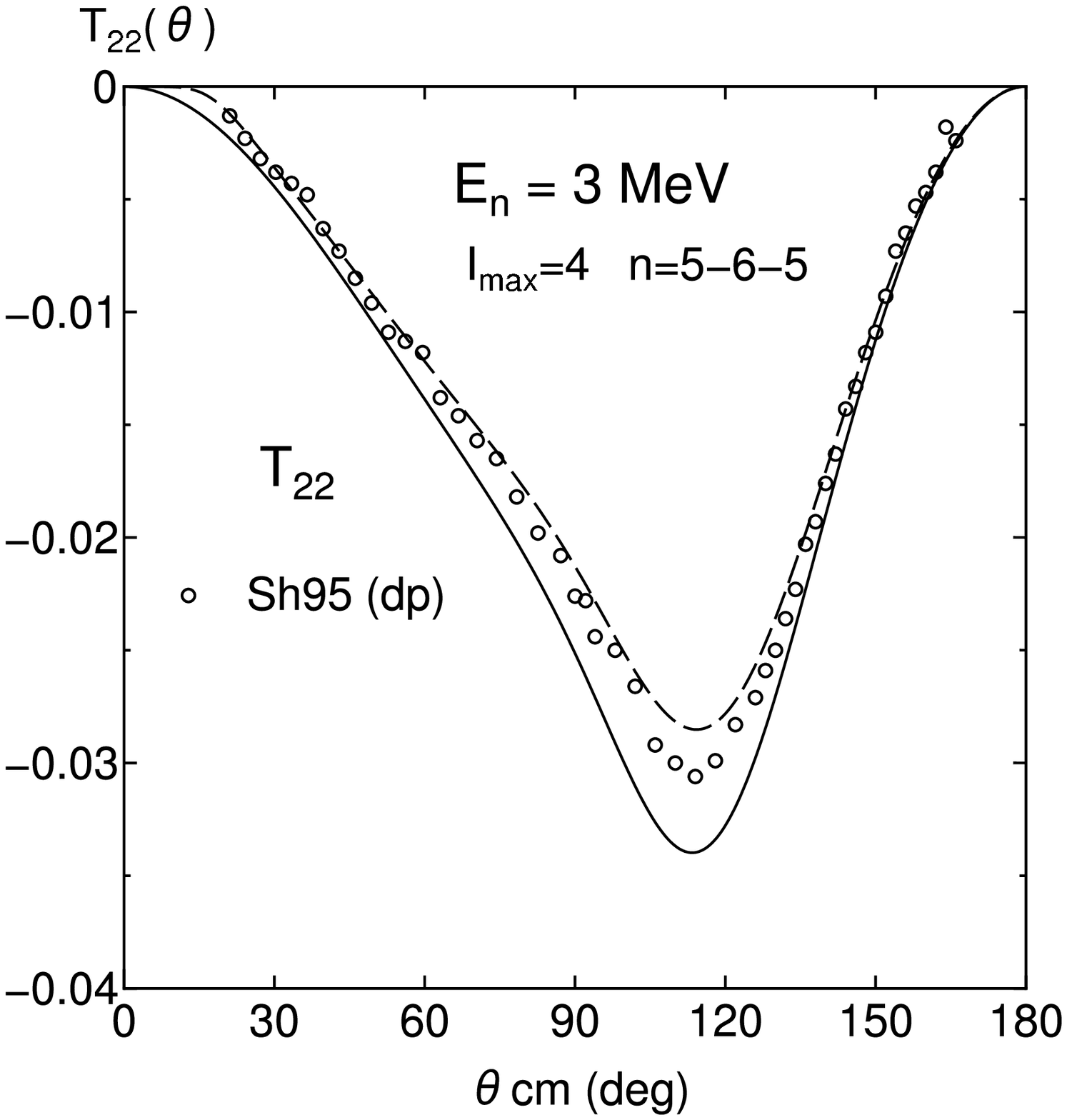}
\end{minipage}~%
\hfill~%
\begin{minipage}{0.48\textwidth}
\includegraphics[angle=0,width=60mm]
{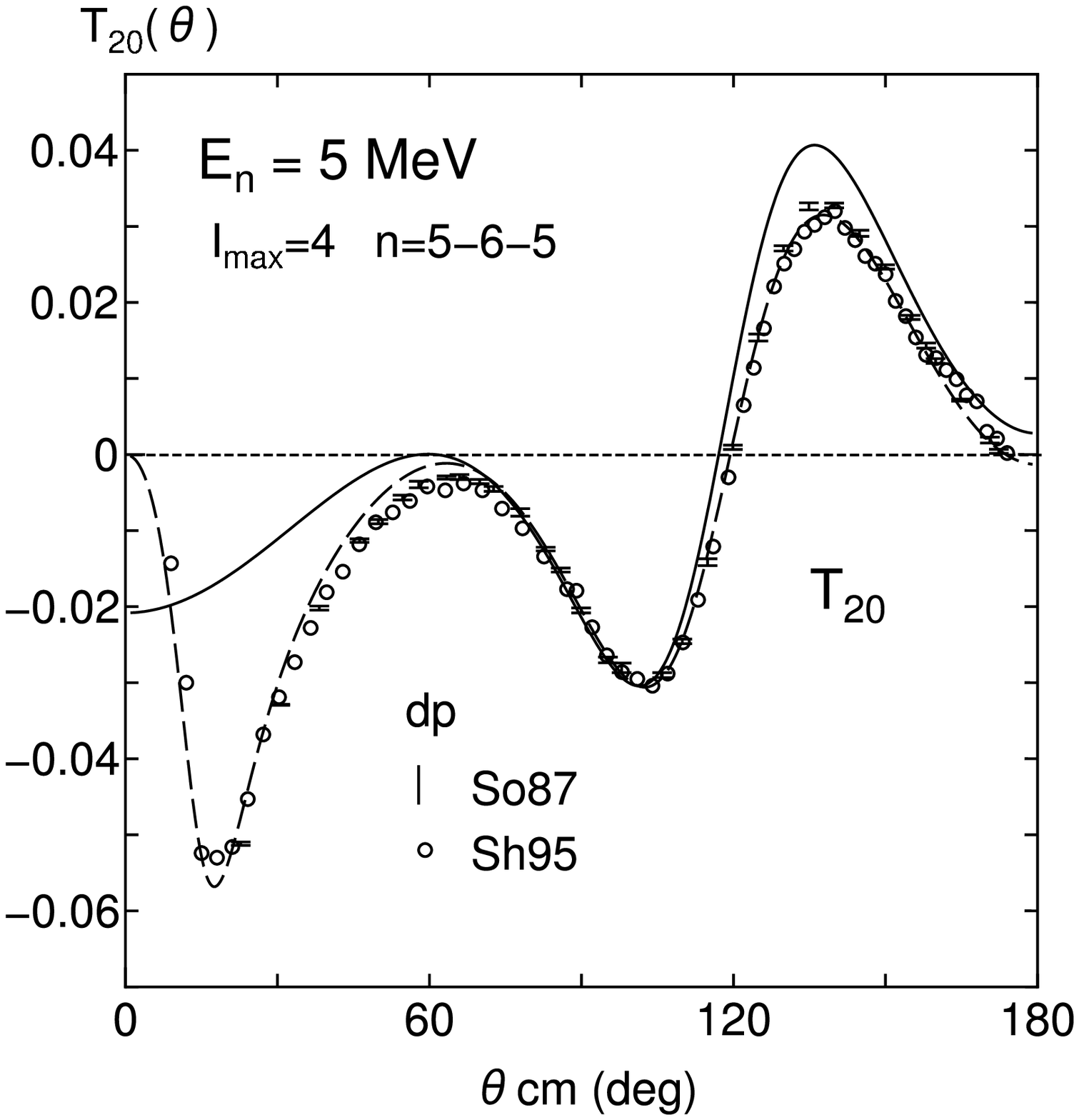}

\includegraphics[angle=0,width=60mm]
{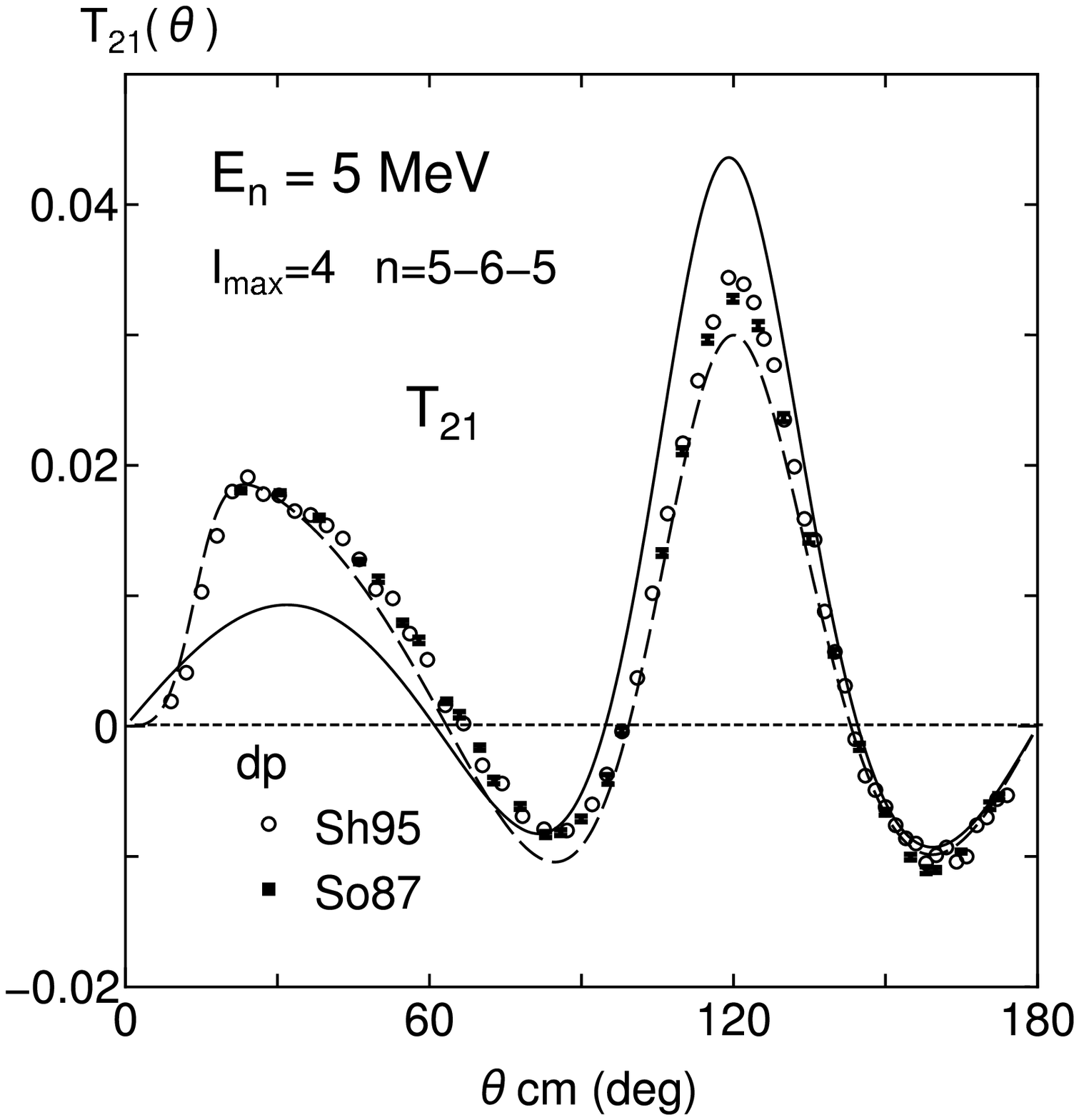}

\includegraphics[angle=0,width=60mm]
{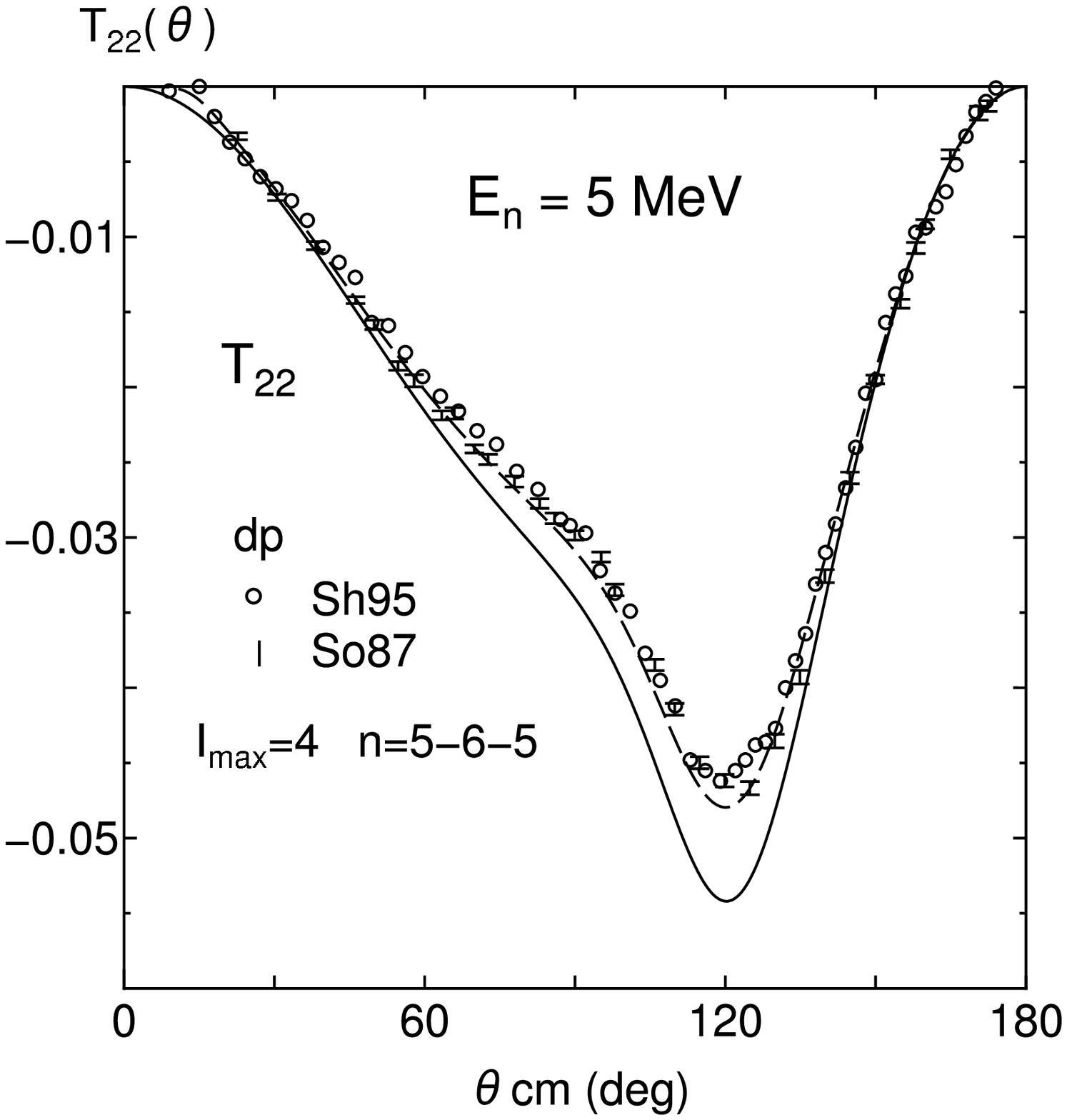}
\end{minipage}
\end{center}
\caption{
The tensor-type deuteron analyzing-powers $T_{2m}(\theta)$ ($m=0$, 1, 2) 
for the $nd$ elastic scattering (solid curve) for $E_n=3$ and 5 MeV.
All the experimental data are for the $dp$ scattering, for which
the $pd$ results with the cut-off Coulomb force are also shown
by dashed curves.
The experimental data are taken from Refs.\,\citen{Sh95} for
Sh95 and \citen{So87} for So87.
}
\label{T2m1}       
\end{figure}

\begin{figure}[htb]
\begin{center}
\begin{minipage}{0.48\textwidth}
\includegraphics[angle=0,width=55mm]
{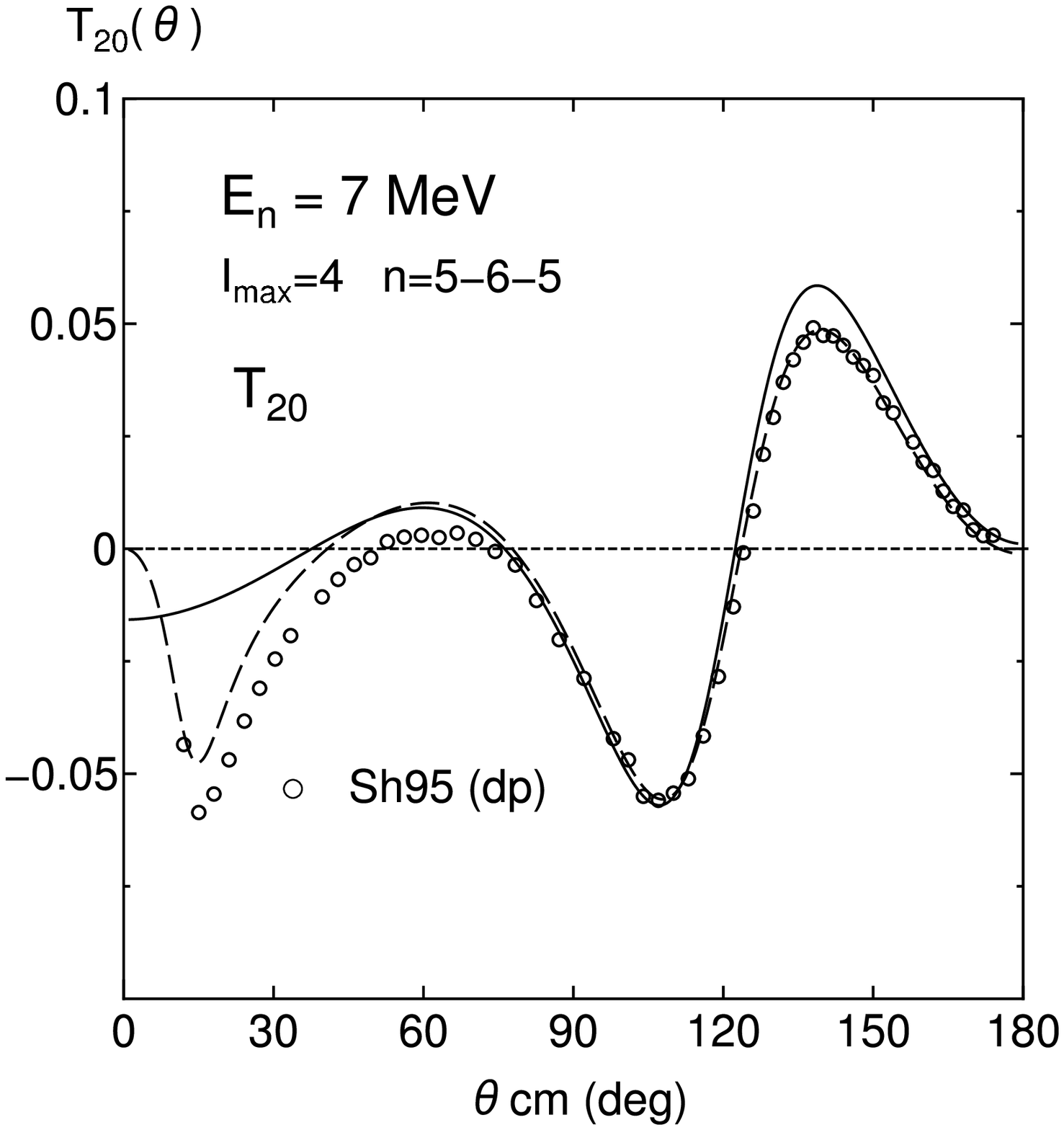}

\includegraphics[angle=0,width=55mm]
{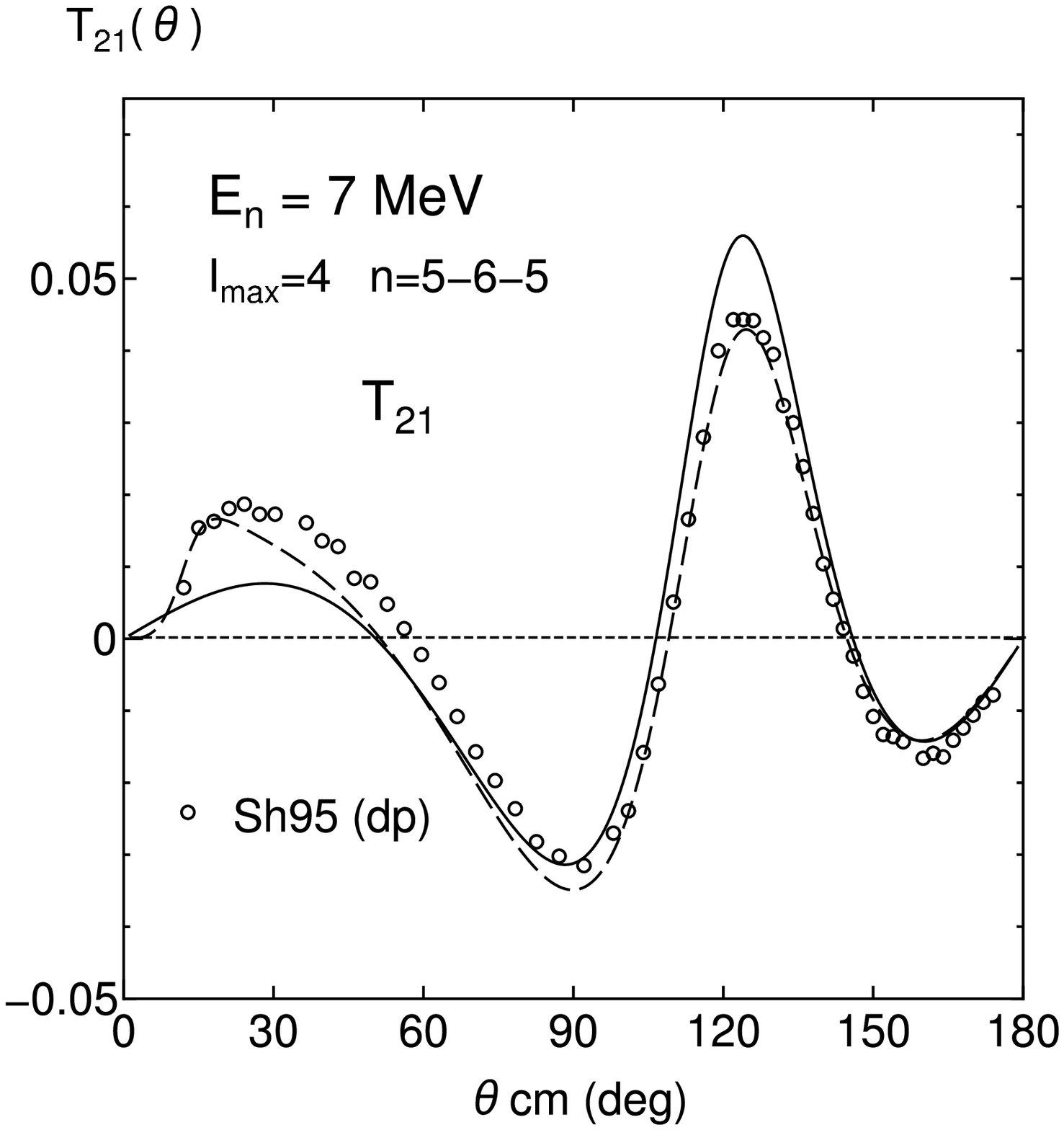}

\includegraphics[angle=0,width=55mm]
{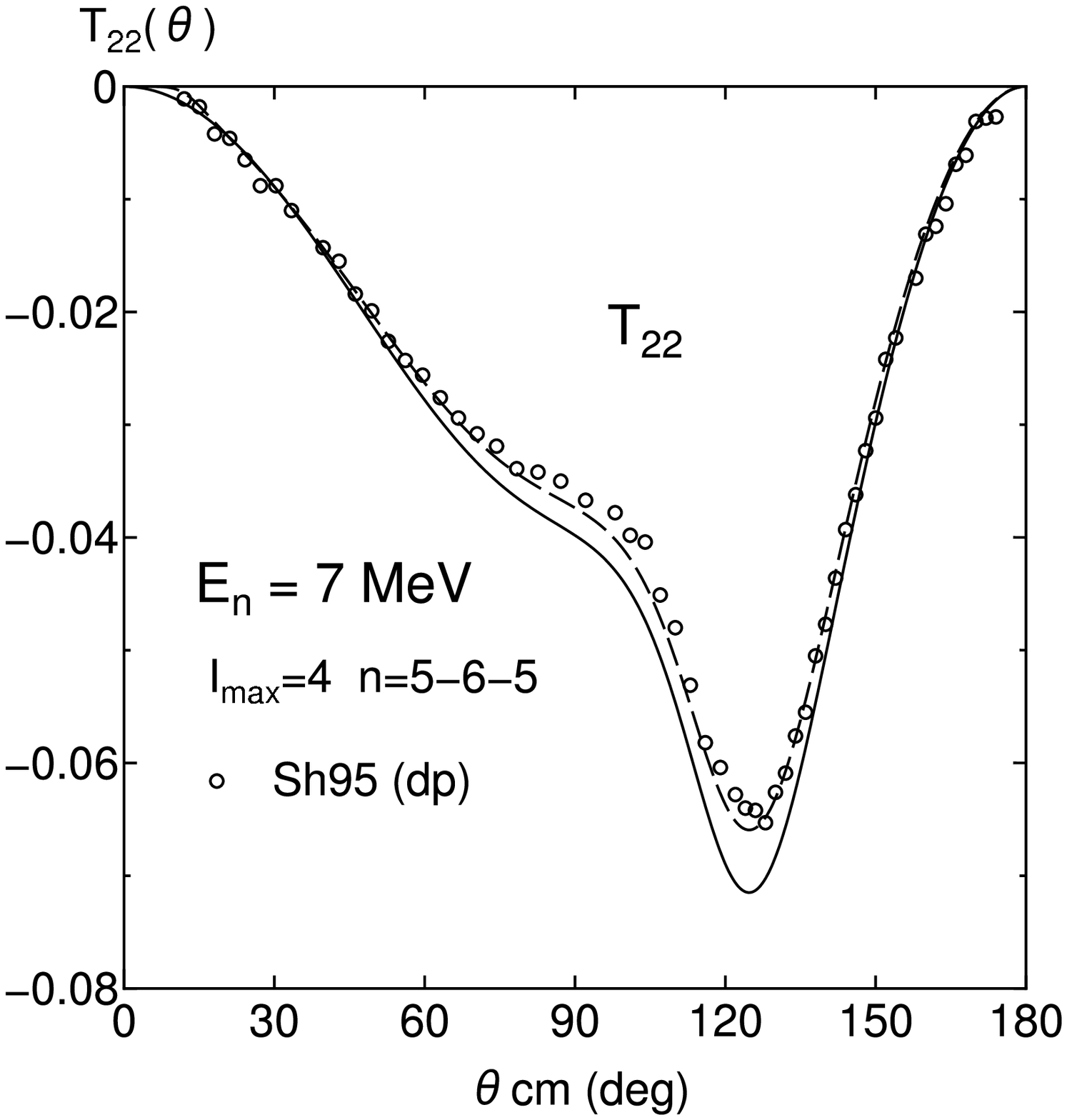}


\end{minipage}~%
\hfill~%
\begin{minipage}{0.48\textwidth}
\includegraphics[angle=0,width=55mm]
{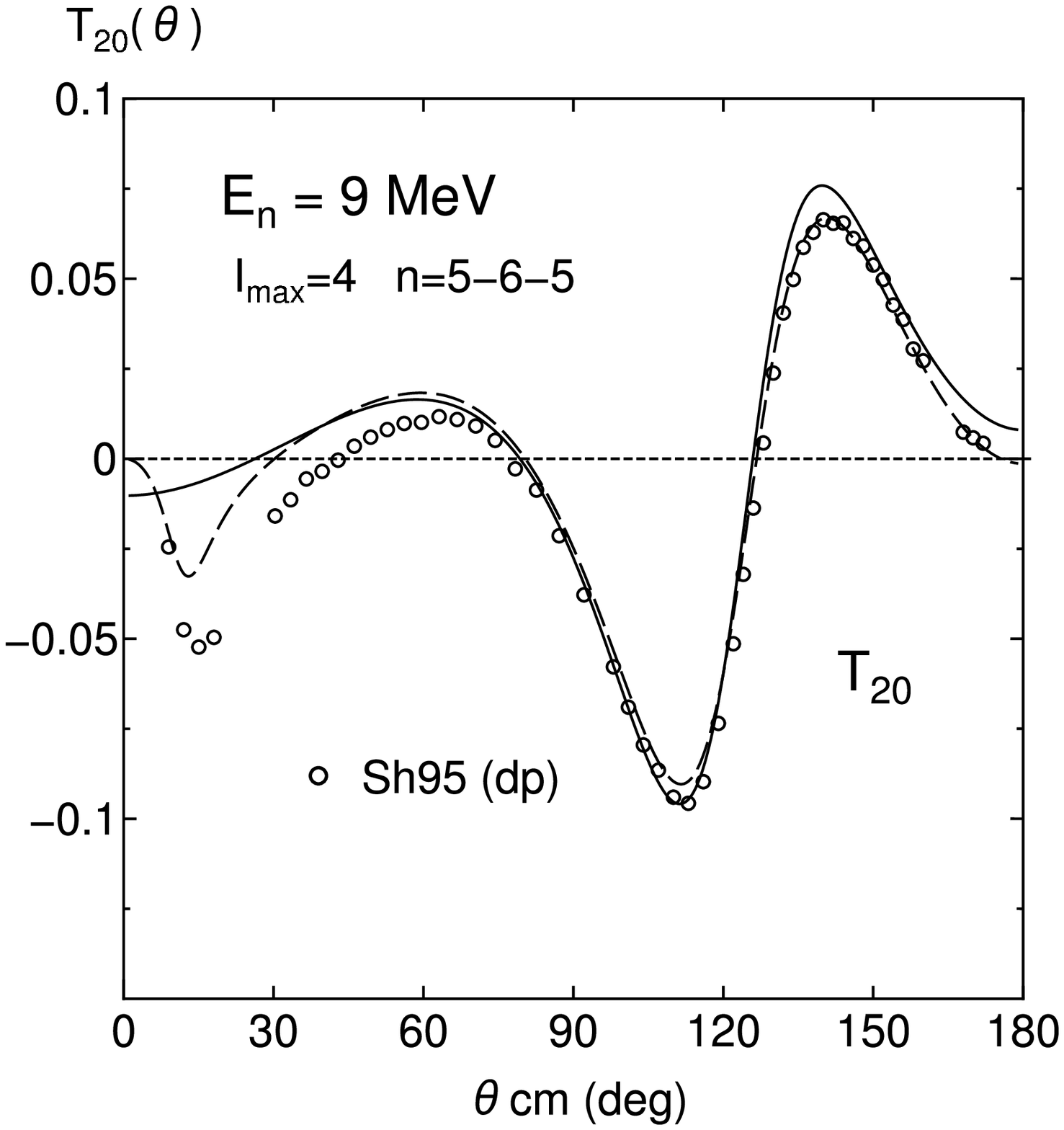}

\includegraphics[angle=0,width=55mm]
{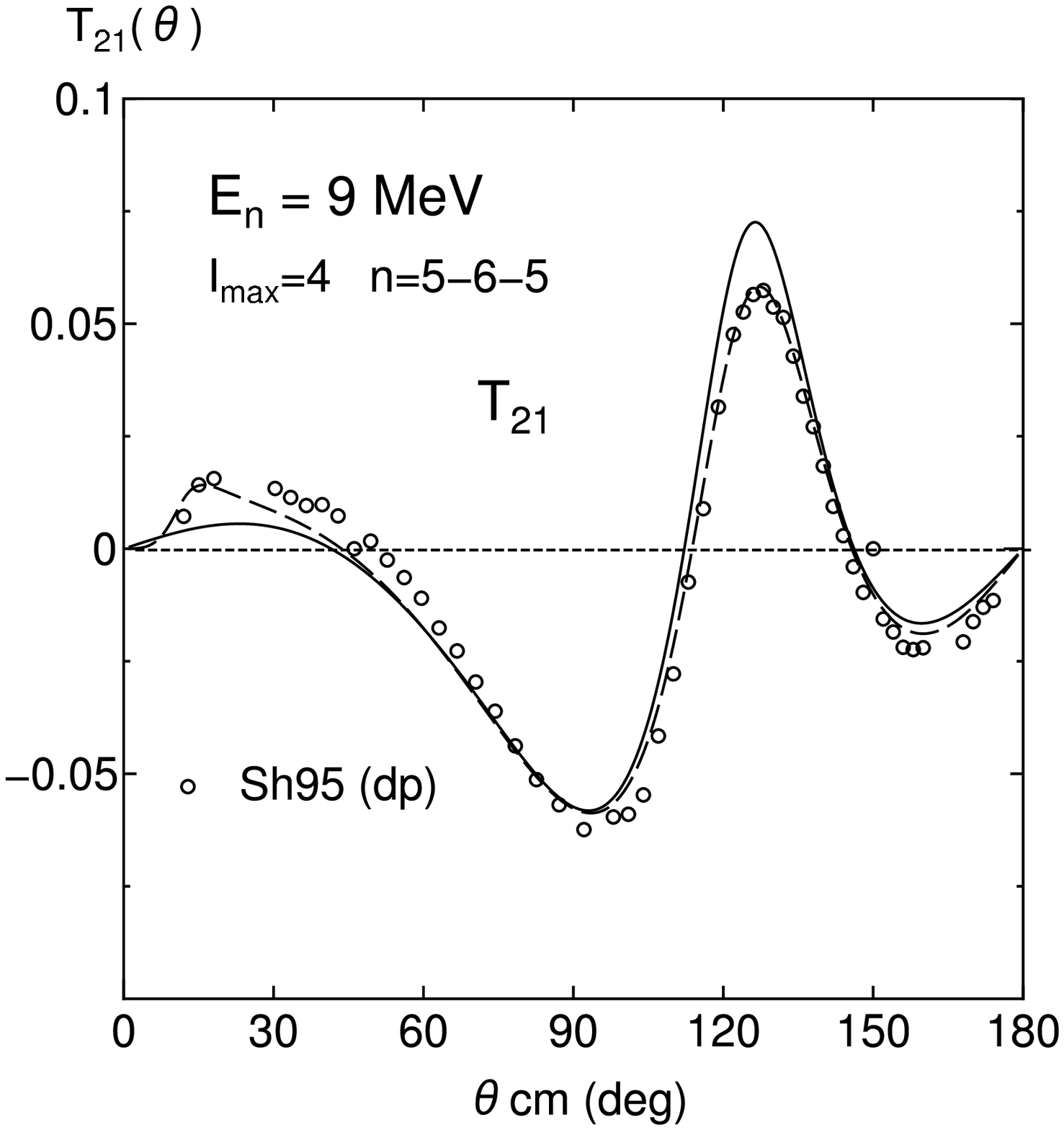}

\includegraphics[angle=0,width=55mm]
{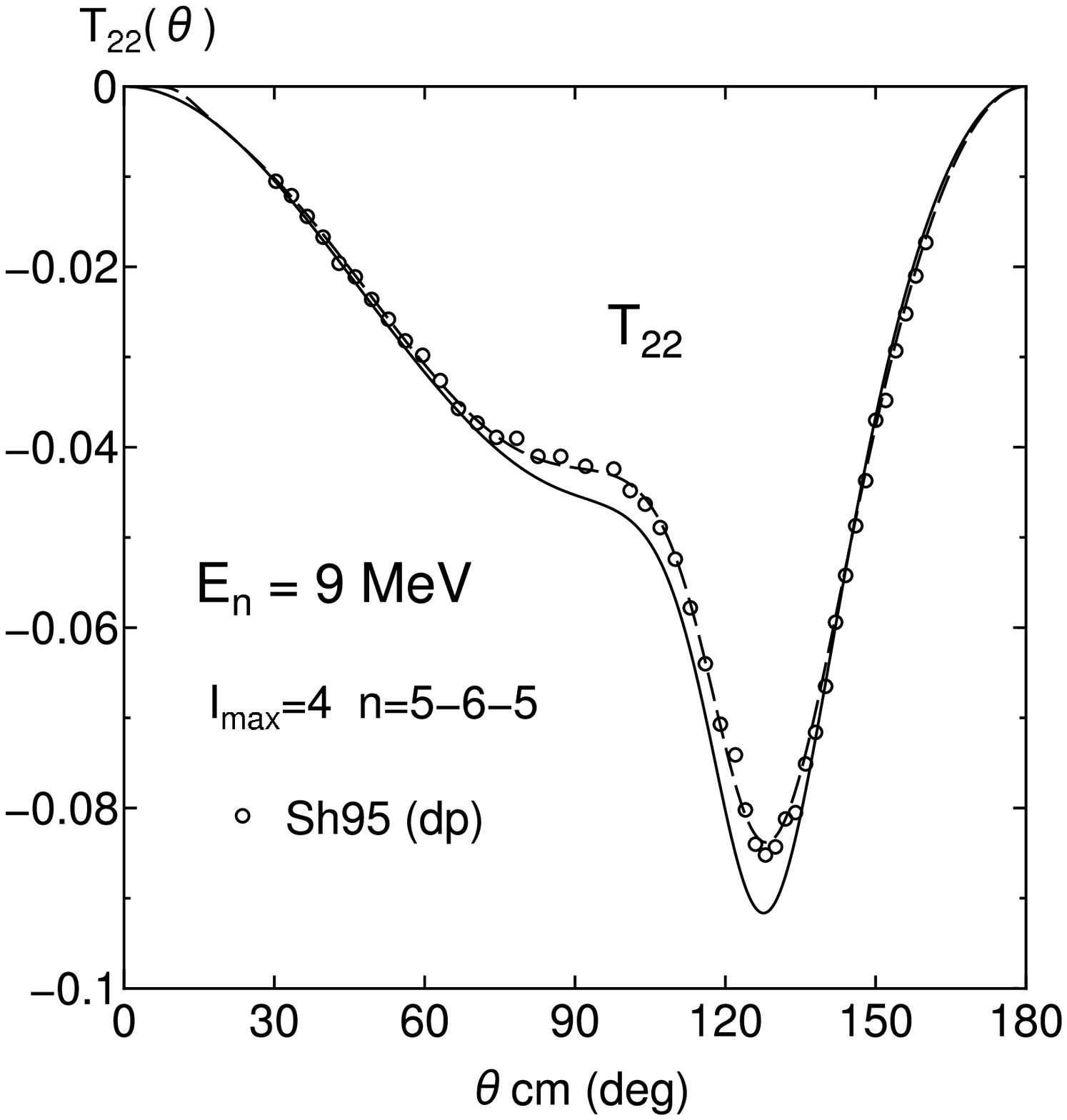}
\end{minipage}
\end{center}
\caption{
The same as Fig.\,\ref{T2m1}, but for the energies $E_n=7$ and 9 MeV.
}
\label{T2m2}       
\end{figure}

\begin{figure}[htb]
\begin{center}
\begin{minipage}{0.48\textwidth}
\includegraphics[angle=0,width=55mm]
{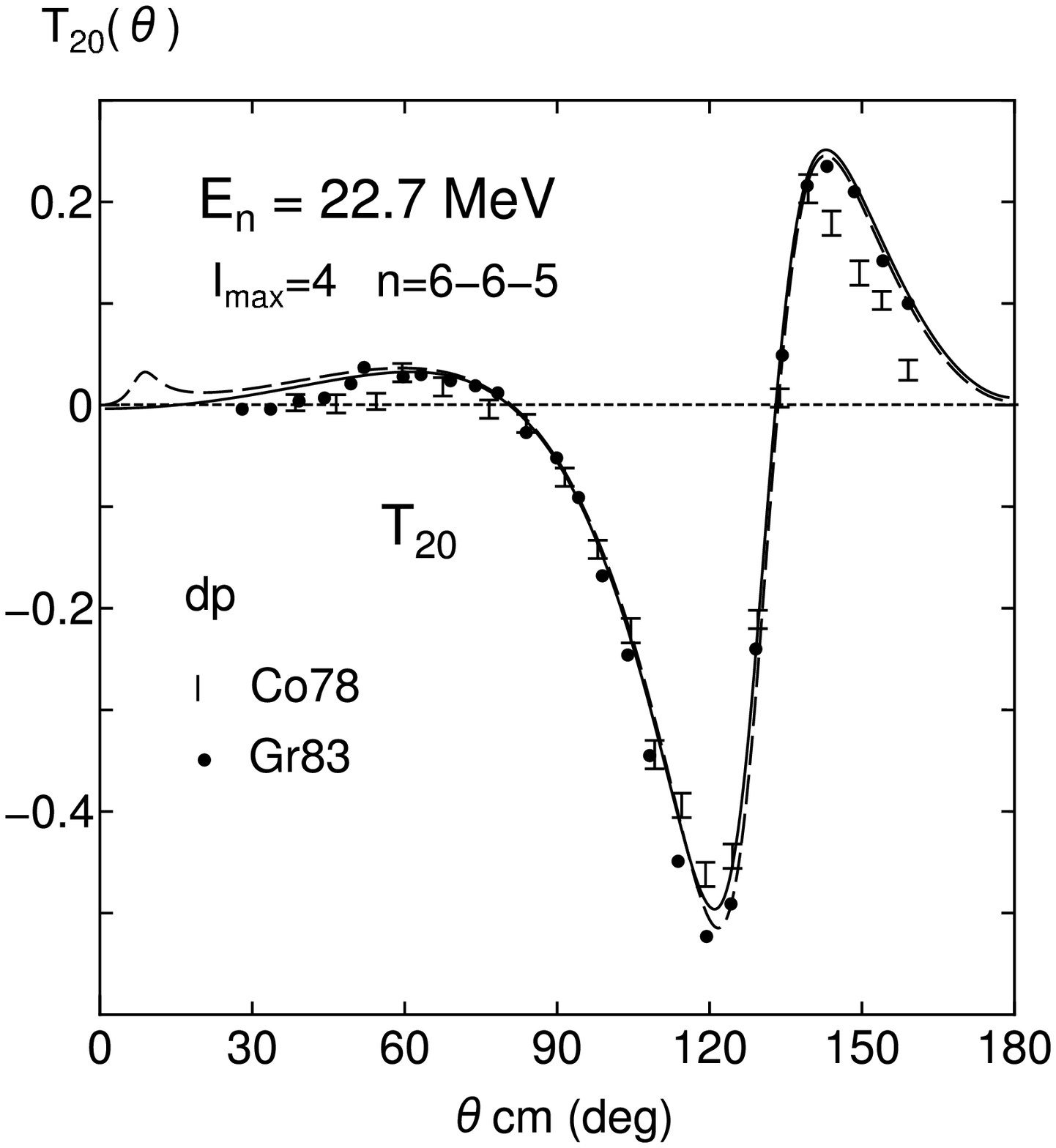}

\includegraphics[angle=0,width=55mm]
{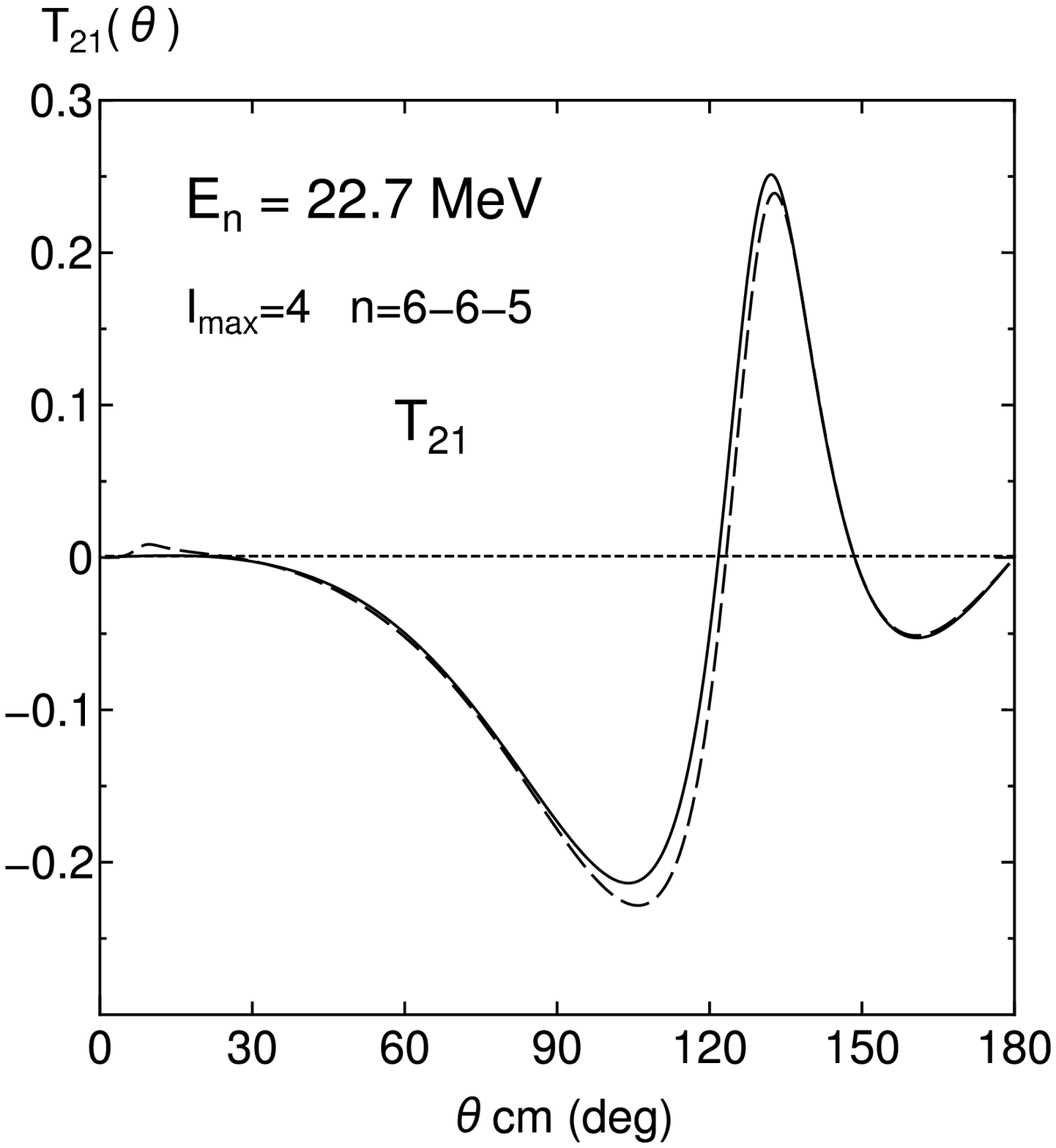}

\includegraphics[angle=0,width=55mm]
{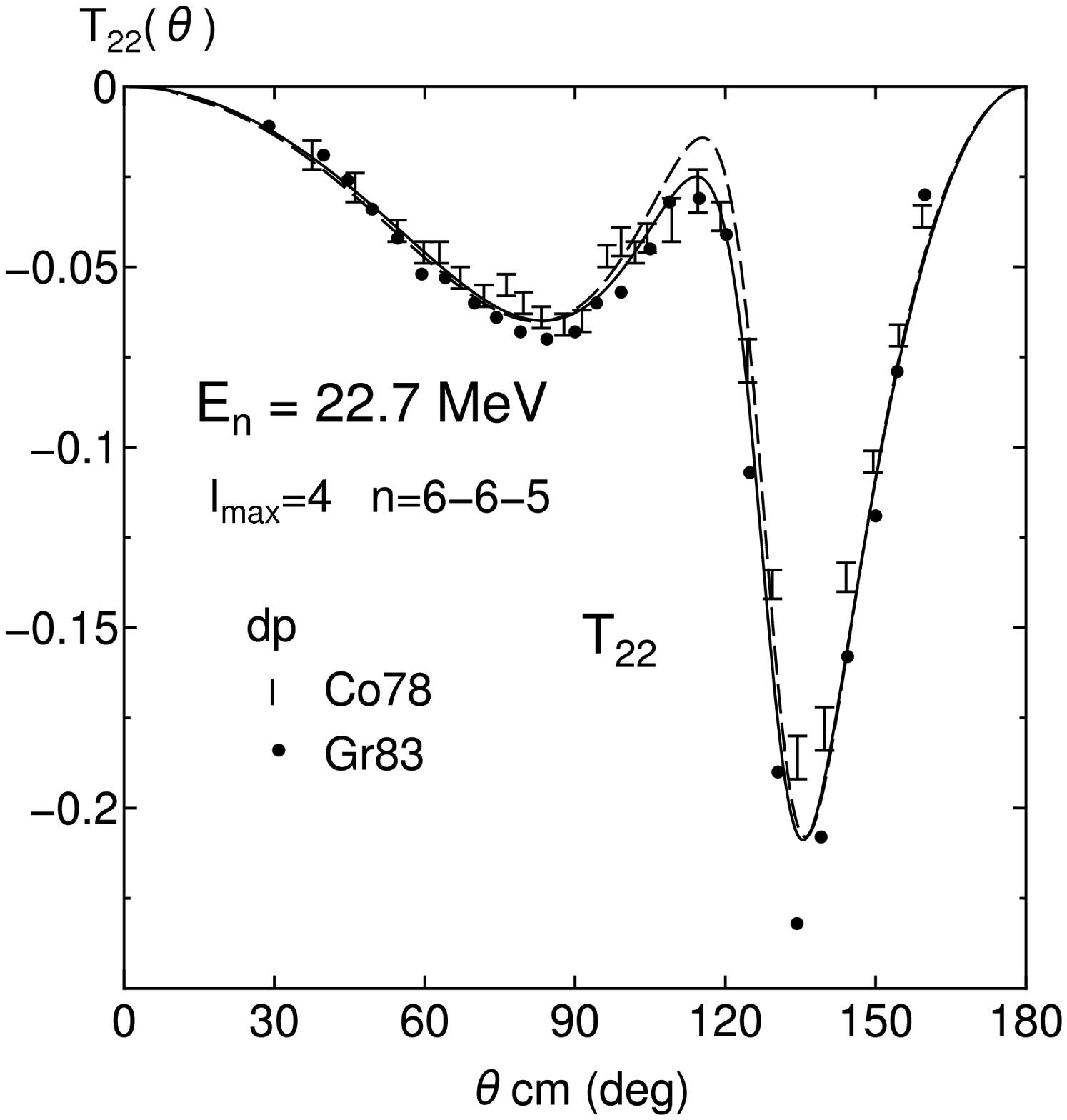}
\end{minipage}~%
\hfill~%
\begin{minipage}{0.48\textwidth}
\includegraphics[angle=0,width=55mm]
{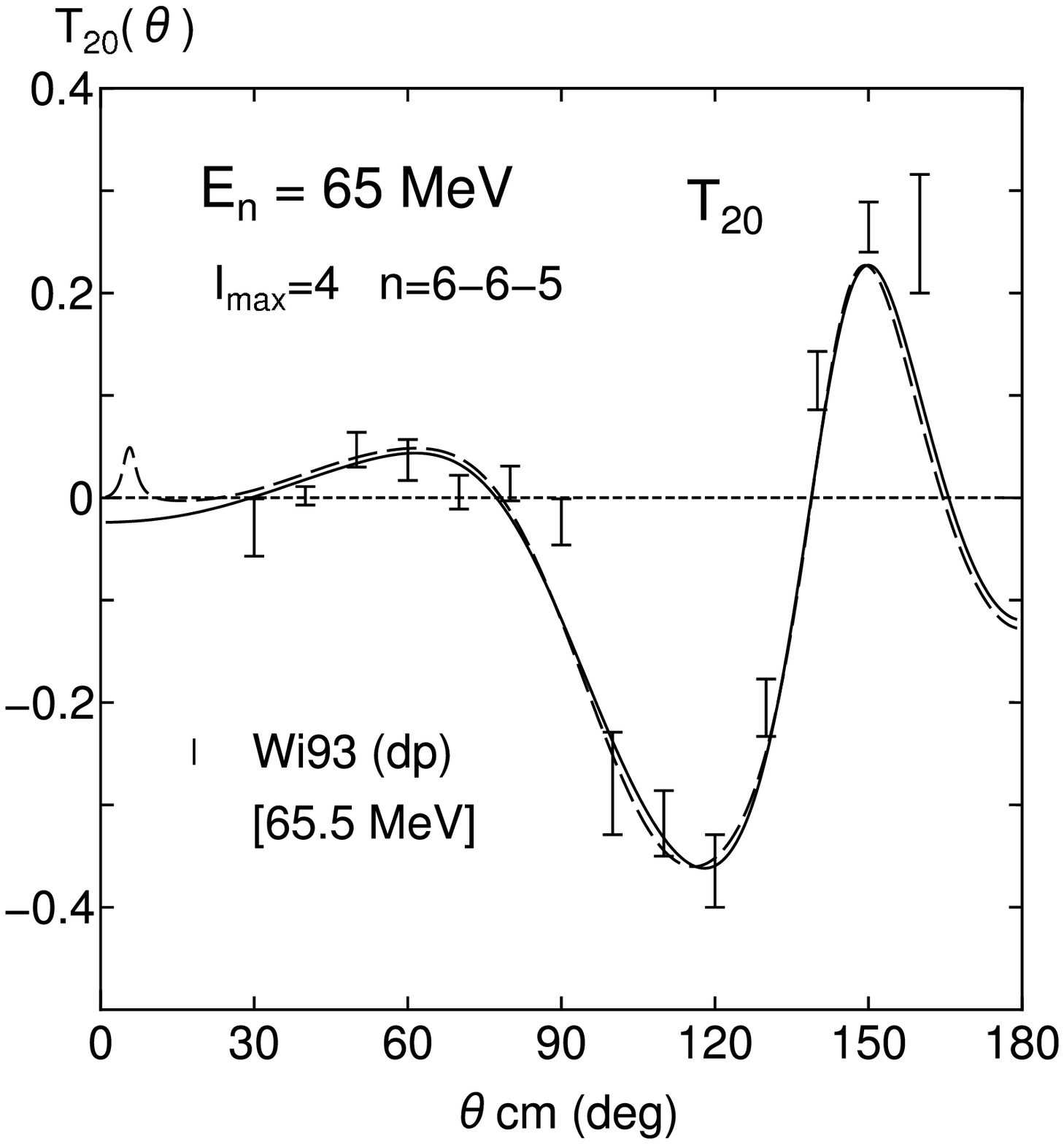}

\includegraphics[angle=0,width=55mm]
{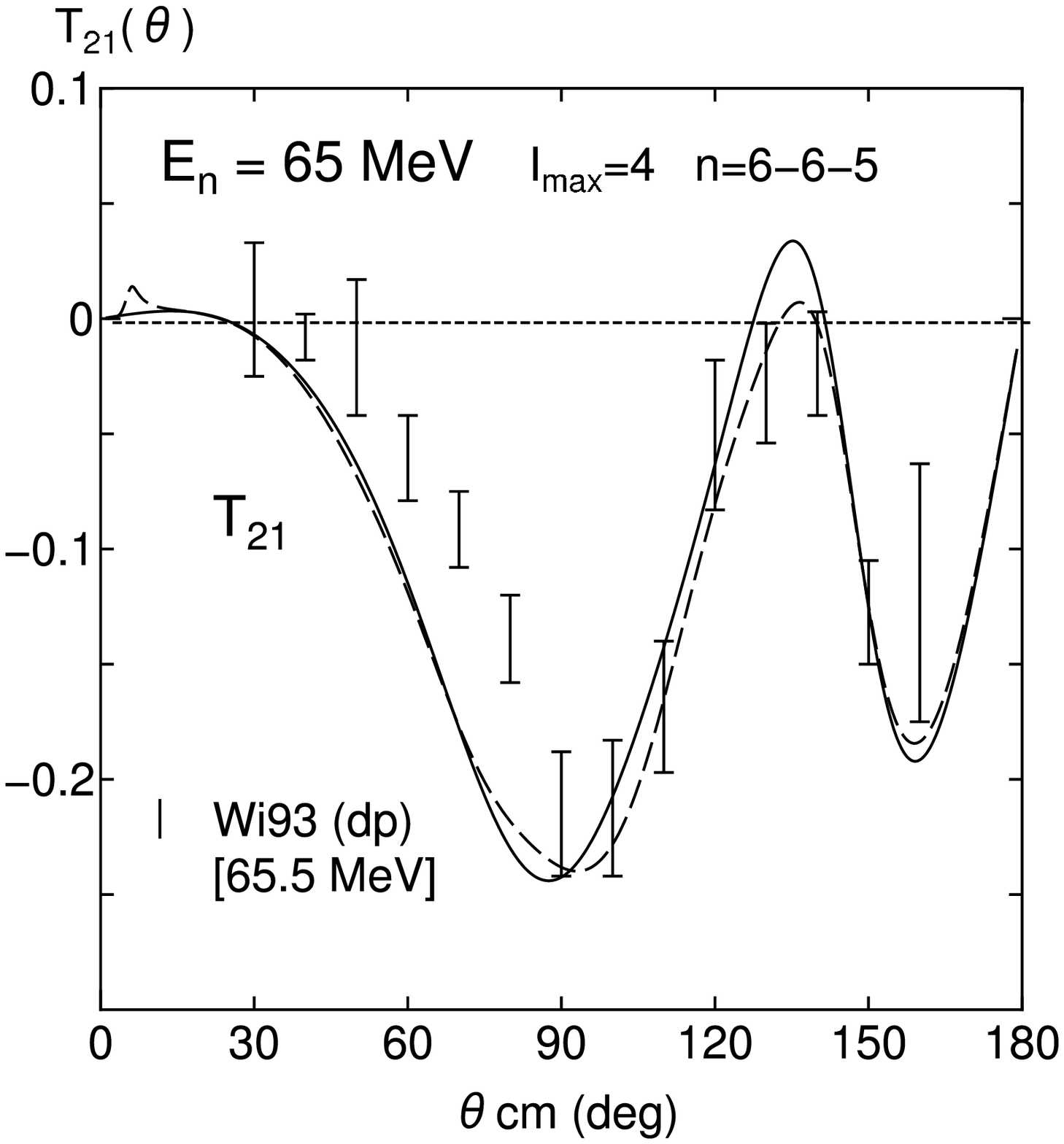}

\includegraphics[angle=0,width=55mm]
{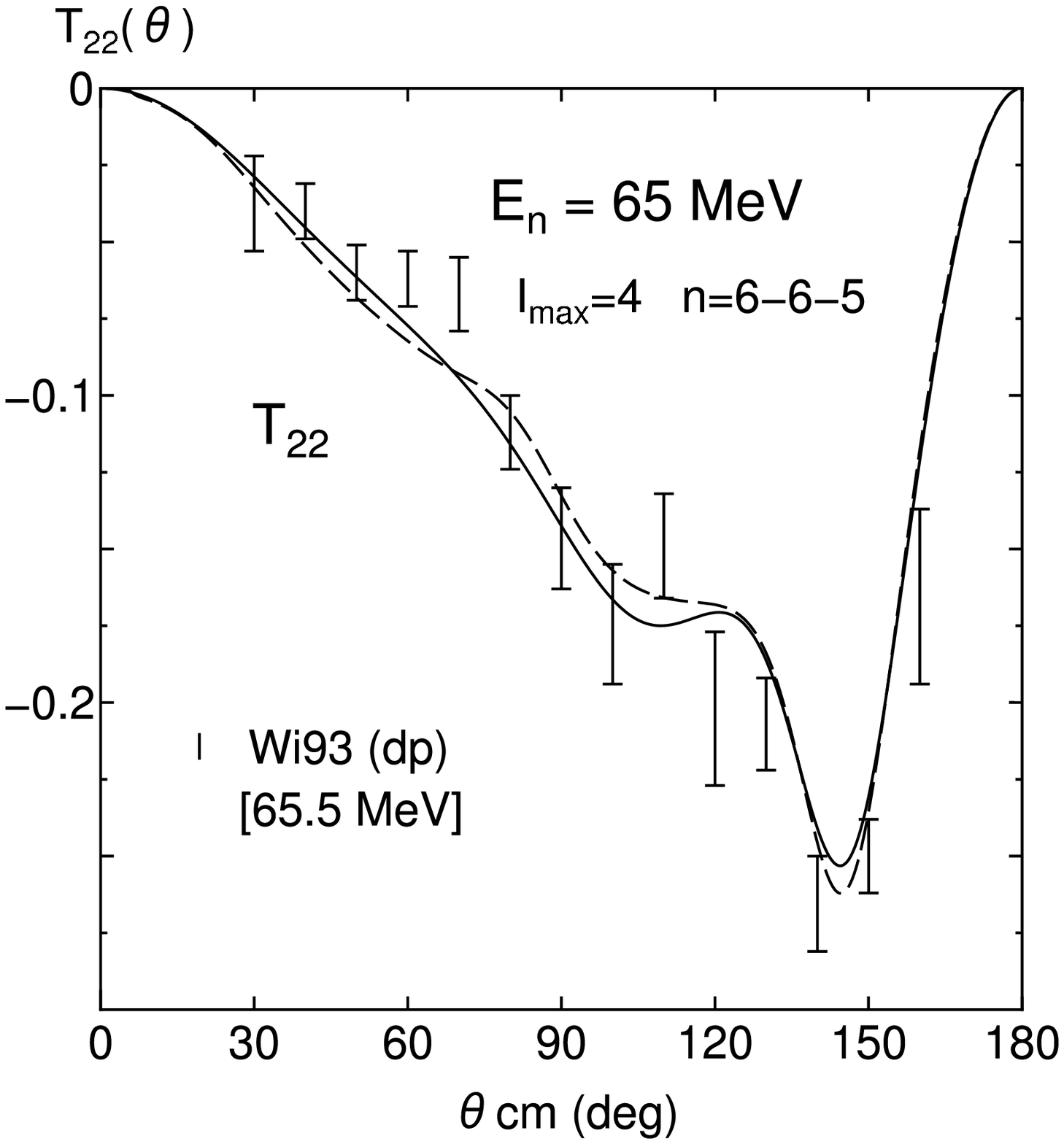}
\end{minipage}
\end{center}
\caption{
The same as Fig.\,\ref{T2m1}, but for the energies $E_n=22.7$ and 65 MeV.
The experimental data are taken from Refs.\,\citen{Co78} for Co78,
\citen{Gr83} for Gr83, and \citen{Wi93} for Wi93.
}
\label{T2m3}       
\end{figure}

We show the vector analyzing-powers of the deuteron
in Fig.\,\ref{T11}, and the tensor analyzing-powers
in Figs.\,\ref{T2m1} - \ref{T2m3}.
All of these observables are measured from the $dp$ elastic
scattering using the polarized deuteron, and discussion
including the Coulomb force is definitely necessary for the
comparison between the theory and experiment.
The introduction of the Coulomb force to the vector
analyzing-power $iT_{11}$ in Fig.\,\ref{T11}, shown with the
dashed curves leads to the enhancement at $\theta_{\rm cm} \leq 30^\circ$
and the reduction at $\theta_{\rm cm}=80^\circ$ - $120^\circ$
for the low-energies $E_p=3$ - 9 MeV, that are very similar to
the situation of $A_y$. 
We have unpleasant rise of the hill at the angular region 
$\theta_{\rm cm}=20^\circ$ - $90^\circ$ for the energies
$E_p=3$ - 10 MeV, which is probably related to the inadequacy
of the nuclear-Coulomb interference term.
The peak height around $\theta_{\rm cm} = 120^\circ$ is
somewhat too low at $E_p=3$ - 9 MeV,
although it is not so serious as in Fig.\,5 of Ref.\,\citen{Ki01}.
Except for these, there is no clear discrepancies
between the theory and experiment. 
As to the first problem, the enhancement by the Coulomb effect
is also seen in other calculations by Berthold et al.\cite{Be90},
Alt et al.\cite{Al02}, Kievsky et al.\cite{Ki01} and  
Deltuva et al.\cite{De05,De05c}, although to less extent
in the last two cases. It is possible that the corresponding
$nd$ quantities are too large.

The tensor-type analyzing-powers $T_{20},~T_{21}$ and $T_{22}$
for the energies $E_p=3$ and 5 MeV in Fig.\,\ref{T2m1} show 
a fairly large Coulomb effect in the whole angles.
In particular, the forward behavior of $T_{20}$ and $T_{21}$ 
in $\theta_{\rm cm} \leq 60^\circ$ for the no-Coulomb calculation
(solid curves) is entirely modified by the Coulomb 
effect, resulting in a good agreement with the $dp$ experimental data.
We have some problems with $T_{20}$ and $T_{21}$ in the angular region
$\theta_{\rm cm}=20^\circ$-$70^\circ$ for the energies $E_p=7$ - 9 MeV,
which might again related to the nuclear-Coulomb interference.
In $T_{22}$, too large negative
values at the minimum points are shifted to the smaller direction,
giving better reproduction of the experimental data.
On the low-energy side with $E_p \le 3$ MeV, the minimum points
of the dips are slightly too high.
Related to this, the perfect fit of $T_{22}$ in Ref.\,\citen{PREP}
at $E_{\rm lab} \leq 10$ MeV is fortuitous, because this calculation
does not include the Coulomb force.
The Coulomb modification diminishes smaller and smaller for
the higher energies $E_p=7$ and 9 MeV in Fig.\,\ref{T2m2},
and for $E_p=22.7$ and 65 MeV in Fig.\,\ref{T2m3}.
The big difference from the experiment in 65 MeV $T_{21}$ at $\theta_{\rm cm}=
50^\circ$ - $90^\circ$ is a common feature with other calculations.\cite{PREP} 
In order to examine the adequacy of our Coulomb treatment,
we have compared our Coulomb effects with the results of calculations
by other authors.\cite{Be90,Al02,De05,De05c,Ki01,Is09}
For example, Ref.\,\citen{De05} shows the differential cross sections
and analyzing-powers of $E_p=3$ and 9 MeV in Figs.\,9 and 10, respectively,
for the coupled-channel potentials of CD Bonn, 
including the



\bigskip

\clearpage

\noindent
$\Delta$ degree of freedom. We find that the general tendency is common,
although some quantitative differences exist because of the different
treatments of the Coulomb force.

\bigskip

\subsection{Analysis of the Coulomb effect
below the deuteron breakup threshold}


In our previous paper,\cite{scl10} we have examined the Coulomb
effect on the $pd$ differential cross sections
for the incident energies below the deuteron breakup threshold. 
In this low-energy region, the channel spin $S_c$ is an almost good
quantum number, and the $J$-averaging of the eigenphase shifts
with respect to the definite $S_c$ and the orbital angular
momentum $\ell$ between the nucleon and the deuteron is very convenient
to reduce the number of phase shift parameters,
at least in discussing the angular distribution of differential
cross sections. 
However, a simple prescription adding the Coulomb amplitude
to the $nd$ scattering amplitude only with the Coulomb phase-shift factors,
which is called the ``Coulomb externally corrected'' approximation
in Ref.\,\citen{De05}, does not work for the differential cross sections
in the low-energy region. This prescription largely overestimates the
differential cross sections. This implies that the modification
of the nuclear phase shifts by the Coulomb force is very important
below the deuteron breakup threshold.
In Ref.\,\citen{scl10}, the Coulomb modification to the nuclear 
phase shifts is therefore incorporated, 
using the difference of the $J$-averaged eigenphase shifts for the
$nd$ and $pd$ scatterings first from the AV18 potential
in Ref.\,\citen{Ki96}, secondly from the fss2 calculated here.
In both cases, we have obtained an almost complete reproduction 
of the differential cross sections by this prescription. 

\begin{figure}[htb]
\begin{center}
\begin{minipage}{0.48\textwidth}
\includegraphics[angle=0,width=54mm]
{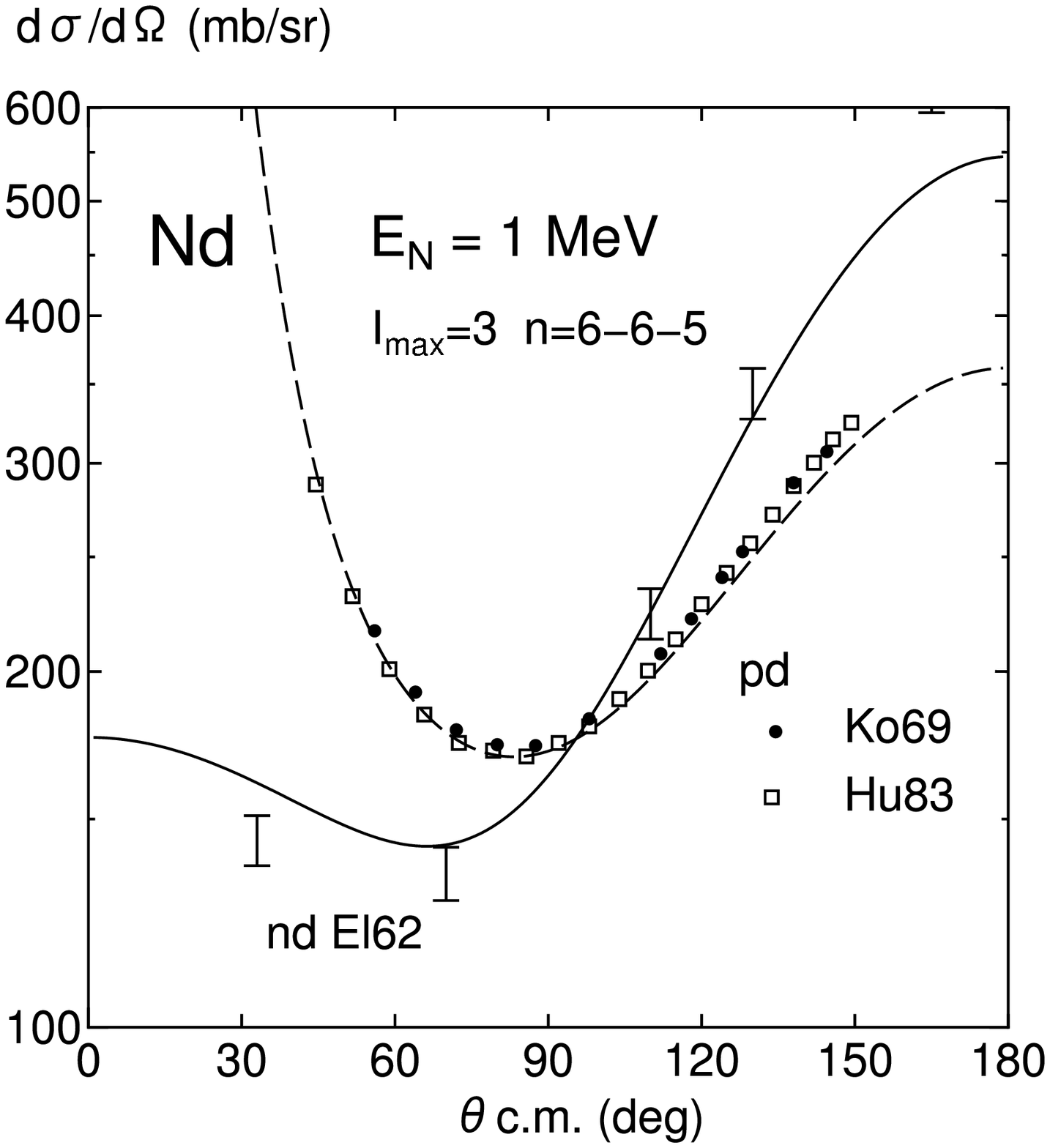}

\includegraphics[angle=0,width=54mm]
{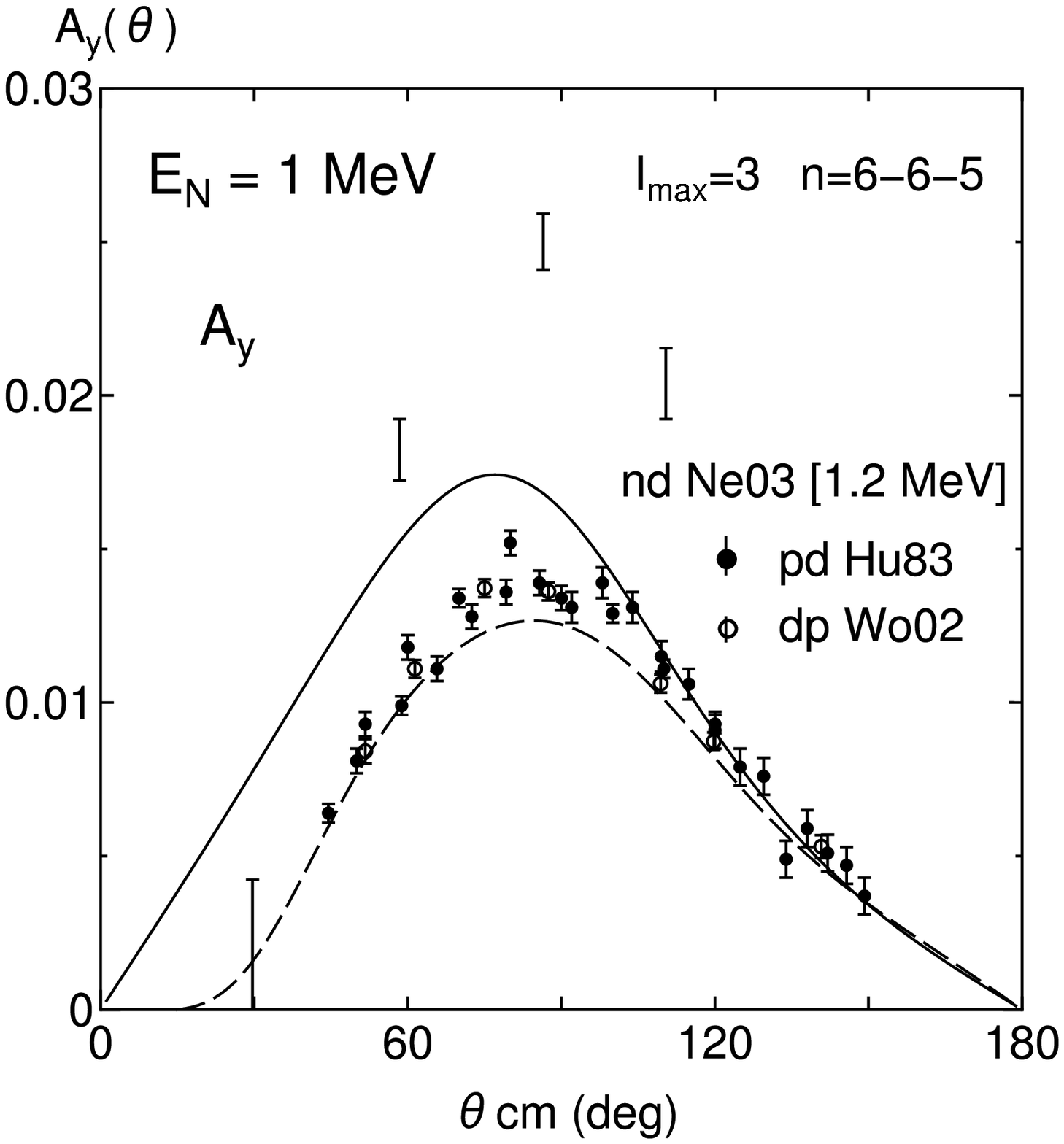}

\includegraphics[angle=0,width=54mm]
{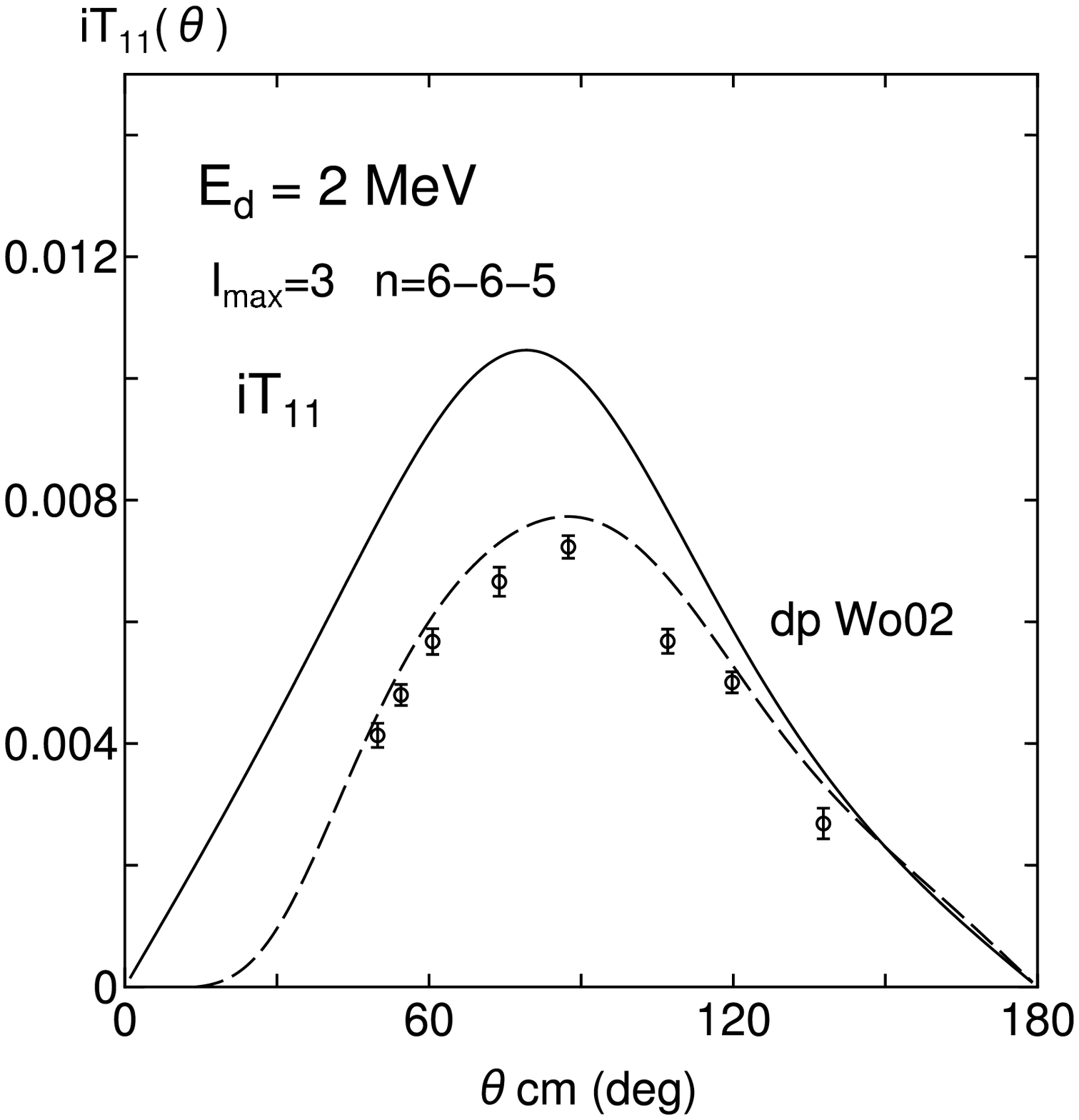}
\end{minipage}~%
\hfill~%
\begin{minipage}{0.48\textwidth}
\includegraphics[angle=0,width=54mm]
{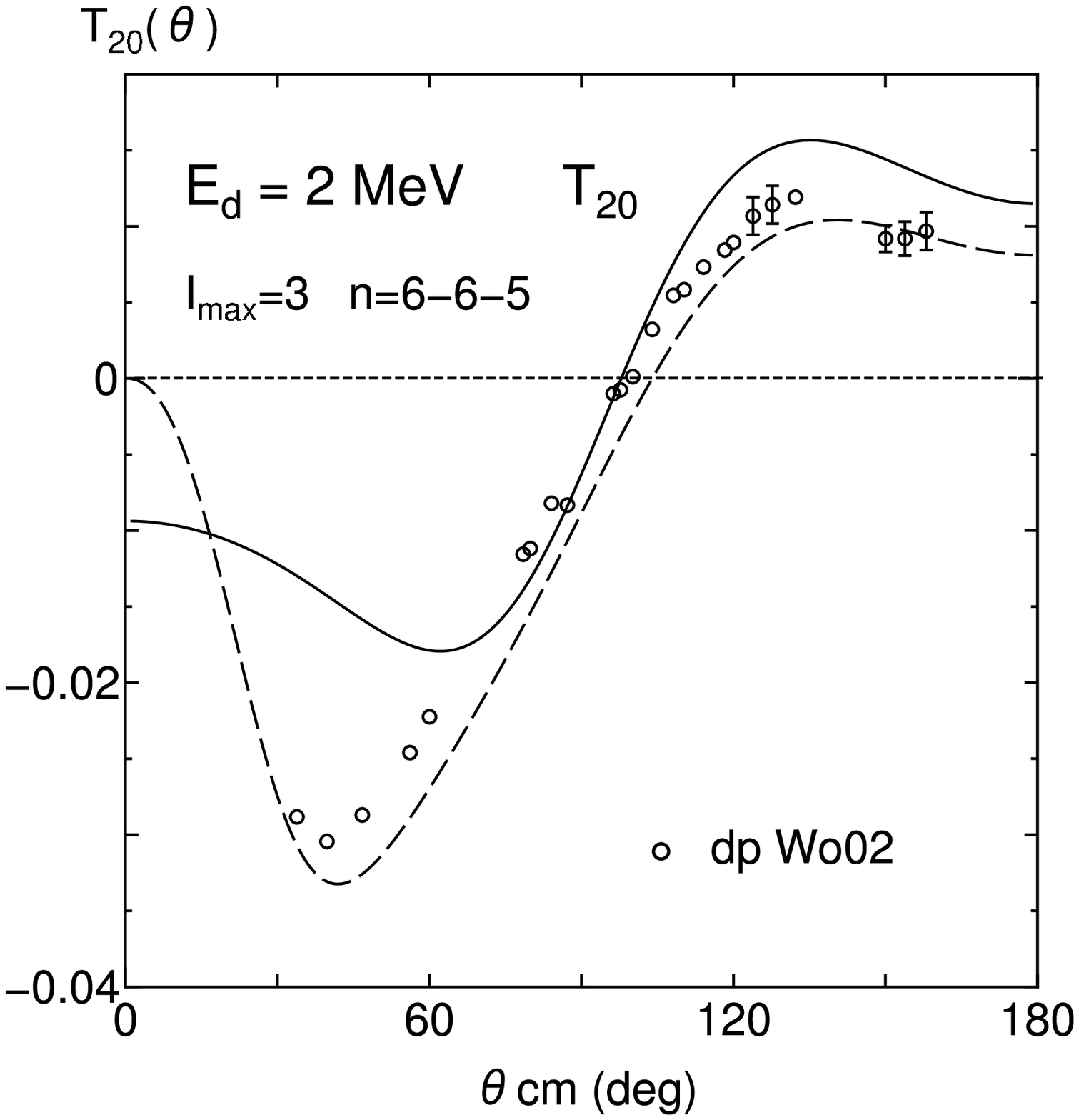}

\includegraphics[angle=0,width=54mm]
{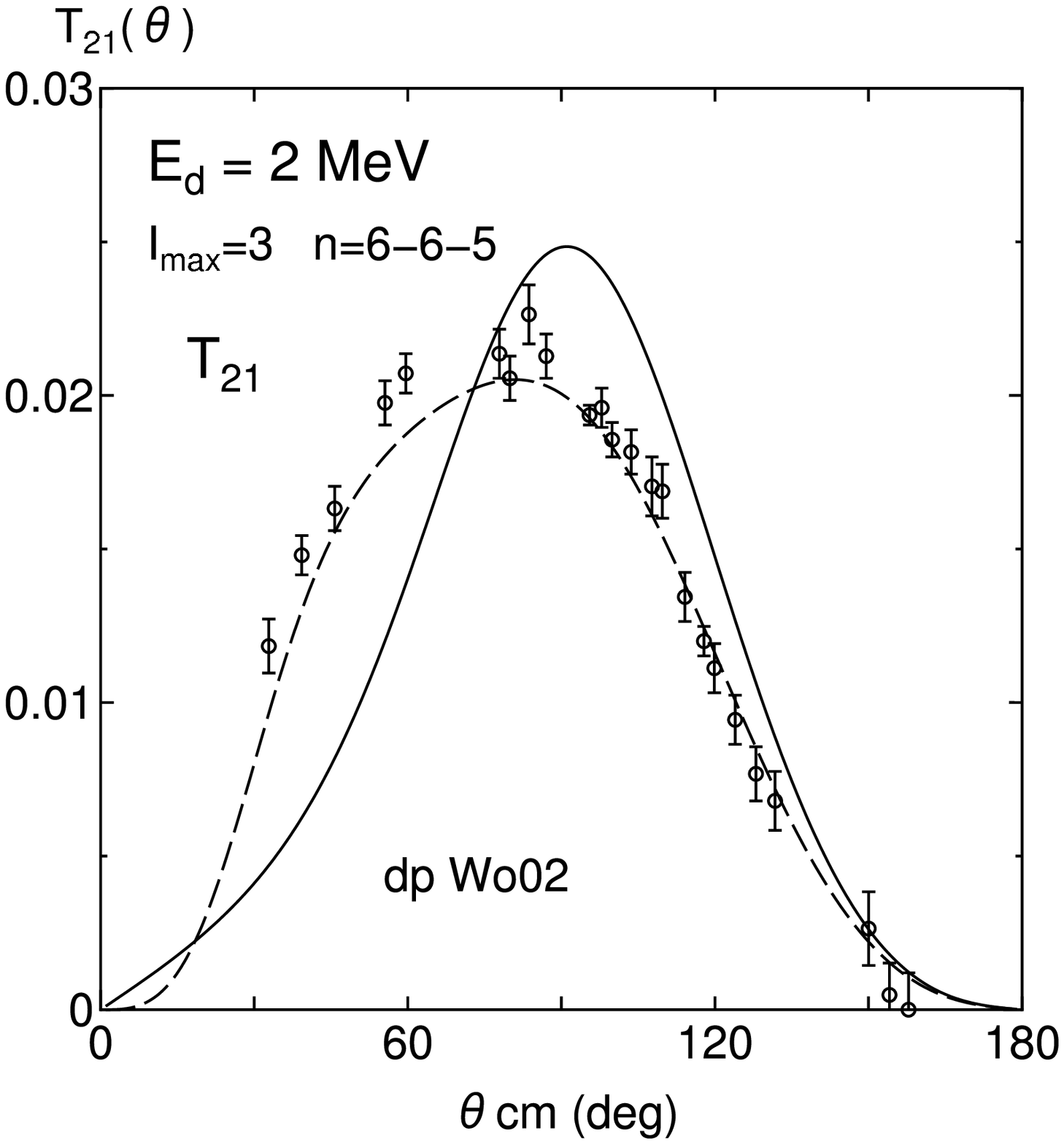}

\includegraphics[angle=0,width=54mm]
{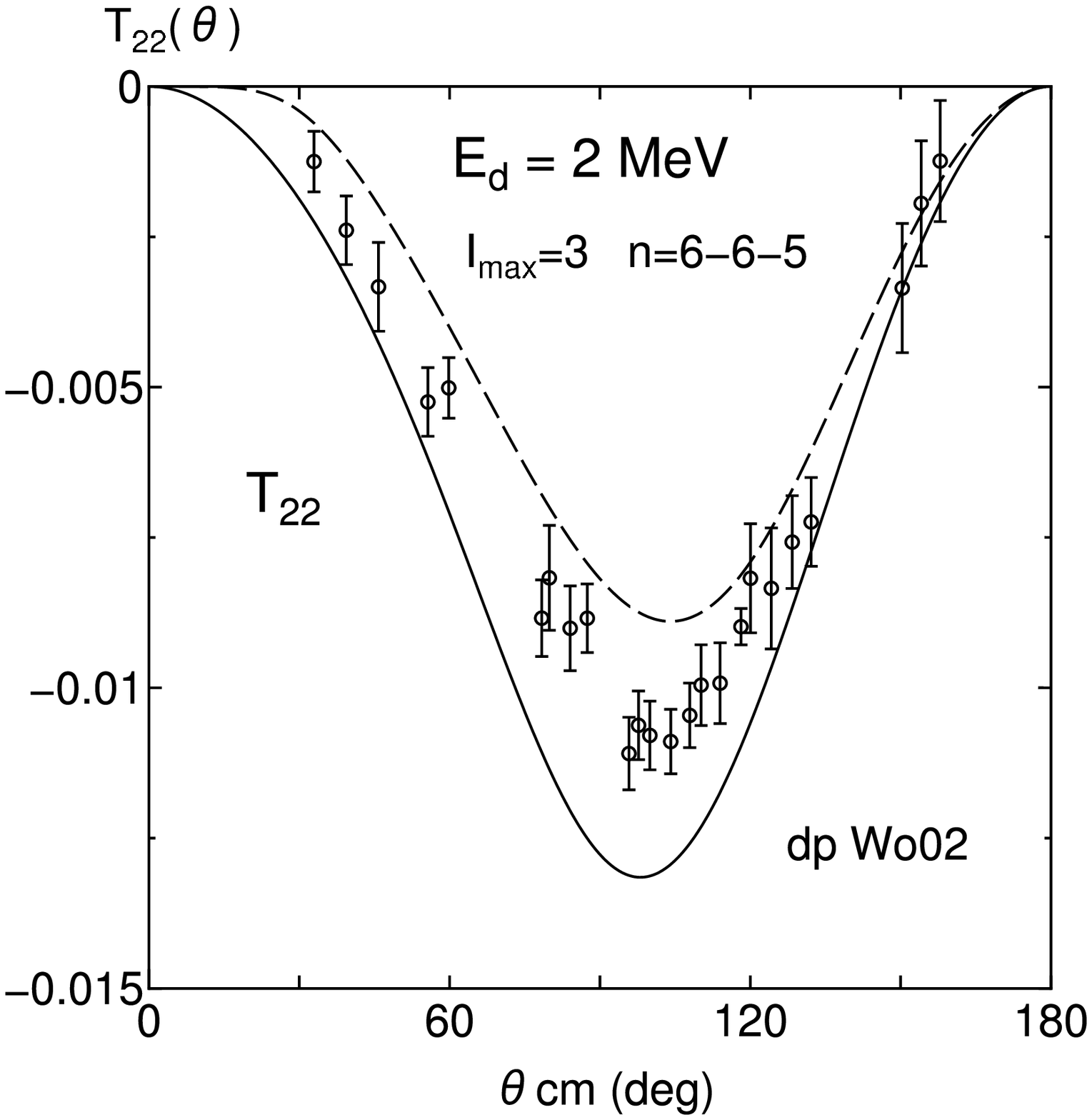}
\end{minipage}
\end{center}
\caption{
The Coulomb effects on the differential cross sections and analyzing-powers
for the $pd$ elastic scattering with the incident energy 
$E/{\rm nucleon}=1$ MeV.
The solid curves denote no-Coulomb calculation and
the dashed curves the results of the cut-off Coulomb force, which
should be compared with the empty circles or the filled circles
with bars.
The experimental data are taken from Ref.\,\citen{El62} for El62 ($nd$)
and from Refs.\,\citen{Ko69} for Ko69 ($pd$), \citen{Hu83} for Hu83 ($pd$), 
\citen{Ne03} for Ne03 ($nd$), and \citen{Wo02} for Wo02 ($dp$). 
}
\label{EN1}       
\end{figure}
\begin{figure}[htb]
\begin{center}
\begin{minipage}{0.48\textwidth}
\includegraphics[angle=0,width=56mm]
{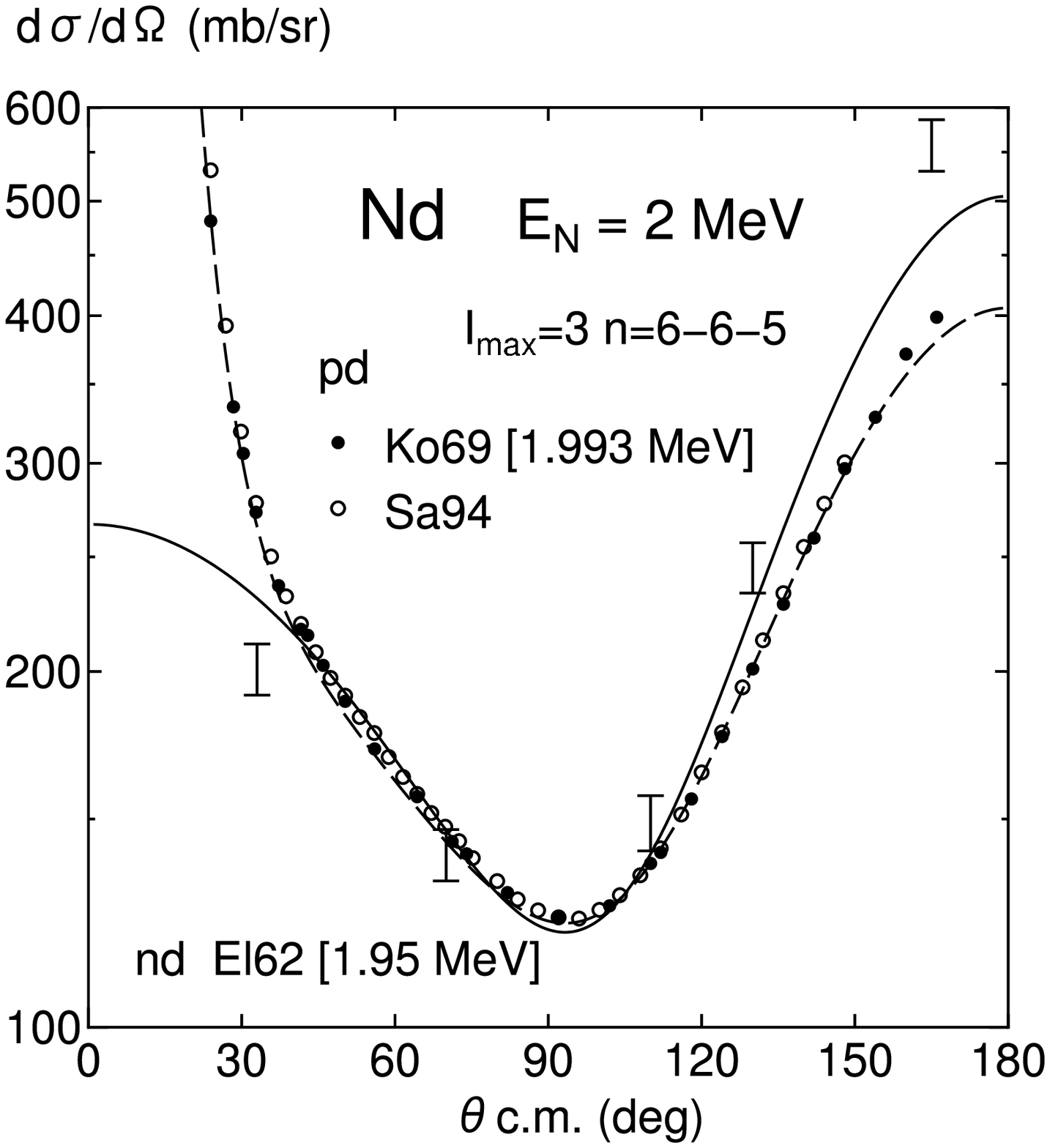}

\vspace{2mm}
\includegraphics[angle=0,width=56mm]
{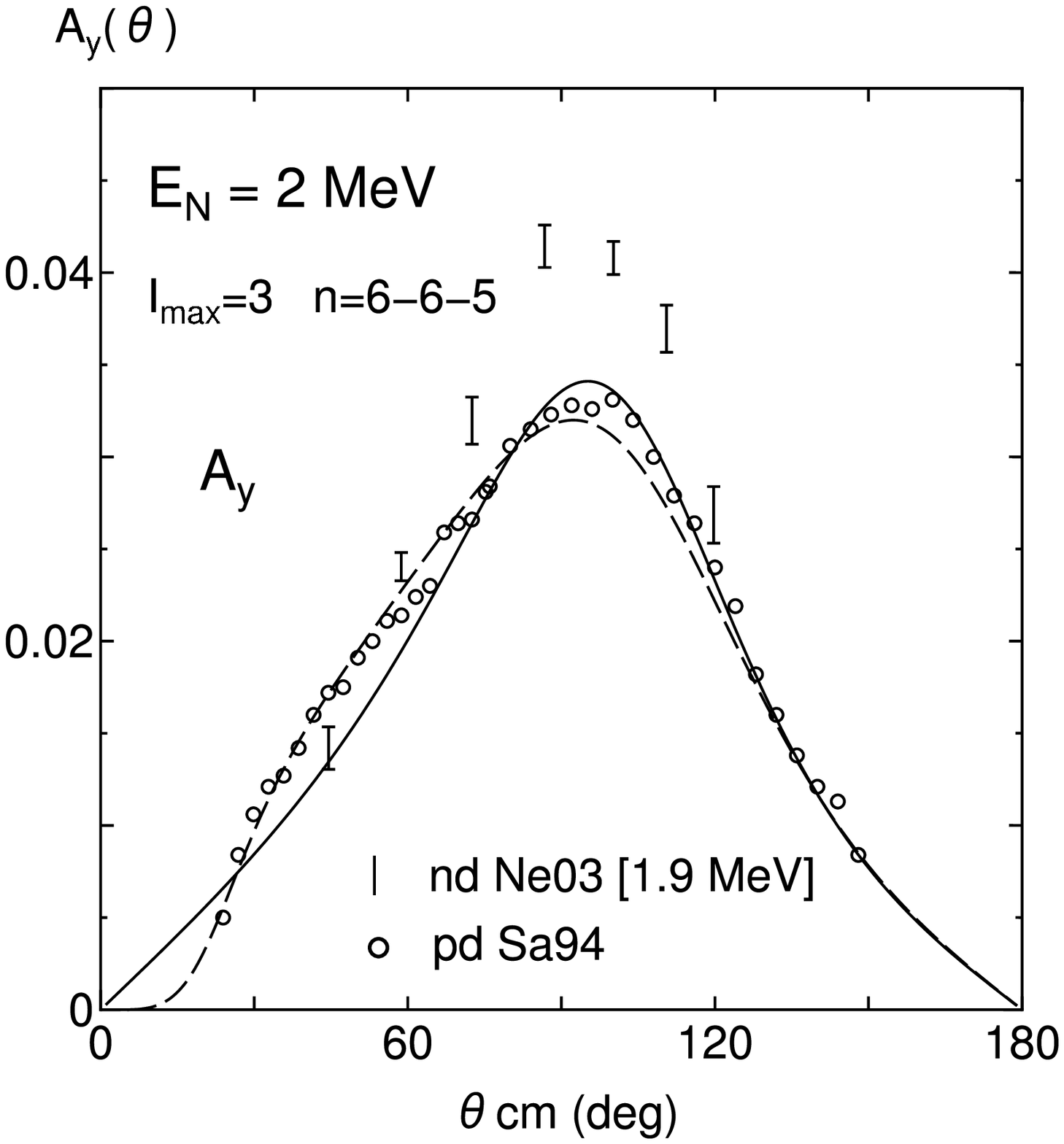}

\vspace{2mm}
\includegraphics[angle=0,width=56mm]
{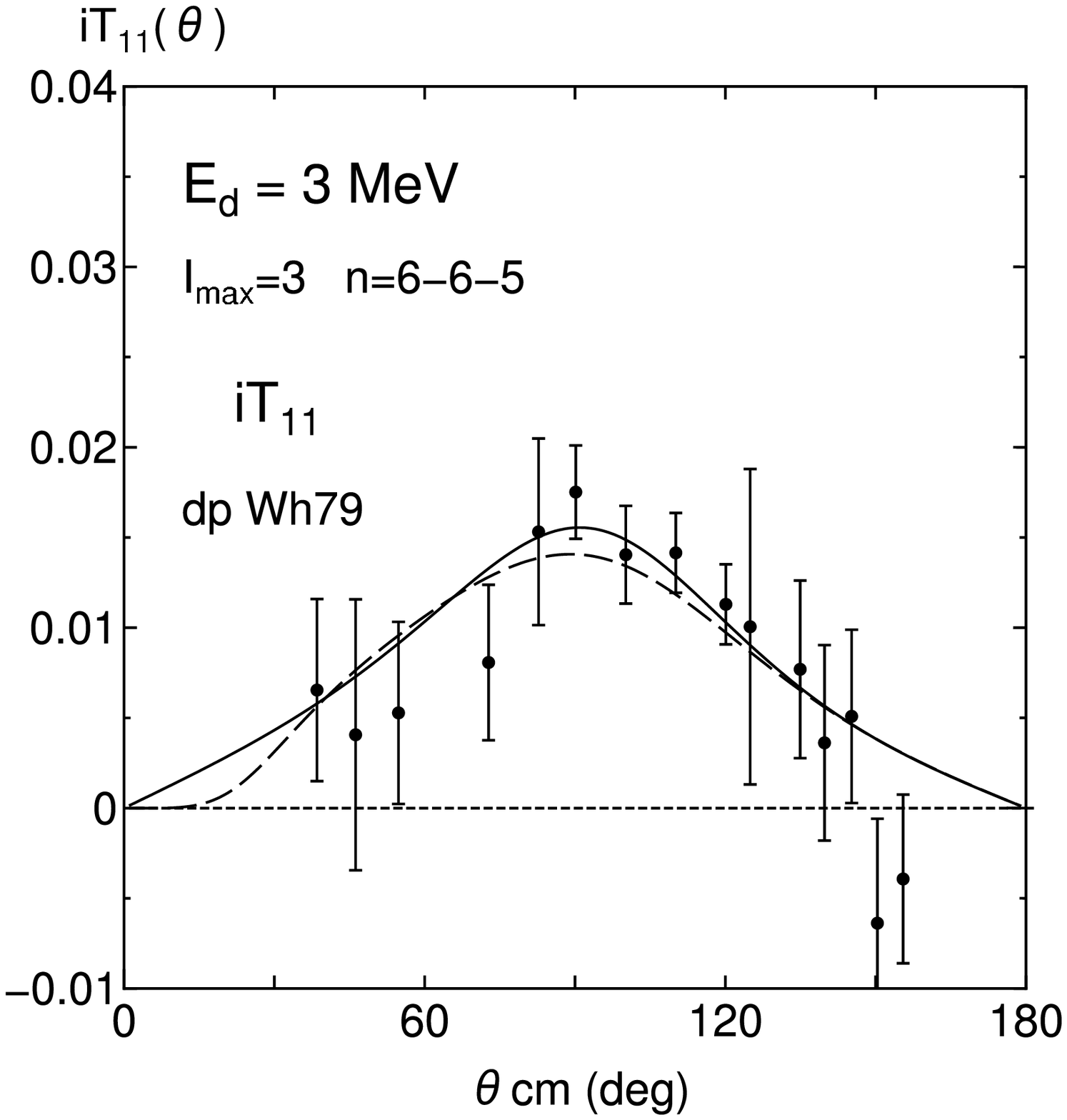}
\end{minipage}~%
\hfill~%
\begin{minipage}{0.48\textwidth}
\includegraphics[angle=0,width=56mm]
{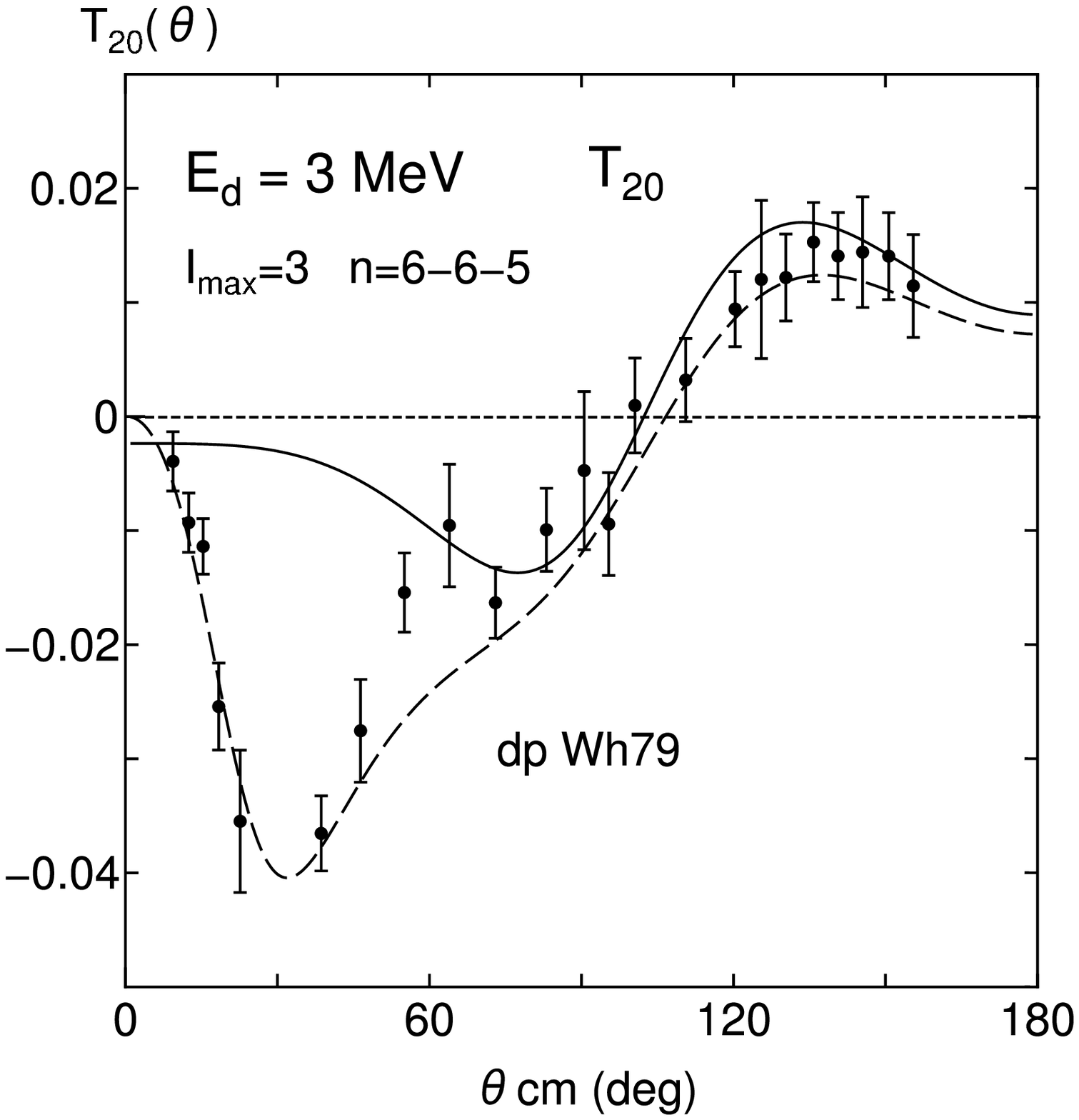}

\vspace{2mm}
\includegraphics[angle=0,width=56mm]
{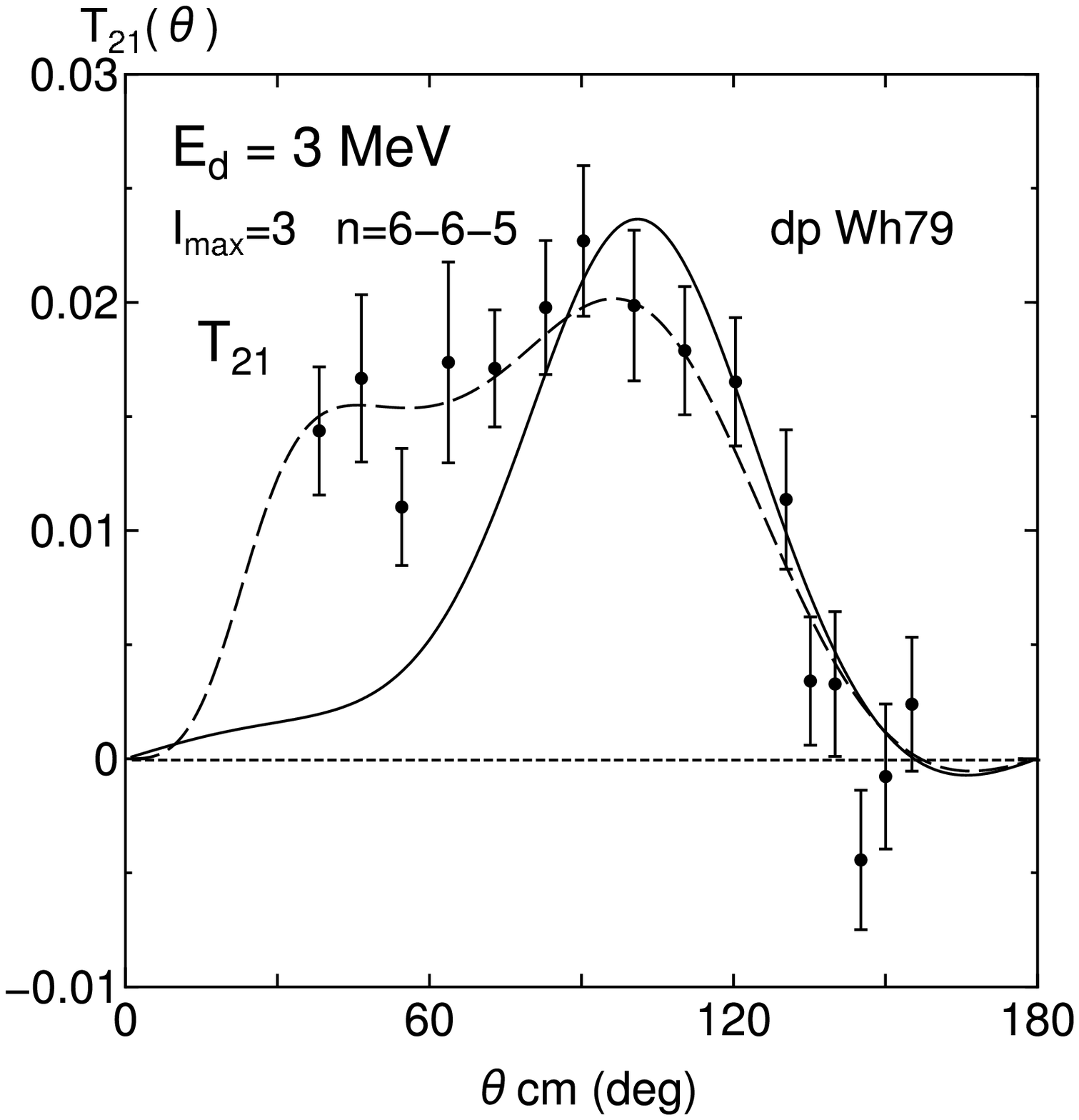}

\vspace{2mm}
\includegraphics[angle=0,width=56mm]
{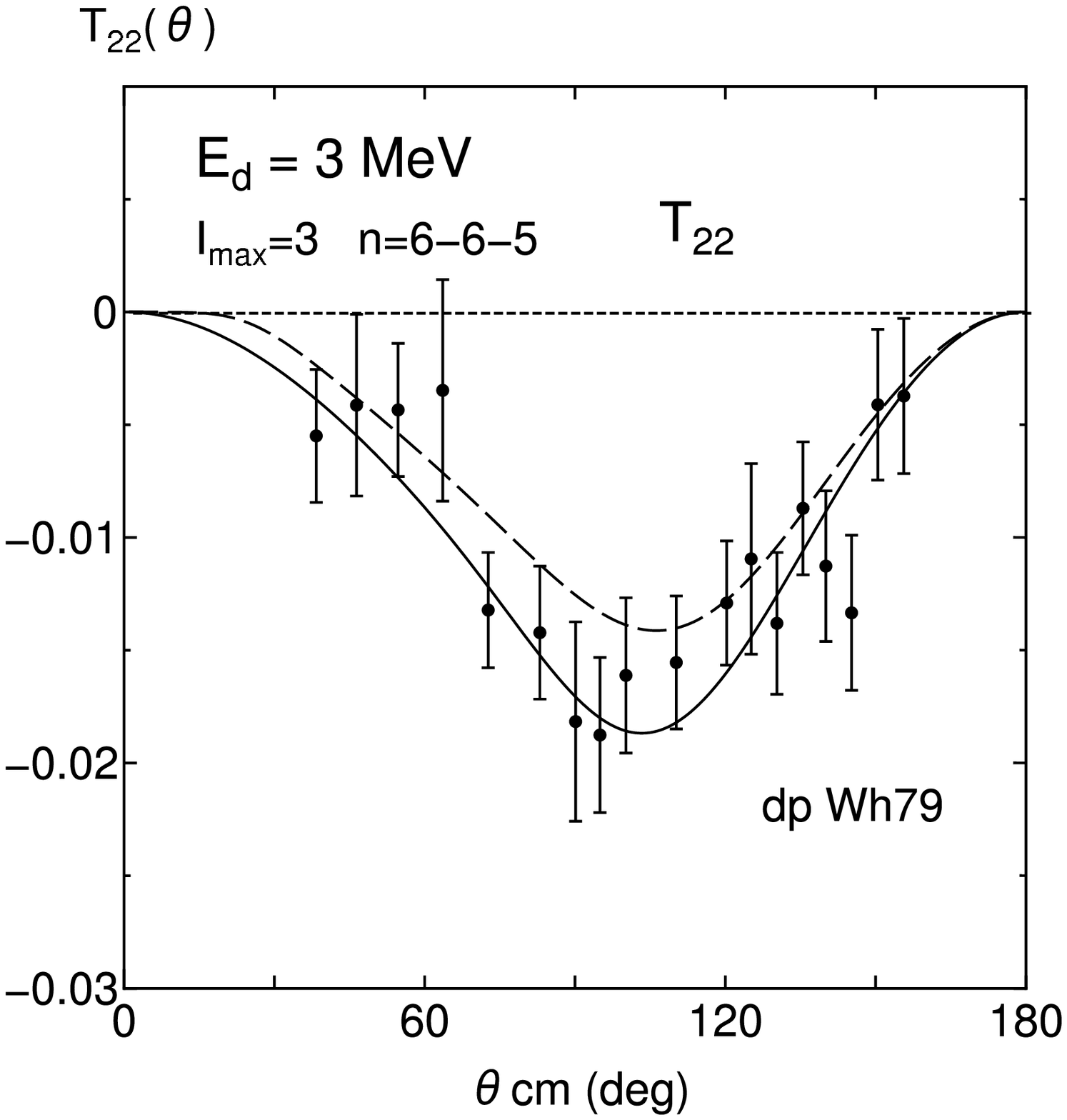}
\end{minipage}
\end{center}
\caption{
The same as Fig.\,\ref{EN1}, but for $(d\,\sigma/d\,\Omega)$ and
$A_y$ for $E_N=2$ MeV, and $iT_{11},~T_{2m}$ for the incident
energy $E/{\rm nucleon}=1.5$ MeV.
The experimental data are taken from Refs.\,\citen{Sa94} for Sa94 ($pd$) 
and \citen{Wh79} for Wh79 ($dp$).
The others are the same as in Fig.\,\protect\ref{EN1}.
}
\label{EN2}       
\end{figure}
\begin{figure}[htb]
\begin{center}
\begin{minipage}{0.48\textwidth}
\includegraphics[angle=0,width=56mm]
{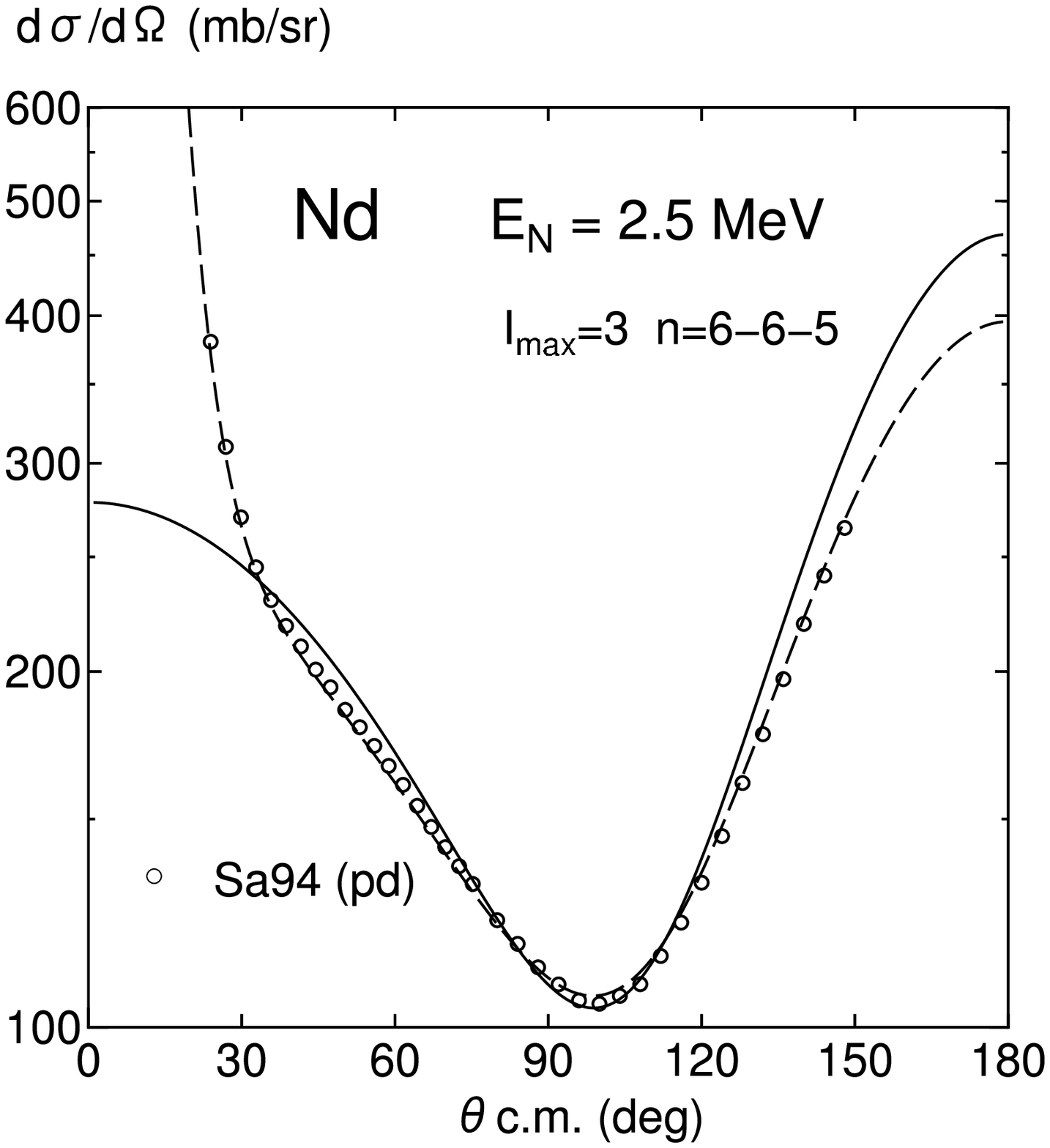}

\vspace{4mm}
\includegraphics[angle=0,width=56mm]
{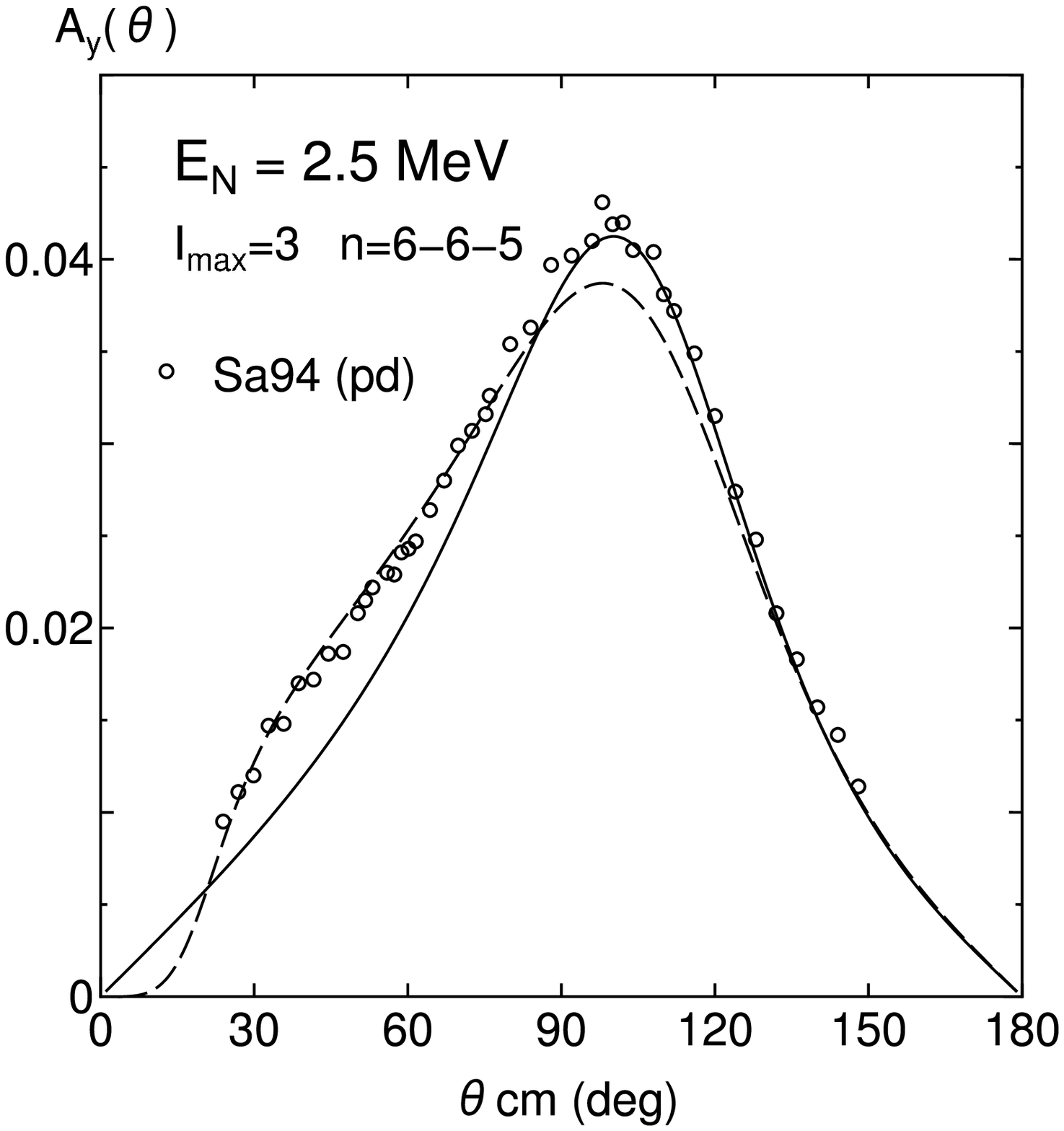}

\vspace{4mm}
\includegraphics[angle=0,width=56mm]
{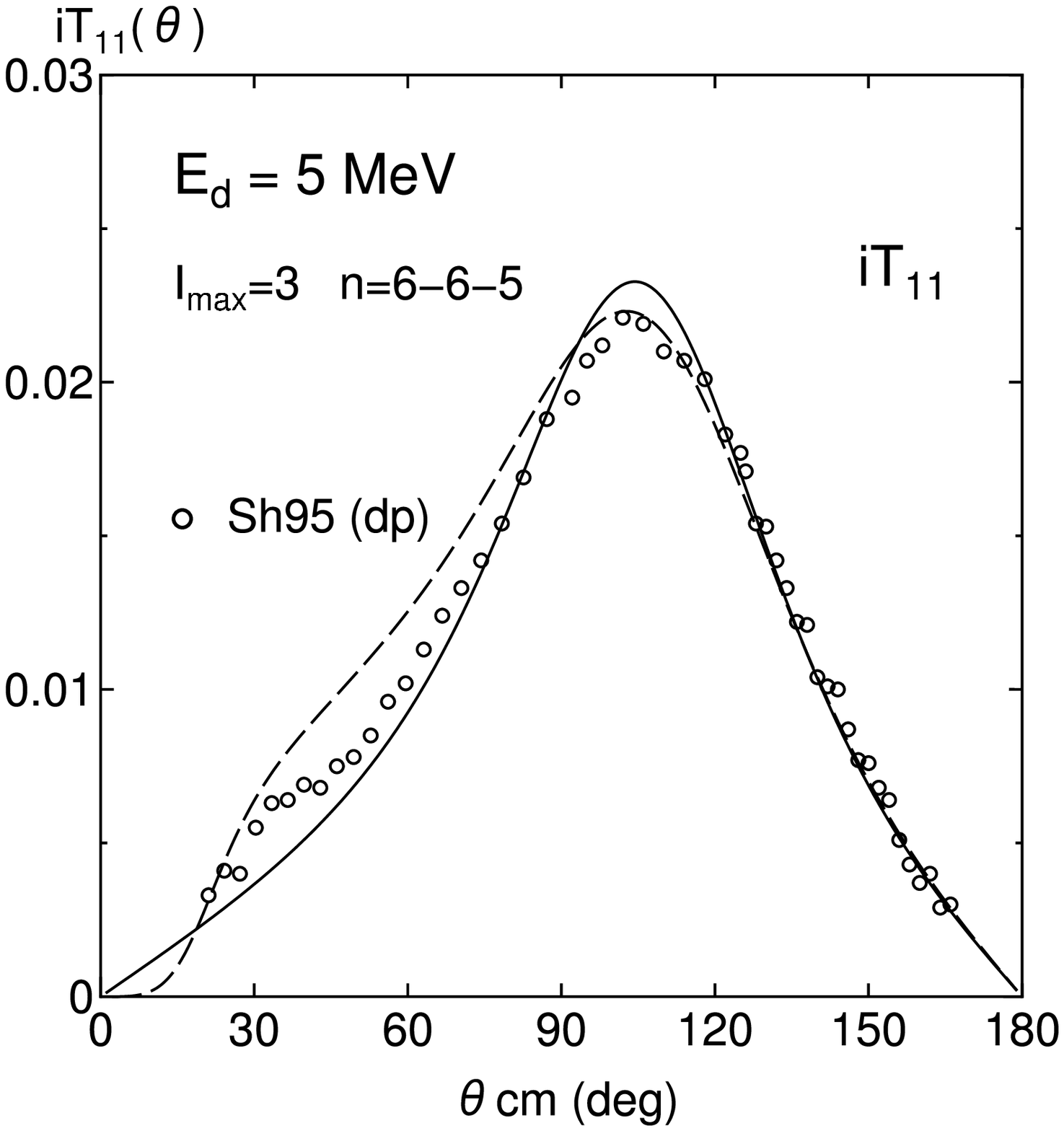}
\end{minipage}~%
\hfill~%
\begin{minipage}{0.48\textwidth}
\includegraphics[angle=0,width=56mm]
{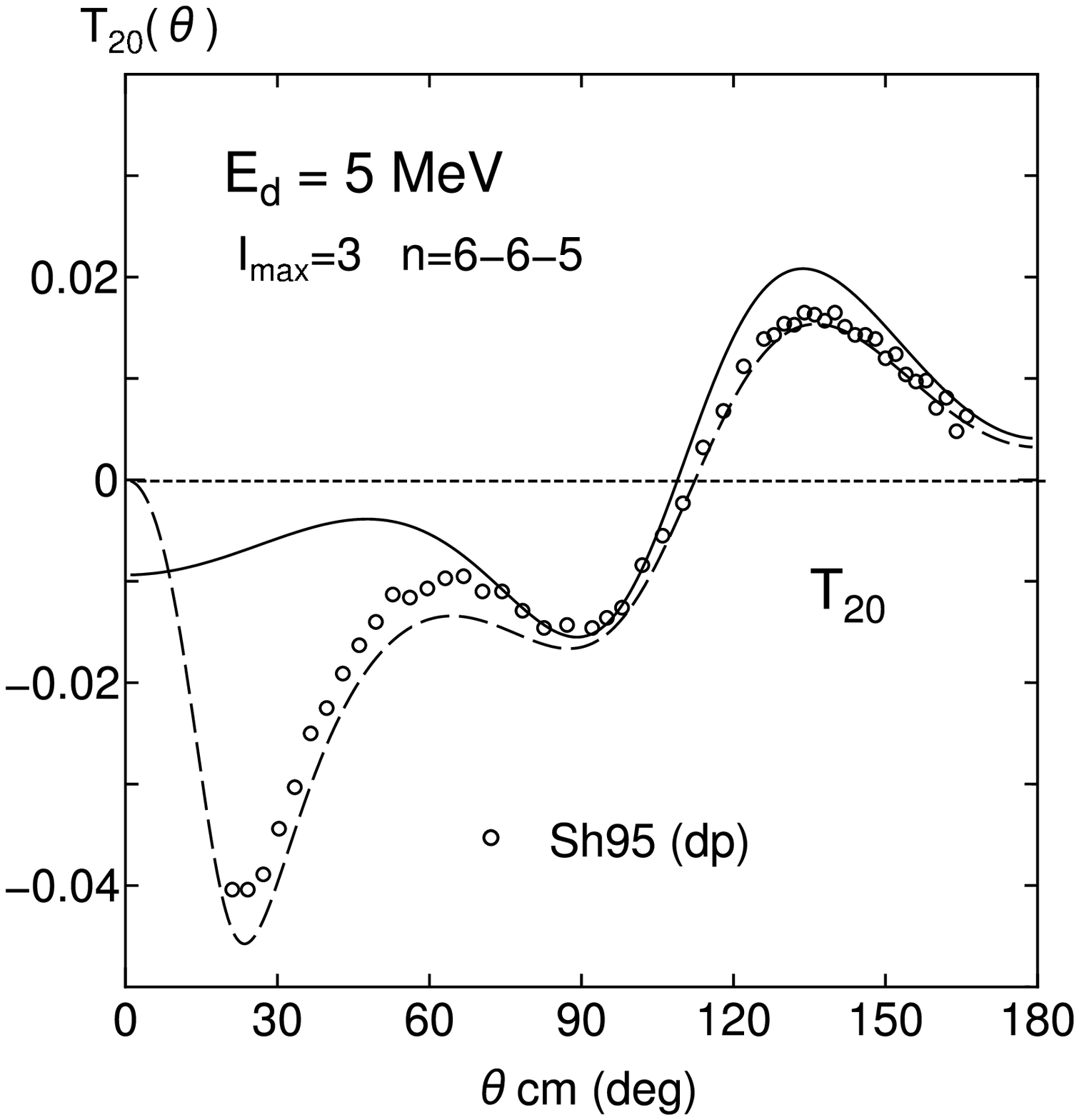}

\vspace{4mm}
\includegraphics[angle=0,width=56mm]
{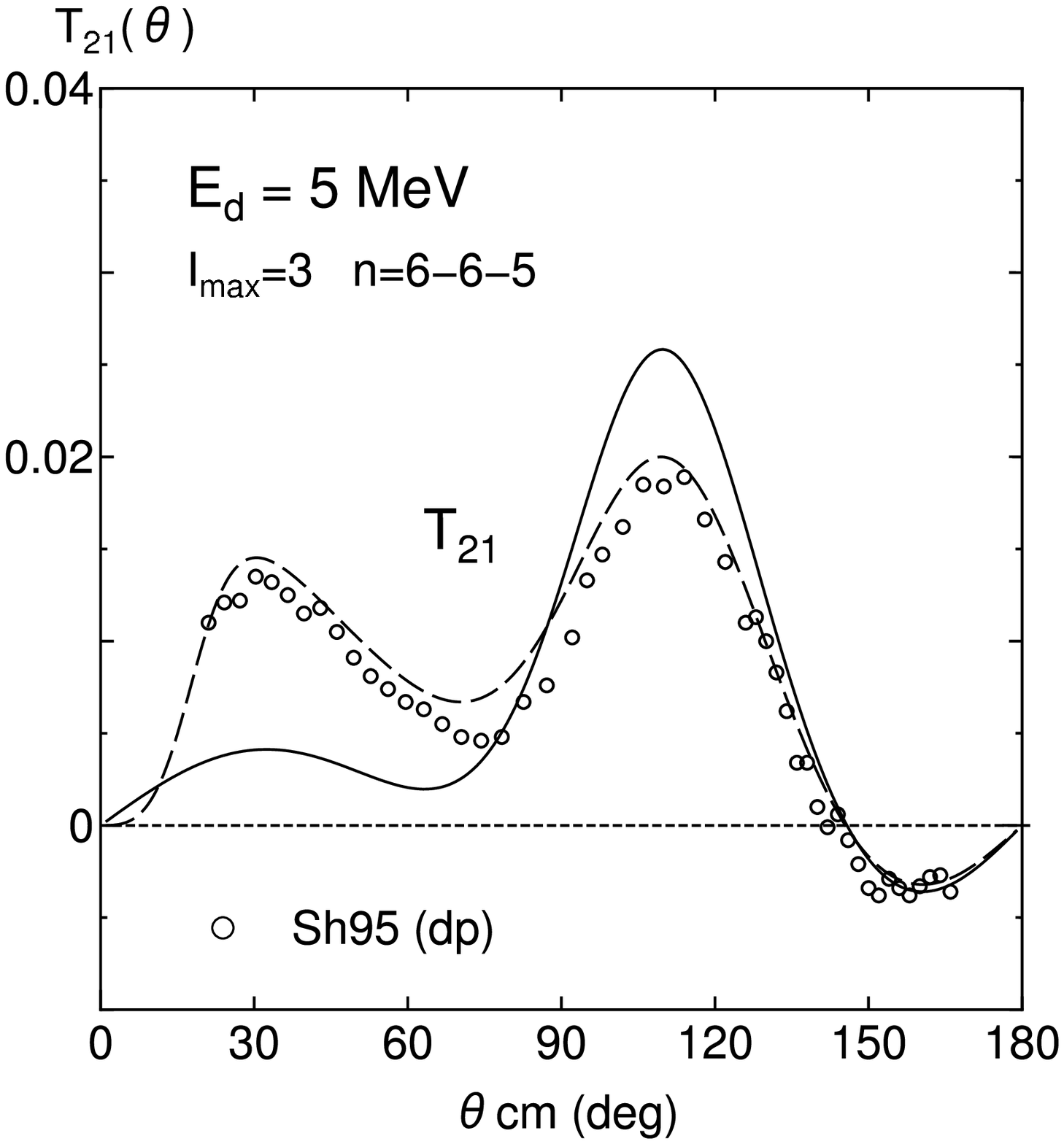}

\vspace{4mm}
\includegraphics[angle=0,width=56mm]
{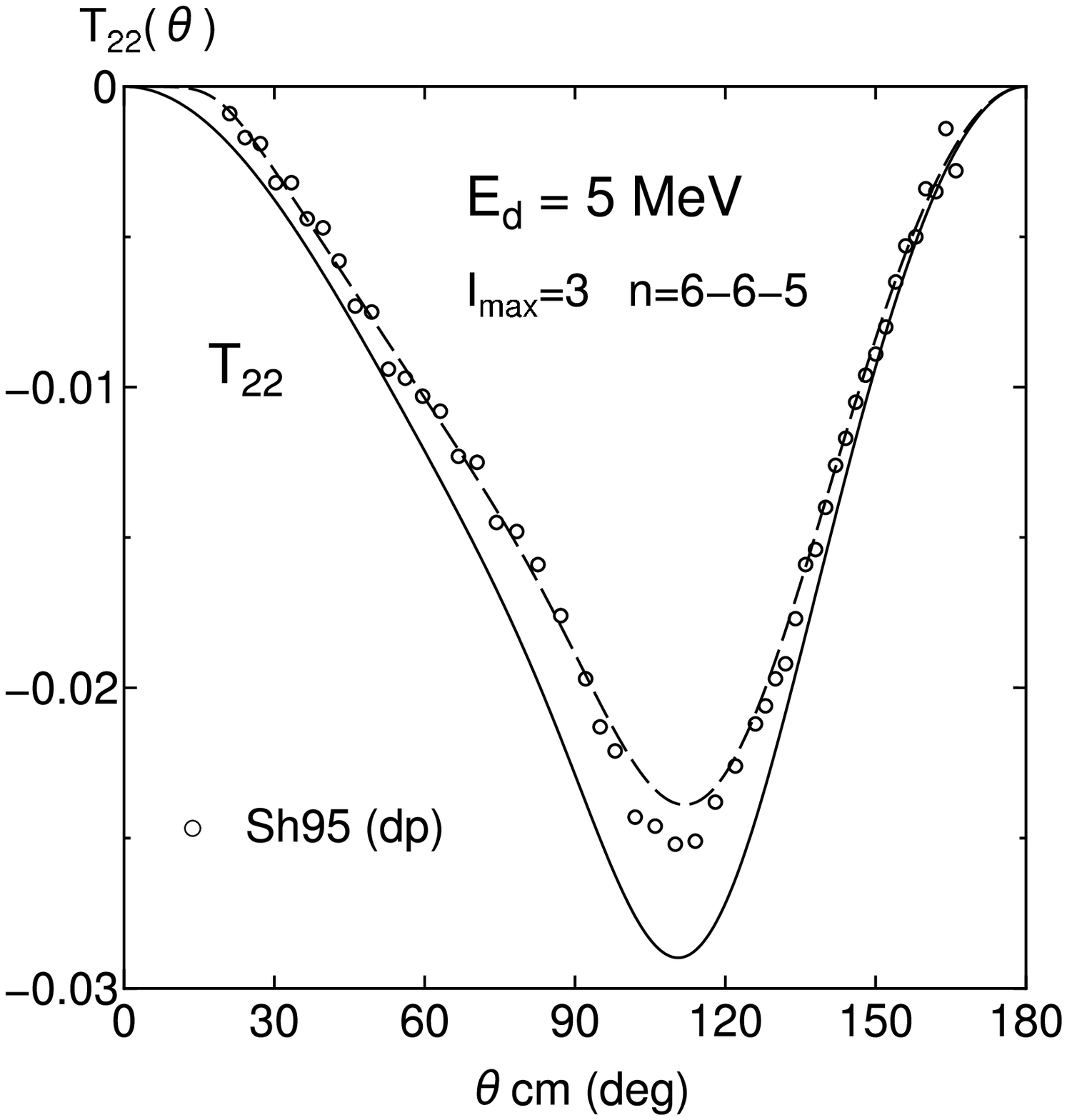}
\end{minipage}
\end{center}
\caption{
The same as Fig.\,\ref{EN1}, but for the incident energy
$E/{\rm nucleon}=2.5$ MeV.
The experimental data are taken from Ref.\,\citen{Sh95} for Sh95 ($dp$).
The others are the same as in Figs.\,\protect\ref{EN1}
and \protect\ref{EN2}.
}
\label{EN3}       
\end{figure}

Here, we incorporate the Coulomb force by the Vincent and Phatak method,
directly to the nuclear scattering amplitudes.
The results are shown by dashed curves
in Figs.\,\ref{EN1} - \ref{EN3}.
We find that the $pd$ or $dp$ experimental data are generally
well reproduced.
In more detail, we find the following two problems:

\bigskip

\begin{enumerate}
\item[1)] The $A_y$ puzzle is more serious below the deuteron breakup
threshold, especially for the $nd$ data.
For instance, the recent $A_y$ measurement of the low-energy
$nd$ scattering in Ref.\,\citen{Ne03} shows the discrepancy
of more than $30\%$, although the difference in the $pd$ case is
not so serious.
\item[2)] The minimum points in the deuteron tensor 
analyzing-power $T_{22}$ are too high,
which is a common feature with the 3 MeV result in Fig.\,\ref{T2m1}.
\end{enumerate}

\bigskip

We can compare our results with those by the variational approach
in Ref.\,\citen{Ki96}, which incorporates the complete Coulomb force
to the AV18 potential and the Urbana three-nucleon force. 
The AV18 potential yields particularly large discrepancy
of more than $30\%$ in this energy region
not only for $A_y$ but also for $i T_{11}$, as seen in
Figs.\,1 and 2 in Ref.\,\citen{Ki96}.
The differential cross sections and the tensor-type deuteron
analyzing-powers $T_{2m}$ are well reproduced.
As to the difference in $T_{22}$ in 2) above,
the comparison with their results indicates
that the three-body force might be important to reproduce
the magnitude at the minimum points.
The deuteron analyzing-powers, $iT_{11}$ and $T_{2m}$, at $E_d=3$ MeV
in Fig.\,\ref{EN2} are very similar to their results in Fig.\,12
of Ref.\,\citen{Ki96}, although the comparison with the experimental data
is not easy because of the experimental errorbars.

\bigskip

\subsection{Polarization transfer coefficients}

\begin{figure}[htb]
\begin{center}
\begin{minipage}{0.48\textwidth}
\includegraphics[angle=0,width=55mm]
{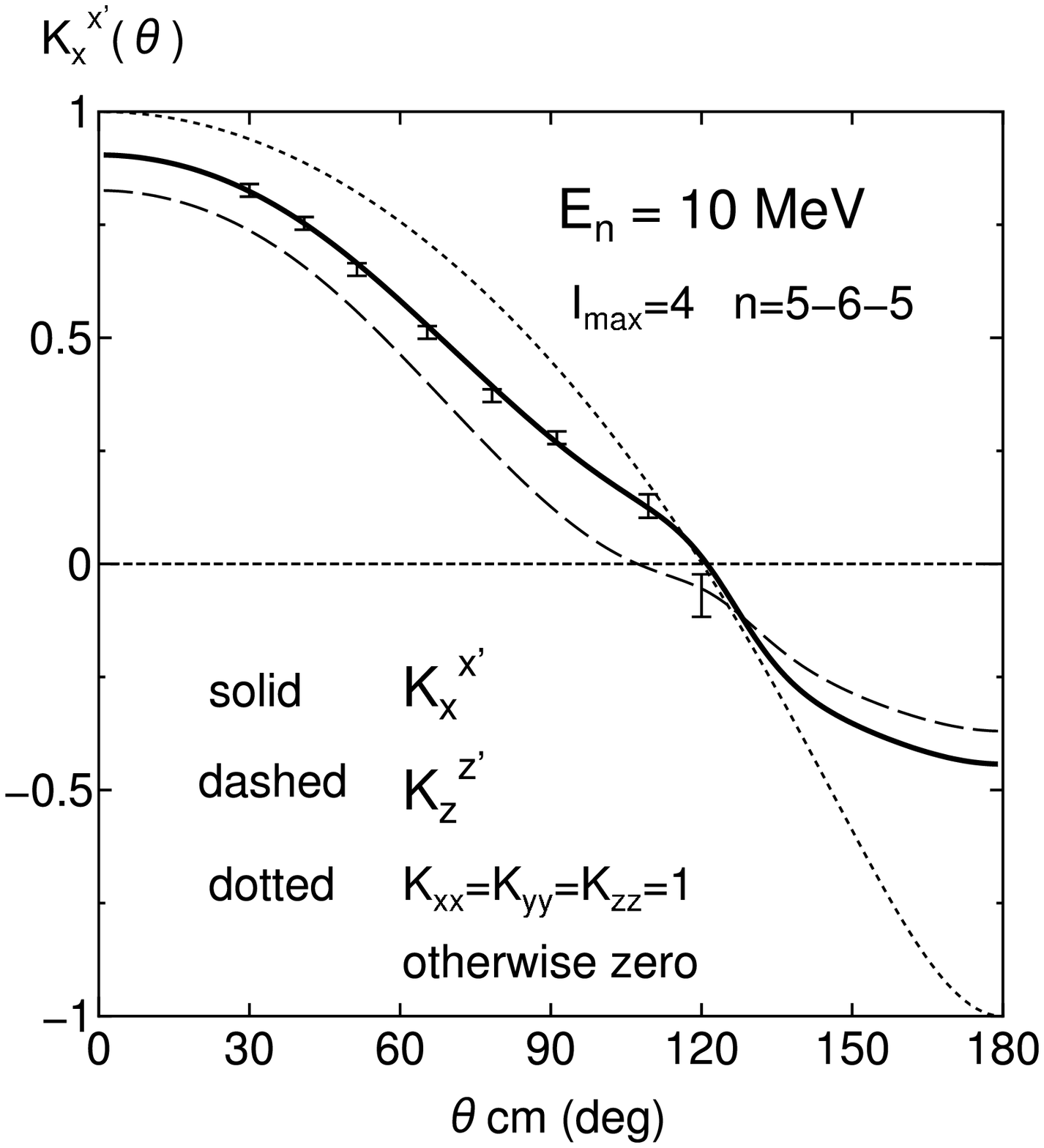}

\includegraphics[angle=0,width=55mm]
{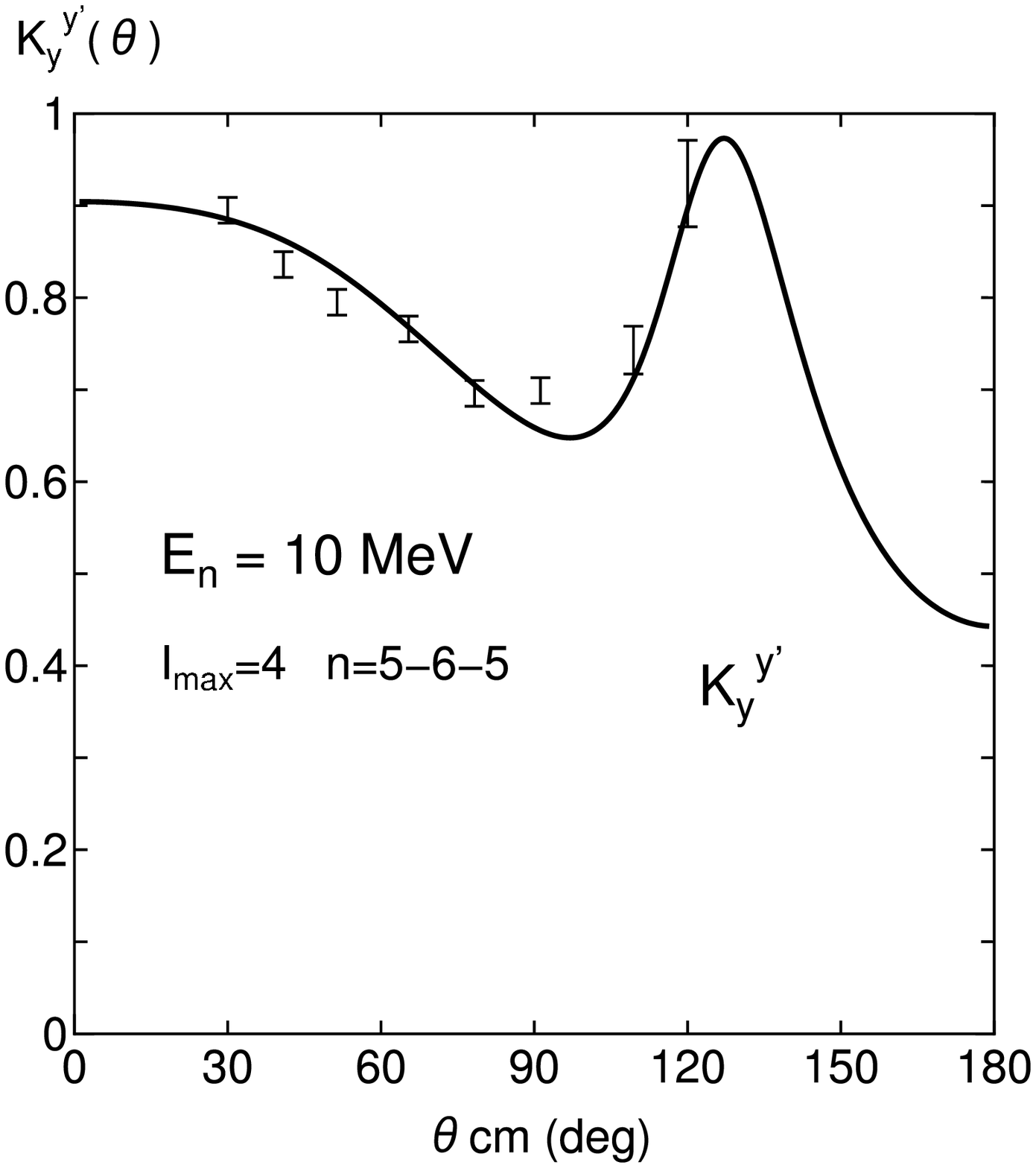}

\includegraphics[angle=0,width=55mm]
{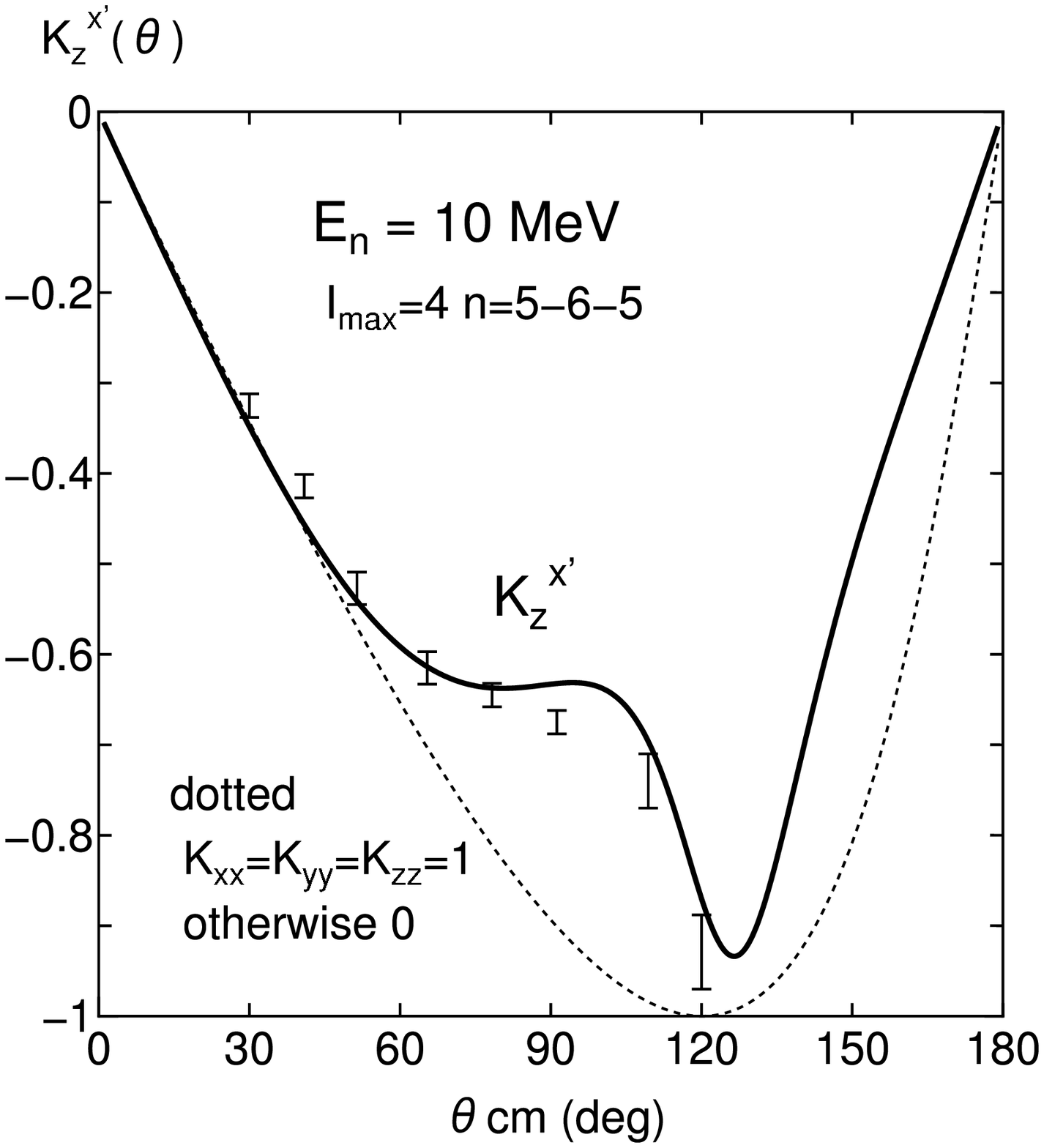}
\end{minipage}~%
\hfill~%
\begin{minipage}{0.48\textwidth}
\includegraphics[angle=0,width=55mm]
{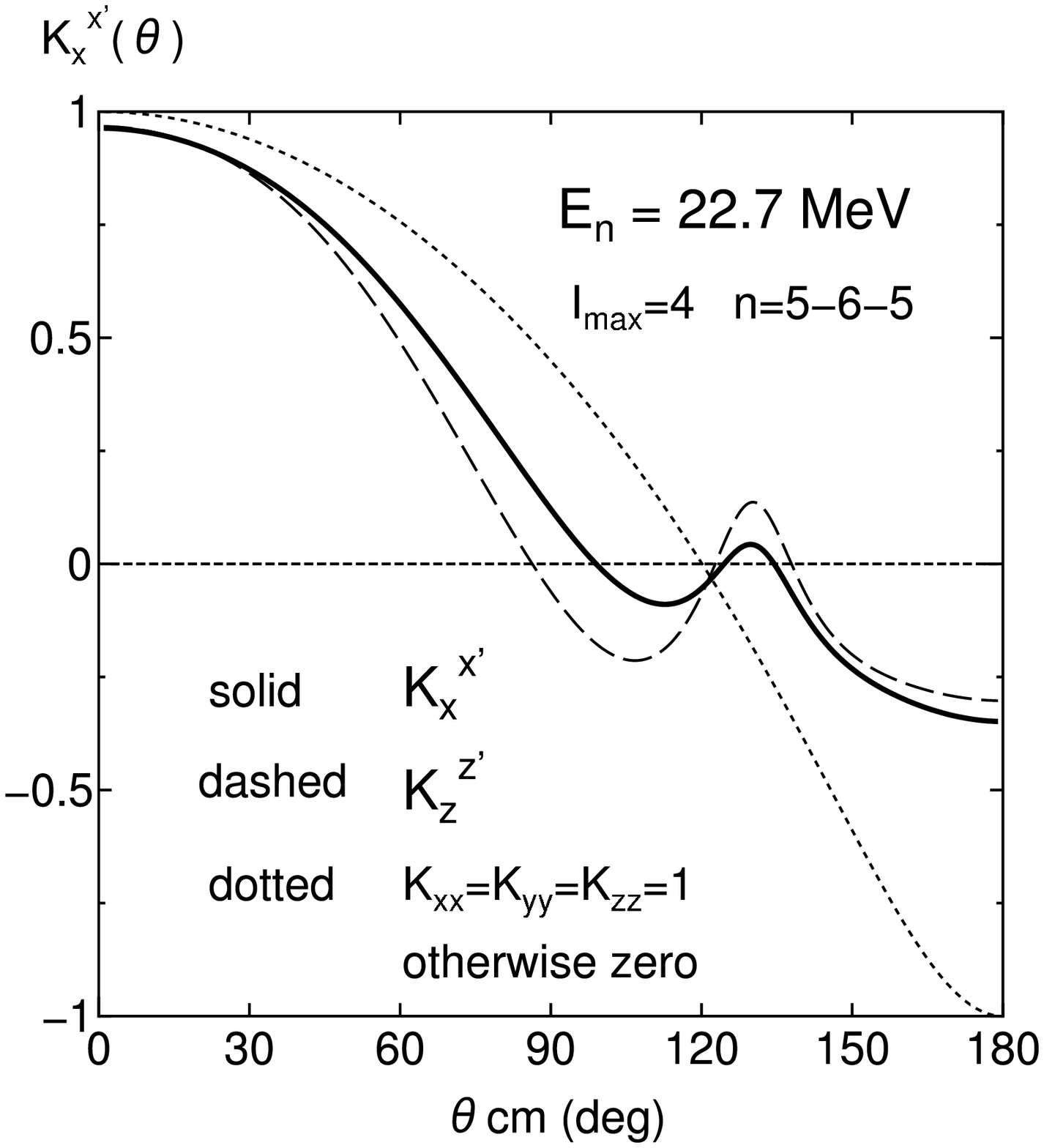}

\includegraphics[angle=0,width=55mm]
{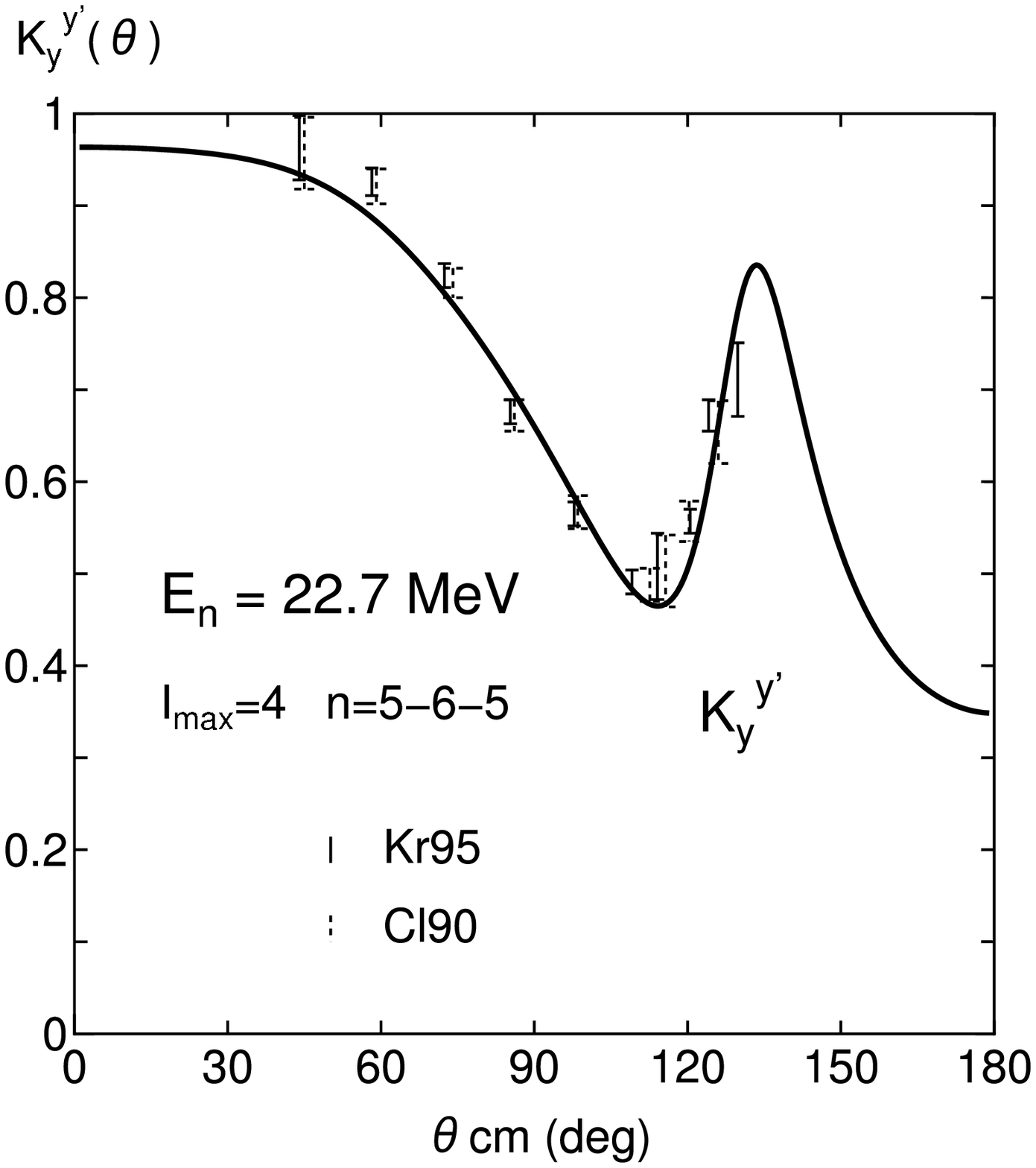}

\includegraphics[angle=0,width=55mm]
{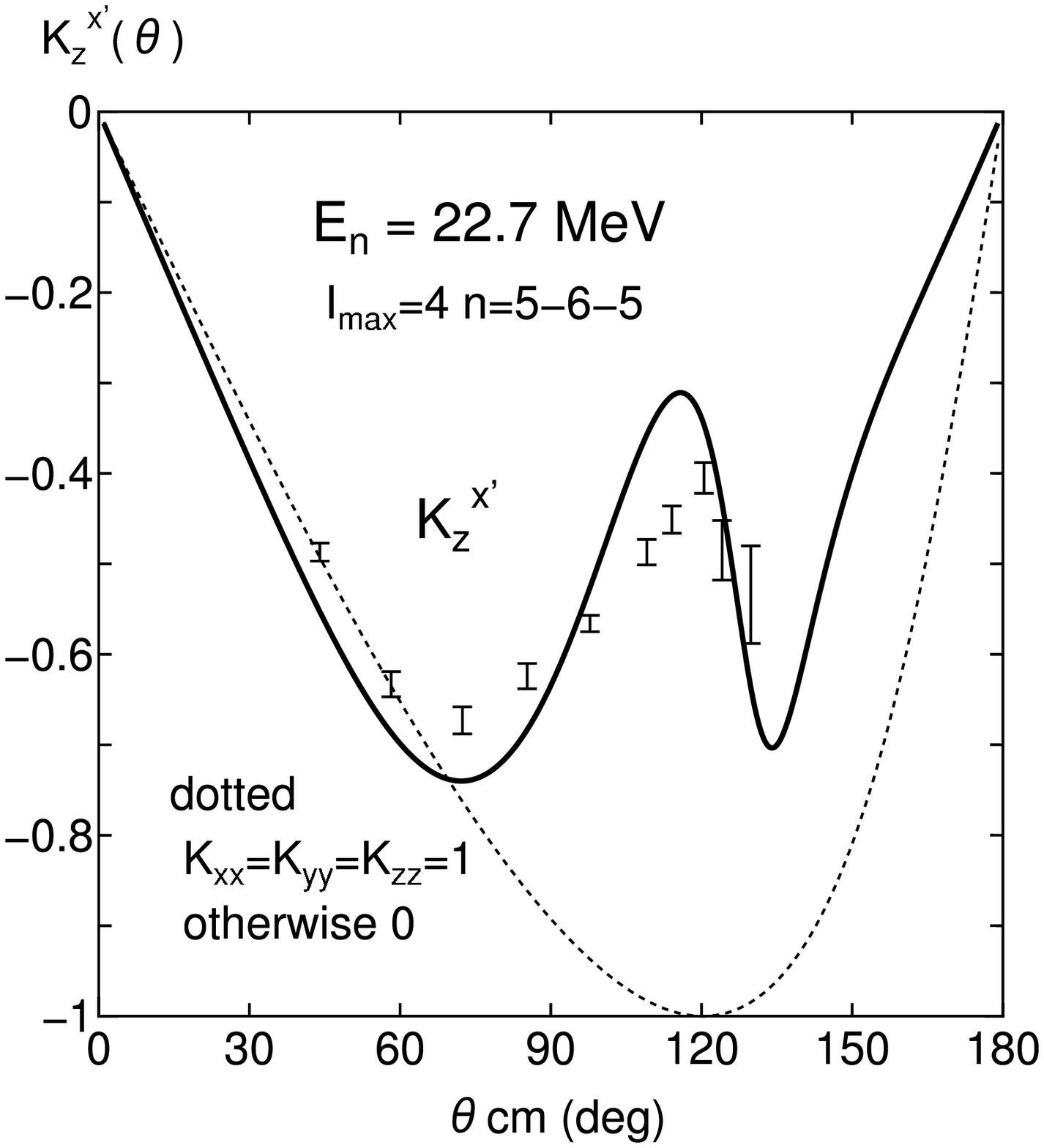}
\end{minipage}
\end{center}
\caption{
The nucleon polarization transfer coefficients
$K^{\beta^\prime}_\alpha(\theta)$ of 
the $nd$ elastic scattering for $E_n=10$ MeV and 22.7 MeV, 
compared with the experiment.
The experimental $pd$ data are taken from Ref.\,\citen{Sp84}
for the 10 MeV data and from Ref.\,\citen{Kr95} for the
22.7 MeV data. The dashed bars in the panel of 22.7 MeV ${K_y}^{y^\prime}$
are from Ref.\,\citen{Cl90b}. 
}
\label{Kn}       
\end{figure}

\begin{figure}[htb]
\begin{center}
\begin{minipage}{0.48\textwidth}
\includegraphics[angle=0,width=55mm]
{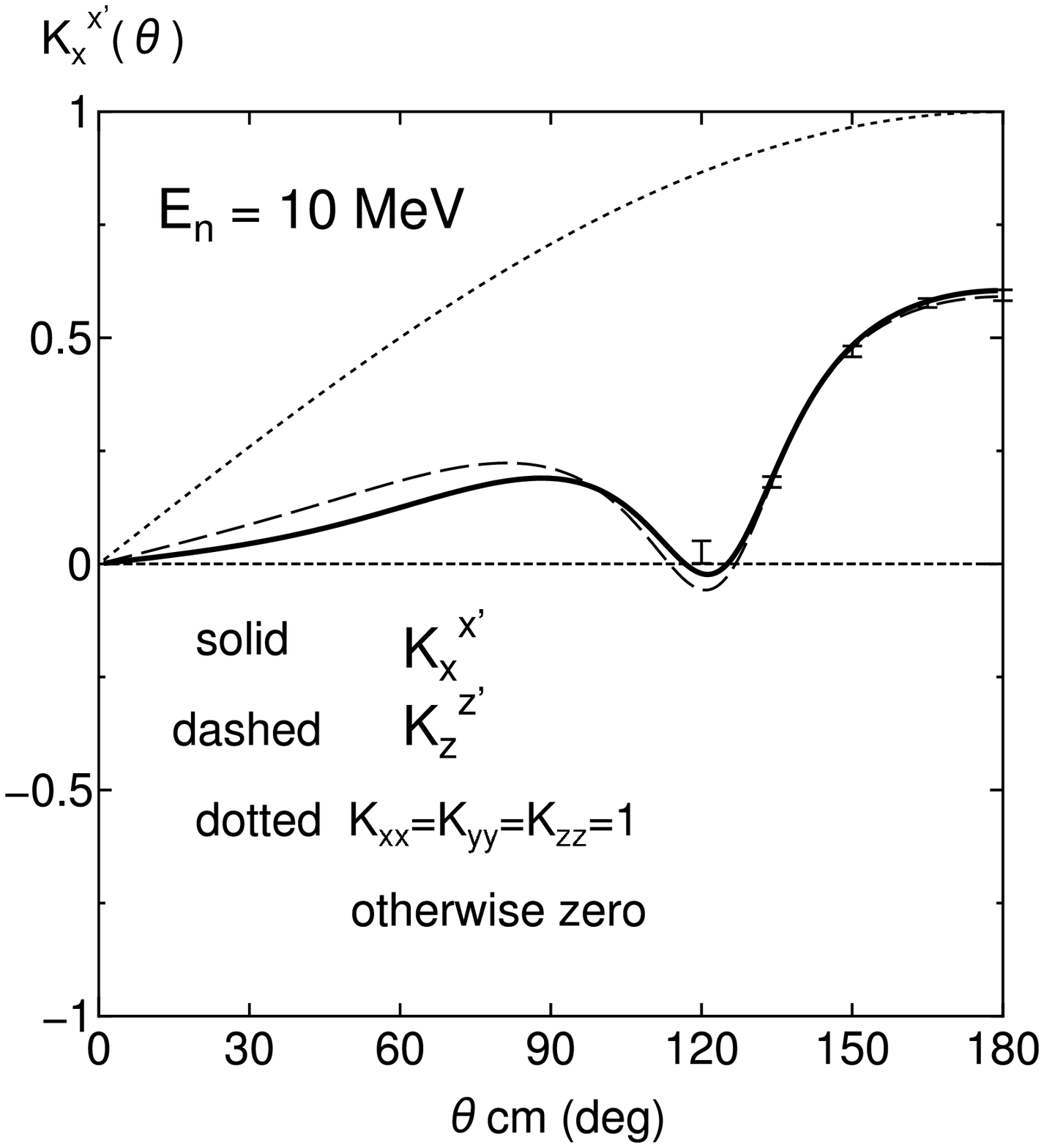}

\includegraphics[angle=0,width=55mm]
{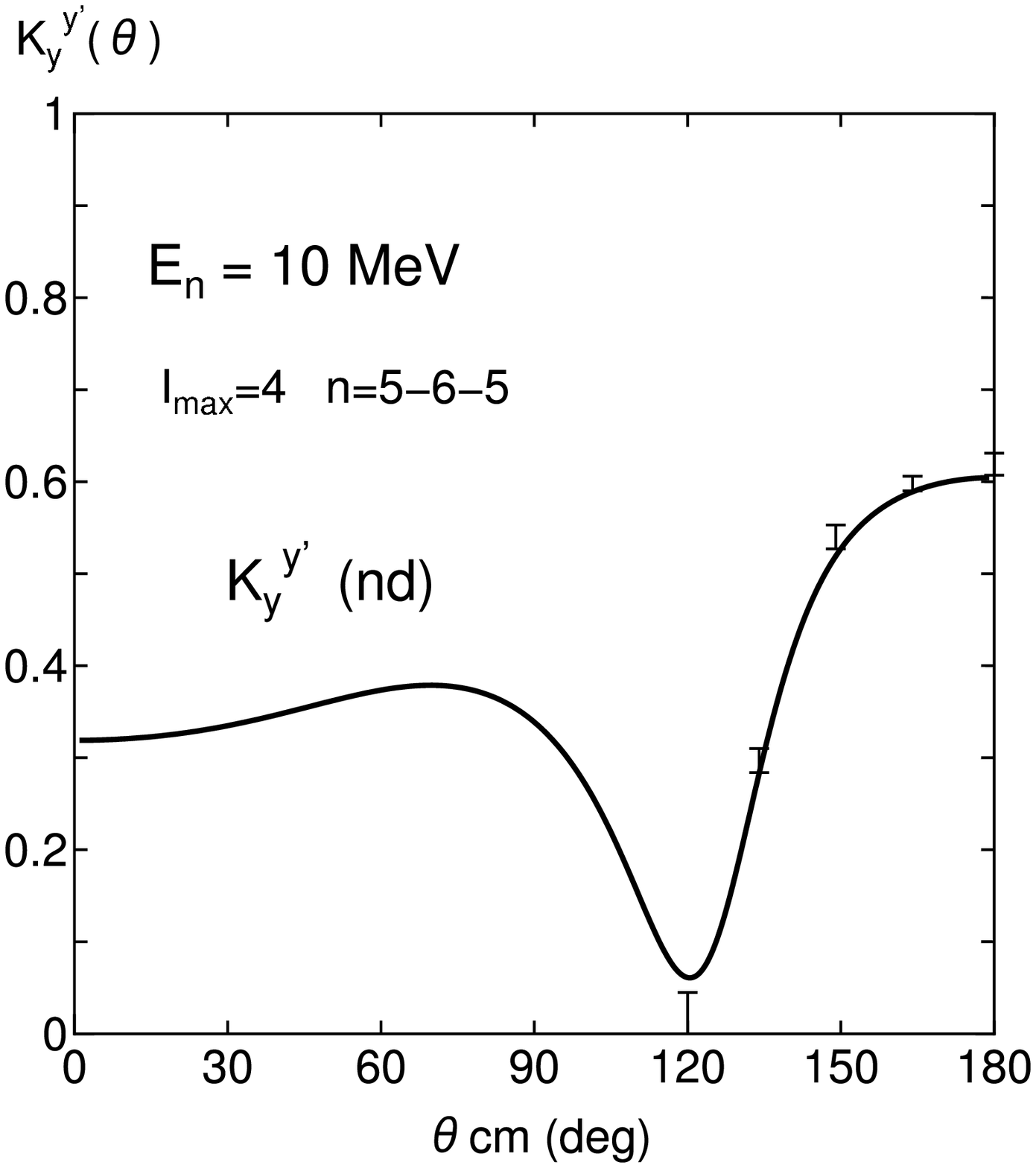}

\includegraphics[angle=0,width=55mm]
{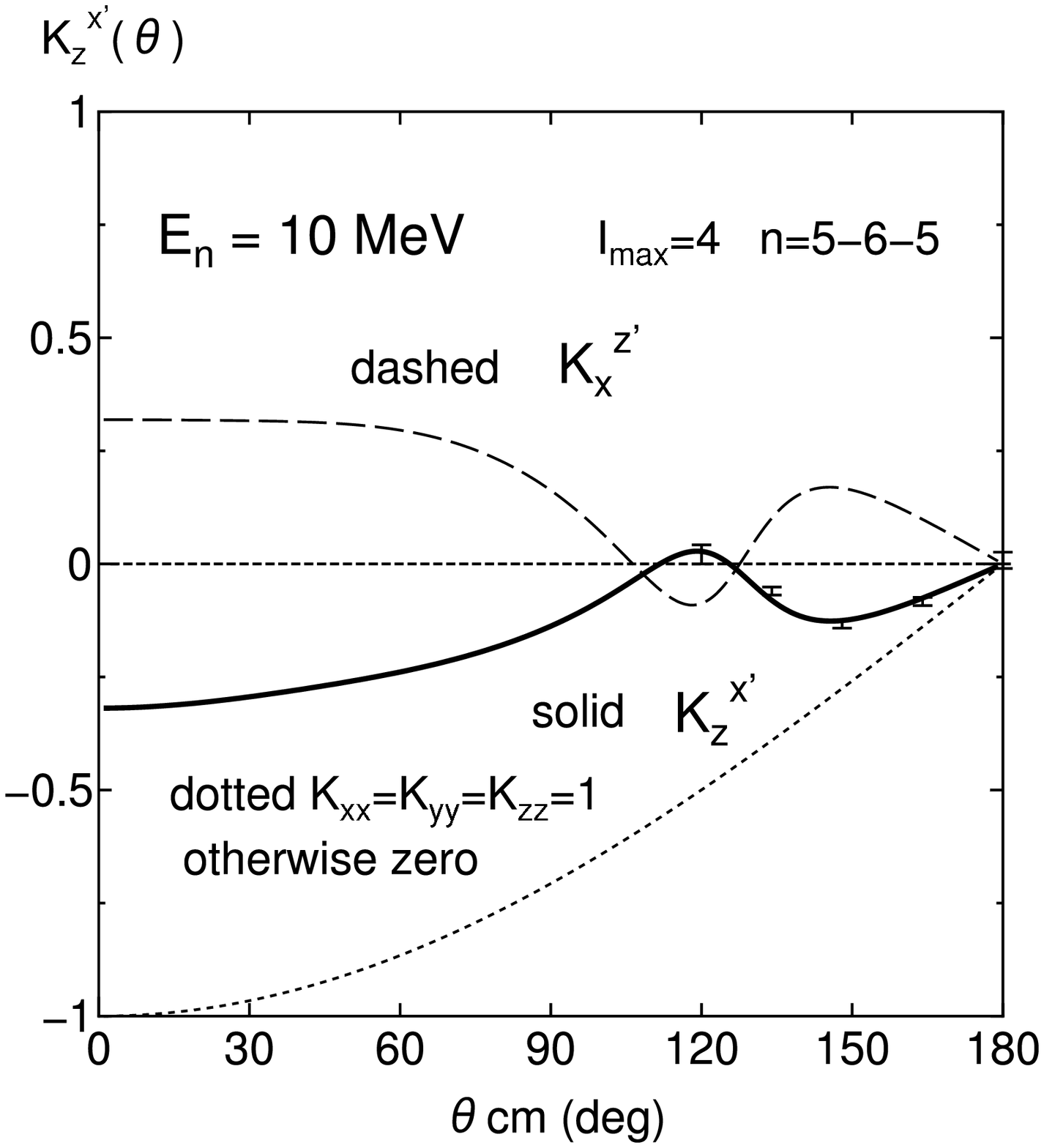}
\end{minipage}~%
\hfill~%
\begin{minipage}{0.48\textwidth}
\includegraphics[angle=0,width=55mm]
{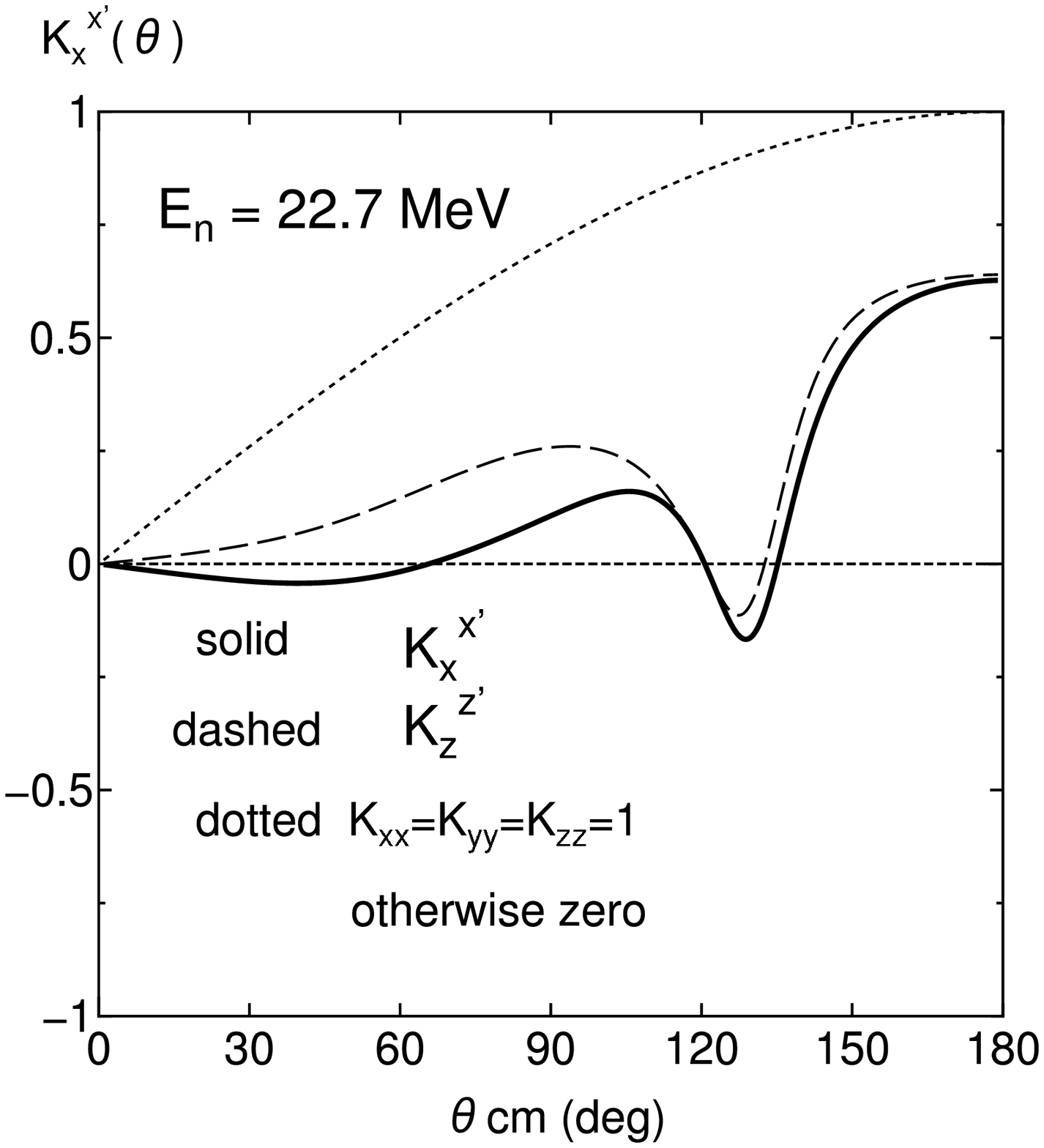}

\includegraphics[angle=0,width=55mm]
{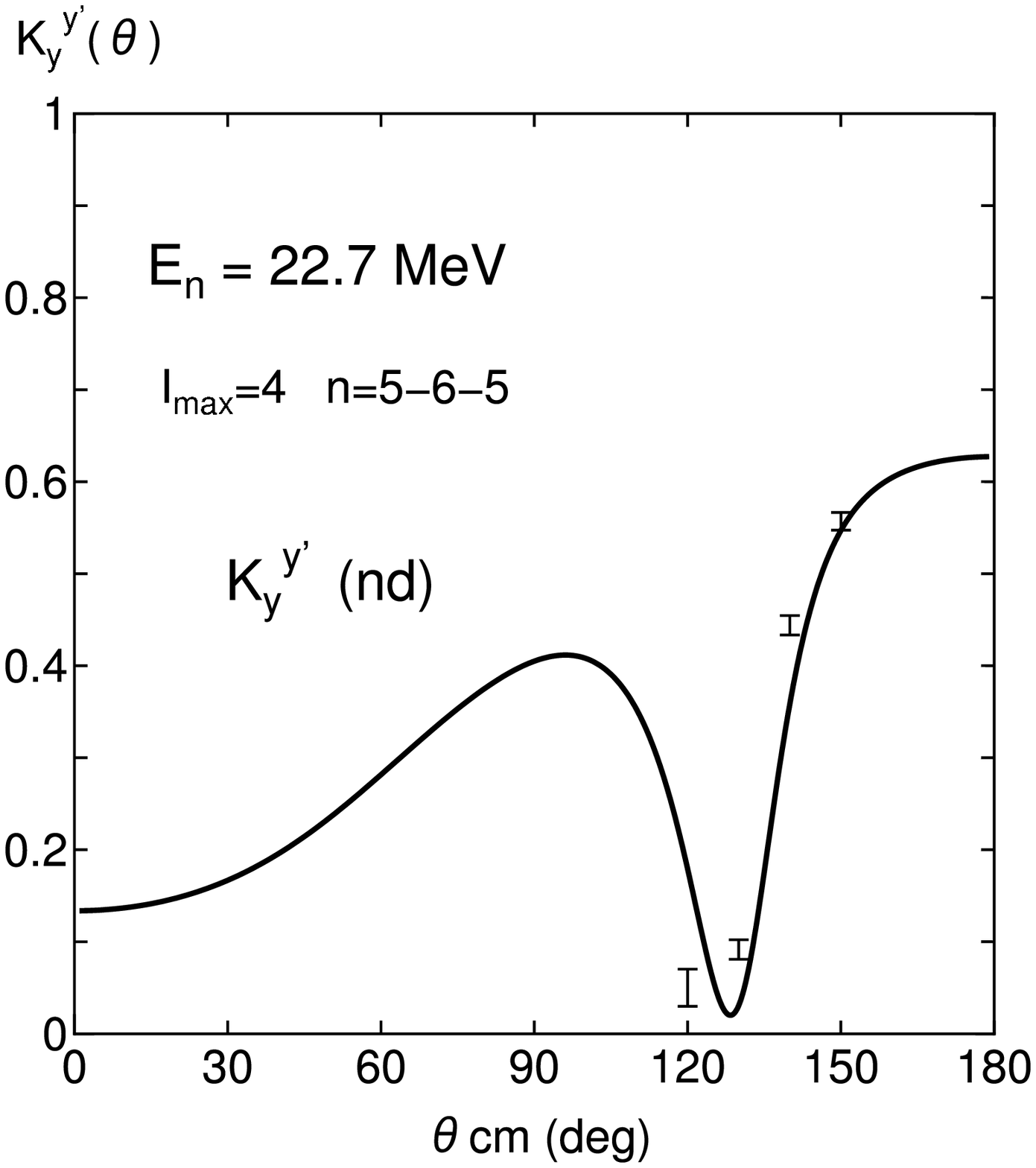}

\includegraphics[angle=0,width=55mm]
{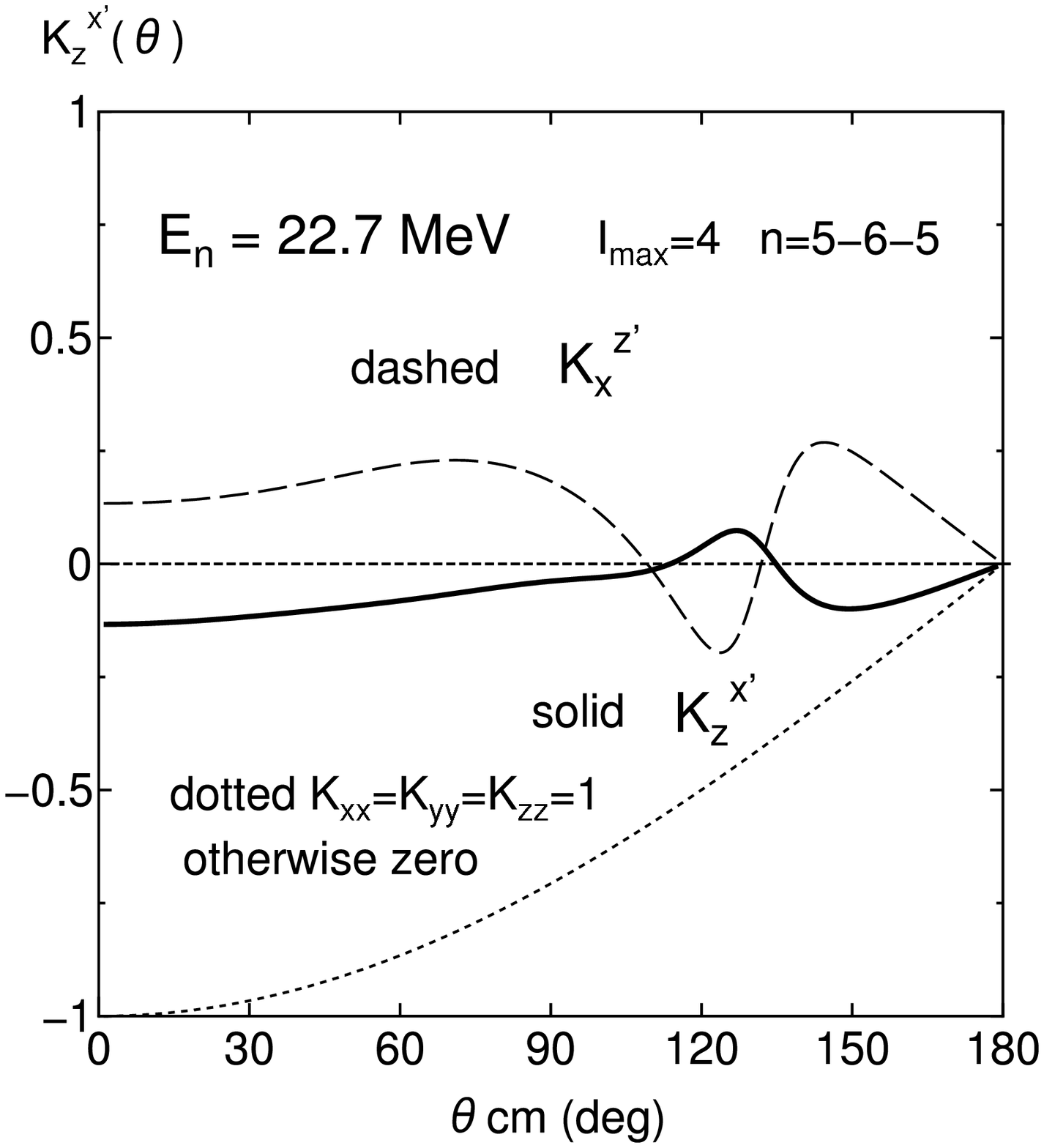}
\end{minipage}
\end{center}
\caption{
The vector-type nucleon to deuteron polarization transfer coefficients
$K^{\beta^\prime}_\alpha(\theta)$ of 
the $nd$ elastic scattering for $E_n=10$ MeV and 22.7 MeV, 
compared with the $pd$ experimental data.
The experimental data are taken from Ref.\,\citen{Sp84}
for the 10 MeV data and from Ref.\,\citen{Gl95} for the
22.7 MeV data. 
}
\label{Kd1}       
\end{figure}

\begin{figure}[htb]
\begin{center}
\begin{minipage}{0.48\textwidth}
\includegraphics[angle=0,width=56mm]
{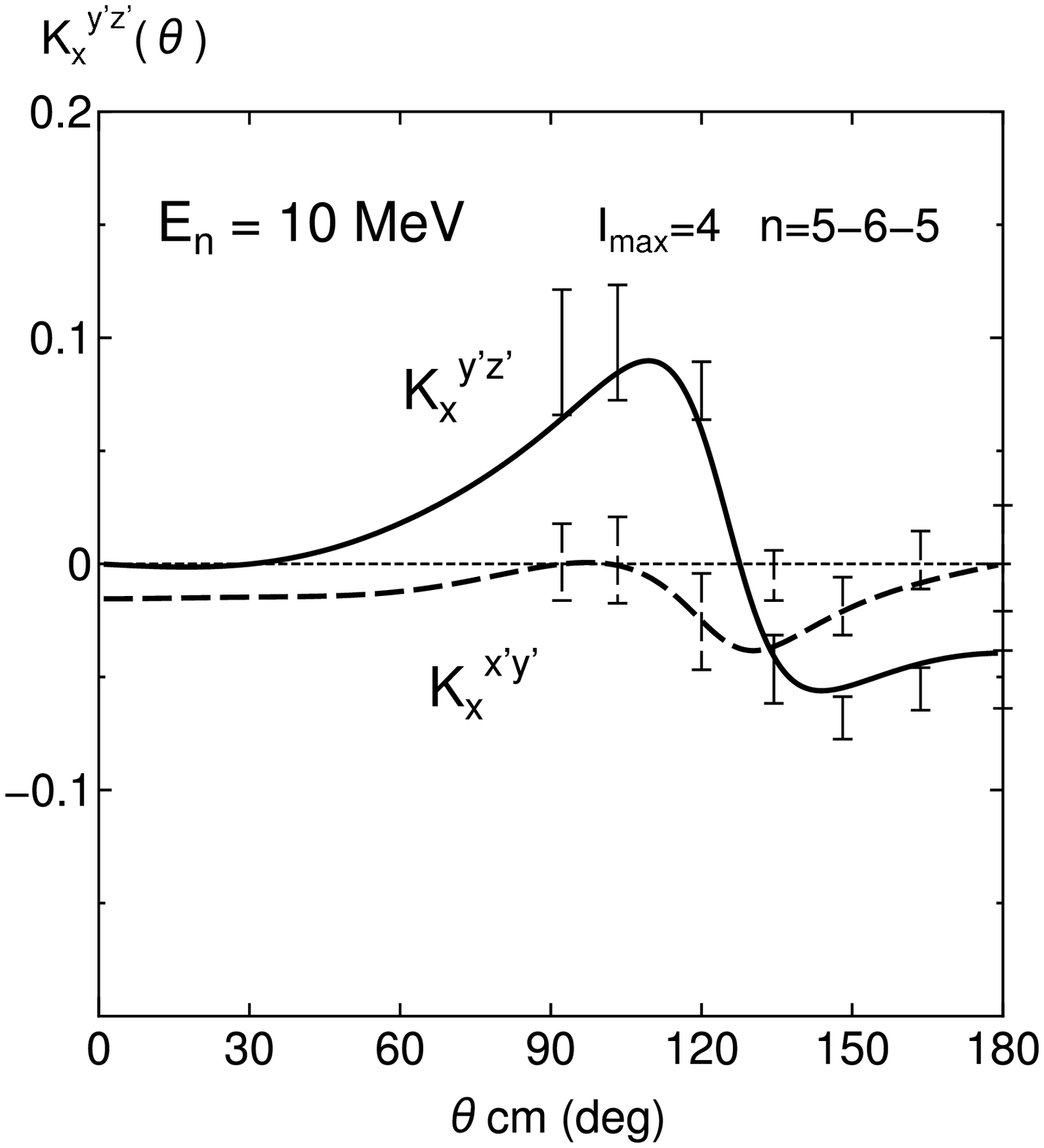}

\includegraphics[angle=0,width=56mm]
{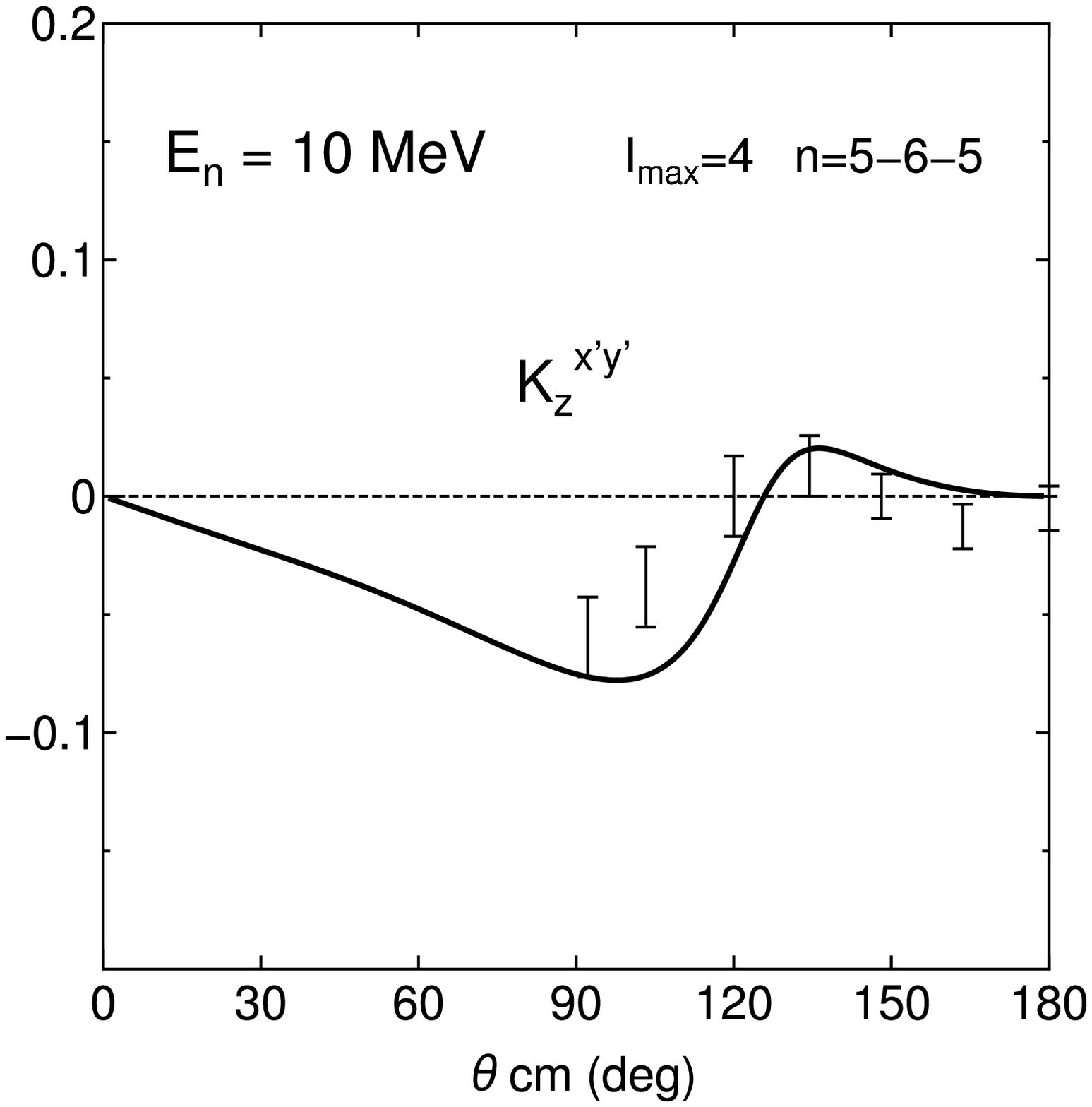}

\includegraphics[angle=0,width=56mm]
{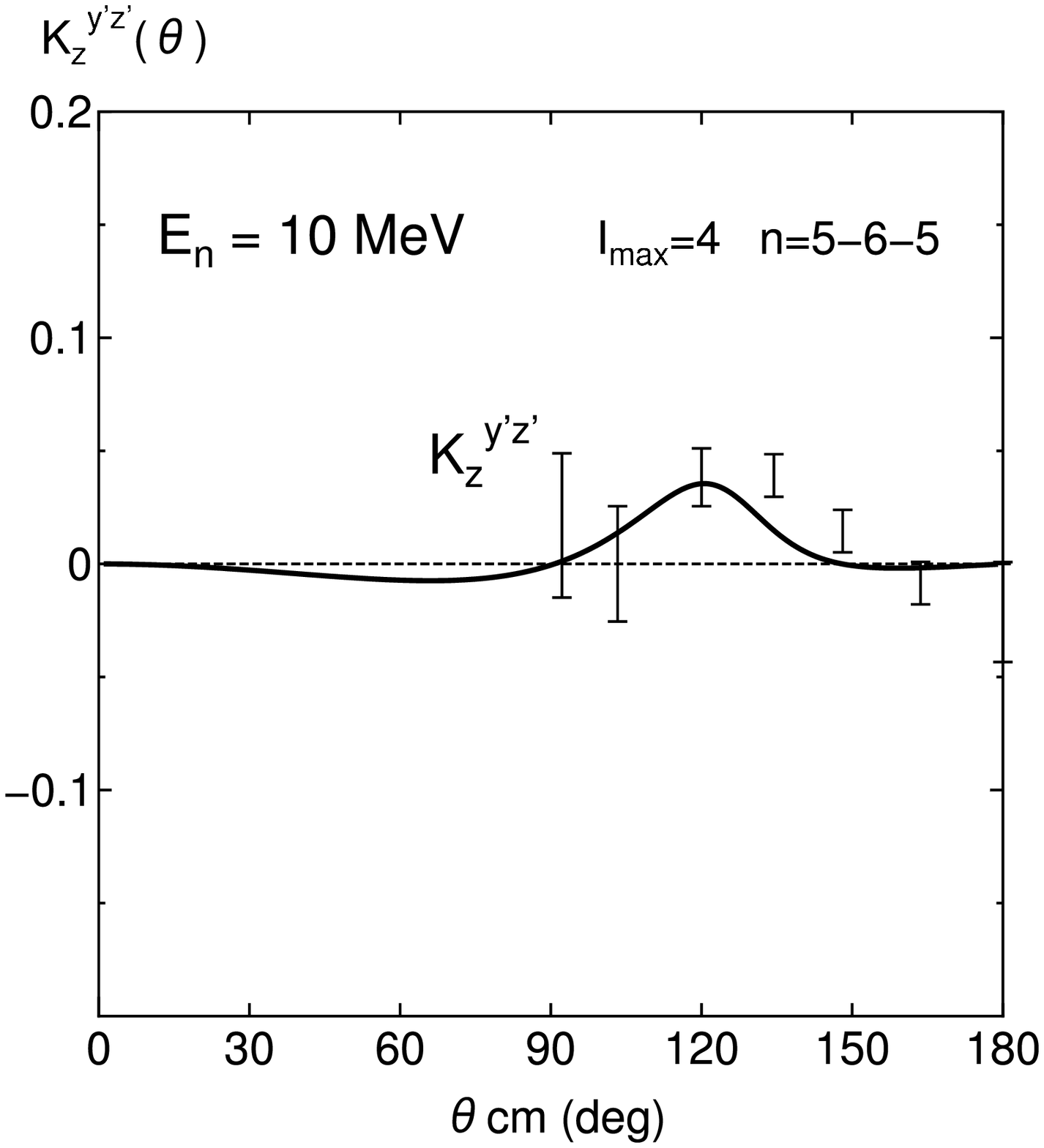}
\end{minipage}~%
\hfill~%
\begin{minipage}{0.48\textwidth}
\includegraphics[angle=0,width=56mm]
{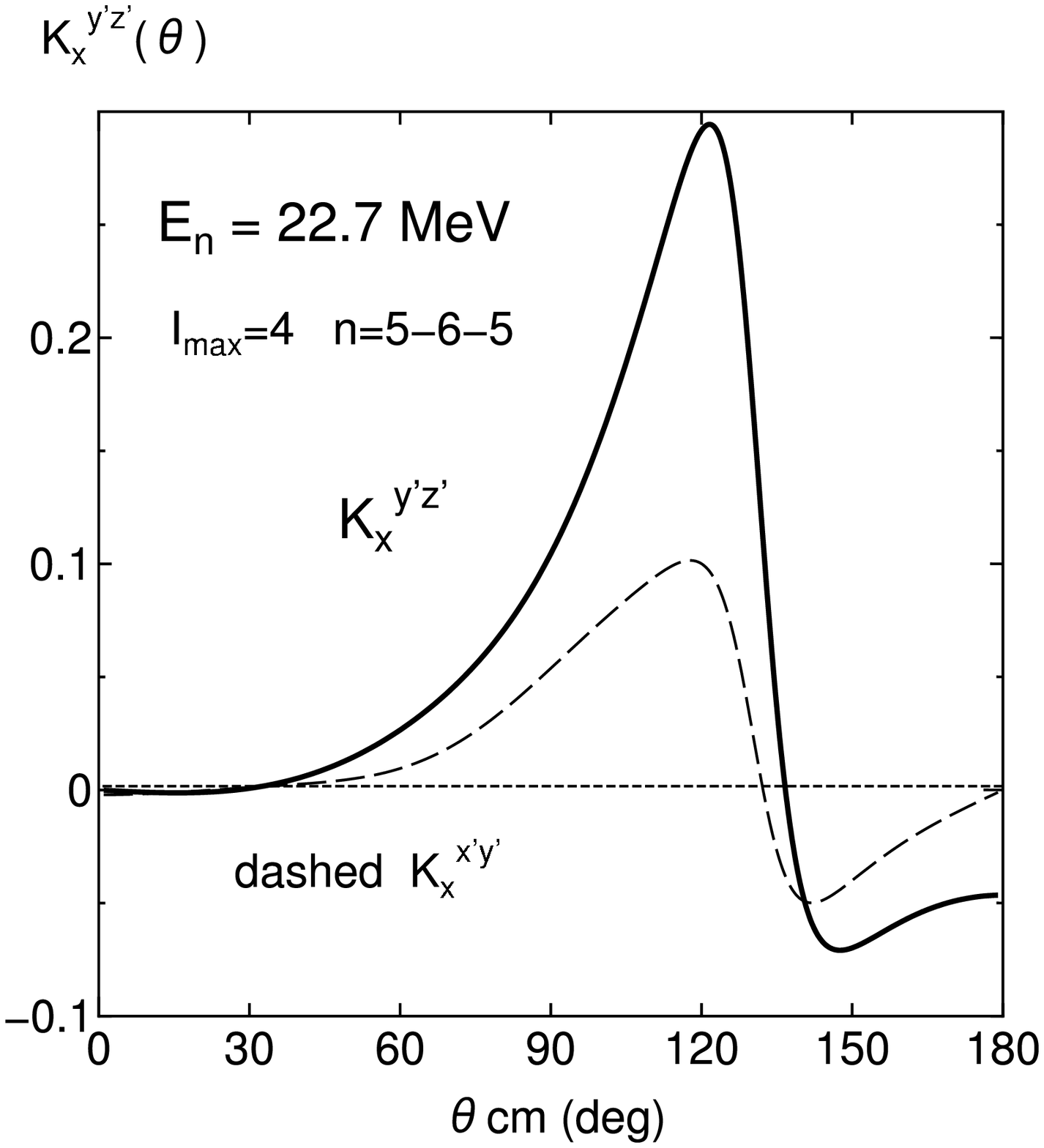}

\includegraphics[angle=0,width=56mm]
{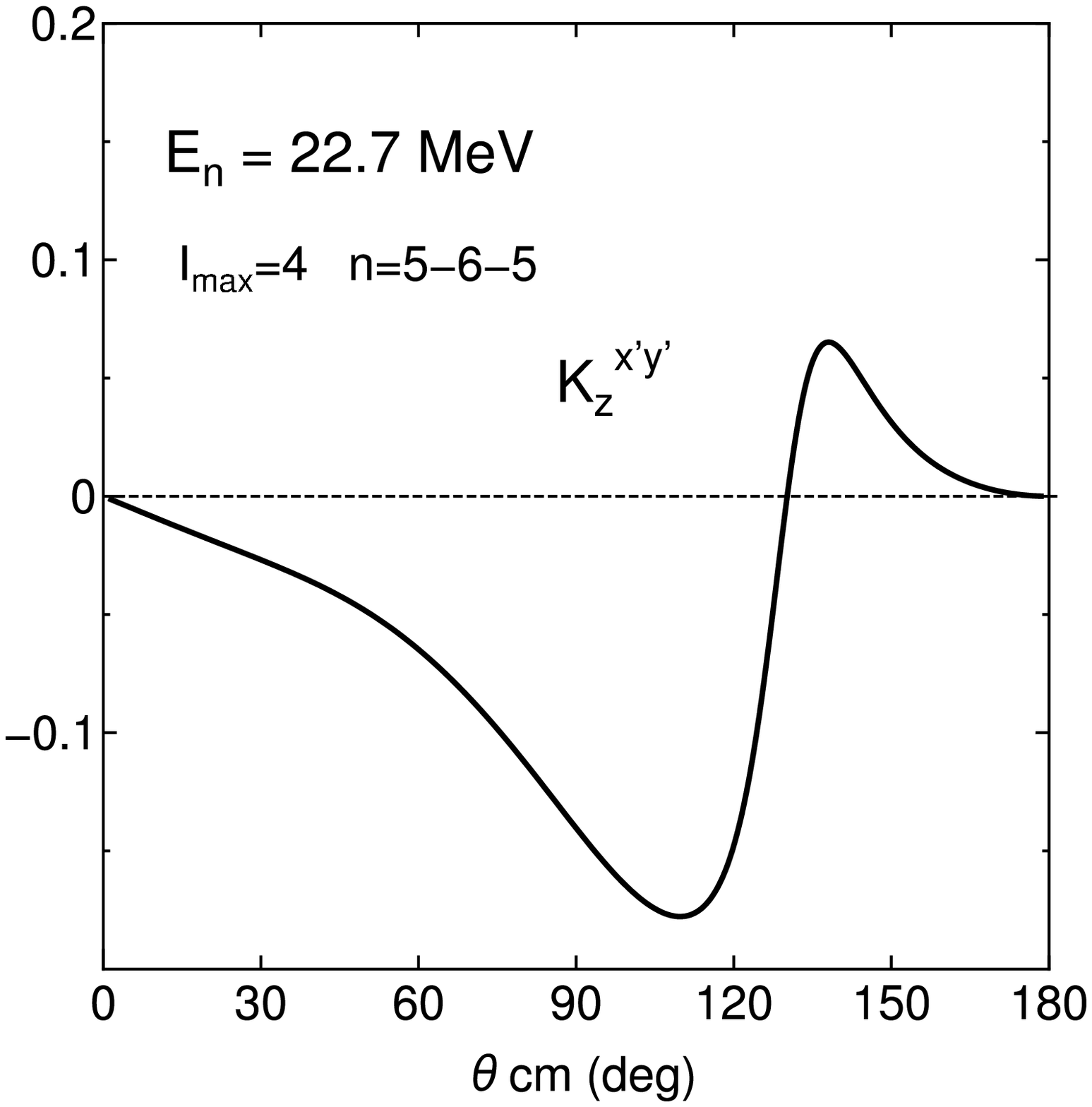}

\includegraphics[angle=0,width=56mm]
{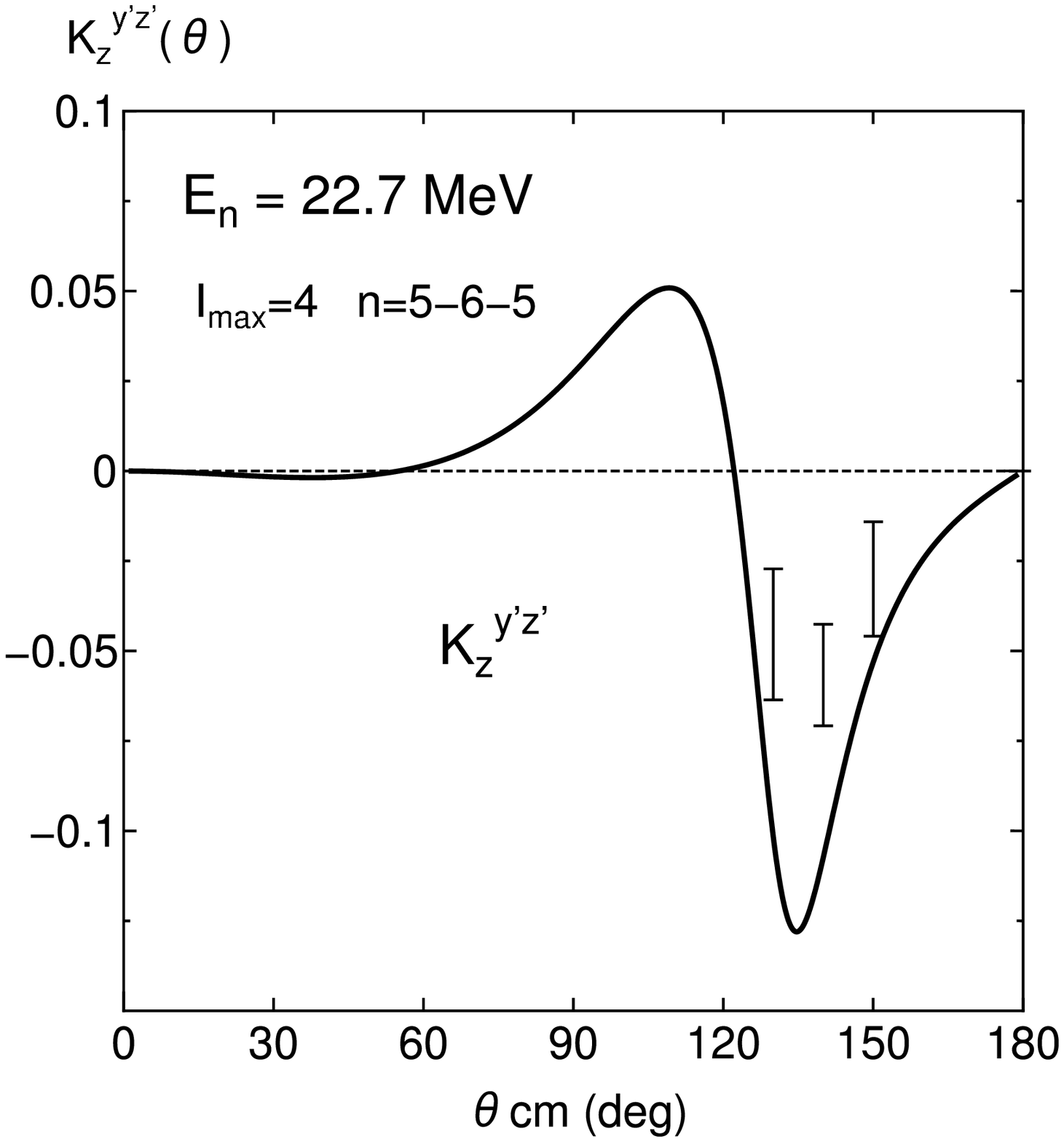}
\end{minipage}
\end{center}
\caption{
The tensor-type nucleon to deuteron polarization transfer coefficients
$K^{\beta^\prime \gamma^\prime}_\alpha(\theta)$ of 
the $nd$ elastic scattering for $E_n=10$ MeV and 22.7 MeV, 
compared with the $pd$ experimental data.
The experimental data are taken from Ref.\,\citen{Sp84}
for the 10 MeV data and from Refs.\,\citen{Gl95,Kr95} for the
22.7 MeV data. 
}
\label{Kd2}       
\end{figure}

\begin{figure}[htb]
\begin{center}
\begin{minipage}{0.48\textwidth}
\includegraphics[angle=0,width=58mm]
{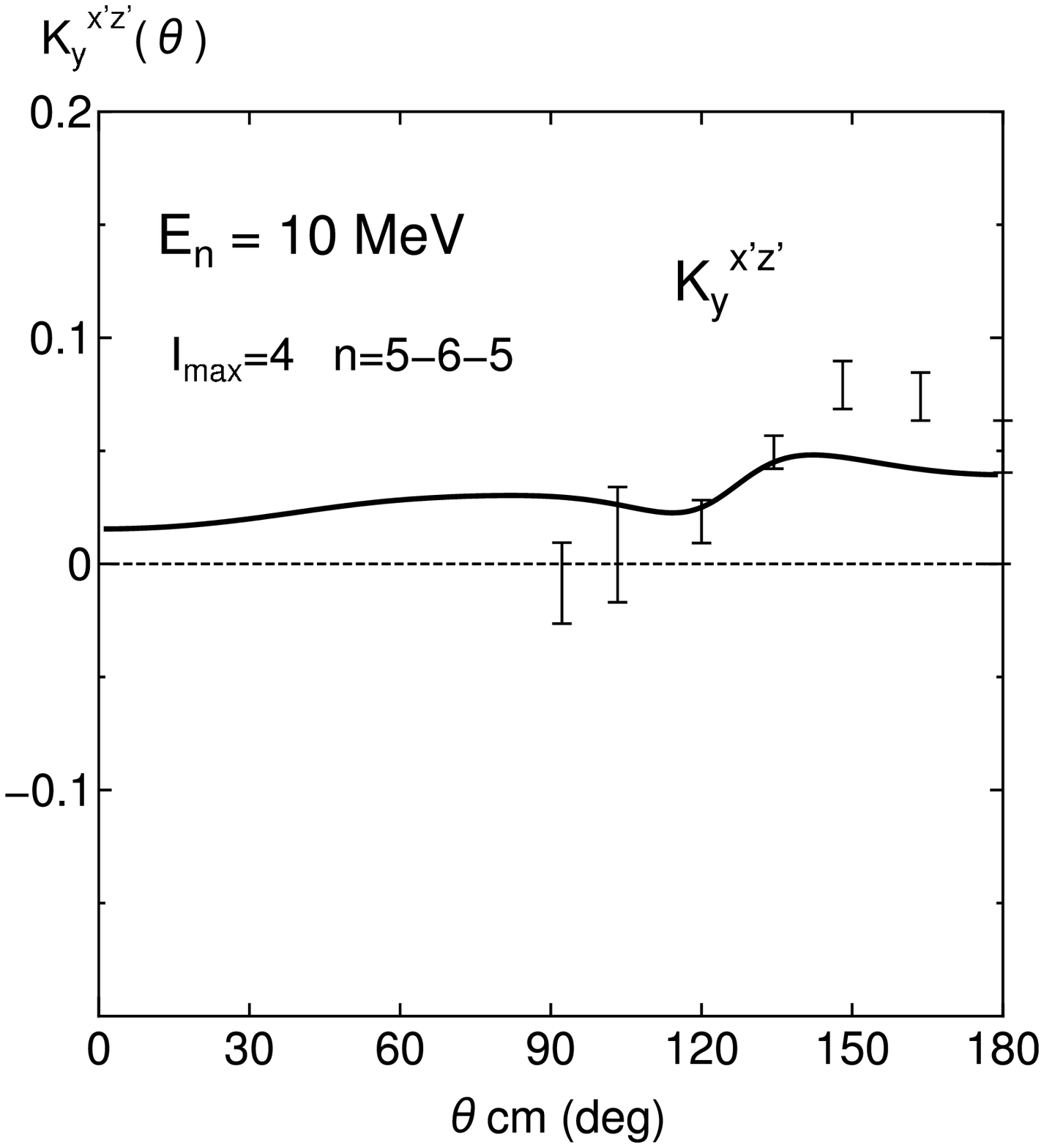}

\vspace{2mm}
\includegraphics[angle=0,width=58mm]
{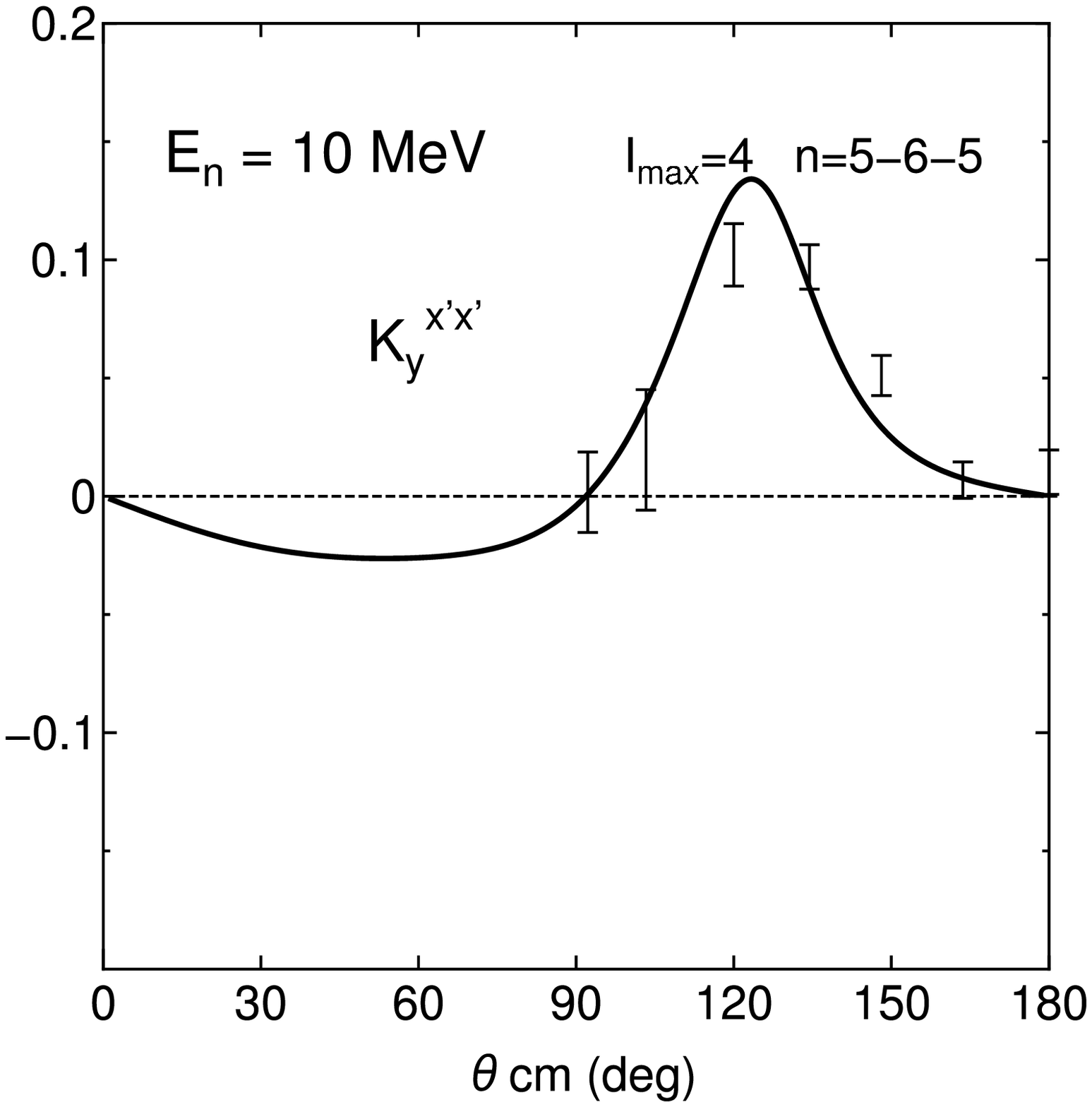}

\vspace{2mm}
\includegraphics[angle=0,width=58mm]
{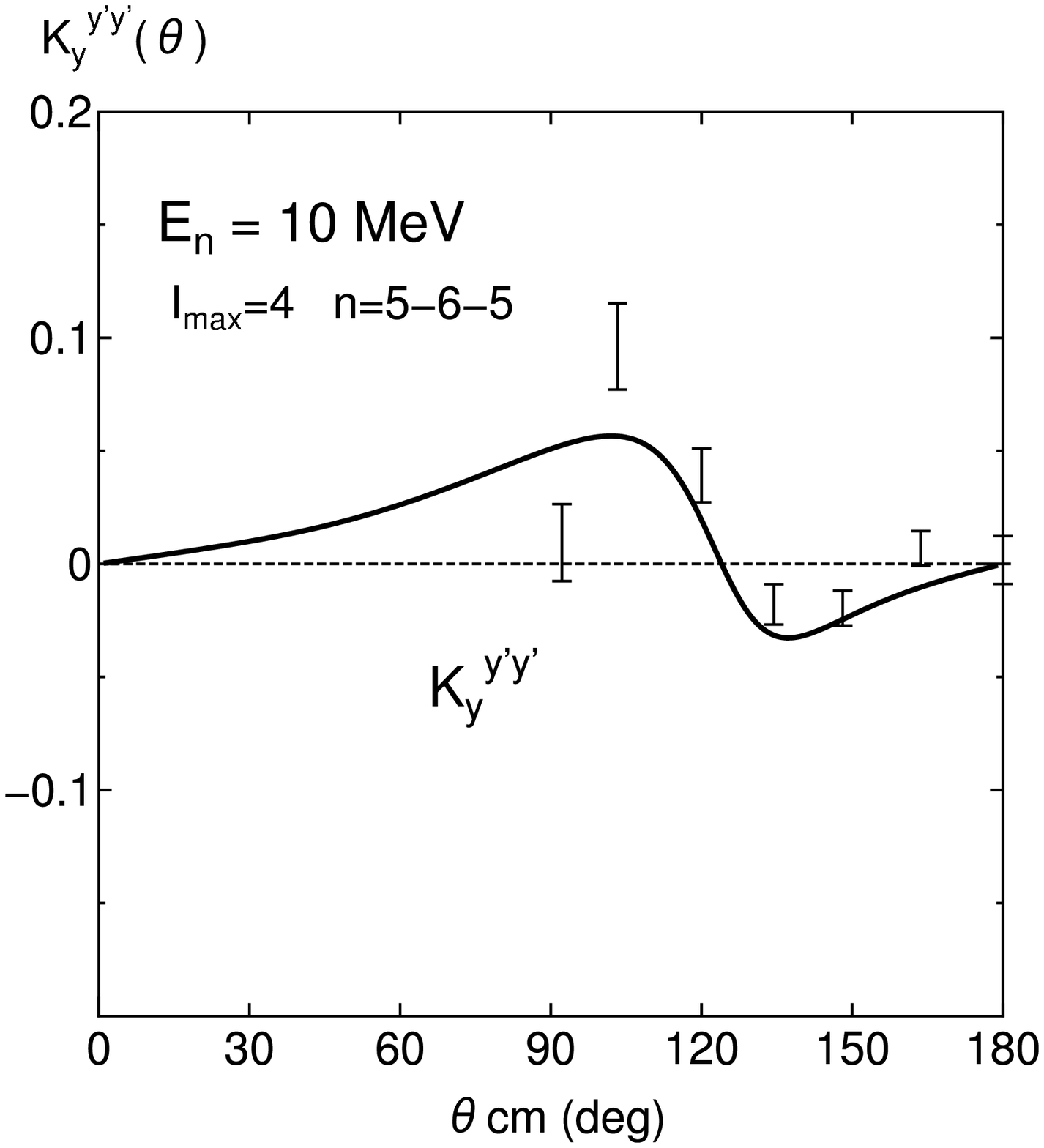}
\end{minipage}~%
\hfill~%
\begin{minipage}{0.48\textwidth}
\includegraphics[angle=0,width=58mm]
{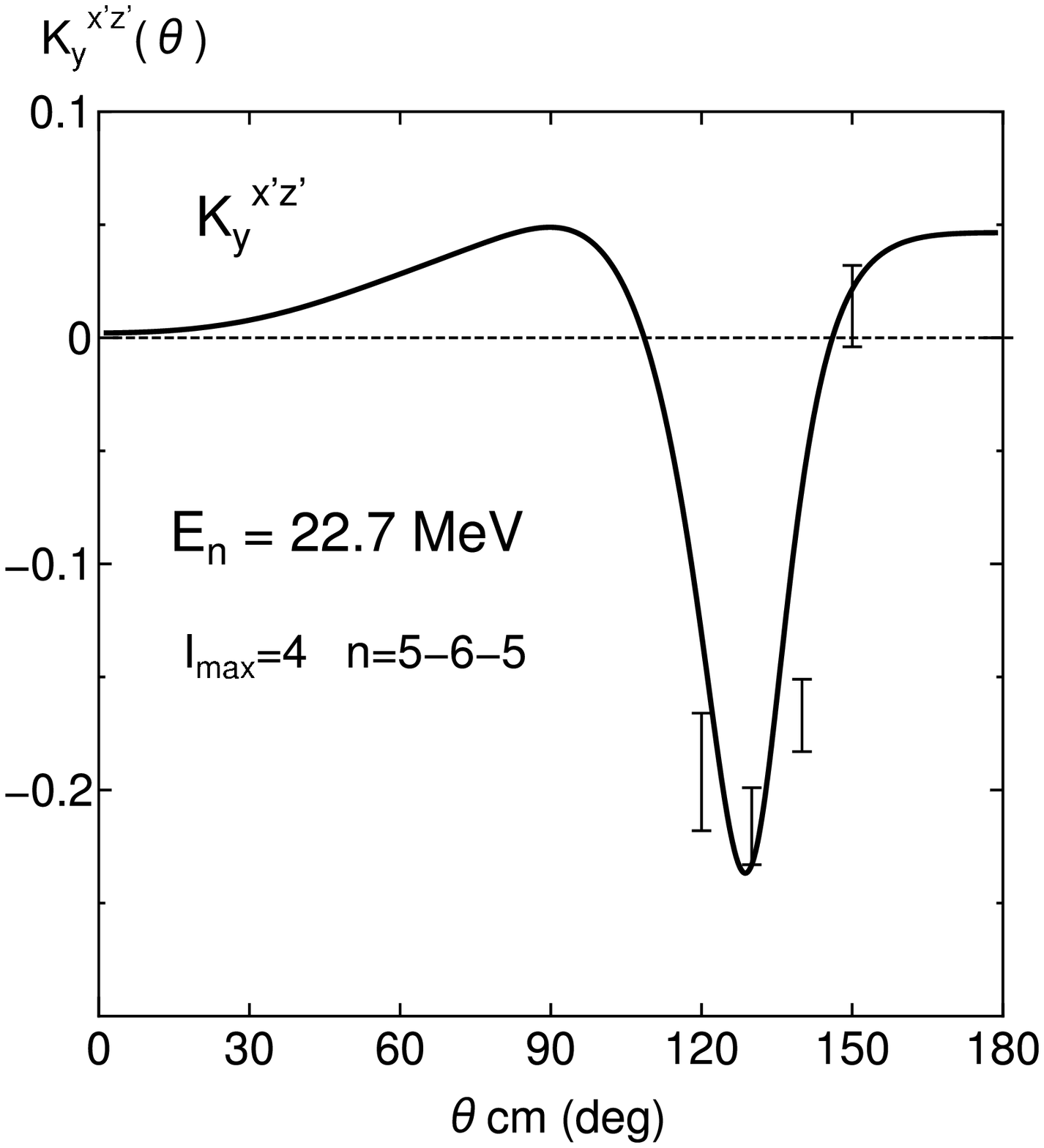}

\vspace{2mm}
\includegraphics[angle=0,width=58mm]
{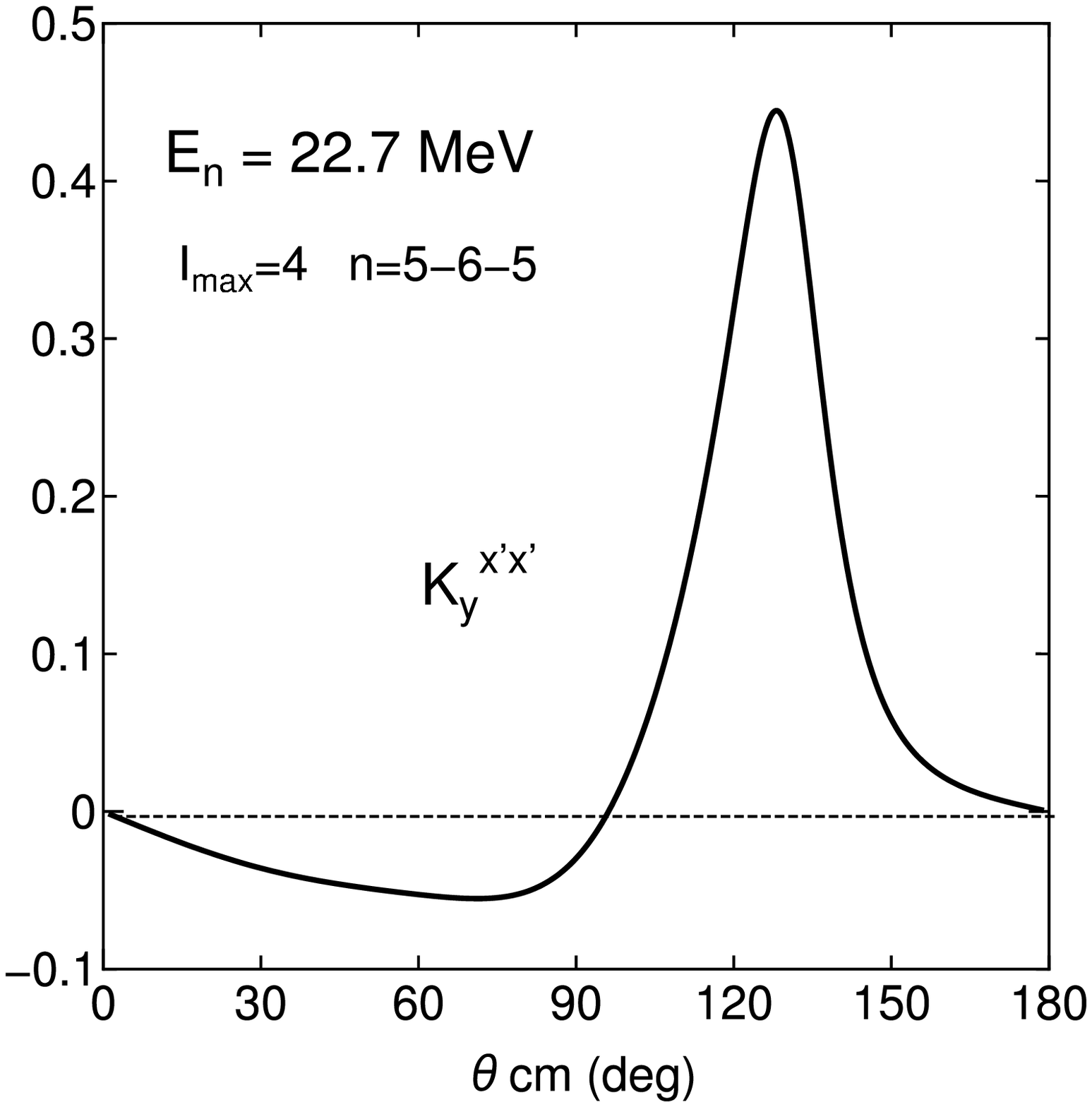}

\vspace{2mm}
\includegraphics[angle=0,width=58mm]
{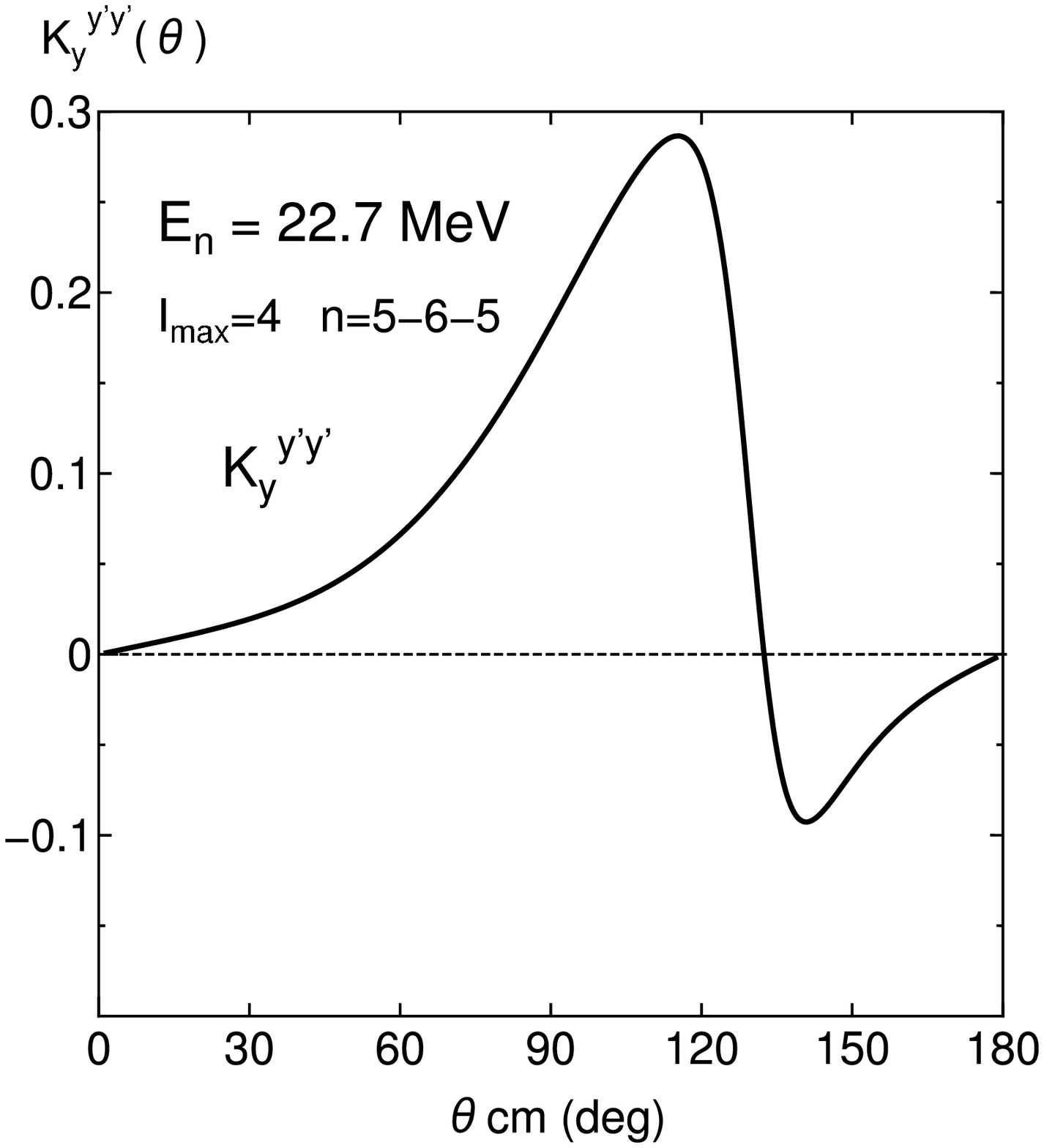}
\end{minipage}
\end{center}
\caption{
The same as Fig.\,\ref{Kd2}, but for the other types of tensor-type
nucleon to deuteron polarization transfer coefficients.
}
\label{Kd3}       
\end{figure}

\begin{figure}[b]
\begin{center}
\begin{minipage}{0.48\textwidth}
\includegraphics[angle=0,width=56mm]
{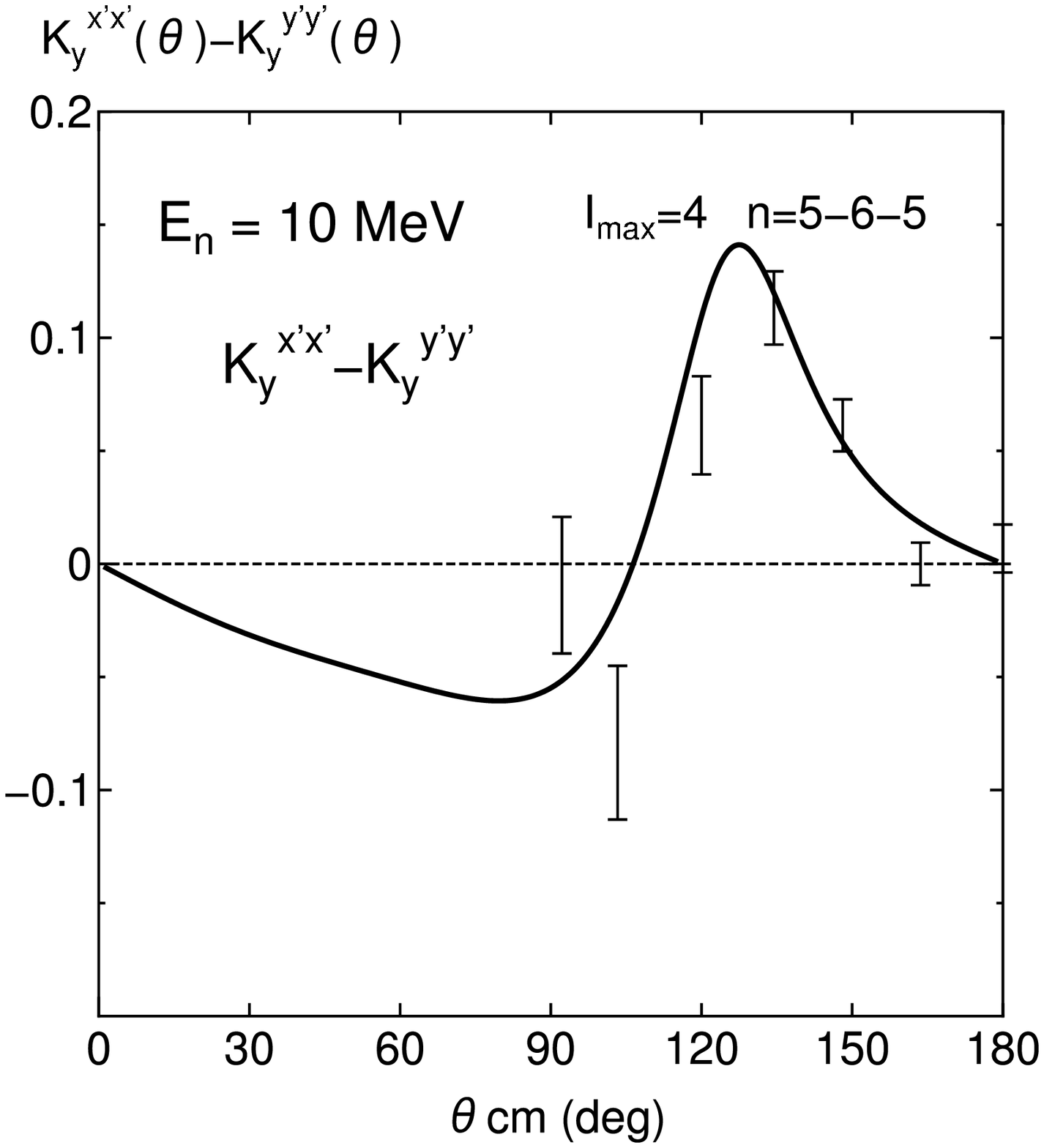}

\includegraphics[angle=0,width=56mm]
{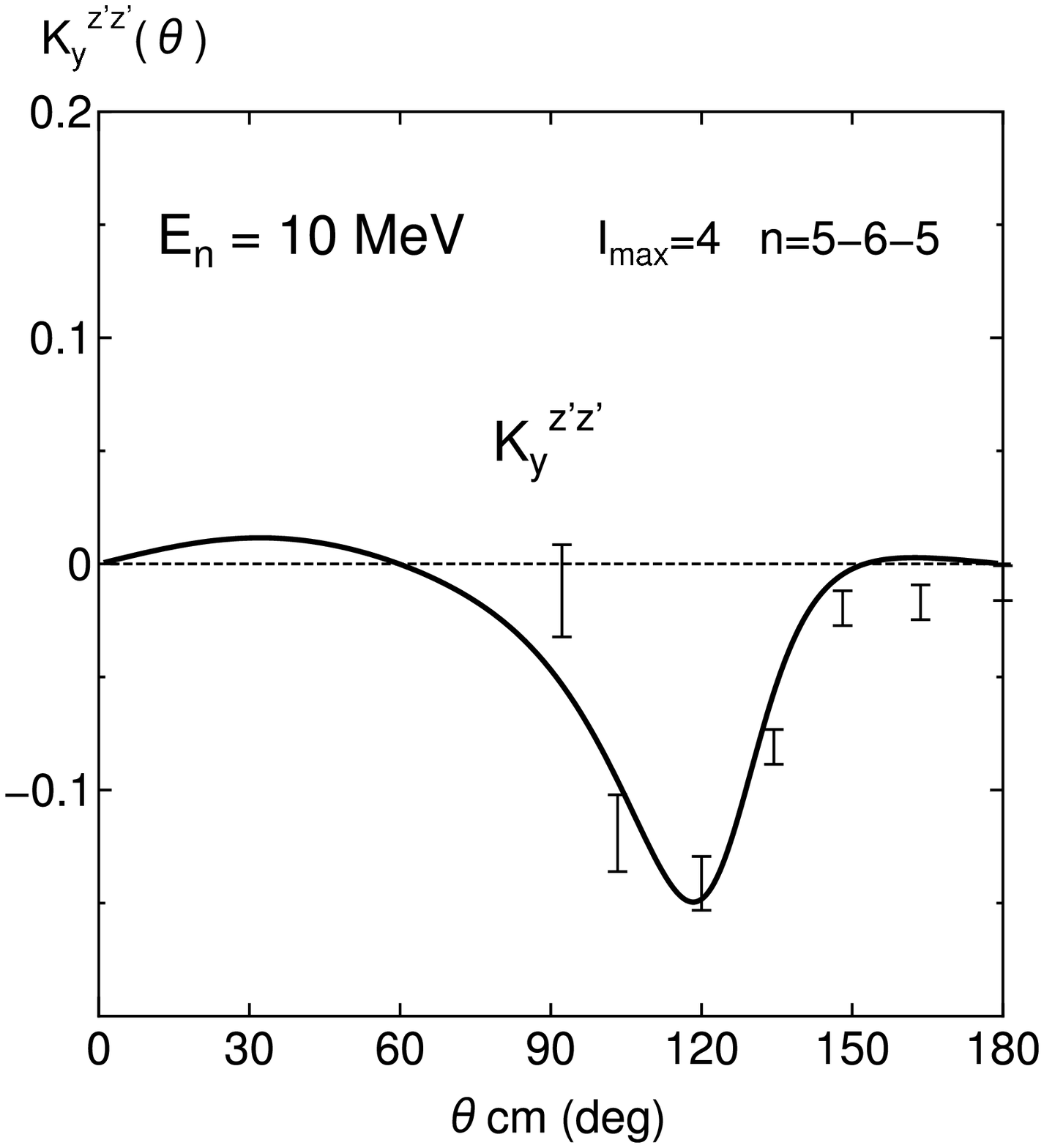}
\end{minipage}~%
\hfill~%
\begin{minipage}{0.48\textwidth}
\includegraphics[angle=0,width=56mm]
{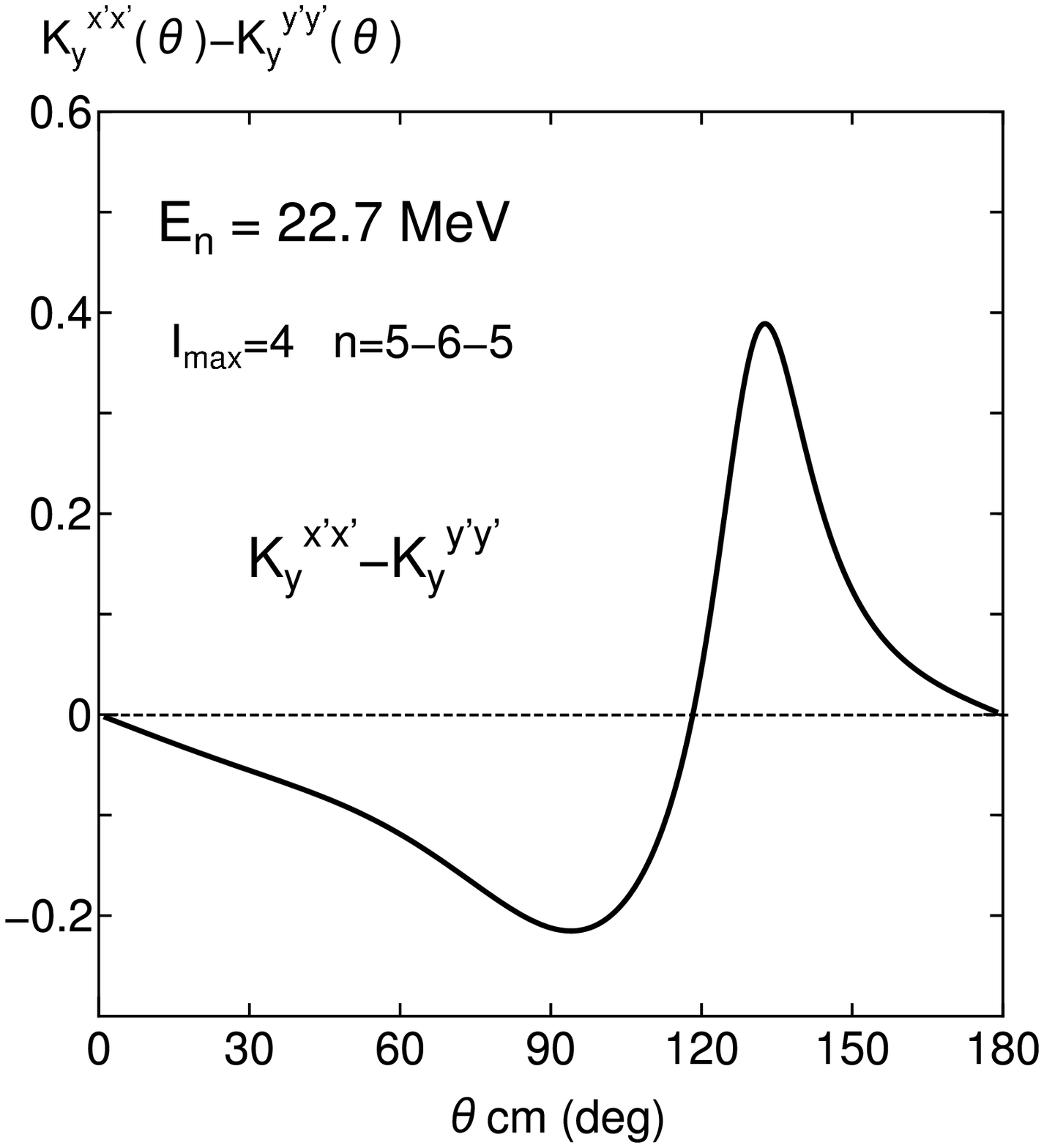}

\includegraphics[angle=0,width=56mm]
{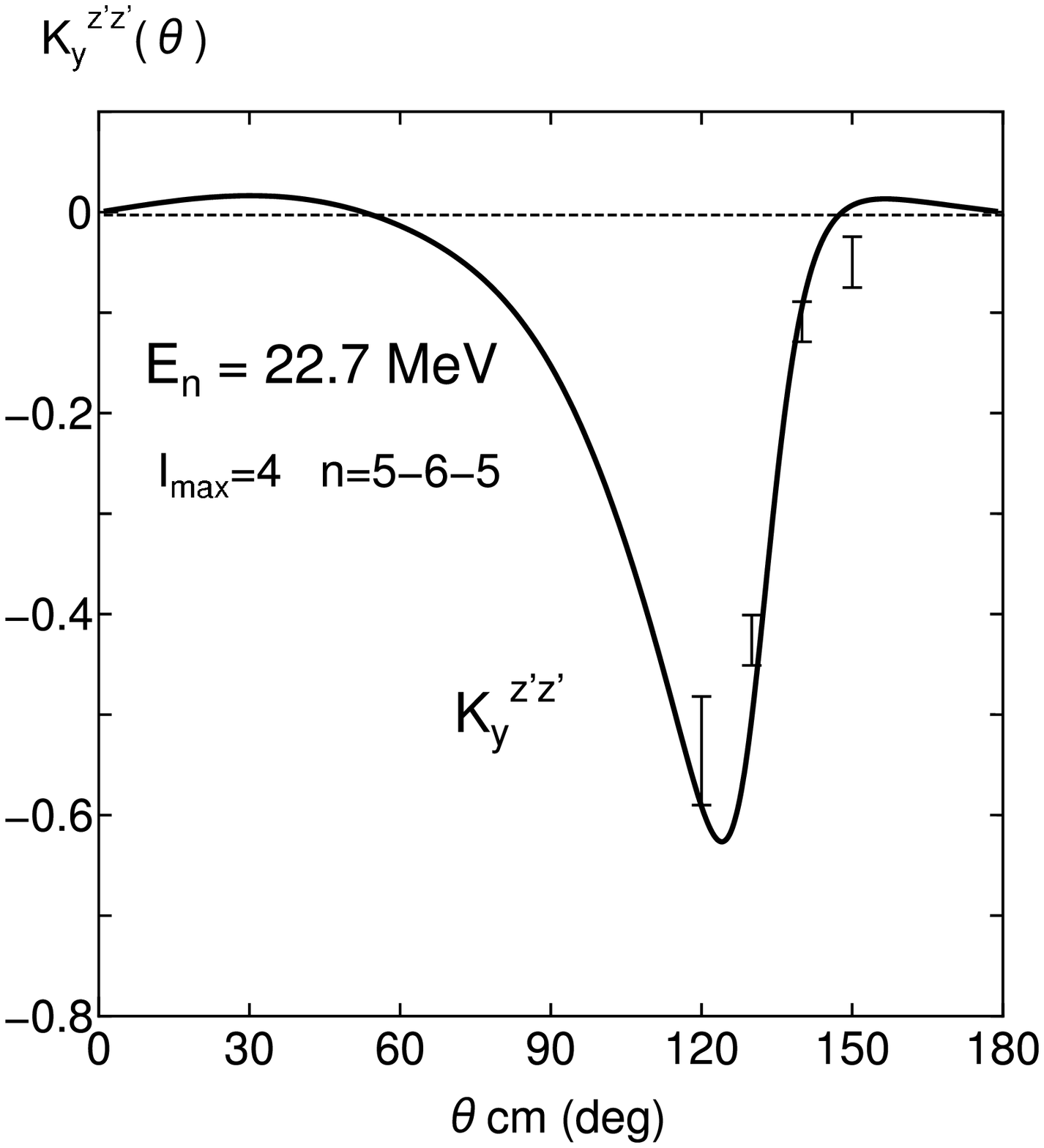}
\end{minipage}
\end{center}
\caption{
The same as Fig.\,\ref{Kd2}, but for the other types of tensor-type
nucleon to deuteron polarization transfer coefficients.
}
\label{Kd4}       
\end{figure}

The nucleon polarization transfer coefficients 
$K^{\beta^\prime}_\alpha(\theta)$ of 
the $nd$ elastic scattering are compared with the $pd$ data
for $E_n=10$ MeV and 22.7 MeV in Fig.\,\ref{Kn}.
The dotted curves represent the kinematical rotation
in \eq{sec4-5-68-5} when $K_{\alpha, \beta}=\delta_{\alpha, \beta}$
is assumed. From here on, we neglect the Coulomb effect and
compare the $nd$ results directly with the $pd$ data. This would be
permissible since the experimental errorbars are still large
for these polarization observables. We find satisfactory agreement
between the theory and experiment. 
The vector-type nucleon to deuteron polarization transfer
coefficients $K^{\beta^\prime}_\alpha(\theta)$ are compared
in Fig.\,\ref{Kd1} at the same energies $E_n=10$ and 22.7 MeV.
The dotted curves again represent the kinematical rotation
in \eq{sec4-5-68-9} when $K_{\alpha, \beta}=\delta_{\alpha, \beta}$
is assumed.
Various types of the nucleon to deuteron polarization transfer 
coefficients are compared with the $pd$ data 
in Figs.\,\ref{Kd2} - \ref{Kd4}.
Here, again we obtain satisfactory agreement although the
experimental errorbars are rather large.

\bigskip

\subsection{Spin correlation coefficients}

\begin{figure}[b]
\begin{center}
\begin{minipage}{0.48\textwidth}
\includegraphics[angle=0,width=54mm]
{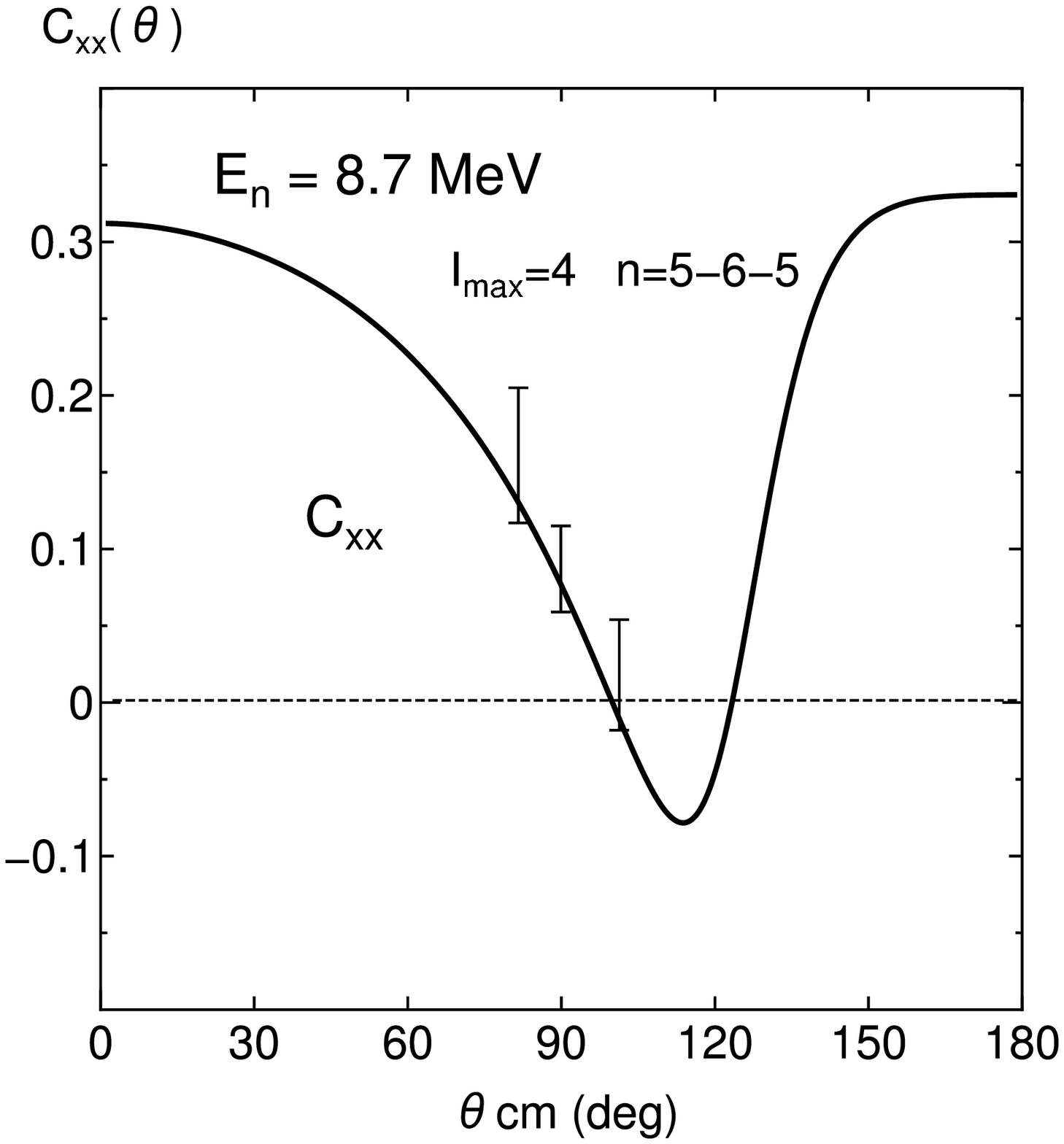}

\includegraphics[angle=0,width=54mm]
{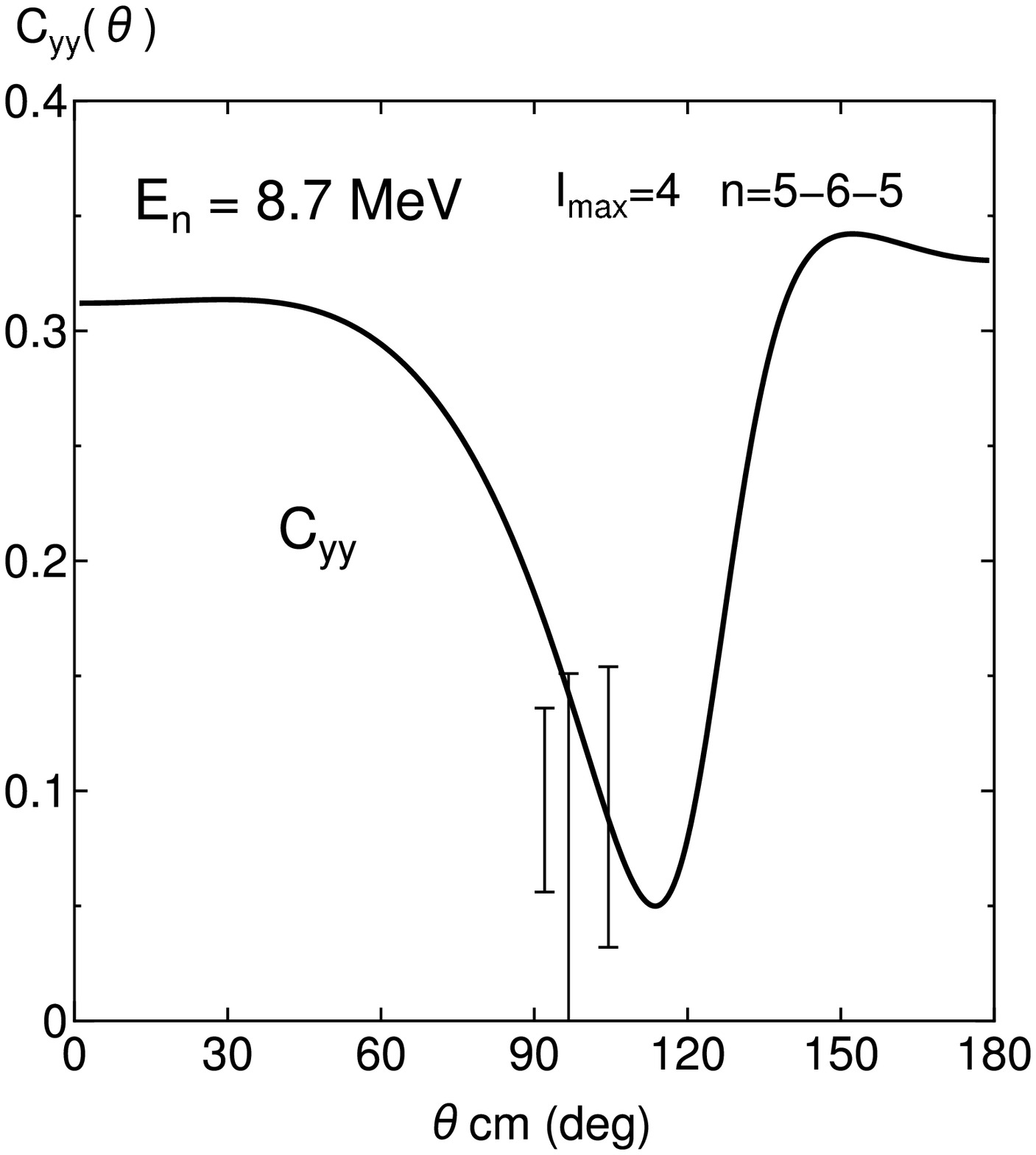}

\includegraphics[angle=0,width=54mm]
{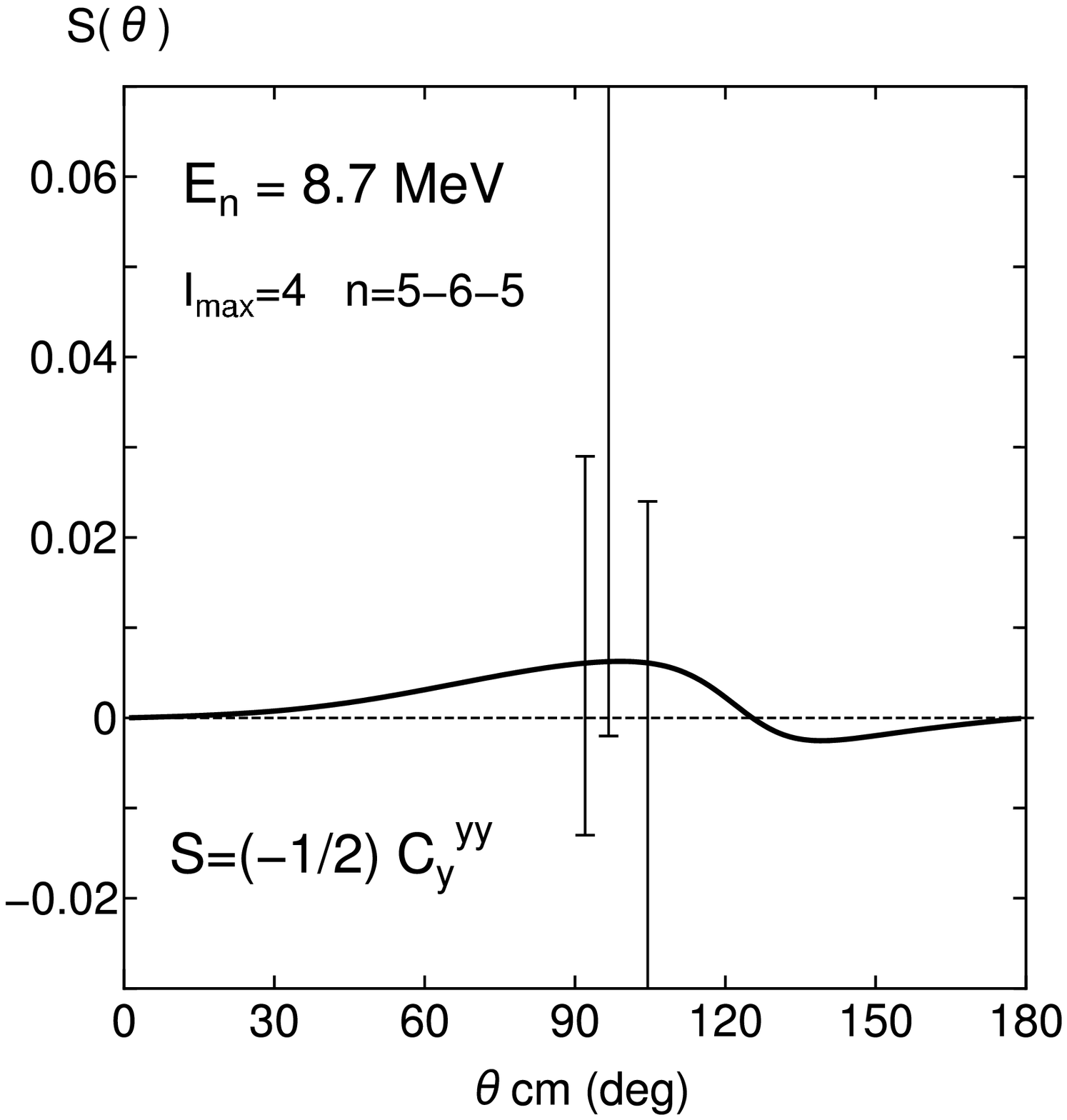}
\end{minipage}~%
\hfill~%
\begin{minipage}{0.48\textwidth}
\includegraphics[angle=0,width=54mm]
{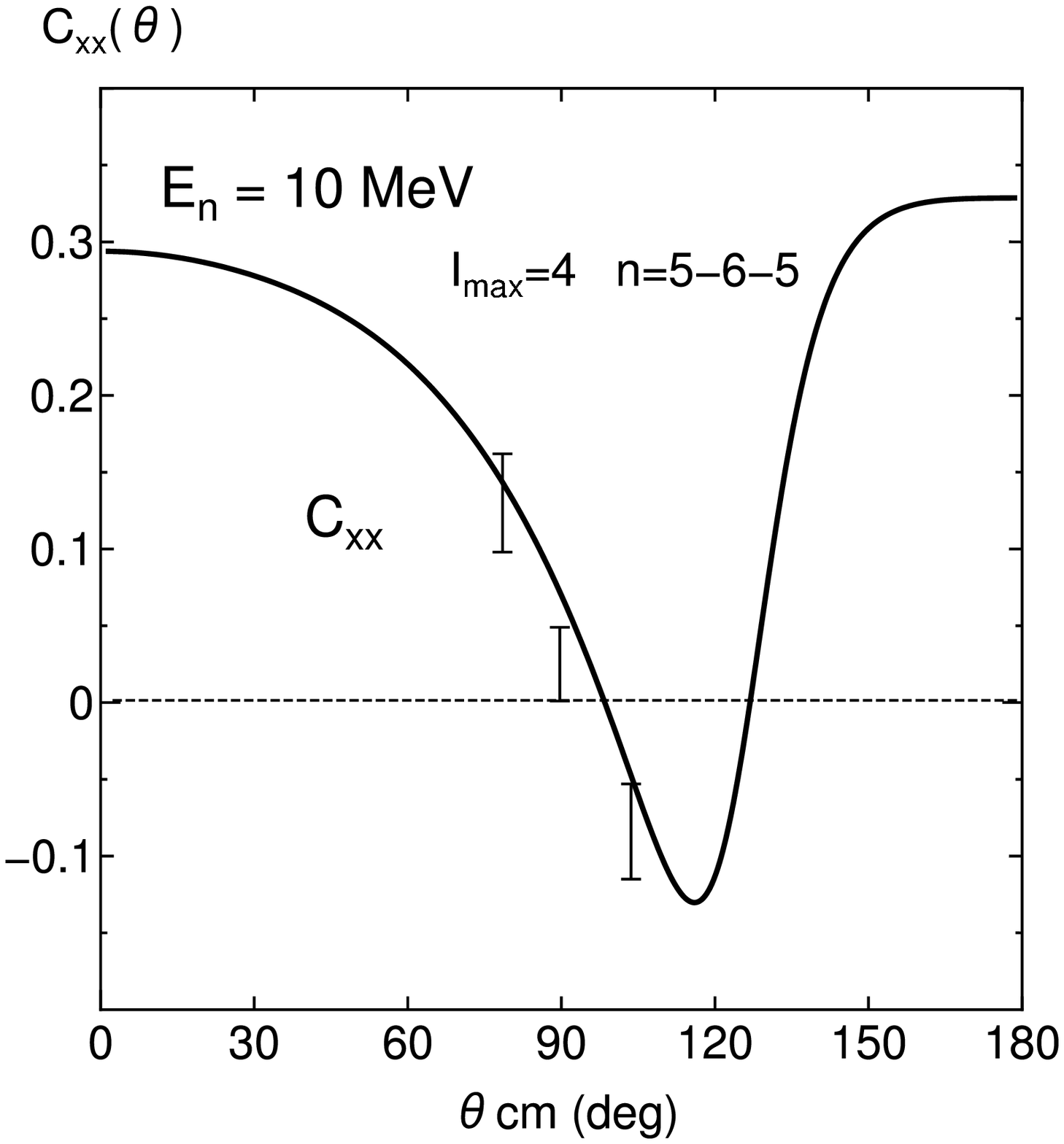}

\includegraphics[angle=0,width=54mm]
{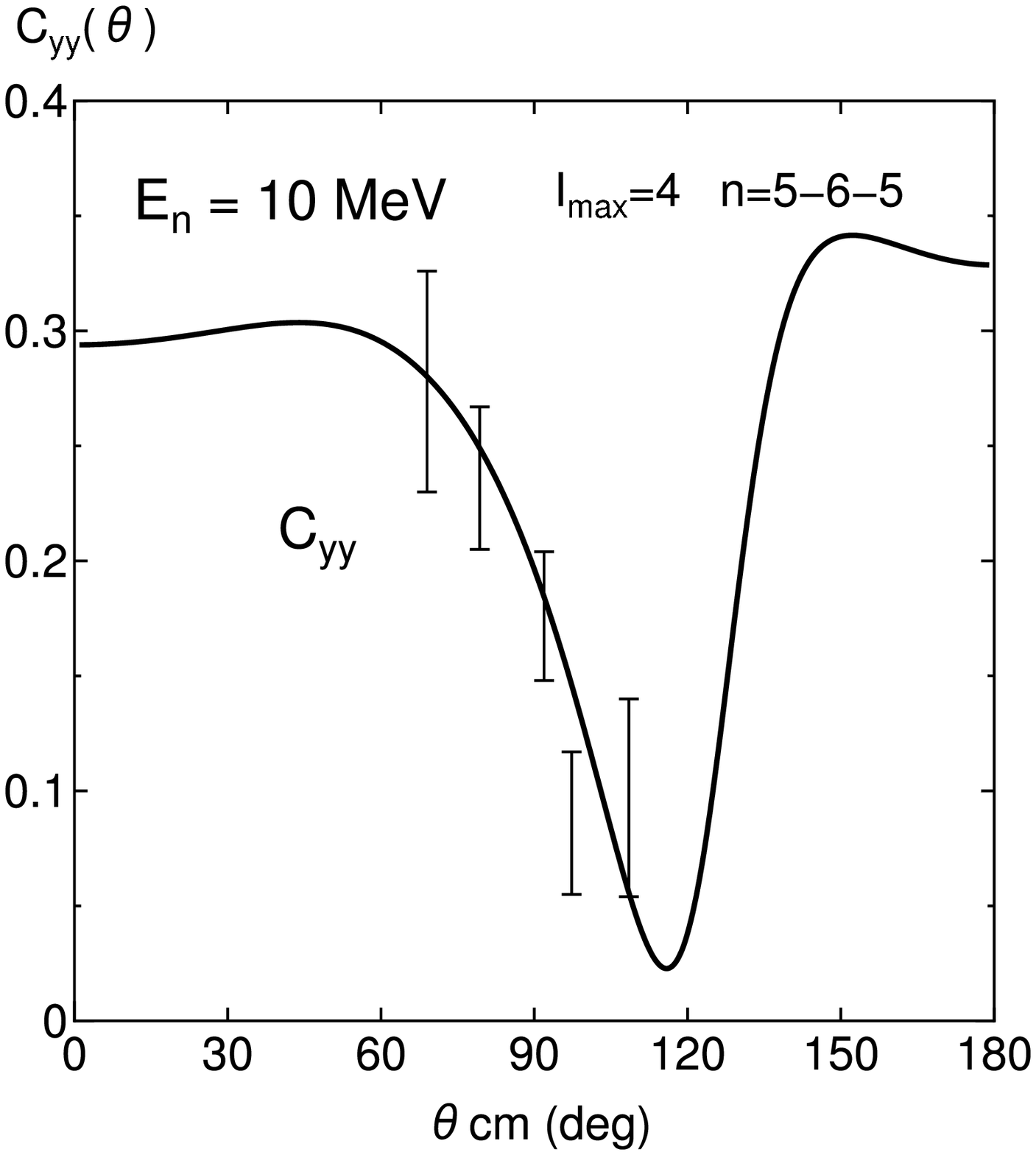}

\vspace{25mm}
\caption{
The spin correlation coefficients,
$C_{xx}$, $C_{yy}$, and $S=(-1/2)C^{yy}_{y}$, of 
the $nd$ elastic scattering for $E_n=8.7$ MeV and 10 MeV, 
compared with the $dp$ experimental data in Ref.\,\citen{Ch75}.
}
\label{Sc1}  
\vspace{10mm}
\end{minipage}
\end{center}
\end{figure}

\begin{figure}[b]
\begin{center}
\begin{minipage}{0.48\textwidth}
\includegraphics[angle=0,width=54mm]
{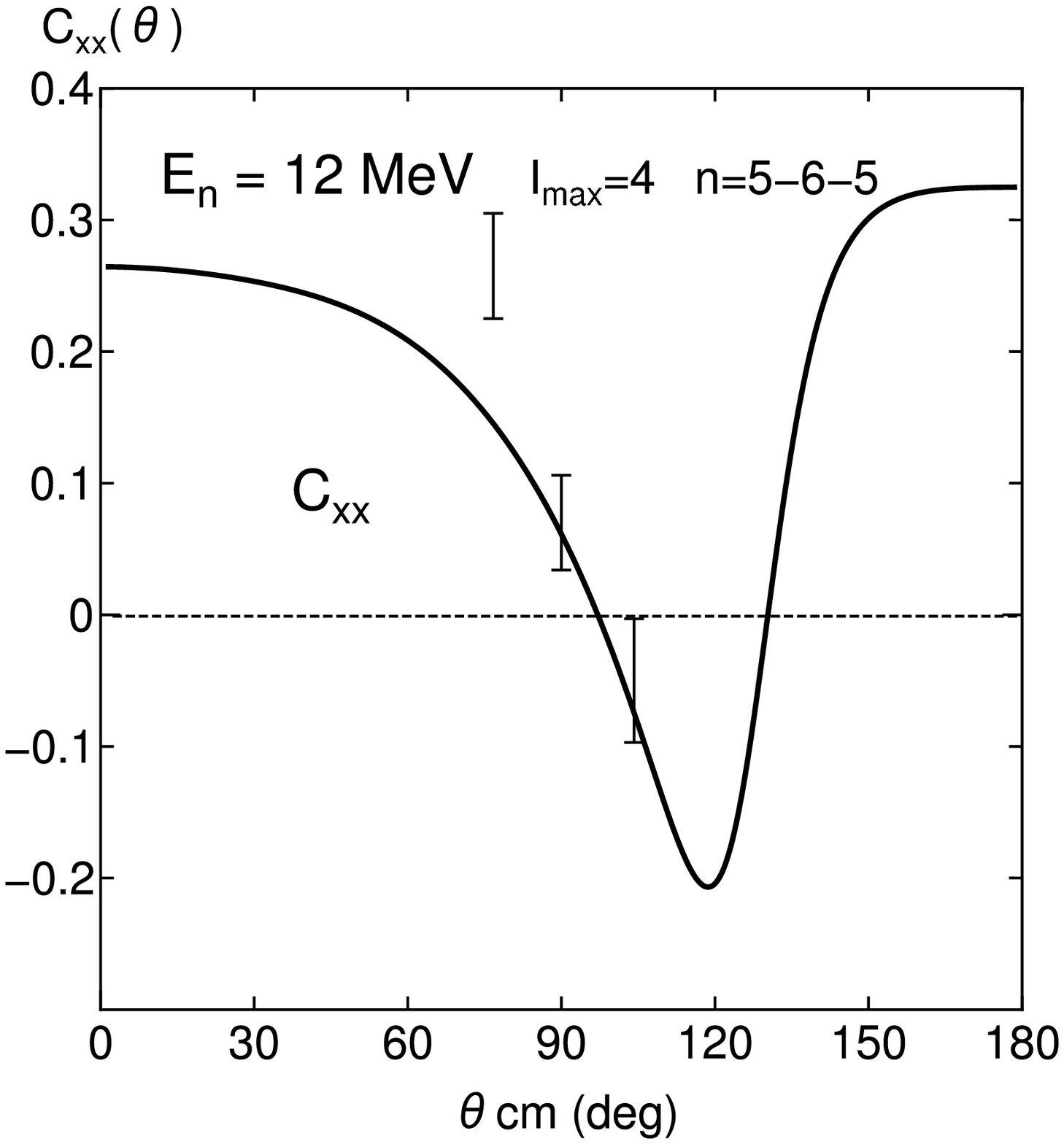}

\includegraphics[angle=0,width=54mm]
{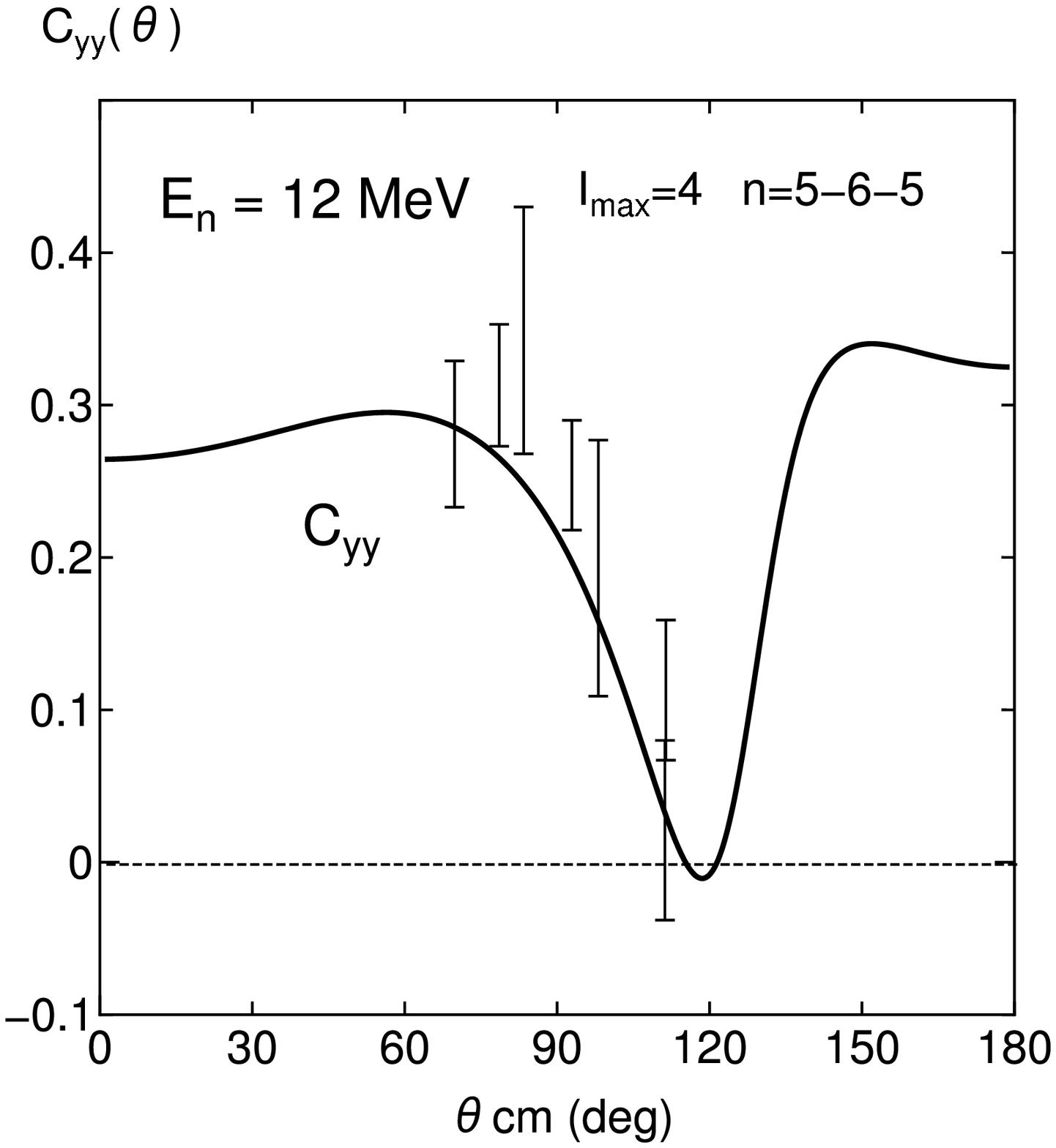}

\includegraphics[angle=0,width=54mm]
{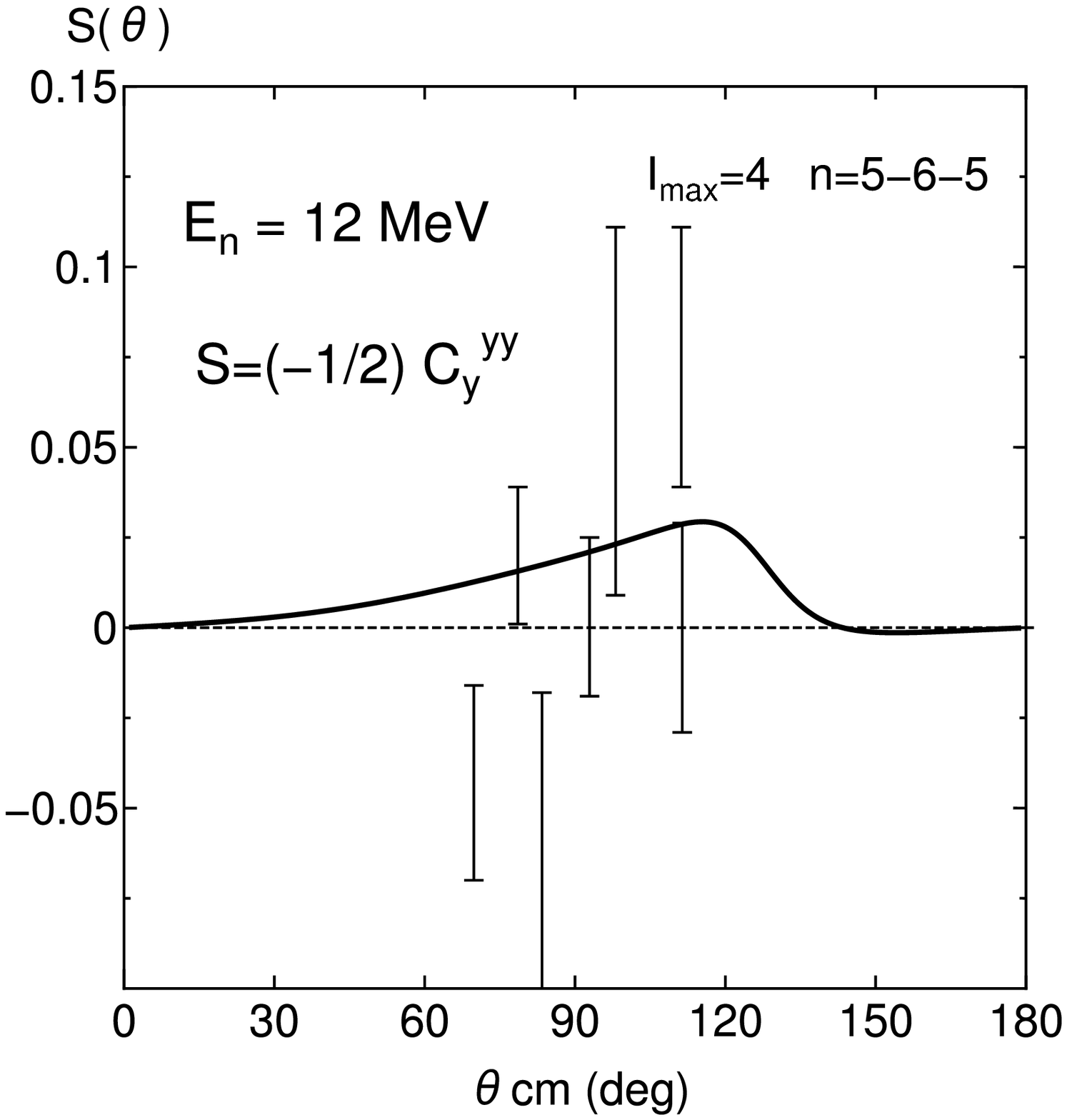}
\end{minipage}~%
\hfill~%
\begin{minipage}{0.48\textwidth}
\includegraphics[angle=0,width=54mm]
{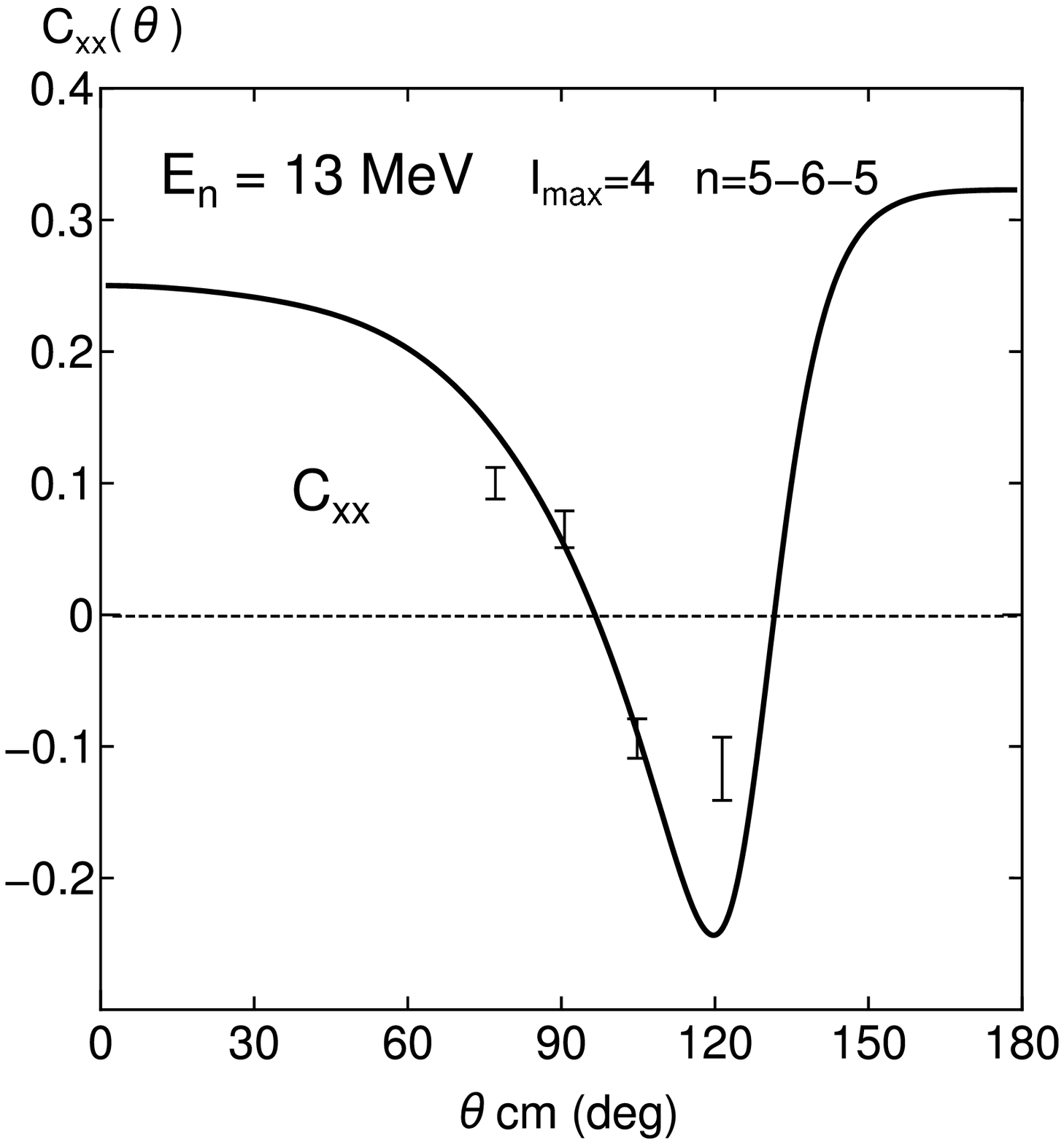}

\includegraphics[angle=0,width=54mm]
{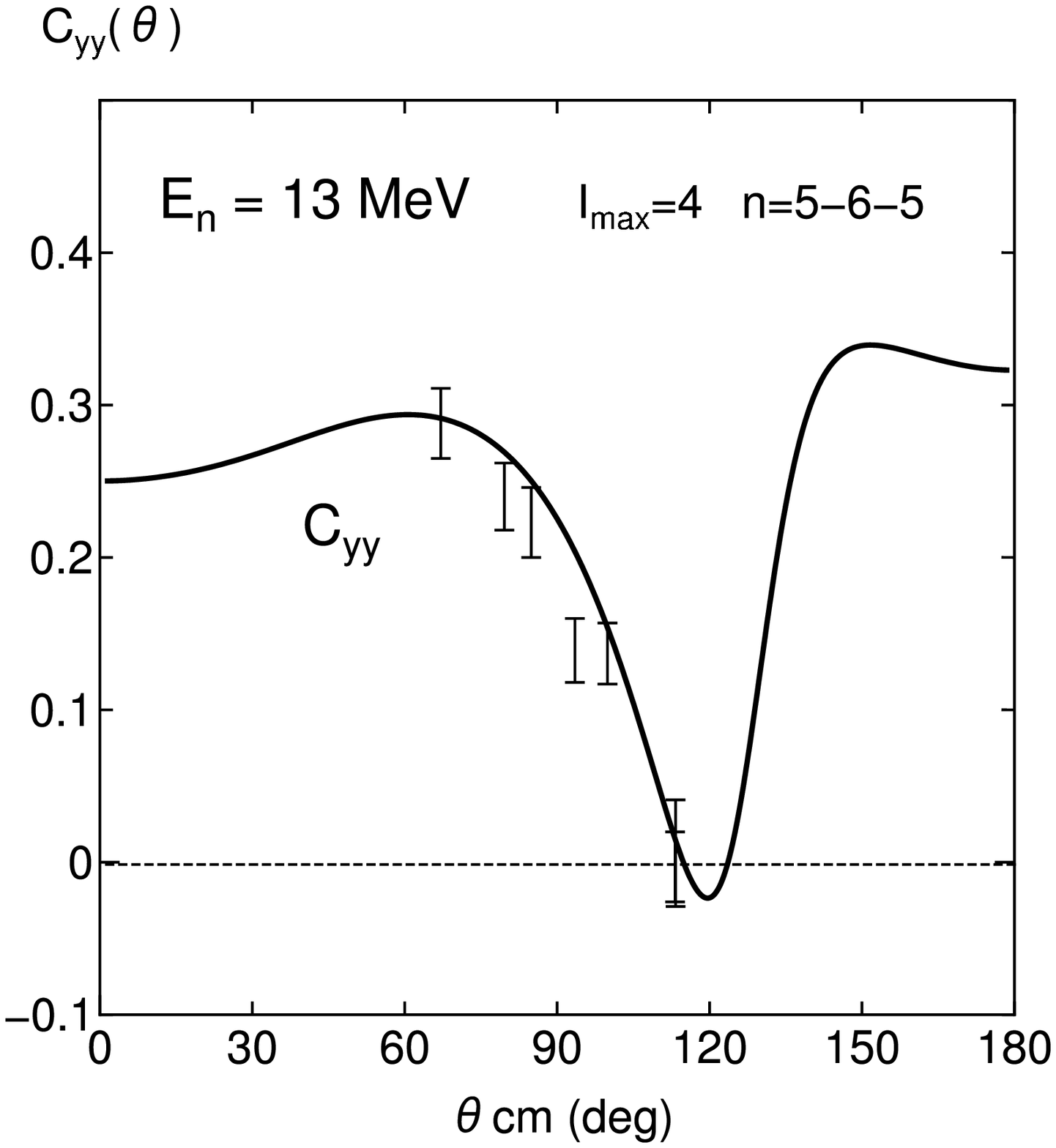}

\includegraphics[angle=0,width=54mm]
{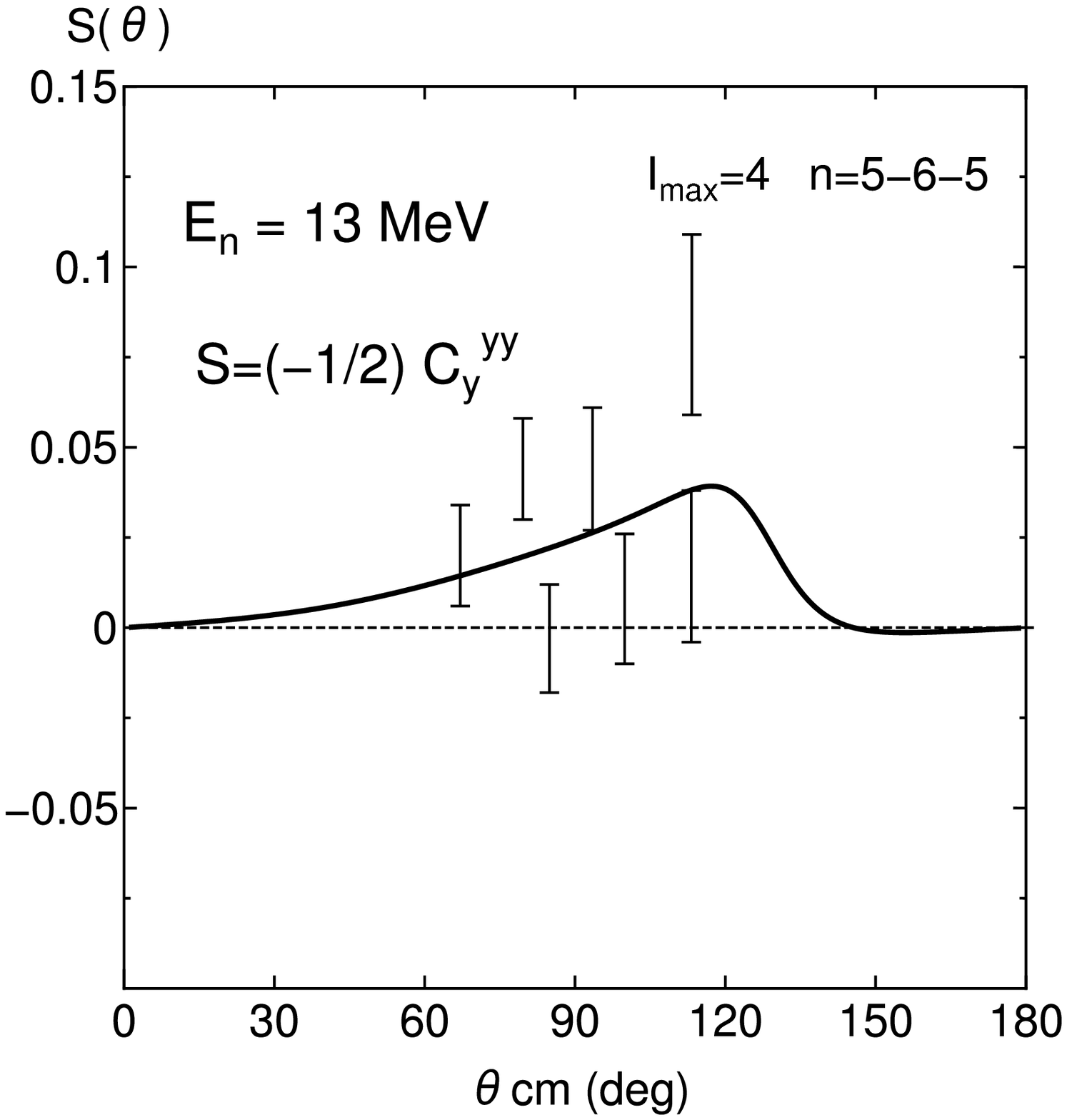}
\end{minipage}
\end{center}
\caption{
The same as Fig.\,\ref{Sc1}, but for $E_n=12$ MeV and 13 MeV. 
}
\label{Sc2}       
\end{figure}

The spin correlation coefficients,
$C_{xx},~C_{yy}$, and $S=(-1/2)C^{yy}_{y}$, for 
the $nd$ elastic scattering at $E_n=8.7$, 10, 12 and 13 MeV
are compared with the $pd$ experimental data
in Figs.\,\ref{Sc1} and \ref{Sc2}.
The experimental data are actually for the
$dp$ scattering at the energies $E_d=17.4$, 19.5, 23.8
and 26.1 MeV.\cite{Ch75} 
We find a reasonable agreement between the theory and experiment,
although the experimental errorbars are rather large.

\bigskip

\section{Summary}

One of the most important purposes of studying 
the three-nucleon ($3N$) system in terms of realistic 
nucleon-nucleon ($NN$) interactions is to clarify
the off-shell properties of the interaction, which can never be 
known from the physical observables of the two-nucleon system.
It is therefore very crucial that the $NN$ interaction 
to start with can reproduce the deuteron properties and
the $NN$ phase shifts very accurately. 
Our quark-model $NN$ interaction fss2 satisfies this criterion 
by taking into account the naive three-quark structure of the nucleon.
It is formulated in the framework of the resonating-group method (RGM),
in which the antisymmetrization of quarks generates a strong nonlocality
and characteristic energy dependence of the interaction between
two three-quark clusters. When this interaction is applied to
the $3N$ system, the short-range repulsion originating 
from this nonlocality behaves quite differently from the meson-exchange
potentials in favor of the deuteron distortion in the spin-doublet
channel. We have found that fss2 yields sufficient attraction
comparable to the AV18 plus Urbana $3N$ force,
leading to the nearly correct triton binding energy and the empirical
value of the spin-doublet effective length 
for the $nd$ interaction.\cite{scl10} 
Note that this strong deuteron distortion effect is only for the 
$J^\pi=1/2^+$ channel. On the other hand, the distortion effect
is marginal in the spin quartet channel or the $J^\pi=3/2^+$ channel,
owing to the Pauli principle on the nucleon level.
As the result, the eigenphase shift
of the dominant $\hbox{}^4S_{3/2}$ state
predicted by fss2 is very similar to the AV18 potential,
in which the effect of $3N$ force is very small.
In the low-energy region, the effect of the $3N$ force hardly appears
since the differential cross sections are dominated by the partial
waves of the spin-quartet channel. It is important to note that
fss2 can also reproduce the differential cross sections at higher
energies, including the sufficient magnitude of the cross sections
at the diffraction minima.\cite{ndsc1}

In this paper, we have extended the previous studies,\cite{ndsc1,scl10}
applying fss2 to the $nd$ elastic scattering,
to various types of polarization observables.
The long-standing $A_y$-puzzle \cite{To98a,To98b,To02,To08} for 
the nucleon vector analyzing-powers in the low-energy 
region $E_n \le 25$ MeV is largely improved
in comparison with the predictions by the AV18 potential.
Although the Coulomb effect obscures the definite conclusion,
the shortage of the maximum peak is about 15$\%$, which is 
somewhat similar to the predictions by old realistic separable
potentials.\cite{Ra88,Ko87a,Ko87b}
The detailed comparison with the phase shift analysis
is not possible since the empirical phase shifts are
not uniquely determined.
Nevertheless, the conclusion drawn in Ref.\,\citen{To02}
based on the solutions derived from the AV18 potential should be 
reconsidered, since $\hbox{}^2P_J$ phase shifts at the energy 
$E_p \sim 10$ MeV are quite different between fss2 and the AV18 potential.
The difference is of the order of several degrees.
It is possible that this difference is caused by the nonlocality
of the quark-model $NN$ interaction. 
Since all the experimental data for the deuteron analyzing-powers
are for the $pd$ or $dp$ scattering, we have introduced 
the cut-off Coulomb force by the Vincent and Phatak method.
This method works very well to reproduce the behavior of the
differential cross sections and analyzing-powers at the forward angles.
In particular, the low-energy observables below the deuteron breakup
threshold are well reproduced except for the $A_y$-puzzle 
and a slight underestimation of the dips in $T_{22}(\theta)$
at the minimum points around $\theta_{\rm cm}=100^\circ$ - $120^\circ$.
A large discrepancy of the vector-type analyzing-power of the deuteron,
$iT_{11}(\theta)$, found in Ref.\,\citen{Ki01} is not observed in our
calculations, although the peak height
around $\theta_{\rm cm} = 120^\circ$ is somewhat too low at $E_p=3$ - 9 MeV.
Instead, we have unpleasant rise of the hill at the angular region 
$\theta_{\rm cm}=20^\circ$ - $90^\circ$ for the energies
$E_p=5$ - 10 MeV, which is probably related to the inadequacy
of the nuclear-Coulomb interference term in our Coulomb treatment.
A similar problem is also seen in $T_{20}$ and $T_{21}$ in the angular rigion
$\theta_{\rm cm}=20^\circ$ - $70^\circ$ for the energies $E_p=7$ - 9 MeV.
Except for these, there is no clear discrepancies 
between the theory and experiment.
We have carefully examined the Coulomb effect on the observables
between our results and other calculations using the modern
meson exchange potentials like AV18 and CD Bonn
potentials.\cite{Ki01,De05,De05c,Is09}
The direction of the Coulomb modification is always the same, although some
quantitative diference appears at the forward angles.
Our calculation somehow overestimates the Coulomb effect and
nuclear-Coulomb interference term is not precisely reproduced.
On the other hand, the behavior of the observables in the backward
angles $\theta_{\rm cm} \gg 90^\circ$ is hardly influenced especially
at higher energies $E_p > 10$ MeV.  
We have also examined various types of polarization transfer coefficients
and the spin correlation coefficients by neglecting the Coulomb force.
They are reasonably reproduced, although the experimental errorbars
are still very large.

In conclusion, we find no apparent disagreement between the theory
and experiment as far as $nd$ elastic scattering with the
energies $E_n \leq 65$ MeV is concerned.
It should be stressed that this conclusion is valid only when 
we deal with the energy dependence of the quark-model RGM kernel
properly.\cite{FK10a,FK10b} As discussed in our previous paper,\cite{ndsc1}
this energy dependence is eliminated by the standard off-shell transformation
utilizing the square root of the normalization kernel.
This procedure yields an extra nonlocal kernel
which is not extremely small and affects various $3N$ observables
in different ways. 
In the next paper, we will discuss the deuteron breakup processes.\cite{ndsc3}

\section*{Acknowledgements}

The authors would like to thank Professors K. Miyagawa,
H. Wita\l a, H. Kamada and S. Ishikawa for many useful comments.
We also thank Professor K. Sagara for providing us
the $pd$ experimental data by the Kyushu university group.
This work was supported by the Grant-in-Aid for Scientific
Research on Priority Areas (Grant No.~20028003), and
by the Grant-in-Aid for the Global COE Program
``The Next Generation of Physics, Spun from Universality and Emergence'' 
from the Ministry of Education, Culture, Sports, Science 
and Technology (MEXT) of Japan. 
It was also supported by the core-stage backup subsidies of Kyoto University.
The numerical calculations were carried out on Altix3700 BX2 at YITP 
in Kyoto University and on the high performance computing system,
Intel Xeon X5680, at RCNP in Osaka University.

\bigskip

\appendix

\section{Spin operators of the deuteron}

In this appendix, we summarize various notations for the
spin operators of the deuteron.
The deuteron spin operators 
are most easily defined from the Wigner-Eckart theorem
\begin{eqnarray}
\langle 1 m^\prime |S_\mu| 1 m \rangle
=\langle 1 m 1 \mu |1 m^\prime \rangle \langle 1 ||\bS|| 1 \rangle_{\rm unc}
\ ,
\label{sec4-5-3}
\end{eqnarray}
with the standard value of the reduce matrix element
$\langle 1 ||\bS|| 1 \rangle_{\rm unc}=\sqrt{1\cdot 2}$.
Here, the subscript ``unc'' stands for the
unconventional reduced matrix element 
with $\langle S ||\bS|| S \rangle_{\rm unc}=\sqrt{S(S+1)}$.
This gives the rank-one spin operator of the deuteron as
\begin{eqnarray}
S_1=-\left( \begin{array}{ccc}
0 & 1 & 0 \\
0 & 0 & 1 \\
0 & 0 & 0 \\
\end{array} \right),\quad
S_0=\left( \begin{array}{ccc}
1 & 0 & 0 \\
0 & 0 & 0 \\
0 & 0 & -1 \\
\end{array} \right),\quad
S_{-1}=\left( \begin{array}{ccc}
0 & 0 & 0 \\
1 & 0 & 0 \\
0 & 1 & 0 \\
\end{array} \right)=-{S_1}^\dagger .
\nonumber \\
\label{sec4-5-4}
\end{eqnarray}
The transformation to the Cartesian representation is carried out by
the standard spherical vector representation:
\begin{eqnarray}
\begin{array}{l}
S_1=-\frac{1}{\sqrt{2}}\left(S_x+iS_y\right) \\ [2mm]
S_0 =S_z \\ [2mm]
S_{-1}=\frac{1}{\sqrt{2}}\left(S_x-iS_y\right) \\
\end{array}\ \ ,\qquad
\begin{array}{l}
S_x=\frac{1}{\sqrt{2}}\left(-S_1+S_{-1}\right) \\ [2mm]
S_y =i\,\frac{1}{\sqrt{2}}\left(S_1+S_{-1}\right) \\ [2mm]
S_z=S_0 \\
\end{array}\ ,
\label{sec4-5-5}
\end{eqnarray}
which yields
\begin{eqnarray}
S_x=\frac{1}{\sqrt{2}}\left( \begin{array}{ccc}
0 & 1 & 0 \\
1 & 0 & 1 \\
0 & 1 & 0 \\
\end{array} \right),\quad
S_y=\frac{i}{\sqrt{2}}\left( \begin{array}{ccc}
0 & -1 & 0 \\
1 & 0 & -1 \\
0 & 1 & 0 \\
\end{array} \right),\quad
S_z=\left( \begin{array}{ccc}
1 & 0 & 0 \\
0 & 0 & 0 \\
0 & 0 & -1 \\
\end{array} \right).
\nonumber \\
\label{sec4-5-6}
\end{eqnarray}
Note that $\bS^\dagger=\bS$ and $\bS^2=2$, 
which is consistent with the reduced matrix element.
The rank-two spin operator of the deuteron is defined by
$S^{(2)}_m=[S S]^{(2)}_m$ and the reduced matrix element
is calculated from
\begin{eqnarray}
\langle 1 m^\prime |S^{(2)}_\mu| 1 m \rangle
=\langle 1 m 2 \mu |1 m^\prime \rangle \langle 1 ||S^{(2)}
|| 1 \rangle_{\rm unc}\ .
\label{sec4-5-7}
\end{eqnarray}
We note that the spin operator $\bS$ does not change the spin value $S=1$,
when it is operated on the deuteron spin wave function
$\chi_1=\chi_1(12)$:
\begin{eqnarray}
S|\chi_1\rangle_{(11)1\mu}=-[S \chi_1]_{1\mu}=\sqrt{2}\chi_{1\mu}\ .
\label{sec4-5-8}
\end{eqnarray}
(Note the phase change by the angular-momentum coupling order.)
We find
\begin{eqnarray}
& & S^{(2)}|\chi_1\rangle_{(12)1\mu}=[S^{(2)}\chi_1]_{1\mu}
=[[SS]_2 \chi_1]_{1\mu}=\sum_{\lambda=0,1,2}\left[ \begin{array}{ccc}
1 & 1 & 2 \\ [2mm]
0 & 1 & 1 \\ [2mm]
1 & \lambda & 1 \\
\end{array}\right]
[S [S\chi_1]_\lambda]_{1\mu}
\nonumber \\
& & =\left[ \begin{array}{ccc}
1 & 1 & 2 \\ [2mm]
0 & 1 & 1 \\ [2mm]
1 & 1 & 1 \\
\end{array}\right]
(-\sqrt{2})[S\chi_1]_{1\mu}
=\sqrt{15}\left\{\begin{array}{ccc}
2 & 1 & 1 \\
1 & 1 & 1 \\
\end{array}\right\}
(-\sqrt{2})(-\sqrt{2}) \chi_{1\mu}
\nonumber \\
& & =\sqrt{15}\frac{1}{6}(-\sqrt{2})(-\sqrt{2})
=\sqrt{\frac{5}{3}} \chi_{1\mu}\ ,
\label{sec4-4-9}
\end{eqnarray}
which yields
\begin{eqnarray}
\langle 1 ||S^{(2)} || 1 \rangle
=\sqrt{\frac{5}{3}}\ .
\label{sec4-5-10}
\end{eqnarray}
This value and \eq{sec4-5-7} gives
\begin{eqnarray}
\langle 1 m^\prime |S^{(2)}_\mu| 1 m \rangle
=(-)^{1-m} \langle 1 m^\prime 1 -m | 2 \mu \rangle\ ,
\label{sec4-5-11}
\end{eqnarray}
owing to the symmetry property of the CG coefficients.
The explicit values of the CG coefficients
%
%
\begin{eqnarray}
\langle 1111|22 \rangle=1\ ,\quad
\langle 1110|21 \rangle=\frac{1}{\sqrt{2}}\ ,\quad
\langle 1010|20 \rangle=\sqrt{\frac{2}{3}}\ ,\quad
\langle 11~1,\,-1 |20 \rangle=\frac{1}{\sqrt{6}}\ ,
\nonumber \\
\label{sec4-5-12}
\end{eqnarray}
and other non-zero coefficients obtained by the symmetries, yield
\begin{eqnarray}
& & S^{(2)}_2=\left( \begin{array}{ccc}
0 & 0 & 1 \\
0 & 0 & 0 \\
0 & 0 & 0 \\
\end{array} \right)\ ,\quad
S^{(2)}_{-2}=\left( \begin{array}{ccc}
0 & 0 & 0 \\
0 & 0 & 0 \\
1 & 0 & 0 \\
\end{array} \right)={S^{(2)}_2}^\dagger ,\nonumber \\
& & S^{(2)}_1=\frac{1}{\sqrt{2}}\left( \begin{array}{ccc}
0 & -1 & 0 \\
0 & 0 & 1 \\
0 & 0 & 0 \\
\end{array} \right)\ ,\quad
S^{(2)}_{-1}=\frac{1}{\sqrt{2}}\left( \begin{array}{ccc}
0 & 0 & 0 \\
1 & 0 & 0 \\
0 & -1 & 0 \\
\end{array} \right)=-{S^{(2)}_1}^\dagger ,\nonumber \\
& &  
S^{(2)}_0=\frac{1}{\sqrt{6}}
\left( \begin{array}{ccc}
1 & 0 & 0 \\
0 & -2 & 0 \\
0 & 0 & 1 \\
\end{array} \right)\ .
\label{sec4-5-13}
\end{eqnarray}
The transformation to the Cartesian representation is carried out by
using a general formula which is obtained from the CG coefficients
\eq{sec4-5-12} and the vector-tensor transformation \eq{sec4-5-5}.
Namely, for two vectors $\ba$ and $\bb$ we have
\begin{eqnarray}
& & [ab]_{22}+[ab]_{2 -2}=a_xb_x-a_y b_y\ \,\qquad
[ab]_{22}-[ab]_{2 -2}=i(a_x b_y+a_y b_x)\ ,\nonumber \\
& & [ab]_{21}+[ab]_{2 -1}=-i(a_y b_z+a_z b_y)\ \,\qquad
[ab]_{21}-[ab]_{2 -1}=-(a_x b_z+a_z b_x)\ ,\nonumber \\
& & [ab]_{20}=\frac{1}{\sqrt{6}}\left(2 a_z b_z-a_x b_x-a_y b_y\right)\ .
\label{sec4-5-14}
\end{eqnarray}
If we set $\ba=\bb=\bS$, we find
\begin{eqnarray}
& & S^{(2)}_2+S^{(2)}_{-2}={S_x}^2-{S_y}^2\ ,\qquad
S^{(2)}_2-S^{(2)}_{-2}=i(S_x S_y+S_y S_x)\ ,\nonumber \\
& & S^{(2)}_1+S^{(2)}_{-1}=-i(S_y S_z+S_z S_y)\ \,\qquad
S^{(2)}_1-S^{(2)}_{-1}=-(S_x S_z+S_z S_x)\ ,\nonumber \\
& & S^{(2)}_0=\frac{1}{\sqrt{6}}\left(2 {S_z}^2-{S_x}^2-{S_y}^2\right)\ .
\label{sec4-5-15}
\end{eqnarray}
Here we should note that the deuteron spin operator $\bS$ satisfies
the commutation relations
\begin{eqnarray}
\left[ S_\alpha, S_\beta \right]=i \sum_\gamma 
\varepsilon_{\alpha \beta \gamma} S_\gamma\ .
\label{sec4-5-16}
\end{eqnarray}
We use this relationship and define the spin operators of the deuteron
in the Cartesian representation $\CP_{\alpha \beta}$, 
following the definition by Ohlsen (see Eq.\,(2.17) of Ref.\,\citen{Oh72}.)
Namely, we define $\CP_\alpha=S_\alpha$ and
\begin{eqnarray}
& & \CP_{xy}=\frac{3}{2}\left(S_x S_y+S_y S_x\right)\ \ ,\qquad
\CP_{xz}=\frac{3}{2}\left(S_x S_z+S_z S_x\right)\ ,\nonumber \\
& & \CP_{yz}=\frac{3}{2}\left(S_y S_z+S_z S_y\right)\ .
\label{sec4-5-17}
\end{eqnarray}
For the diagonal part $\CP_{\alpha \alpha}$, we define
$\CP_{\alpha \alpha}=3 {S_\alpha}^2-2$ symmetrically, which
leads to $\CP_{xx}+\CP_{yy}+\CP_{zz}=0$, owing to $\bS^2=2$.
We also use the normalized tensor operators
of the deuteron $T_{1\mu}$ and $T_{2\mu}$ defined by
\begin{eqnarray}
T_{1 \mu}=\sqrt{\frac{3}{2}}~S_{\mu}\ \ ,\qquad
T_{2 \mu}=\sqrt{3}~S^{(2)}_\mu\ ,
\label{sec4-5-18}
\end{eqnarray}
which satisfy
\begin{eqnarray}
{T_{\kappa \nu}}^\dagger=(-)^\nu T_{\kappa -\nu} \ \ ,\qquad
\frac{1}{3} Tr\left(T_{\kappa \nu}\,
T^\dagger_{\kappa^\prime \nu^\prime} \right)
=\delta_{\kappa, \kappa^\prime} \delta_{\nu, \nu^\prime}\ .
\label{sec4-5-19}
\end{eqnarray}
The Gl{\"o}ckle et al.'s review article \cite{PREP} also uses the
notation $S^{(i)}$ ($i=0$ - 8) for $T_{\kappa \nu}$, which satisfy
\begin{eqnarray}
{S^{(i)}}^\dagger = S^{(i)}\ \ ,\qquad
\frac{1}{3} Tr\left(S^{(i)}\,S^{(j)}\right)=\delta_{i,j}\ .
\label{sec4-5-20}
\end{eqnarray}
All of these different notations are related by
\begin{eqnarray}
& & S^{(0)}=1\ ,\nonumber \\
& & S^{(1)}=\sqrt{\frac{3}{2}}\CP_x
=\sqrt{\frac{3}{2}}S_x
=\frac{\sqrt{3}}{2}
\left( \begin{array}{ccc}
0 & 1 & 0 \\
1 & 0 & 1 \\
0 & 1 & 0 \\
\end{array} \right)\ ,\nonumber \\
& & S^{(2)}=\sqrt{\frac{3}{2}}\CP_y
=\sqrt{\frac{3}{2}}S_y
=i\,\frac{\sqrt{3}}{2}
\left( \begin{array}{ccc}
0 & -1 & 0 \\
1 & 0 & -1 \\
0 & 1 & 0 \\
\end{array} \right)\ ,\nonumber \\
& & S^{(3)}=\sqrt{\frac{3}{2}}\CP_z
=\sqrt{\frac{3}{2}}S_z
=\sqrt{\frac{3}{2}}
\left( \begin{array}{ccc}
1 & 0 & 0 \\
0 & 0 & 0 \\
0 & 0 & -1 \\
\end{array} \right)\ ,\nonumber \\
& & S^{(4)}=\sqrt{\frac{2}{3}}\CP_{xy}
=\sqrt{\frac{2}{3}}\frac{3}{2} \left(S_x S_y+S_y S_x\right)
=i\,\sqrt{\frac{3}{2}}
\left( \begin{array}{ccc}
0 & 0 & -1 \\
0 & 0 & 0 \\
1 & 0 & 0 \\
\end{array} \right)\ ,\nonumber \\
& & S^{(5)}=\sqrt{\frac{2}{3}}\CP_{yz}
=\sqrt{\frac{2}{3}}\frac{3}{2} \left(S_y S_z+S_z S_y\right)
=i\,\frac{\sqrt{3}}{2}
\left( \begin{array}{ccc}
0 & -1 & 0 \\
1 & 0 & 1 \\
0 & -1 & 0 \\
\end{array} \right)\ ,\nonumber \\
& & S^{(6)}=\sqrt{\frac{2}{3}}\CP_{xz}
=\sqrt{\frac{2}{3}}\frac{3}{2} \left(S_x S_z+S_z S_x\right)
=\frac{\sqrt{3}}{2}
\left( \begin{array}{ccc}
0 & 1 & 0 \\
1 & 0 & -1 \\
0 & -1 & 0 \\
\end{array} \right)\ ,\nonumber \\
& & S^{(7)}=\frac{1}{\sqrt{6}} \left(\CP_{xx}-\CP_{yy}\right)
=\frac{1}{\sqrt{6}} 3\left({S_x}^2-{S_y}^2\right)
=\sqrt{\frac{3}{2}}
\left( \begin{array}{ccc}
0 & 0 & 1 \\
0 & 0 & 0 \\
1 & 0 & 0 \\
\end{array} \right)\ ,\nonumber \\
& & S^{(8)}=\frac{1}{\sqrt{2}} \CP_{zz}
=\frac{1}{\sqrt{2}} \left(3{S_z}^2-2\right)
=\frac{1}{\sqrt{2}}
\left( \begin{array}{ccc}
1 & 0 & 0 \\
0 & -2 & 0 \\
0 & 0 & 1 \\
\end{array} \right)\ .\nonumber \\
\label{sec4-5-21}
\end{eqnarray}
The relationship between the tensor notation and $S^{(i)}$ is
\begin{eqnarray}
& & S^{(2)}_2=\frac{1}{\sqrt{6}}\left(S^{(7)}+iS^{(4)}\right)\ \ ,\qquad
S^{(2)}_{-2}=\frac{1}{\sqrt{6}}\left(S^{(7)}-iS^{(4)}\right)\ ,\nonumber \\
& & S^{(2)}_1=-\frac{1}{\sqrt{6}}\left(S^{(6)}+iS^{(5)}\right)\ \ ,\qquad
S^{(2)}_{-1}=\frac{1}{\sqrt{6}}\left(S^{(6)}-iS^{(5)}\right)\ ,\nonumber \\
& & S^{(2)}_0=\frac{1}{\sqrt{3}} S^{(8)}\ ,
\label{sec4-5-22}
\end{eqnarray}
or
\begin{eqnarray}
& & T_{00}=S^{(0)}=1\ ,\nonumber \\
& & T_{11}=-\frac{1}{\sqrt{2}}\left(S^{(1)}+iS^{(2)}\right)\ \ ,\qquad
T_{1,-1}=\frac{1}{\sqrt{2}}\left(S^{(1)}-iS^{(2)}\right)\ ,\nonumber \\
& & T_{10}=S^{(3)}\ ,\nonumber \\
& & T_{22}=\frac{1}{\sqrt{2}}\left(S^{(7)}+iS^{(4)}\right)\ \ ,\qquad
T_{2,-2}=\frac{1}{\sqrt{2}}\left(S^{(7)}-iS^{(4)}\right)\ ,\nonumber \\
& & T_{21}=-\frac{1}{\sqrt{2}}\left(S^{(6)}+iS^{(5)}\right)\ \ ,\qquad
T_{2,-1}=\frac{1}{\sqrt{2}}\left(S^{(6)}-iS^{(5)}\right)\ ,\nonumber \\
& & T_{20}=S^{(8)}\ .
\label{sec4-5-23}
\end{eqnarray}
If we express these by $\CP_\alpha$ and $\CP_{\alpha \beta}$, we obtain
\begin{eqnarray}
\ \hspace{-10mm}
& & i\frac{1}{2}\left(T_{11}+T_{1,-1}\right)=\frac{\sqrt{3}}{2} \CP_y
\ \ ,\qquad 
T_{20}=\frac{1}{\sqrt{2}} \CP_{zz}\ ,\nonumber \\
\ \hspace{-10mm}
& & \frac{1}{2}\left(T_{21}-T_{2,-1}\right)=-\frac{1}{\sqrt{3}}
\CP_{xz}\ \ ,\qquad
\frac{1}{2}\left(T_{22}+T_{2,-2}\right)=\frac{1}{2\sqrt{3}}
\left(\CP_{xx}-\CP_{yy}\right)\ ,
\label{sec4-5-24}
\end{eqnarray}
where the other components of the deuteron analyzing-powers
are zero as in $\S\,2.2$.

\bigskip

\newcommand{\etal}{{\em et al.}}


\begin{thebibliography}{99}
\bibitem{PPNP}
Y. Fujiwara, Y. Suzuki and C. Nakamoto,
\JL{Prog.~Part.~Nucl.~Phys.,58,2007,439}.
\bibitem{renRGM} Y. Suzuki, H. Matsumura, M. Orabi, Y. Fujiwara,
P. Descouvemont, M. Theeten and D. Baye,
\PLB{659,2008,160}.
\bibitem{ren} Y. Fujiwara, K. Miyagawa, M. Kohno,
Y. Suzuki and H. Nemura,
\PRC{66,2002,021001(R)};
Y. Fujiwara, K. Miyagawa, M. Kohno and Y. Suzuki,
\PRC{70,2004,024001};
Y. Fujiwara, Y. Suzuki, M. Kohno and K. Miyagawa,
\PRC{77,2008,027001}.
\bibitem{ndsc1} Y. Fujiwara and K. Fukukawa,
\PTP{124,2010,433}.
\bibitem{AGS}
E. O. Alt, P. Grassberger and W. Sandhas, \NPB{2,1967,167}.
\bibitem{NO65} H. P. Noyes,
\PRL{15,1965,538}.
\bibitem{KO65} K. L. Kowalski,
\PRL{15,1965,798}; Errata {\bf 15} (1965), 908.
\bibitem{spline82} W. Gl{\"o}ckle, G. Hasberg and A. R. Neghabian,
\JL{Z.~Phys.~A,305,1982,217}.
\bibitem{Wi03} H. Wita\l a, Th. Cornelius and W. Gl{\"o}ckle,
\JL{Few-Body Systems,3,1988,123}.
\bibitem{PREP} W. Gl{\"o}ckle, H. Wita\l a, D. H{\"u}ber, H. Kamada
and J. Golak, \JL{Physics Reports,274,1996,107}.
\bibitem{Liu05} 
H. Liu, Ch. Elster and W. Gl{\"o}ckle,
\PRC{72,2005,054003}.
\bibitem{apfb08} K. Fukukawa, Y. Fujiwara and Y. Suzuki, 
\JL{Mod. Phys. Lett. A,24,2009,1035}.
%
\bibitem{scl10} K. Fukukawa and Y. Fujiwara,
arXiv: nucl-th1010.2024, submitted to Prog. Theor. Phys. (2010).
%
\bibitem{To98a} W. Tornow and H. Wita\l a,
\NPA{637,1998,280}.
\bibitem{To98b} W. Tornow, H. Wita\l a and A. Kievsky,
\PRC{57,1998,555}.
\bibitem{To02} W. Tornow, A. Kievsky and H. Wita\l a,
\JL{Few-Body~Systems,32,2002,53}.
\bibitem{To08} W. Tornow, J. H. Esterline and G. J. Weisel,
\JL{J. Phys. G: Nucl. Part. Phys.,35,2008,125104}.
\bibitem{Vi74} C. M. Vincent and S. C. Phatak,
\PRC{10,1974,391}.
\bibitem{apfb05} Y. Fujiwara, in
{\em Proceedings of the Third Asia-Pacific Conference on Few-Body
Problems in Physics} (APFB05),
World Scientific Co. Ltd., 2007, pp. 221 - 225.
\bibitem{Be90} G. H. Berthold, A. Stadler and H. Zankel,
\PRC{41,1990,1365}.
\bibitem{Al02} E. O. Alt, A. M. Mukhamedzhanov, M. M. Nishonov
and A. I. Sattarov,
\PRC{65,2002,064613}
\bibitem{Ki96} A. Kievsky, S. Rosati, W. Tornow and M. Viviani,
\NPA{607,1996,402}.
\bibitem{Ki01} A. Kievsky, M. Viviani and S. Rosati,
\PRC{64,2001,024002}.
\bibitem{De05} A. Deltuva, A. C. Fonseca and P. U. Sauer, 
\PRC{71,2005,054005}.
\bibitem{De05c} A. Deltuva, A. C. Fonseca, A. Kievsky, S. Rosati,
P. U. Sauer and M. Viviani,
\PRC{71,2005,064003}.
\bibitem{Is09} S. Ishikawa,
\PRC{80,2009,054002}, and private communications.
\bibitem{Bl52} J. M. Blatt and L. C. Biedenharn,
\JL{Rev.~Mod.~Phys.,24,1952,258}.
\bibitem{Fu98} T. Fujita, Y. Fujiwara, C. Nakamoto and Y. Suzuki,
\PTP{100,1998,931}.
%
\bibitem{Oh72} G. G. Ohlsen,
\JL{Rep.~Prog.~Phys.,35,1972,717}.
\bibitem{Mc94} J. E. McAninch, L. O. Lamm and W. Haeberli,
\PRC{50,1994,589}.
\bibitem{Sa94} K. Sagara, H. Oguri, S. Shimizu, K. Maeda,
H. Nakamura, T. Nakashima and S. Morinobu,
\PRC{50,1994,576}.
\bibitem{To91} W. Tornow, C. R. Howell, M. Alohali, Z. P. Chen,
P. D. Felsher, J. M. Hanly, R. L. Walter, G. Weisel, G. Mertens,
I. {\v S}laus, H. Wita\l a and G. Gl{\"o}ckle,
\PLB{257,1991,273}.
\bibitem{Ho87} C. R. Howell, W. Tornow, K. Murphy, H. G. Pf{\"u}tzner,
M. L. Roberts, Anli Li, P. D. Felsher, R. L. Walter, I. {\v S}laus,
P. A. Treado and Y. Koike,
\JL{Few-Body~Systems,2,1987,19}.
\bibitem{Ni98} N. Nishimori, K. Sagara, T. Fujita, F. Wakamatsu,
T. Bussaki, K. Maeda, H. Akiyoshi, K. Tsuruta, H. Nakamura
and T. Nakashima,
\NPA{631,1998,697c}.
\bibitem{Ne03} E. M. Neidel, W. Tornow, D. E. Gonz{\'a}lez Trotter,
C. R. Howell, A.S. Crowell, R. A. Macri, R. L. Walter, G. J. Weisel,
J. Esterline, H. Wita\l a, B. J. Crowe III, R. S. Pedroni and D. M. Markoff, 
\PLB{552,2003,29}.
\bibitem{To82} W. Tornow, C. R. Howell, R. C. Byrd, R. S. Pedroni
and R. L. Walter,
\PRL{49,1982,312}.
\bibitem{To83} W. Tornow, R. C. Byrd, C. R. Howell, R. S. Pedroni
and R. L. Walter,
\PRC{27,1983,2439}.
\bibitem{We10} G. J. Weisel, W. Tornow, B. J. Crowe III,
A. S. Crowell, J. H. Esterline, C. R. Howell, J. H. Kelley, 
R. A. Macri, R. S. Pedroni, R. L. Walter and H. Wita\l a,
\PRC{81,2010,024003}.
\bibitem{Do82} P.~Doleschall, W.~Gr\"uebler, V.~K\"onig, P.~A.~Schmelzbach,
F.~Sperisen, and B.~Jenny, \NPA{380,1982,72}.
\bibitem{Sa83} M. Sawada, S. Seki, K. Furuno, Y. Tagishi, Y. Nagashima, 
J. Schimizu, M. Ishikawa, T. Sugiyama, L. S. Chuang, W. Gr{\"u}ebler,
J. Sanada, Y. Koike and Y. Taniguchi,
\PRC{27,1983,1932}.
\bibitem{Ko87a} Y. Koike and J. Haidenbauer,
\NPA{463,1987,365c}.
\bibitem{Ko87b} Y. Koike, J. Haidenbauer and W. Plessas,
\PRC{35,1987,396}.
\bibitem{Ra88} G. Rauprich, H. J. H{\"a}hn, M. Karus, P. Nie{\ss}en,
K. R. Nyga, H. Oswald, L. Sydow, H. Paetz gen. Schieck and Y. Koike,
\JL{Few-Body~Systems,5,1988,67}.
\bibitem{Co78} H. E. Conzett,
\JL{Lecture~Note~in~Physics,87,1978,477}.
\bibitem{Do78} H. Dobiasch, R. Fischer, B. Haesner, H. O. Klages,
P. Schwarz, B. Zeitnitz, R. Maschuw, K. Sinram and K. Wick,
\PLB{76,1978,195}.
\bibitem{Jo65} A. R. Johnston, W. R. Gibson, J. H. P. C. Megaw, 
R. J. Griffiths and R. M. Eisberg,
\PL{19,1965,289}.
\bibitem{Za73} J. Zamudio-Cristi, B. E. Bonner, F. P. Brady,
J. A. Jungerman and J. Wang,
\PRL{31,1973,1009}.
\bibitem{Bu68} S. N. Bunker, J. M. Cameron, R. F. Carlson,
J. Reginald Richardson, P. Tomas, W. T. H. Van Oers and J. W. Verba,
\NPA{113,1968,461}.
\bibitem{Ro82} J. L. Romero, J. L. Ullmann, F. P. Brady, J. D. Carlson,
D. H. Fitzgerald, A. L. Sagle, T. S. Subramanian, C. I. Zanelli,
N. S. P. King, M. W. McNaughton and B. E. Bonner,
\PRC{25,1982,2214}.
\bibitem{Sh82} H. Shimizu, K. Imai, N. Tamura, K. Nisimura,
K. Hatanaka, T. Saito, Y. Koike and Y. Taniguchi,
\NPA{382,1982,242}.
\bibitem{Sh95} S. Shimizu, K. Sagara, H. Nakamura, K. Maeda,
T. Miwa, N. Nishimori, S. Ueno, T. Nakashima and S. Morinobu,
\PRC{52,1995,1193}.
\bibitem{So87} J. Sowinski, D. D. Pun Casavant and L. D. Knutson,
\NPA{464,1987,223}.
\bibitem{Sp84} F. Sperisen, W. Gr{\"u}ebler, V. K{\"o}nig, P. A. Schmelzbach,
K. Elsener, B. Jenny, C. Schweizer, J. Ulbricht and P. Doleschall,
\NPA{422,1984,81}.
\bibitem{Gr83} W. Gr{\"u}ebler, V. K{\"o}nig, P. A. Schmelzbach,
F. Sperisen, B. Jenny, R. E. White, F. Seiler and H. W. Roser,
\NPA{398,1983,445}.
\bibitem{Wi93}
H. Wita\l a, W. Gl{\"o}ckle, L. E. Antonuk, J. Arvieux, D. Bachelier,
B. Bonin, A. Boudard, J. M. Cameron, H. W. Fielding, M. Gar\c{c}on, F. Jourdan,
C. Lapointe, W. J. McDonald, J. Pasos, G. Roy, I. The, J. Tinslay,
W. Tornow, J. Yonnet and W. Ziegler,
\JL{Few-Body~Systems,15,1993,67}.
\bibitem{El62} A. J. Elwyn, R. O. Lane and A. Langsdorf Jr.,
\JL{Phys.~Rev.,128,1962,779}.
\bibitem{Ko69} D. C. Kocher and T. B. Clegg,
\NPA{132,1969,455}.
\bibitem{Hu83} E. Huttel, W. Arnold, H. Berg,
H. H. Krause, J. Ulbricht and G. Clausnitzer,
\NPA{406,1983,435}.
\bibitem{Wo02} M. H. Wood, C. R. Brune, B. M. Fisher, H. J. Karwowski,
D. S. Leonard, E. J. Ludwig, A. Kievsky, S. Rosati and M. Viviani,
\PRC{65,2002,034002}.
\bibitem{Wh79} R. E. White, W. Gr{\"u}ebler, B. Jenny, V. K{\"o}nig,
P. A. Schmelzbach and H. R. B{\"u}rgi,
\NPA{321,1979,1}.
\bibitem{Kr95} W. Kretschmer and POLAR collaboration,
\JL{AIP~Conference~Proc.,339,1995,355}.
%
\bibitem{Cl90b} M. Clajus, P.M. Egun, W. Gr{\"u}ebler, P. Hautle,
I. {\v S}laus, B. Vuaridel, F. Sperisen, W. Kretschmer, A. Rauscher,
W. Schuster, R. Weidmann, M. Haller, M. Bruno, F. Cannata, M. D'Agostino,
H. Wita\l a, Th. Cornelius, W. Gl{\"o}ckle and P.A. Schmelzbach,
\PLB{245,1990,333}
\bibitem{Gl95} A. Glombik, B. Aum{\"u}ller, W. Kretschmer, G. Martin,
K. M{\"u}mmler, G. Suft, R. Weidmann,
I. {\v S}laus, M. Bruno, P. Milazzo, M. Clajus, G. Mertens, W. Gr{\"u}ebler,
P.A. Schmelzbach, W. Gl{\"o}ckle and H. Wita\l a,
\JL{AIP~Conference~Proc.,334,1995,486}.
\bibitem{Ch75} J. Chauvin, D. Garreta and M. Fruneau,
\NPA{247,1975,335}.
\bibitem{FK10a} Y. Fujiwara and K. Fukukawa,
\JL{Mod.~Phys.~Lett.~A,25,2010,1759}.
\bibitem{FK10b} Y. Fujiwara and K. Fukukawa,
to be published in {\em International Journal of Modern Physics E} (2010).
\bibitem{ndsc3} Y. Fujiwara and K. Fukukawa,
submitted to Prog. Theor. Phys. (2011).
%
\end{thebibliography}
\end{document}